\documentclass[11pt]{article}
\usepackage{amsmath,amssymb}
\usepackage{pictex}
\font\dm=cmr9

\numberwithin{equation}{section}
\let\eq\equiv
\let\ot\otimes
\let\ov\overline
\let\os\overrightarrow
\let\pa\partial
\let\q\quad
\def\qh#1{\quad\hbox{#1}\quad}
\let\ul\underline
\let\wh\widehat
\let\wt\widetilde
\let\a\alpha
\let\b\beta
\let\d\delta
\let\D\varDelta
\let\ve\varepsilon
\let\g\gamma
\let\G\varGamma
\let\k\kappa
\let\l\lambda
\let\La\Lambda
\let\o\omega
\let\O\varOmega
\let\vf\varphi
\let\si\sigma
\let\Si\varSigma
\let\t\theta
\let\T\varTheta
\let\z\zeta
\def\m{{\mu\nu}}
\def\cD{\mathcal D}
\def\cL{\mathcal L}
\def\C{\mathbb C}
\def\dg #1,#2,#3,{{{#1}_{#2}}^{\!#3}}
\def\gd #1,#2,#3,{{#1^{#2}}_{\!#3}}
\def\nad#1#2{\overset{\rm #1}{#2}}
\def\nadd#1#2{\overset{#1}{#2}}
\def\falg{{\mathrel{\lower5pt\hbox{${\scriptstyle\sim}$}\hskip-5pt g}}}
\def\falH{{\mathrel{\lower4pt\hbox{${\scriptstyle\sim}$}\hskip-7.5pt H}}}
\def\tl{\mathopen{\hbox{\dm[}}}
\def\tp{\mathclose{\hbox{\dm]}}}
\def\hor{\operatorname{hor}}
\def\ver{\operatorname{ver}}
\def\id{\operatorname{id}}
\def\div{\operatorname{div}}
\def\ad#1{\operatorname{ad}_{#1}}
\def\Spin{{\rm Spin}}
\def\SO{{\rm SO}}
\def\pp#1#2{\frac{\pa #1}{\pa #2}}
\catcode`@=11
\def\X#1#2{\mathopen{\hbox{$\left#2\vbox to\ifcase#1\or9.5\p@\or11.5\p@\or
14.5\p@\or17.5\p@\fi{}\right.\n@space$}}}
\def\Y#1#2{\mathclose{\hbox{$\left.\vbox to\ifcase#1\or9.5\p@\or11.5\p@\or
14.5\p@\or17.5\p@\fi{}\right#2\n@space$}}}
\def\twl#1{\vbox{\m@th\ialign{##\crcr
      =\crcr\noalign{\kern-\p@\nointerlineskip}
      $\hfil\displaystyle{#1}\hfil$\vrule height6pt width0pt \crcr}}}
\catcode`@=12

\let\ti\textit
\def\beq{\begin{equation}\label}
\def\bea{\begin{eqnarray}\label}
\def\bml{\begin{multline}\label}
\def\bg{\begin{gather}\label}
\let\bal\aligned \let\eal\endaligned
\let\bga\gathered \let\ega\endgathered
\let\nn\nonumber
\let\la\lambda
\def\U{\text{U}}
\let\Ps\varPsi
\let\PS\varPsi
\def\(#1){{(#1)}}
\def\[#1]{{[#1]}}
\def\gdv{\nad{gauge}d}
\def\pl{_{\rm pl}}

\def\ap{approximat}
\def\ct{constant}
\def\cd{coordinate}
\def\co{cosmolog}
\def\di{dimension}
\def\JT{(Jordan--Thiry)}
\def\spt{space-time}
\def\pt{potential}
\def\up#1{\uppercase{#1}}
\def\eu{\expandafter\up}
\def\cf{coefficient}
\def\cfn{confinement}
\def\cn{connection}
\def\dv{derivative}
\def\elm{electromagnetic}
\def\e{equation}
\def\f{function}
\def\gr{gravitation}
\def\nos{nonsymmetric}
\def\NK#1{Nonsymmetric Kaluza--Klein #1Theory}
\def\rf{reflection}
\def\so{solution}
\def\st{such that }
\def\s{symmetric}
\def\tf{transformation}
\def\wrt{with respect to }
\def\Pl{Planck}
\let\TM\texttrademark

\newdimen\jedn
\catcode`@=11
\newcount\@multicnt
\newdimen\@xdim
\newdimen\@ydim
\long\def\multi(#1,#2)(#3,#4)#5#6{\@multicnt=#5\relax
\@xdim=#1\jedn
\@ydim=#2\jedn
\loop\ifnum \@multicnt>\z@
\raise\@ydim\hbox to\z@{\kern\@xdim\hbox{#6}\hss}\advance\@multicnt by-1 \advance\@xdim
#3\jedn\advance\@ydim #4\jedn \repeat \ignorespaces}
\catcode`@=12
\def\grub #1 {\linethickness=#1}
\catcode`!=11
\def\MT  #1 #2 {\!start (#1,#2)}
\def\LT  #1 #2 {\!ljoin (#1,#2)}
\catcode`!=12
\def\cput(#1,#2,#3){\put {$#3$} at #1 #2 }
\def\lput(#1,#2,#3){\put {$#3$}[r] at #1 #2 }
\def\rput(#1,#2,#3){\put {$#3$}[l] at #1 #2 }
\def\linia(#1,#2)(#3,#4){\MT #1 #2 \LT #3 #4 }
\def\bps{\beginpicture\setcoordinatesystem units <\jedn,\jedn>}
\let\epict\endpicture
\newdimen\sz
\newcount\nr
\newcount\nd
\newcount\nf
\newcount\ng
\newdimen\hs
\sz=\textwidth
\divide\sz by4000
\jedn\sz

\def\obraz#1 {\global\nr=#1
\ifodd\nr \vtop to1050\jedn \bgroup \vfill \bps \fi 
\hfil \vbox \bgroup \hsize=\hs
\leftline \bgroup \hskip-0.7\hs
\nd=\nr \global\divide\nd by8 \nf=\nd \global\multiply\nf by8
\ng=\nr \advance\ng by-\nf
\lput(-150,770,{\rm\ifcase\ng H\or A\or B\or C\or D\or E\or F\or G\fi})}
\def\koniec#1 #2 {\hfil \egroup \vskip4pt
\centerline{\hfil $a=#1$, $b=#2$\hfil }\egroup \hfil
\ifodd\nr \else \epict \egroup \vfil \fi
\nd=\nr \divide\nd by8 \nf=\nd \multiply\nf by8
\ifnum\nr=\nf
\begin{center}Fig. \the\nd.\ \opis\end{center}\fi }
\def\kon#1;{\hfil \egroup \vskip8pt
\centerline{\hfil $#1$\hfil }\egroup \hfil
\ifodd\nr \else \epict \egroup \vfil \fi
\nd=\nr \divide\nd by8 \nf=\nd \multiply\nf by8
\ifnum\nr=\nf
\begin{center}Fig. \the\nd.\ \opis\end{center}\fi }

\author{M. W. Kalinowski\\Faculty of Physics, Warsaw University, ul. Ho\.za
69, 00-681 Warsaw, Poland;\\
Pracownia Bioinformatyki, Instytut Medycyny
Do\'swiadczalnej i~Klinicznej PAN,\\
ul. Pawi\'nskiego 5, 02-106 Warszawa, Poland\\
e-mail: markwkal@bioexploratorium.pl, mkalinowski@imdik.pan.pl}
\title{On some developments\\
in the Nonsymmetric Kaluza--Klein Theory}
\date{}
\begin{document}
\maketitle
\begin{abstract}We consider a condition for a charge \cfn\ and
gravito-electromagnetic wave \so s in the \NK{}. We consider also an
influence of a cosmological constant on a static, spherically \s\ \so.
We remind to the reader some fundamentals of the \NK{} and a geometrical
background behind the theory. Simultaneously we give some remarks concerning
misunderstanding connected to several notions of Kaluza--Klein Theory,
Einstein Unified Field Theory, geometrization and unification
of physical interactions. We reconsider Dirac field in the \NK{}.
\end{abstract}

\section*{Introduction}
The \NK{(Jordan--Thiry) } has been developed (see Refs \cite1, \cite2,
\cite3, \cite4). The theory unifies \gr al theory described by NGT (\eu\nos\
\eu\gr al Theory) (see Ref.~\cite5) and Electrodynamics. The theory has been
extended to Nonabelian Gauge Fields, Higgs Fields, scalar field with
applications to cosmology. Some possibilities to get a \cfn\ of colour have
been suggested. A~nonsingular spherically symmetric \so\ has been derived.
The \NK{} can be obtained from the \eu\nos\ Jordan--Thiry Theory by putting
the scalar field into zero. In this way it is a limit of the \eu\nos\
Jordan--Thiry Theory. The \eu\nos\ Jordan--Thiry Theory has several physical
applications in \co y, e.g.: 1.~\co ical \ct, 2.~inflation, 3.~quintessence,
and some possible relations to a dark matter problem. Simultaneously the
theory unifies gravity with gauge fields in a nontrivial way via geometrical
unification of two fundamental invariance principles in physics: 1.~\cd\
invariance principle, 2.~gauge invariance principle. Unification on the level
of invariance principles is more important than on a level of interactions
for from invariance principles we get conservation laws (via Noether
theorem). In some sense Kaluza--Klein theory unifies an energy-momentum
conservation law with an electric charge conservation law.
This unification has
been achieved in more than 4-\di al world. It is nontrivial for we can get
some additional effects unknown in conventional theories of gravity and gauge
fields (\elm\ or Yang--Mills fields). All of these effects which we call
``interference effects'' between gravity and gauge fields are testable in
principle in an experiment or an observation. The formalism of this
unification has been described in references \cite1--\cite4. The \NK{} is an
example of a geometrization of \gr al and \elm\ interactions according to the
Einstein programme. In
this paper we consider conditions for a charge \cfn\ in the theory and three
\so s of \NK{} \e s describing gravito--\elm\ waves. We consider also an influence of a \co ical \ct\
on a static, spherically \s\ \so. This \so\ can be considered as a model of
an electron for it has remarkable properties being nonsingular in electric
and \gr al fields. Simultaneously this \so\ has been built from elementary
fields. The properties of the \so\ can be considered as ``interference
effects'' between \elm\ and \gr al fields in our unification. In this way the
theory achieves an old dream of Einstein, Weyl, Kaluza, Eddington and
Schr\"odinger on a {\bf unitary classical field theory} by having particles
as spherically \s\ singularity-free \so s of the field \e s.

The \NK{} should be called a Unified Field Theory according to the definition
which we quote here (see Ref.~\cite{dict}): ``\ti{Unified Field Theory: Any
theory which attempts to express \gr al theory and \elm\ theory within a
single unified framework. Usually, an attempt to generalize Einstein's
general theory of relativity from a theory of \gr\ alone to a theory of
gravity and classical
electromagnetism}''. In our case this single unified framework is a multi\di
al analogue of geometry from Einstein Unified Field Theory (treated as a
generalized gravity) defined on an \elm\ bundle.

Summing up the \NK{} connects old ideas of unitary field theories (unified
field theories) with some modern applications.

The paper has been divided into four sections.
In the first section we give some elements of the \NK{} in some new setting.
We give also a condition for a dielectric \cfn\ of a charge. In the second
section we give three \so s of the field \e s describing gravito--\elm\
waves. In the third section we deal with a spherically-\s\ \so\ in a presence
of a cosmological constant. In the fourth section we give a theory of a Dirac
field in the \NK{} getting CP-violation and EDM (Electric Dipole Moment) for
a fermion. We reconsider some notion known from our previous papers.
In Conclusions we give also some remarks
concerning some misunderstanding concerning Kaluza--Klein Theory, Einstein
Unified Field Theory commonly met. We put our investigations on a wider
background. In Appendix A we give some notions of differential
geometry used in the paper and in Appendix~B some details of calculations.
In Appendix~C we give some elements of Clifford algebra and spinor theory.
Appendix~D is devoted to a redefinition of the \NK{} in terms of~GR (General
Relativity) and additional ``matter fields''.

In this paper we use the following convention. Capital Latin indices
$A,B,C=1,2,3,4,5$ (Kaluza--Klein indices), lower Greek cases
$\a,\b,\g=1,2,3,4$ (\spt\ indices), lower Latin cases
$a,b,c=1,2,3$ (space indices).  In Appendix~A we use Latin lower indices in a
Lie algebra $\mathfrak G$ of a Lie group $G$, $a,b,c=1,2,\dots,n=\dim
\mathfrak G=\dim G$. This cannot cause any misunderstanding. In Appendix~B we
use capital Latin indices $A,B,C,W,N,M=1,2,\dots,n$, where $n$ is the
dimension of a manifold equipped with nonsymmetric tensor and a nonsymmetric
connection. This also does not result in any misunderstanding.

\section{Elements of the \NK{}}
The basic logic of a construction is as follows. We define a \NK{} as a
five-\di al analogue of NGT using our extension of natural metrization of an
\elm\ fibre bundle achieving in this way a unification of two fundamental
principles of invariance (i.e.\ a \cd\ invariance principle
and a gauge invariance principle) reducing both to
the \cd\ invariance principle in 5-\di al world (see Ref.~\cite4 for details).

Let us notice that our construction from Ref.~\cite4 is more general for it
contains a scalar field $\rho$ (or~$\Ps$) which here is put to $\rho=1$
(${\Ps=0}$).

Let $\ul P$ be a principal fibre bundle with a structural group $G=U(1)$ over
a space-time $E$ with a projection $\pi$ and let us define on this bundle a
connection~$\a$. We call this bundle an {\it \elm\ bundle} and $\a$ an {\it
\elm\ connection} (see Appendix A for details).
We define a curvature $2$-form for the connection~$\a$
\beq{1.1}
\O=d\a\eq \frac12\,\pi^*(F_\m\ov\t^\mu\wedge\ov\t^\mu),\quad \mu,\nu=1,2,3,4,
\end{equation}
where
\beq{1.2}
F_\m=\pa_\mu A_\nu- \pa_\nu A_\mu, \q e^*\a=A_\mu\ov\t^\mu.
\end{equation}
$A_\mu$ is a 4-potential of the \elm\ field, $e$ is a local section of~$\ul
P(e:E\supset U\to P)$, $F_\m$ is an \elm\ field strength, and $\ov\t^\mu$ is a frame
on~$E$. Bianchi identity is
\beq{1.3}
d\O=0,
\end{equation}
so the 4-potential exists. This is of course simply the first pair of Maxwell
\e s. On the space-time $E$ we define a \nos\ metric tensor $g_{\a\b}$ \st
\beq{1.4}
\bal {}&g_{\a\b}=g_{(\a\b)}+g_{[\a\b]}\\
&g_{\a\b}g^{\g\b}=g_{\b\a}g^{\b\g}=\d^\g_\a,
\eal
\end{equation}
where the order of indices is important. In such a way we suppose that
\beq{1.5}
g=\det g_{\a\b}\ne0.
\end{equation}
We suppose also that
\beq{1.6}
\wt g=\det g_{(\a\b)}\ne0,
\end{equation}
defining an inverse tensor ${\wt g}^{(\a\b)}$ for $g_{(\a\b)}$ \st
${\wt g}^{(\a\b)}g_{(\b\mu)}=\d_\mu^\a$.
The combination of \s\ and anti\s\ tensor Eq.\eqref{1.4} is going to new insides
in an inverse tensor.
We define also on $E$ two connections ${{\ov w}^\a}_{\!\b}$ and ${{\ov W}^\a}_{\!\b}$
\beq{1.7}
{{\ov w}^\a}_{\!\b}={{\ov \G}^\a}_{\!\b\g}{\ov \t}^\g
\end{equation}
and
\beq{1.8}
{{\ov W}^\a}_{\!\b}={{\ov W}^\a}_{\!\b\g}\ov \t^\g,\quad \a,\b,\g=1,2,3,4,
\end{equation}
\st
\beq{1.9}
{{\ov W}^\a}_{\!\b}={{\ov w}^\a}_{\!\b}-\tfrac23\,{\d^\a}_{\!\b}\ov W,
\end{equation}
where
$$
\ov W=\ov W_\g\ov \t^\g = \frac12\bigl({{\ov W}^\si}_{\!\g\si}-
{{\ov W}^\si}_{\!\si\g}\bigr)\ov \t^\g.
$$
For the connection ${{\ov w}^\a}_{\!\b}$ we suppose the following conditions
\beq{1.10}
\bga
\ov D g_{\a+\b-}=\ov Dg_{\a\b}-g_{\a\d}{\ov Q}^\d_{\b\g}(\ov\G){\ov\t}^\g=0\\
\gd \ov Q,\a,\b\a,(\ov\G)=0,
\ega
\end{equation}
where $\ov D$ is the exterior covariant \dv\ \wrt $\gd \ov w,\a,\b,$ and $\gd
\ov Q,\a,\b\a,(\ov \G)$ is a torsion of $\gd\ov w,\a,\b,$. $\gd \ov W,\a,\b,$
is called an unconstrained \cn\ and $\gd \ov w,\a,\b,$ a constrained \cn.
Thus we have defined
on space-time all quantities present in Moffat's theory of \gr\ (NGT, see
Ref.~\cite{5a}). In this approach we consider test particles moving along
geodesics \wrt a Levi--Civita \cn\ generated by a tensor $g_\(\a\b)$ on~$E$,
i.e.\ $\gd{\wt{\ov w}}{},\a,\b,=\gd{\wt{\ov\G}}{},\a,\b\g,\ov \t{}^\g$.
Let us introduce on~$P$ a frame (a lift horizontal base)
\beq{1.11}
\t^A=\bigr(\pi^*({\ov\t}^\a),\l\a=\t^5\bigr), \q \l={\rm const}.
\end{equation}

Now we turn to the natural \nos\ metrization of the bundle $P$. We have
\bg{1.12a}
\ov \g=\pi^\ast \ov g-\t^5\ot\t^5=\pi^\ast(g_{(\a\b)}\ov\t{}^\a \ot
\ov \t{}^\b)-\t^5\ot\t^5\\
\ul\g=\pi^\ast\ul g=\pi^\ast(g_{[\a\b]}\ov\t{}^\a\wedge\ov\t{}^\b)\label{1.12b}
\end{gather}
where $\ov g$ is a \s\ tensor on~$E$, $\ov g=g_{(\a\b)}\ov\t{}^\a\ot\ov\t
{}^\b$ and $\ul g$ is a 2-form on~$E$, $\ul
g=g_{[\a\b]}\ov\t{}^\a\wedge\ov\t{}^\b$. Taking both parts together we get
\beq{1.12}
\g=\pi^*g-\t^5\ot\t^5=\pi^*(g_{\a\b}{\ov\t}^\a\ot{\ov\t}^\b)-\t^5\ot\t^5.
\end{equation}

The \nos\ metric $\g$ is biinvariant \wrt the action of a group $\U(1)$ on
$P$.
From the classical Kaluza--Klein theory we know that $\l=2\frac{\sqrt{
G_N}}{c^2}$. We work with such a system of units that $G_N=c=1$ and $\l=2$. Thus
we have in a matrix form
\beq{1.13}
\bga
\g_{AB}=\left(\begin{array}{c|c} g_{\a\b} &  0 \\
\hline 0 & -1 \end{array} \right),\\
\g=\g_{AB}\t^A\ot\t^B.
\ega
\end{equation}

The tensor $\g_{AB}$ has this shape in a lift horizontal base, which is of
course nonholonomic (${d\t^5\ne0}$). We can find it in a holonomic system of \cd s. Let us
take a section $e:E\supset U\to P$ and attach to it a \cd~$x^5$, selecting $x^\mu={\rm
const}$ on the fibre in such a way that $e$ is given by the condition $x^5=0$
and $\z_5=\pa/\pa x^5$. Then we have $e^\ast \,dx^5=0$ and
$$
\a=\frac1\l \,dx^5+\pi^\ast(A_\mu\ov\t{}^\mu), \quad\hbox{where}\quad
A=A_\mu \ov\t{}^\mu=e^\ast \a.
$$
Taking $\ov\t{}^\mu=dx^\mu$ one gets $\t^5=dx^5+\pi^\ast(\l A_\mu\,dx^\mu)$.
Putting the last result into Eq~\eqref{1.12} one finds
\bml{1.16n}
\g=\pi^\ast\Bigl(\bigl(g_{\a\b}-\l^2A_\a A_\b\bigr)\,dx^\a\ot dx^\b\Bigr)\\
-\pi^\ast(\l A_\a\,dx^\a)\ot dx^5
-dx^5\ot(\l A_\b\,dx^\b)-dx^5\ot dx^5.
\end{multline}

In this \cd\ system the tensor $\g$ takes a matrix form
\beq{1.14n}
\g_{AB}=\left(\begin{array}{c|c}
g_{\a\b}-\l^2A_\a A_\b&-\l A_\a \vrule width0pt depth5pt height0pt\\
\hline
-\l A_\b&-1
\end{array}\right).
\end{equation}
In order to have the correct dimension of a four-\pt\ we should rather write
$e^\ast \a=(q/\hbar c)A=\mu A$, where $q$ is an elementary charge and $\hbar$
is \Pl's \ct. The same is true for the curvature of \cn\ on the \elm\ bundle
$\O=\l\mu \pi^\ast(F)$, $F=\frac12 F_{\mu\nu}\ov\t{}^\mu \ov\t{}^\nu$.
Moreover, it can be absorbed by a \ct~$\l$ (we have only one \ct\ as in
classical Kaluza theory and a Planck's \ct\ appearence is illusory for the
theory is classical (not quantum) and eventually demands quantization).

In this way Eq.~\eqref{1.14n} gives us a classical Kaluza--Klein
approach with a five-\di al metric tensor and with a Killing vector~$\z_5$.
Even if in Eq.~\eqref{1.13} we have not any four-\pt\ $A_\mu$, an \elm\ field
exists as an \elm\ \cn~$\a$.

The \cn\ contains (\pt ly) all possible four-\pt\ $A_\mu$ (it means, in any
possible gauge). To choose a gauge means here to take a section of a bundle.
An \elm\ \cn\ $\a$ is really an \elm\ field. We can also consider $e^\ast
\O$. Moreover, the structural group of an \elm\ bundle $\U(1)$ is abelian. It
means $\O=d\a+\frac12[\a,\a]=d\a$. This means we can use Eq.~\eqref{1.1}. In
this theory we have two more fibre bundles. Two fibre bundles of frames
over~$E$ (a~\spt) and over~$P$ (a bundle manifold). Moreover, in order to
simplify a formalism we do not refer explicitly to those fibre bundles.

Tensor \eqref{1.13} in a lift horizontal base
looks simpler than \eqref{1.14n}, moreover, we do not loose any information.
Introducing nonholonomic frames we can write very complicated tensors as
diagonal and sometimes it causes some misunderstanding. In this way
$F_{\mu\nu}$ tensor is connected to $A_\mu$---four-\pt\ via a curvature of an
\elm\ fibre bundle and not via a metric \eqref{1.14n} which has more
mathematical sound. Eq.~\eqref{1.12b} is equivalent to Eq.~\eqref{1.14n}. The
first one is written in a lift horizontal frame and the second in $dx^A=
(dx^\a,dx^5)$.

Now we define on $P$ a \cn\ $\gd w,A,B,$ \st
\beq{1.14}
\bga
\gd w,A,B,=\gd\G,A,BC,\t^C,\quad A,B,C=1,2,3,4,5,\\
D\g_{A+B-}=D\g_{AB}-\g_{AD}\gd Q,D,BC,(\G)\t^C=0,
\ega
\end{equation}
which is invariant \wrt an action of the group $U(1)$ on $P$. $D$~is an
exterior covariant \dv\ \wrt the \cn\ $\gd w,A,B,$ and $\gd Q,D,BC,(\G)$ is
its torsion. Let us notice that for $\gd w,A,B,$ we do not suppose any
constraints on its torsion.
In Refs \cite1, \cite2,~\cite3 it is shown that
\beq{1.15}
\gd w,A,B,=\left( \begin{array}{c|c}
\pi^*(\gd \ov w,\a,\b,)+g^{\g\a}H_{\g\b}\t^5\  & \ H_{\b\g}\t^\g
\vrule height0pt depth 5pt width0pt \\
\hline
g^{\a\b}(H_{\g\b}+2F_{\b\g})\t^\g\ &\ 0
\vrule height10pt depth 0pt width0pt
\end{array} \right)
\end{equation}
where $H_{\b\g}$ is a tensor on $E$ \st
\beq{1.16}
g_{\d\b}g^{\g\d}H_{\g\a}+g_{\a\d}g^{\d\g}H_{\b\g}=
2g_{\a\d}g^{\d\g}F_{\b\g}.
\end{equation}
It is possible to prove that (see Appendix B)
\beq{1.17}
H_{\g\b}=-H_{\b\g} \q (\hbox{if }F_\m=-F_{\nu\mu}).
\end{equation}
Tensors $H_{\mu\nu}$ and $F_{\mu\nu}$ define a five-\di al \cn\ on~$P$. In
the case of a \s\ tensor $g_{\a\b}$, $H_{\mu\nu}=F_{\mu\nu}$ and the theory
reduces to ordinary Kaluza--Klein theory.

We define on $P$ a second \cn
\beq{1.17a}
\bga
\gd W,A,B,=\gd w,A,B,-\tfrac49\gd \d,A,B,\ov W\\
\ov W=\hor\ov W.
\ega
\end{equation}
\eu\cn\ $W^A{}_B$ is a five-\di al analogue of the \cn\ $\ov W{}^\a{}_\b$
known in Einstein Unified Field Theory and NGT (Moffat theory of \gr, see
Ref.~\cite{5a}). According to our notation ``$\ov{\ \ \vphantom{a}}$''
over a symbol means the
quantity is defined on~$E$ (a~\spt), ``$\wt{\ \ }$'' over a symbol means the
quantity is defined \wrt Levi--Civita \cn s, i.e.\ $\gd {\wt{\ov \G}}{},\a,\b\g,$
mean \cf s of Levi--Civita \cn s on~$E$.

The \cn s \eqref{1.15} and \eqref{1.17a} unify \elm\ and \gr al interactions in the
\NK{}. In the theory we can also consider a dual frame $\z_A=(\z_\a,\z_5)$
\st $\t^A(\z_B)=\gd\d,A,B,$. In this way $[\z_\a,\z_\b]=\frac\la 2F_{\a\b}
\z_5$ and the remaining commutators of vector fields vanish. In the classical
Kaluza--Klein Theory the geodetic \e s describe a motion of a charged
particle (test particle), i.e.\ we get a Lorentz force term.
Moreover, in the case of classical (Riemannian)
Kaluza--Klein Theory  we have to do with only one (Levi--Civita) \cn\ on~$P$.
Here we have to do with several possibilities (see Ref.~\cite4). For we have
$H_\m=-H_{\nu\mu}$ it seems now that we should choose a Riemannian part
of \eqref{1.15}. It means Levi--Civita \cn\ generated by $\g_{(AB)}$.

This means we have $u^B\wt\nabla_B u^A=0$, where $\wt\nabla$ means a covariant
\dv\ \wrt ${{\wt w}^A}{}_B=\gd \wt\G{},A,BC,\t^C$
(a~Riemannian part of a \cn\ $\gd w,A,B,$),
$u^A(\tau)$ is a tangent vector to a geodetic line. Eventually one gets
\beq{1.23a}
\frac{\wt{\ov D}u^\a}{d\tau}+\frac q{m_0}\,\wt g{}^{(\a\mu)}\,F_{\mu\b}u^\b=0,
\quad \frac q{m_0}=2u^5={\rm const}.
\end{equation}
$2u^5$ has an interpretation as $q/m_0$ for a test particle, where $q$ is a
charge and $m_0$ is a rest mass of a test particle. $\frac{\wt{\ov
D}}{d\tau}$ means a covariant \dv\ \wrt ${\wt{\ov w}{}^\a}\!_\b$ along a curve
to which $u(\tau)$ is tangent. ${{\wt{\ov w}}{}^\a}\!_\b=
{{\wt{\ov\G}}{}^\a}\!_{\b\g}\wt\t{}^\g$ is of course a Riemannian part of
$\gd\ov w,\a,\b,
=\gd\ov \G,\a,\b\g,\ov\t{}^\g$ (a~Levi--Civita \cn\ generated by
$g_{(\a\b)}$).

Let us calculate a Moffat--Ricci curvature scalar for $\gd W,A,B,$,
$R(W)$, $R(W)\sqrt{\det \g_{AB}}$ is a five-dimensional Lagrangian density.
One gets
\beq{1.18}
R(W)=\ov R(\ov W)+\bigl(2(g^{[\m]}F_\m)^2-H^\m F_\m\bigr)
\end{equation}
where
\beq{1.19}
\ov R(\ov W)=g^\m\ov R_\m(\ov \G)+\tfrac23 g^{[\mu\nu]}\ov W_{[\mu,\nu]}
\end{equation}
is a Moffat--Ricci scalar for the \cn\ $\gd \ov W,\a,\b,$ and $\ov R_{\a\b}
(\ov\G)$ is a Moffat--Ricci tensor for the \cn\ $\gd\ov w,\a,\b,$. In
particular
\beq{1.20}
\ov R_\m(\ov\G)=\gd\ov R,\a,\m\a,(\ov\G)+\tfrac12\gd\ov R,\a,\a\m,
(\ov\G),
\end{equation}
where $\gd \ov R,\a,\m\rho,(\ov\G)$ are components of the ordinary curvature
tensor  for $\ov\G$. In addition
\beq{1.21}
H^{\mu\a}=g^{\b\mu}g^{\g\a}H_{\b\g}.
\end{equation}
The action of the theory simply reads
$$
S=\int_V \sqrt{\det \g_{AB}}\,R(W)\,d^5x.
$$
Using Palatini variation principle \wrt $g_{\a\b}$, $A_\mu$, $\gd \ov
W,\a,\b,$, i.e.\
\beq{1.21a}
0=\d\int_V \sqrt{\det\g_{AB}}\,R(W)\,d^5x=
2\pi\d\int_U\sqrt{-g}\Bigl(\ov R(\ov W)+\bigl(2(g^{[\m]}F_\m)^2-H^\m F_\m\bigr)
\Bigr)\,d^4x
\end{equation}
(Eqs \eqref{1.18} and \eqref{1.21a} give us 5-\di al action in terms of 4-\di al
quantities which can be compared to the standard field theory action),
where $V=U\times U(1)$, $U\subset E$,
one gets from \eqref{1.21a} fields \e s
\bg{1.22}
\ov R_{\a\b}(\ov W)-\tfrac12g_{\a\b}\ov R(\ov W)=8\pi\nad{em}T_{\a\b}\\
{\falg^{[\m]}}_{,\nu}=0 \label{1.23}\\
g_{\m,\si}-g_{\z\nu}\gd\ov\G,\z,\mu\si,-g_{\mu\z}\gd\ov\G,\z,\si\nu,=0
\label{1.24}\\
\pa_\mu(\falH^{\a\mu})=2\falg^{[\a\b]}\pa_\b(g^{[\m]}F_\m) \label{1.25}
\end{gather}
where
\bea{1.26}
\!\nad{em}T_{\a\b}&=&\frac1{4\pi}\Bigl\{g_{\g\b}g^{\tau\mu}g^{\ve\g}H_{\mu\a}
H_{\tau\ve}-2g^{[\m]}F_\m F_{\a\b}-\tfrac14\,g_{\a\b}\bigl(H^\m
F_\m-2(g^{[\m]}F_\m)^2\bigr)\Bigr\}\ \\
\!\falg^{[\m]}&=&\sqrt{-g}\,g^{[\m]} \label{1.27}\\
\!\falH^\m&=&\sqrt{-g}\,g^{\b\mu}g^{\g\a}H_{\b\g}=\sqrt{-g}\,H^{\mu\a}
\label{1.28}
\end{eqnarray}
One can prove
\bg{1.29}
H^\m H_\m=F^\m H_\m,\\
g^{[\m]}F_\m=g^{[\m]}H_\m\label{1.30}
\end{gather}
and
$$
g^{\si\nu}g^{\a\nu}H_{\si\a}F_\m+g^{\mu\si}g^{\nu\b}H_{\b\si}F_\m=2g^{\mu\si}
g^{\nu\b}F_\m F_{\b\si}.
$$
We have also
\beq{1.31}
g^{\a\b}\nad{em}T_{\a\b}=0.
\end{equation}

Equations \eqref{1.22}--\eqref{1.25} can be written in the form
\bg{1.32}
\ov R_{(\a\b)}(\ov\G)=8\pi \nad{em}T_{(\a\b)}\\
\ov R_{[[\a\b],\g]}(\ov\G)=8\pi\nad{em}T_{[[\a\b],\g]} \label{1.33}\\
{\ov\G}_\mu=0 \label{1.34}\\
g_{\m,\si}-g_{\z\nu}{\ov\G}^\z_{\mu\si}-g_{\mu\z}{\ov\G}^\z_{\si\nu}=0
\label{1.35}\\
\pa_\mu\bigl(\falH^{\a\mu}-2\falg^{[\a\mu]}(g^{[\nu\b]}F_\m)\bigr)=0
\label{1.36}
\end{gather}
where $\ov R_{\a\b}(\ov\G)$ is a Moffat--Ricci tensor for the \cn\
$\gd\ov w,\a,\b,=\gd \ov\G,\a,\b\g,{\ov\t}^\g$ and
\beq{1.37}
\ov\G_\mu=\gd\ov\G,\a,[\mu\a],,
\end{equation}
$_{,\g}$ means a partial \dv\ \wrt $x^\g$ (as usual).

Four-\di al quantities in the theory $A_\mu$, $g_\(\m)$ and $g_\[\m]$ are an
\elm\ field, a metric and a skew-\s\ tensor. They correspond to particles:
a~photon (a~spin one), a~graviton (a~spin~2) and a~skewon (a~spin zero).

In the theory we get a current density
\beq{1.38}
{\ov J}^\a=2\pa_\mu\bigl(\sqrt{-g}\,g^{[\a\mu]}(g^{[\nu\b]}F_{\nu\b})\bigr)
\end{equation}
which is conserved by its definition (in this way it is a topological
current). Equation \eqref{1.36} can be written in the form
\bg{1.39}
\ov\nabla_\mu H^{\a\mu}=J^\a\\
J^\a=2\ov\nabla_\mu\bigr(g^{[\a\mu]}(g^{[\nu\b]}F_{\nu\b})\bigr)
=2g^{[\a\b]}\ov\nabla_\mu(g^{[\nu\b]}F_{\nu\b})\label{1.40}
\end{gather}
where $\ov\nabla_\mu$ is a covariant \dv\ for the connection $\gd\ov
w,\a,\b,$.

Equation \eqref{1.16} can be solved \wrt $H_{\nu\mu}$ (see Appendix B)
\beq{1.41}
H_{\nu\mu}=F_{\nu\mu}-{\wt g}^{(\tau\a)}F_{\a\nu}g_{[\mu\tau]}
+{\wt g}^{(\tau\a)}F_{\a\mu}g_{[\nu\tau]}.
\end{equation}
However the form of Eq. \eqref{1.16} is easier to handle from theoretical
point of view. Writing $H_\m$ in the form
\beq{1.42}
H_\m=F_\m -4\pi M_\m
\end{equation}
we get
\beq{1.43}
Q_\m^5=8\pi M_\m=2{\wt g}^{(\tau\a)}\bigl(F_{\a\mu}g_{[\nu\tau]}
-F_{\a\nu}g_{[\mu\tau]}\bigr),
\end{equation}
where $Q_\m^5$ is a torsion in the fifth dimension for the \cn\ $\gd w,A,B,$
on~$P$ and $M_\m$ is an \elm\ polarization tensor induced by a \nos\
tensor $g_{\a\b}$ (if $g_{\a\b}=g_{(\a\b)}$, $F_\m=H_\m$). In this way $H_\m$
can be considered as an induction tensor of an \elm\ field. Moreover, the
second pair of Maxwell \e s \eqref{1.39} suggests that rather $H^\m$ should be
considered as an induction tensor. It is easy to see that if we take $g_\[\m]=
F_\m=0$, $g_\(\a\b)=\eta_{\a\b}$ (a~Minkowski tensor) we satisfy field \e s,
i.e.\ Eqs \eqref{1.32}--\eqref{1.36}. It means an empty Minkowski space is a
\so\ of the \e s.

It is easy to see that the theory contains GR as a limit $g_\[\m]=0$. In this
case we get Einstein \e s with \elm\ sources. If we consider a \NK{} with
external sources (see Ref.~\cite4), we recover GR in the limit $g_\[\m]=0$.
The theory satisfies a Bohr correspondence principle to~GR. Thus we recover
all the achievements of~GR, e.g.\ Newton law, post Newtonian corrections, \gr
al waves. Moreover, in our theory we have gravito-\elm\ waves which are more
general (Section~2). Post Newtonian approximation in the \NK{} with material
(external) sources can be done similarly as in NGT.

Our theory contains an anti\s\ field $g_\[\m]$. Moreover it does not results
in ghosts (see Ref.~\cite{Mann} and references cited therein, especially
Ref.~\cite{9n}). One of these five theories of gravity is Moffat's NGT (see
Ref.~\cite{5a}) in real version. Our approach on the level of ghost
consideration corresponds to NGT. It means after a linearization our theory
does not differ from NGT and we can apply results from Ref.~\cite{9n}.

In the theory we get an \elm\ field lagrangian
\beq{1.43a}
\cL_{\rm em}=-\frac1{8\pi}\bigl(H^\m F_\m-2(g^{[\m]}F_\m)^2\bigr)
\end{equation}
which can be written in the form
\beq{1.43b}
\cL_{\rm em}=-\frac1{8\pi}\bigl((g^{\mu\a}g^{\nu\b}-g^{\nu\b}{\wt g}^{(\mu\a)}
+g^{\nu\b}g^{\mu\o}{\wt g}^{(\tau\a)}g_{\o\tau})
F_{\a\b}F_\m-2(g^{[\m]}F_\m)^2\bigr)
\end{equation}
or
\beq{1.43c}
\cL_{\rm em}=-\frac1{8\pi}\bigl(F^\m F_\m - 2(g^{[\m]}F_\m)^2
+(g^{\nu\b}g^{\mu\o}{\wt g}^{(\tau\a)}g_{\o\tau}-g^{\nu\b}{\wt g}^{(\mu\a)})
F_{\a\b}F_\m\bigr)
\end{equation}
where
$$
F^\m=g^{\mu\a}g^{\nu\b}F_{\a\b}.
$$

Let us consider energy--momentum tensor of an \elm\ field in the \NK{}, i.e.\
$\nad{em}T_{\a\b}$. Using Eq.~\eqref{1.41} one gets
\beq{1.48c}
\nad{em}T_{\a\b}=\nad{o}T_{\a\b}+\frac1{4\pi}\,t_{\a\b}
\end{equation}
where
\beq{1.49c}
\nad{o}T_{\a\b}=\frac1{4\pi}\Bigl(\gd F,\tau,\a,F_{\tau\b}-\frac14\,g_{\a\b}
F^{\m}F_\m\Bigr)
\end{equation}
is an energy--momentum tensor of an \elm\ field in N.G.T.,
\beq{1.50c}
\gd F,\tau,\a,=g^{\tau\g}F_{\g\a}=-\dg F,\a,\tau,
\end{equation}
and
\bml{1.51c}
t_{\a\b}=g_{\g\b}\dg F,\nu,\tau,F_{\o\tau}g^{\ve\g}{\wt g}^{(\rho\nu)}
{\wt g}^{(\d\o)}g_{[\a\rho]}g_{[\ve\d]}
-g_{\g\b}{\wt g}^{(\rho\nu)}\bigl(F^{\mu\g}F_{\nu\mu}g_{[\a\rho]}
+F_{\mu\ve}\dg F,\nu,\mu,g^{\ve\g}g_{[\a\rho]}\bigr)\\
{}-2g^{[\m]}F_\m F_{\a\b}
+\frac14\,g_{\a\b}\Bigl(2\bigl(g^{[\m]}F_\m\bigr)^2
-\bigl(g^{\nu\d}g^{\mu\o}{\wt g}^{(\tau\ve)}g_{\o\tau}-g^{\nu\d}{\wt g}^{(\mu\ve)}
\bigr)F_{\ve\d}F_\m\Bigr)
\end{multline}
is a correction coming from the \NK{}.

Let us consider the second pair of Maxwell \e s in the \NK{}. One writes them
in the following form (using \eqref{1.41}):
\beq{1.52c}
\ov\nabla_\mu F^{\a\mu}=\gd J,\a,p,+J^\a
\end{equation}
where $J^\a$ is a topological current and $\gd J,\a,p,$ is a polarization
current
\bml{1.53c}
\gd J,\a,p,=4\pi \ov\nabla_\mu M^{\a\mu}=\frac{4\pi}{\sqrt{-g}}\,
\pa_\mu\bigl(\sqrt{-g}\,M^{\a\mu}\bigr)=
\ov\nabla_\mu\Bigl(g^{\a\b}g^{\mu\g}{\wt g}^{(\tau\rho)}
\bigl(F_{\rho\g}g_{[\b\tau]}-F_{\rho\b}g_{[\g\tau]}\bigr)\Bigr)\\
{}=\frac1{\sqrt{-g}}\,\pa_\mu\Bigl(\falg^{\a\b}g^{\mu\g}{\wt g}^{(\tau\rho)}
\bigl(F_{\rho\g}g_{[\b\tau]}-F_{\rho\b}g_{[\g\tau]}\bigr)\Bigr).
\end{multline}

The energy momentum tensor in the form \eqref{1.48c} and the second pair of
Maxwell \e s can be obtained directly from Palatini variation principle \wrt
$\gd \ov W,\a,\b,$, $g_\m$ and $A_\mu$ for
\beq{1.54c}
R(W)=R(\ov W)+8\pi\cL_{\rm em}
\end{equation}
where $\cL_{\rm em}$ is given by Eq.~\eqref{1.43c}.

Writing as usual
\bea{1.44}
F_\m&=&\left(\begin{matrix}
0 &\ &-B_3 &\ &B_2 &\ &-E_1\\
B_3&&0&&-B_1&&-E_2\\
-B_2&&B_1&&0&&-E_3\\
E_1&&E_2&&E_3&&0 \end{matrix} \right)\\
H^\m&=&\left(\begin{matrix}
0 &\ &-H^3&\ &H^2&\ &-D^1\\
H^3&&0&&-H^1&&-D^2\\
-H^2&&H^1&&0&&-D^3\\
D^1&&D^2&&D^3&&0 \end{matrix} \right) \label{1.45}
\end{eqnarray}
and introducing Latin indices $a,b=1,2,3$ we get
\bea{1.46}
E_a=F_{4a}, &\q& D^a=H^{4a}\\
\os D=(D^1,D^2,D^3), &\q& \os E=(E_1,E_2,E_3) \label{1.47}\\
\os B=-(F_{23},F_{31},F_{12}), &\q& \os H=-(H^{23},H^{31},H^{12}) \label{1.48}
\end{eqnarray}
or
\bea{1.49}
B_a=-\frac12\,{\ve_a}^{bc}F_{bc}, &\q& F_{cm}=-\dg\ve,cm,e,B_e\\
H^a=-\frac12\,\gd\ve,a,bc,H^{bc}, &\q& H^{cm}=-\gd\ve,cm,e,H^e. \label{1.50}
\end{eqnarray}
$\ve_{abc}$ is a usual 3-dimensional antisymmetric symbol. $\ve_{123}=1$ and it
is unimportant for it if its indices are in up or down position. We keep those
indices in up and down position only  for a convenience.

Using Eq. \eqref{1.41} one gets
\bea{1.51a}
D^a&=&A^{ac}E_c+C^{ad}B_d \\
H^a&=&{\ov A}^{ac}E_c+{\ov C}^{ad}B_d. \label{1.51b}
\end{eqnarray}
In formulae \eqref{1.51a} and \eqref{1.51b} $A^{ac}$ can be identified with
$\ve_{ac}$ (a~dielectric constant tensor) and ${\ov C}^{ab}$ with $(\mu^{-1})
_{ab}$ (an inverse of magnetic constant tensor). Remaining \cf s have more
complex interpretation. Moreover, it is possible to think about them as on
material properties of some kind generalized medium.
A medium with nonzero $C^{ad}$ and $\ov C{}^{ad}$ is called {\it
bianisotropic}. In our case they are induced
by the \nos\ tensor $g_{\a\b}$, where
\begin{eqnarray}
\kern-20pt
A^{me}&=&g^{\mu e}g_{[\d\mu]}\wt g^{(4\d)}g^{m4}
-g^{44}\bigl(g_{[\d\mu]}g^{\mu e}\wt g^{(m\d)}-g^{me}\bigr)
+g^{\o4}g_{[\d\o]}g^{4e}\wt g^{(m\d)}-g^{m4}g^{4e}
\label{1.52}\\
\kern-20pt
C^{pe}&=&\ve_{mz}{}^p\bigl(g^{z4}g^{me}+g^{\mu e}g_{[\d\mu]}g^{z4}\wt g^{(m\d)}
-g^{\o4}g_{[\d\o]}g^{me}\wt g^{(z\d)}\bigr)
\label{1.53}\\
\kern-20pt
\ov A^{pm}&=&\tfrac12 \ve^p{}_{ek}\bigl(g^{4k}g^{me}-g^{mk}g^{4e}
-(g^{mk}\wt g^{(4\d)}
+g^{4k}\wt g^{(m\d)})g^{\mu e}g_{[\d\mu]}
-g^{\o k}g_{[\d\o]}g^{me}\wt g^{(4\d)}\bigr)
\label{1.54}\\
\kern-20pt
\ov C^{pf}&=&\tfrac12\ve^p{}_{ek}\ve_{wm}{}^f\bigl(g^{wk}g^{me}-g^{wk}g^{\mu e}
g_{[\d\mu]}\wt g^{(m\d)}-g^{\o k}g_{[\d\o]}g^{me}\wt g^{(w\d)}\bigr)
\label{1.55}
\end{eqnarray}

Let us come back to the formula \eqref{1.16} and transform it to the form
\beq{1.74}
g_{\si\rho}g_{\d\b}g^{\g\d}g^{\a\si}H_{\g\a}+H_{\b\rho}=2F_{\b\rho}.
\end{equation}
Eq.~\eqref{1.74} gives us $F_{\b\rho}$---a strength of an \elm\ field in term
of $H_{\b\rho}$---an induction tensor. Moreover, we know that an induction
tensor is rather given by Eq.~\eqref{1.21}. Thus we get
\beq{1.75}
F_{\b\rho}=\frac12\bigl(g_{\mu\b}g_{\nu\rho}+g_{\si\rho}g_{\d\b}
g^{\g\d}g^{\a\si}g_{\mu\g} g_{\nu\a}\bigr)H^\m.
\end{equation}
After some calculations one gets
\bea{1.76}
B_a&=&K_{ae}H^e + L_{an}D^n\\
E_r&=&\ov K_{re}H^e+\ov L_{rn}D^n,\label{1.77}
\end{eqnarray}
where
\bea{1.78}
K_{ae}&=&\frac14 \,\dg\ve,a,br, \bigl(g_{mb}g_{nr}+g_{\si r}g_{\d b}
g^{\g\d}g^{\a\si}g_{m\g}g_{n\a}\bigr)\gd\ve,mn,e,\\
L_{an}&=&\frac14 \,\gd\ve,br,a, \Bigl[\bigl(g_{nb}g_{4r}+g_{\si r}g_{\d b}
g^{\g\d}g^{\a\si}g_{n\g}g_{4a}\bigr)-\bigl(g_{4b}g_{nr}+
g_{\si r}g_{\d b}g^{\g\d}g^{\a\si}g_{4\g}g_{n\a}\bigr)\Bigr] \label{1.79}\\
\ov K_{re}&=&\frac12 \, \X1(g_{m4}g_{nr}+g_{\si r}g_{\d4}g^{\g\d}g^{\a\si}
g_{m\g}g_{n\a}\Y1)\gd\ve,mn,e, \label{1.80}\\
\ov L_{rn}&=&\frac12 \X2[\X1(g_{44}g_{nr}+g_{\si r}g_{\d4}g^{\g\d}g^{\a\si}
g_{4\g}g_{n\a}\Y1) - \X1(g_{n4}g_{4r}+g_{\si r}g_{\d4}g^{\g\d}g^{\a\si}
g_{n\g}g_{4\a}\Y1)\Y2]. \label{1.81}
\end{eqnarray}
In this way we get $\os B$ and $\os E$ in terms of $\os H$ and~$\os D$,
similarly to the formulae \eqref{1.52}--\eqref{1.55}. They are in some sense
dual to those formulae.

Moreover, in Ref.~\cite{3Ch} one can find an induction between electric and
magnetic fields caused by topological effects. Another \cn\ between our work
and Ref.~\cite{3Ch} is using a dimensional reduction which is close to the
Kaluza--Klein idea.

In our theory the \nos\ tensor $g_\m$ (a~\nos\ \gr al field) results as a
bianisotropic linear medium. Due to this we get Eqs
\eqref{1.51a}--\eqref{1.51b}. Moreover, in classical \elm\ theory there are
some formalisms for linear bianisotropic phenomena (see Refs \cite{11n},
\cite{12n}, \cite{13n}, \cite{14n}). The authors consider classical
electrodynamics of continuous media in several forms including a covariant
form, i.e.
$$
\bal
H^\m&=\k^{\m\a\b} F_{\a\b}\\
\hbox{or}\q F_{\a\b}&=\ov\k_{\a\b\m}H^\m
\eal
$$
(see Ref.~\cite{11n}), similar to our Eqs \eqref{1.41} and \eqref{1.16}.

In the theory of spatially dispersive materials one can find the following
formulae written in dyadic formalism,
$$
\os D=\twl\ve \cdot\os E+\twl\xi\cdot \os H \qh{and}
\os B=\twl\mu \cdot\os H-\twl\xi{}^T \cdot \os E,
$$
where $\twl\ve$ is the permittivity and $\twl\mu$ is the permeability
depending on space (and time). $\twl\xi$ is one more parameter for
mirror-a\s\ structure (chiral materials). For magnetoelectric coupling media
one gets
$$
\os D=\twl\ve \cdot\os E+\twl\z\cdot \os H \qh{and}
\os B=\twl\mu \cdot\os H+\twl\z{}^T \cdot \os E,
$$
where $\twl\z$ is one more parameter which has to do with parity property of the
material \wrt time inversion. These \e s are similar to our Eqs
\eqref{1.51a}--\eqref{1.51b} (see Ref.~\cite{12n}). In Refs \cite{13n},
\cite{14n} the authors are using differential forms in classical
electrodynamics also for bianisotropic media (see Ref.~\cite{14n}, also
\cite{15n}), getting similar formulae in covariant form. In Ref.~\cite{13n}
the authors consider also nonlinear electrodynamics by Born--Infeld (see
Ref.~\cite{16n}) and also in a more general Pleba\'nski's form (see
Ref.~\cite{17n}). In our approach constitutive relations are linear (in
nonlinear electrodynamics they are nonlinear), but field \e s are nonlinear.
In Ref.~\cite{12n} one considers several currents, e.g.\ external,
polarization. In our case we have also several currents, i.e.\ a polarization
current and topological current, see Eq.~\eqref{1.52c} and \eqref{1.53c} and
\eqref{1.40}. The covariant form of the linear constitutive \e s has been
considered in a general form in Ref.~\cite{11n}.

In GR we have also a tensor $H^\m$ given by the formula
$$
H^\m=g^{\mu\a}g^{\nu\b}F_{\a\b}
$$
where $g^{\mu\a}$ is an inverse tensor of a \s\ metric in~GR. In this way we
can have an induction tensor $H^\m$ in curvilinear \cd s in space
(e.g.~spherical). Someone can define an induction ``tensor'' in a different
way, as a tensor density, i.e.
$$
\mathrel{\lower4pt\hbox{${\scriptstyle\sim}$}\hskip-6.5pt h}^\m
=\sqrt{-g}\,H^\m.
$$
We do not follow this approach. In this way we have to do with bianisotropic
medium in GR and also in flat Minkowski space in curvilinear \cd s, i.e.
$$
\bga
D^a=\X1(g^{44}g^{ab}-g^{4b}g^{a4}\Y1)E_a - \X1(g^{4m}g^{an}-g^{4n}g^{am}\Y1)
\dg\ve,mn,e, B_e\\
H^a=\tfrac12\gd\ve,a,mn,\X1(g^{mb}g^{n4}-g^{m4}g^{nb}\Y1)E_b
+ \tfrac12\gd\ve,a,mn,\dg\ve,cb,e, g^{\tl m[c}g^{n\tp b]}.
\ega
$$
It is easy to see that the ``medium'' is bianisotropic if $g^{4m}\ne0$.

In the \NK{} the situation is more complex:
$$
H^\m = g^{\mu\a}g^{\nu\b}H_{\a\b}
$$
and $H_{\a\b}$ is given by the formula \eqref{1.41}.

One can also consider a tensorial density
$$
\mathrel{\lower4pt\hbox{${\scriptstyle\sim}$}\hskip-6.5pt h}^\m
=\sqrt{-g}\,H^\m.
$$
Moreover, we do not follow this approach. In Refs \cite{18n}, \cite{19n} one
can find some conditions posed on $\ve_{ac}$ (or~$\ve \d_{ac}$) and
$(\mu^{-1})_{ab}$ (or~$\frac{\d_{ab}}\mu$). In the \NK{} such conditions are not
satisfied for our ``generalized medium'' is a \gr al field described by the
\nos\ tensor $g_\m$. Let us notice the following fact: even if $g^{4m}=0
\ne g^{m4}$ (in the \nos\ case) our constitutive relations can still describe
bianisotropic medium.

Let us notice that in the case of a diagonal $g_{(\a\b)}$,  $F_\m=
H_\m$, also in the case of spherically symmetric $g_\m$ we have $F_\m=H_\m$,
which was extensively used in order to find an exact \so\ for field \e s
(see~\cite4).

Formulae \eqref{1.44}--\eqref{1.45} have a formal character. The physical
meaning of $(\os D,\os H)$ and $(\os E,\os B)$ as induction or strength
vectors of electric or magnetic fields is sound only in a stationary case.

Let us consider a \nos\ tensor for axially \s\ and stationary space-time in
cylindrical coordinates (see Ref.~\cite6)
\beq{1.56}
g_\m=\left(\begin{matrix}
-e^{2(n-l)}&\ &0&\ &ade^n&\ &de^n\\
0&&-e^{2(n-l)}&&kae^m&&ke^n\\
-ade^n&&-ake^n&&ca^2e^{2l}-r^2e^{-2l}&&ace^{2l}\\
-de^n&&-ke^n&&ace^{2l}&&ce^{2l} \end{matrix}\right)
\end{equation}
where $c=1+d^2+k^2$, $x^1=r$, $x^2=z$, $x^3=\t$, $x^4=t$, and all the \f s
$n,l,a,b,d,k$ are \f s of~$r$ and~$z$ only,
\beq{1.57}
g=r^2e^{4(n-l)}, \q \wt g=-r^2e^{4(n-l)}(1+d^2+k^2).
\end{equation}

An \elm\ field is described by
\bg{1.58}
F_\m=\left( \begin{matrix}
0&\ &0&\ &p&\ &s\\
0&&0&&q&&u\\
-p&&-q&&0&&0\\
-s&&-u&&0&&0 \end{matrix} \right)\\
p=B_z,\ s=-E_r,\ q=-B_r,\ u=-E_z\nn,
\end{gather}
and all the \f s depend on $r$ and $z$ only.

In this case we can calculate $H_\m$ and $H^\m$. One gets
\beq{1.59}
H_\m=\frac1{\wt g}\left(\begin{matrix}
0&\ &r^2m_1&\ &-pr^2m_2&\ &-r^2sm_2\\
-r^2m_1&&0&&-qr^2m_2&&-r^2um_2\\
pr^2m_2&&qr^2m_2&&0&&r^2m_3\\
r^2sm_2&&r^2um_2&&-r^2m_3&&0 \end{matrix} \right)
\end{equation}
where $m_1=(du-ks)e^{5n-6l}$, $m_2=(1+d^2+k^2)e^{4(n-l)}$,
$m_3=\bigl(d(p-as)+k(q-as)\bigr)e^{3n-2l}$,
\beq{1.60}
H^\m=\frac1P
\left(\begin{matrix}
0&\ &0&\ &r^4m_4&\ &ar^4m_4+r^6m_6\\
0&&0&&r^4m_5&&-ar^4m_5+r^6m_7\\
-r^4m_4&&-r^4m_5&&0&&0\\
-ar^4m_4-r^6m_6&&ar^4m_5-r^6m_7&&0&&0 \end{matrix} \right)
\end{equation}
where
\beq{1.61}
\bal
P=e^{12(n-l)}r^6(1+d^2+k^2),\q\\
m_4=e^{10n-8l}(1+d^2+k^2)(as-p),\q &m_5=e^{10n-8l}(1+d^2+k^2)(au-q), \\
m_6=e^{10n-12l}(s+d^2s+dku),\q &m_7=e^{10n-12l}(dks+u+k^2u).
\eal
\end{equation}

Let us come back to Eq. \eqref{1.25} (the second part of Maxwell \e s) and let
us consider it for $\a=4$
\beq{1.62}
\pa_m\falH^{4m}=2\falg^{[4b]}\pa_b \bigl(g^{[\m]}F_\m\bigr), \q m,b=1,2,3,
\end{equation}
or
\beq{1.63}
\div\bigr(\sqrt{-g}\,\os D\bigr)=\rho\sqrt{-g}.
\end{equation}
In this equation $\rho$ represents a density of an electric charge.

If $\os D=0$ the density of charge equals zero. The problem which we now pose
is as follows. Is it possible to have $\os D=0$ and $\os E\ne0$? This means
that we have a condition
\beq{1.64}
A^{ae}E_e+C^{ae}H_e=0.
\end{equation}
This means we have a \cfn\ of a charge induced by a special properties of
``vacuum'' (i.e.\ a \gr al field described by \nos\ tensor $g_\m$). It means
non-zero electric field and zero charge distribution.

In the case of a stationary, axially \s\ field we have conditions
\bea{1.65a}
as-p&=&\frac{r^2(s+d^2s+dku)}{a(1+d^2+k^2)}\,e^{-4l}\\
au-q&=&\frac{r^2(dks+u+k^2u)}{a(1+d^2+k^2)}\,e^{-4l}\label{1.65b}
\end{eqnarray}
with $u,s\ne0$.

These conditions can be imposed on \f s $n,l,d,k,a$ to be satisfied for any
nonzero $u,s$ with some dependence on $p$ and~$q$. One gets
\beq{1.66}
\bga
e^{4l}a^2(1+d^2+k^2)=r^2(1+d^2)\\
p=-u\,\frac{r^2dk}{a(1+d^2+k^2)}\,e^{-4l}\\
e^{4l}a^2(1+d^2+k^2)=r^2(1+k^2)\\
q=-s\,\frac{dkr^2}{a(1+d^2+k^2)}\,e^{-4l}
\ega
\end{equation}
and finally
\bg{1.67}
d=k, \q a=re^{-2l}\sqrt{\frac{1+d^2}{1+2d^2}}\,,\\
p=-\b u,\q q=-\b s, \label{1.68}
\end{gather}
where
\beq{1.69}
\b=\frac{rde^{2l}}{\sqrt{(1+d^2)(1+2d^2)}}
\end{equation}
or
\bea{1.70}
B_z&=&\b E_z \\
B_r&=&-\b E_r. \label{1.71}
\end{eqnarray}

\def\elm{electromagneti}
We do not give here any quantum version of the theory. It is a classical
field theory and a classical theory of a charge \cfn. Moreover, a \cfn\ is a
nonperturbative effect and cannot be obtained in perturbative quantum field
theory. This dielectric model of a charge \cfn\ can be considered as an
``interference effect'' between gravity and \elm sm in our unified classical
field theory. In order to find a quantum version of the theory we
should consider Ashtekar--Lewandowski canonical quantization procedure of the
theory, which is suitable here for the theory is nonlinear and contains
gravity (see Ref.~\cite{AL}).

According to the Einstein programme of geometrization and unification of
physical interactions we should get \e s where on the left-hand side we have
geometrical quantities and on the right-hand side material quantities. A~full
programme is completed if on the right-hand side we get zero. It means all
quantities have been geometrized. Eqs \eqref{1.22}--\eqref{1.25} give us in
this sense geometrization and unification of gravity and \elm sm if we shift
$8\pi \nad{em}T_{\a\b}$ from the right-hand side to the left in
Eq.~\eqref{1.22} and $2\falg^{[\a\b]}\pa_\b(g^{[\m]}F_\m)$ from the
right-hand side to the left in Eq.~\eqref{1.25}. Having in mind
Eq.~\eqref{1.26} we see that $8\pi \nad{em}T_{\a\b}$ is geometrized.
According to this geometrization and unification programme all quantities
coming from higher dimension should get an interpretation in terms of matter
defined on the \spt. In this
way in the theory we geometrized all quantities completing Einstein
programme for unification of gravity and \elm sm getting ``interference
effects'' between both interactions.

\allowdisplaybreaks
\section{Gravito--\elm c waves \so s in the \NK{}}
Let us consider the following \nos\ metric in cartesian coordinates
\beq{2.1}
g_\m=\left(\begin{matrix} -1&\ &0&\ &r&\ &-r\\
0&&-1&&s&&-s\\
-r&&-s&&e-1&&-e\\
r&&s&&-e&&1+e\end{matrix}\right)
\end{equation}
where
$$
\bal
e&=e(x,y,z-t)\\
s&=s(x,y,z-t)\\
r&=r(x,y,z-t)
\eal
$$
which describes generalized plane wave for $g_\m$ and
\beq{2.2}
F_\m=\left(\begin{matrix}
0&\ &0&\ &k&\ &-k\\
0&&0&&p&&-p\\
-k&&-p&&0&&0\\
k&&p&&0&&0\end{matrix} \right)
\end{equation}
where
$$
\bal
p&=p(x,y,z-t)\\
q&=q(x,y,z-t)
\eal
$$
which describes generalized plane \elm c wave. All the \f  s mentioned here
are subject of field \e s in \NK{}. Using results from Ref.~\cite7 one gets
the following \e s
\beq{2.3}
-\D e+4Q-\biggl[\Bigl(\pp rx-\pp sy\Bigr)^2+\Bigl(\pp ry+\pp sx\Bigr)^2\biggr]
=4\biggl[\Bigl(\pp Ax\Bigr)^2+\Bigl(\pp Ay\Bigr)^2\biggr]
\end{equation}
where
\begin{eqnarray}
{}&&Q=\biggl[\Bigl(\pp rx\Bigr)^2+r\,\pp{^2r}{x^2}+\frac12\,\pp sx\Bigl(\pp ry+
\pp sx\Bigr) +\frac12\,s\Bigl(\pp{^2r}{x\pa y}+\pp{^2s}{x^2}\Bigr)
+\Bigl(\pp sy\Bigr)^2\nn\\
&&\qquad{}+s\,\pp{^2s}{y^2}+\frac12\,\pp ry\Bigl(\pp ry+\pp sx
\Bigr)+\frac12\,r\Bigl(\pp{^2r}{y^2}+\pp{^2s}{x\pa y}\Bigr)\biggr]\label{2.4}\\
&&\D A(x,y,z-t)=0 \label{2.5}\\
&&p=\pp Ax,\q q=-\pp Ay\label{2.6}\\
&&\pp ry=\pp sx+H(x,y,z-t), \q \D H(x,y,z-t)=0 \label{2.7}
\end{eqnarray}
where $\D=\pp{^2}{x^2}+\pp{^2}{y^2}$ is the Laplace operator in two
dimensions. Equation \eqref{2.3} is a Poisson \e\ for~$e$.
Using Eq.~\eqref{1.23} one gets
$$
\pp sy=-\pp rx \qh{or} s=-\pp Bx,\ r=\pp By.
$$
In this way we get
\beq{2.7n}
\D B=H.
\end{equation}

$A$ and $H$ are arbitrary harmonic \f s in two dimensions with arbitrary
dependence on $(z-t)$ of $C^2$ class. $B$~is an arbitrary \so\ of Poisson \e\
and arbitrary \f\ for $(z-t)$ of $C^2$ class.
The \f~$e$ can be obtained from the Poisson \e
\beq{2.9}
\D e=f
\end{equation}
where $f$ is given in terms of $B,A,H$ and takes a simple form
$$
f=-4\biggl[\Bigl(\pp Ax\Bigr)^2+\Bigl(\pp Ay\Bigr)^2\biggr]
+\biggl(\pp{^2B}{y^2}-\pp{^2B}{x^2}\biggr)^2+2\Bigl(\pp Bx\,\pp Hx+
\pp By\,\pp Hy\biggr).
$$
The dependence on $(z-t)$ is
parametric and is given by the dependence on $(z-t)$ of $B,A,H$. This \so\
describes a generalized plane gravito-\elm c wave.

Now we consider the following \nos\ tensor in cartesian coordinates
\beq{2.10}
g_\m=\left(\begin{matrix}
-a&\ &0&\ &r&\ &-r\\
0&&-a&&s&&-s\\
-r&&-s&&-b&&0\\
r&&s&&0&&b\end{matrix}\right),
\end{equation}
$$
\bal
a&=a(x,y), \q r=r(x,y,z-t),\\
b&=b(x,y), \q s=s(x,y,z-t),
\eal
$$
and an \elm c field strength tensor
\bg{2.11}
F_\m=\left(\begin{matrix}
0&\ &0&\ &p&\ &-p\\
0&&0&&q&&-q\\
-p&&-q&&0&&0\\
p&&q&&0&&0\end{matrix}\right),\\
p=p(x,y,z-t), \q q=q(x,y,z-t).
\end{gather}

We put \eqref{2.10} and \eqref{2.11} into the field \e\
\eqref{1.22}--\eqref{1.28}. Using the results from Ref.~\cite8 we get the
following \e s:
\bea{2.12}
{}&&\D(\a/a)=4a\b^2-\frac2a \biggl[\Bigr(\pp \psi x\Bigr)^2
+\Bigl(\pp\psi y\Bigr)^2\biggr]\\
&&p=\pp \psi y, \q q=-\pp \psi x \label{2.13}\\
&&\psi=\psi(x,y,z-t), \q \D\psi=0 \nn\\
&&\b=\frac1{2a} \Bigl(\pp ry-\pp sx\Bigr)\label{2.14}\\
&&\a=r^2+s^2. \label{2.15}
\end{eqnarray}
From the remaining field \e\ we get
\beq{2.16}
\D\b=0,
\end{equation}
where $\D=\pp{^2}{x^2}+\pp{^2}{y^2}$,
\beq{2.17}
a=e^A,
\end{equation}
where $A$ is an arbitrary harmonic \f\ in two variables,
\beq{2.18}
\D A=0.
\end{equation}
The \f\ $b$ is given by
\beq{2.20}
b=G_1(z+t)G_2(z-t)
\end{equation}
where $G_1$ and $G_2$ are arbitrary \f s of $C^2$ class (of on variable). The
\f\ $\b$ should be written in the following form
\beq{2.21}
\b(x,y,z,t)=f_0(z-t)\b_0(x,y)
\end{equation}
where $\b_0$ is an arbitrary harmonic \f\ and $f_0$ is an arbitrary \f\ of
one variable (of $C^2$ class). Let us consider Eq.~\eqref{1.23}. One gets
\beq{2.22}
\pp ry+\pp sx=0.
\end{equation}
From this equation we get
\beq{2.23}
r=\pp \vf x, \q s=-\pp\vf y
\end{equation}
where $\vf$ is a \f\ (arbitrary) of two variables (of~$C^3$) and a \f\ of
$z-t$.

One gets
\beq{2.24}
\D\vf=2e^Af_0\b_0.
\end{equation}
Let $\wt\vf$ be any arbitrary \so\ of Eq.~\eqref{2.24} (remembering that $A$,
$f_0$ and~$\b_0$ are arbitrary). In this way
\beq{2.25}
\a=r^2+s^2=\biggl(\pp{\wt\vf}y \biggr)^2+\biggl(\pp{\wt\vf}x\biggr)^2.
\end{equation}
Let us consider Eq.~\eqref{2.12} in the following form
\beq{2.26}
\D g=4e^A\b_0^2f_0^2-\frac2{e^A}\biggl[\biggl(\pp\psi x \biggr)^2
+\biggl(\pp\psi y \biggr)^2\biggr]
\end{equation}
and let $\wt g$ be any \so\ of this Poisson \e.

Thus
\bg{2.27}
\frac\a a=\wt g\\
\biggl(\pp{\wt\vf}y \biggr)^2+\biggl(\pp{\wt\vf}x \biggr)^2=\wt g
e^A.\label{2.28}
\end{gather}
Eq.\ \eqref{2.28} gives us a consistency condition for an existence of
gravito-\elm c wave.

Let us consider the following \nos\ metric
\beq{2.26a}
g_\m=\left(\begin{matrix}
0&\ &0&\ &0&\ &1\\
0&&b&&0&&l+q\\
0&&0&&b&&m+p\\
1&&l-q&&m-p&&-v
\end{matrix}\right)
\end{equation}
and an \elm c field strength tensor
\beq{2.27a}
F_\m=\left(\begin{matrix}
0&\ &0&\ &0&\ &0\\
0&&0&&0&&s\\
0&&0&&0&&u\\
0&&-s&&-u&&0
\end{matrix}\right)
\end{equation}
in cartesian coordinates
$$
x^1=x, \q x^2=y, \q x^3=z, \q x^4=t.
$$
It is possible to consider $x^4$ as $z-t$. We suppose that
$$
\pp{}{x^1}\,g_{\m}=\pp{}{x^1}\,F_{\m}=0.
$$
Using results from Refs \cite{10}, \cite{11} and field \e\ of the \NK{} we
get the following \e s
\beq{2.28a}
\bal
{}&s=\pp \psi{x^3},\\
&u=\pp \psi{x^2},\\
&\biggl(\pp{^2}{(x^2)^2}+\pp{^2}{(x^3)^2}\biggr)\psi=0, \q
\psi=\psi(x^2,x^3,x^4).
\eal
\end{equation}
Supposing
\beq{2.29}
b=1, \q q=0
\end{equation}
we get further
\beq{2.30}
\bga
p=p(x^2,x^4)\\
\pp{^3}{(x^2)^3}\,p=0\\
\pp l{x^3}=\pp m{x^2}\,, \q \pp l{x^2}=-\pp m{x^3}\\
l=l(x^2,x^3,x^4), \q m=m(x^2,x^3,x^4).
\ega
\end{equation}
Thus $l$ and $m$ are harmonically conjugate.

Writing
\beq{2.31}
v=1+f, \q f=f(x^2,x^3,x^4),
\end{equation}
where $f$ satisfies the equation
\beq{2.32}
\D f=2p^2\biggl(\frac{p_{,22}}p+\Bigl(\frac{p_{,2}}p\Bigr)^2\biggr)
-2\bigl((\psi_{,2})^2+(\psi_{,3})^2\bigr),
\end{equation}
the form of $p$ can be easily found
$$
p=\frac{(x^2)^2}2\,\vf_1(x^4)+x^2\vf_2(x^4)+\vf_3(x^4)
$$
where $\vf_i$, $s=1,2,3$, are arbitrary \f s of one variable of $C^3$ class.

This \so\ represents a gravito-\elm c wave if we interpret $x^4$ as wave front
variable $z-t$. ``,'' means a \dv\ \wrt $x^2$ or~$x^3$, $f$~is a \so\ of a
Poisson \e\ with a parametric dependence on~$x^4$ imposed by $\psi$, $\vf_i$,
$i=1,2,3$. \eu\f s $\psi$, $l$ and $m$ are arbitrary \f s of $x^4$ variable
of $C^2$ class.

For all three \so s considered here $t_\m=M_\m=J^\mu=\gd J,\mu,p,=0$. Moreover,
$\os D\ne \os E$ for all \so s. Those \so s are not \so s of \nos\ theory of
gravity coupled to a Maxwell field. They are examples of wave \so s in the
\NK{}, a unified theory of gravity and an \elm c field.

\section{An influence of a cosmological constant on a \so\ in the \NK{}}
Let us consider Eqs.\ \eqref{1.22}--\eqref{1.28} and let us change
$\nad{em}T_{\a\b}$ into
$$
T_{\a\b}=\nad{em}T_{\a\b}-\frac{\La}{8\pi}\,g_{\a\b}.
$$
It means we introduce a cosmological constant $\La$ to the theory.

In the \NK{} $\La=0$. Moreover, in the \eu\nos\ Nonabelian Kaluza--Klein
Theory this constant is in general non-zero. The Yang--Mills field can be
reduced to a $U(1)$ subgroup of a group $G$ (see Ref.~\cite{3} for more
details). We also consider \NK{} with external sources, and a cosmological
constant term can be added to the external energy momentum tensor (see
Ref.~\cite4).

Let us consider corrected field \e s in a static, spherically \s\ case. Using
results from Ref.~\cite4 we get the following exact \so
\beq{3.1}
g_\m=\left(\begin{matrix}
-\a&\ &0&\ &0&\ &\o\\
0&&-r^2&&0&&0\\
0&&0&&-r^2\sin^2\t&&0\\
-\o&&0&&0&&\g
\end{matrix}\right)
\end{equation}
where
\bea{3.2}
\o&=&\frac{l^2}{r^2}\\
\a^{-1}&=&\biggl(1+\frac{Q^2}{\ov br}\,g\Bigl(\frac r{\ov b}\Bigr)+\frac{\La r^2}3
\biggr) \label{3.3}\\
\g&=&\biggl(1+\frac{l^4}{r^4}\biggr)
\biggl(1+\frac{Q^2}{\ov br}\,g\Bigl(\frac r{\ov b}\Bigr)+\frac{\La r^2}3
\biggr) \label{3.4}\\
F_{14}&=&E=-\frac Q{r^2}\biggl(\frac{r^4}{r^4+{\ov b}^4}\biggr) \label{3.5}\\
{\ov b}^4&=&4l^4 \label{3.6}
\end{eqnarray}
$Q$ is an electric charge and $l$ is an integration constant of length
dimension. Let us notice that a \so\ with a cosmological constant differs
from the previous \so\ (see Ref.~\cite4) only by a term $\frac{\La r^2}3$ similarly as in
General Relativity for a Schwarzschild \so\ with a \co ical \ct.

Similar \so\ is known in Einstein Unified Field Theory and in Schr\"odinger
Theory (see Ref.~\cite{Takeno}) (i.e.\ which generalizes Schwarzschild-like
\so\ from those theories to the case with non-zero \co ical \ct). This means
that \co ical \ct\ results only via term $\frac{\La r^2}{3}$. Thus the \so\
is not asymptotically flat.

It is easy to see that
\beq{3.7n}
F_{14}=E \underset{r\to\infty}{\longrightarrow}-\frac Q{r^2}
\end{equation}
from Eq.~\eqref{3.5}. From Eq.~\eqref{3.4} and the definition of the \f\
$g(x)$ (see Eq.~\eqref{3.10} below) one gets
\beq{3.8n}
\a^{-1}\underset{r\to\infty}{\longrightarrow}1-\frac{2m_N}r
+\frac{Q^2}{r^2}+\frac{\La r^2}3
\end{equation}
where
\beq{3.9n}
m_N=\frac{\pi Q^2}{2\sqrt 2 \ov b}
\end{equation}
(see Ref.~\cite4) is an energy of a \so\ (a~Schwarzschild-like or
Kottler-like asymptotically). It is also easy to see that $\a^{-1}(0)=1$,
$F_{14}(0)=E(0)=0$.

All the details concerning the \so\ without a \co ical \ct\ can be found
in Ref.~\cite4. In Ref.~\cite4 it has been proved that the energy of the \so\
is finite and the total charge is equal to~$Q$. An electric field of the \so\
is plotted in Fig.~3 of Ref.~\cite4 (it is the same as in the case with
non-zero \co ical \ct).

Moreover, we repeat some details important for a reader. This \so\ is not a
\so\ with a \co ical \ct\ in a \nos\ theory of gravity coupled to a Maxwell
field. This \so\ cannot be
obtained in pure NGT with \elm c sources. Thus its remarkable properties
concerning nonsingularity of electric and \gr al fields are ``interference
effects'' between gravity and electromagnetism in our unification of gravity
and electromagnetism. The \so\ asymptotically behaves as
Reissner--Nordstr\"om \so\ in~NGT. Due to this it satisfies Bohr
correspondence principle between our unification (\NK{}) and NGT and General
Relativity. The \so\ achieves an old dream by Einstein, Weyl, Kaluza,
Schr\"odinger, Eddington on a \ti{unitary classical field theory} which has
spherically-\s\ singularity-free \so s of the field \e s treated as particles.

The properties of the solution are traced back to Abraham--Lorentz (see Refs
\cite{A2}, \cite{A3}, \cite{A4}, \cite{A5}) idea and more advanced in
Born--Infeld electrodynamics (see Refs \cite{16n},~\cite{A1}) of an electron
being a particle-like finite-energy field configuration, the soliton, which
energy is of completely field nature. It means it is an energy of field
selfinteraction. The \so\ is in rest. In order to get a moving soliton it is
enough to boost it via a Lorentz \tf.

Our theory is \ti{nonlinear} (nonlinear field \e s). Moreover, our
constitutive \e s between a field strength tensor $F_\m$ and an induction
tensor $H^\m$ are \ti{linear}.

Let us introduce the following notation
\bea{3.7}
a&=&\frac{Q^2}{2l^2} \q \Bigl(a=\frac{G_NQ^2}{2c^2l^2}\Bigr)\\
b&=&\frac{2l^2\La}3 \label{3.8}
\end{eqnarray}
In this notation one gets
\beq{3.9}
\a^{-1}=1+a\,\frac{g(x)}x+bx^2=f(x)
\end{equation}
where
\beq{3.10}
g(x)=\frac1{4\sqrt2}\log\biggl(\frac{x^2+\sqrt2 x+1}{x^2-\sqrt2 x+1}
\biggr)-\frac1{2\sqrt2}\biggl[\arctan(\sqrt2 x+1)+\arctan(\sqrt2 x-1)
\biggr]
\end{equation}
(see Ref.~\cite4),
\beq{3.11}
x=\frac{r}{\sqrt2 l}\,.
\end{equation}
The \f\ $g(x)$ has been plotted in Fig.~2 of Ref.~\cite4.
Similarly as in Ref.~\cite4 one writes $f(x)=1-P(x)$. In this way $P(x)$ has
an interpretation of a generalized Newtonian \gr al field in normalized
radial coordinate, i.e.\ $P(x)=-a\frac{g(x)}x-bx^2$ (see Schwarzschild \so\
in~GR, see also Fig.~6 of Ref.~\cite4 in the case of $b=0$).
The \so\ described here can be considered as a classical model of an
electron. We can try to quantize it using collective \cd\ approach. One can
also apply loop quantum gravity approach (see Ref.~\cite{AL}) similarly as in
Ref.~\cite{GP}.

Now it is interesting to examine an influence of the cosmological constant
$\La$ (via~$b$) on properties of the \so. The most interesting case is to
find horizons for such a \so. It means, to find zeros of the \f
\beq{3.12}
f(x)=\a^{-1}=1+a\,\frac{g(x)}x+bx^2=0.
\end{equation}
(The \f\ $f(x)$ gives us information of \gr\ field of the discussed \so. In
the case of zero \co ical \ct\ we have a plot in Fig.~6 of Ref.~\cite4.)

All the plots of the \f\ $f(x)$ for some values of parameters $a$ and~$b$
give us some information on the behaviour of relativistic \gr al field of the
\so\ for some critical values of $a$ and~$b$ and some typical behaviour of
the \so\ among critical values. For we expect horizons $f(x)$ gives us more
than $P(x)$. We have of course $P(0)=0$ ($f(0)=1$).

In the previous case $b=0$ (see Ref.~\cite4)
we get that there is a critical value of $a=a_{\rm
crt}= 3.17\dots$ such that for $a<a_{\rm crt}$ there is not any horizon,
for $a=a_{\rm crt}$ there is one horizon, and for $a>a_{\rm crt}$ there are
two horizons. Let us suppose that $\La>0$ ($b>0$) and examine $a_{\rm crt}$
in this case. One gets
$$
\bal
\hbox{for }b&=0.001, \q && a_{\rm crt}\cong 3.2\\
\hbox{for }b&=0.01, \q && a_{\rm crt}\cong 3.6\\
\hbox{for }b&=0.1, \q && a_{\rm crt}\cong 4\\
\hbox{for }b&=1, \q && a_{\rm crt}\cong 11.
\eal
$$
It means that as before we have for $a<a_{\rm crt}$ not any horizon,
$a=a_{\rm crt}$ one horizon and for $a>a_{\rm crt}$ two horizons.
Thus the cosmological constant results in a higher value of $a_{\rm crt}$.

It is interesting to consider a negative value of $\La$ ($b<0$). In this case
we have always one more horizon (the so-called de~Sitter horizon, a
cosmological horizon). For example for $a=0.1$, $b=-0.001$ we have only one
horizon. For $a=5$, $b=-0.001$ we have three horizons, two as before for
$b>0$ and one de~Sitter horizon. For $a=3$, $b=-0.01$ we have two horizons,
one de~Sitter horizon and one double as before for $b>0$. In this case $a=3$
is a critical value for $b=-0.001$. For lower negative values of~$b$, i.e.\
$b=-0.1$, we have for $a=0.1$, $0.7$, $3$ only one (de~Sitter) horizon.

Summing up, for $\La>0$ we get two horizons, one horizon or no horizon,
$0<r_{H_1}<r_{H_2}$,
$0<r_H$ (as in the case of Reissner--Nordstr\"om \so). For $\La<0$ we have a
de~Sitter horizon
$$
\bal
0&<r_S\\
0&<r_{H_1}<r_{H_2}<r_S\\
0&<r_H<r_S
\eal
$$
and inside it one or two horizons or a case without an inside horizon.
In Figures 1, 2, 3, 4, 5 we give plots of the \f\ $f(x)$ for several values of
parameters $a$ and $b$.

One gets a value of critical parameter $a_{\rm crt}$ and a critical value of
a horizon radius from the following equations (let us remind to the reader
that $x$ is $r$ (a~radius) in a convenient unit $\sqrt2\,l$,
see Eq.~\eqref{3.11})
\beq{3.13}
\bal
0&=f(x_{\rm crt})=1+a_{\rm crt}\,\frac{g(x_{\rm crt})}{x_{\rm crt}}
+bx_{\rm crt}^2\\
0&=\frac{df}{dx}(x_{\rm crt})=2bx_{\rm crt}+\frac{a_{\rm crt}}{x_{\rm crt}}
\,\frac{dg}{dx}(x_{\rm crt})-\frac{a_{\rm crt}}{x_{\rm crt}^2}\,g(x_{\rm crt}).
\eal
\end{equation}
After some algebra we get
\bg{3.14}
x_{\rm crt}^3(1+bx_{\rm crt}^2)+g(x_{\rm crt})(x_{\rm crt}^4+1)(
3bx_{\rm crt}^2+1)=0\\
a_{\rm crt}=-\frac{(1+bx_{\rm crt}^2)x_{\rm crt}}{g(x_{\rm crt})}
=\frac{(1+x_{\rm crt}^4)}{x_{\rm crt}^2}\,(3bx_{\rm crt}^2+1)\label{3.15}
\end{gather}

\newpage
\def\opis{Plots of the \f\ $f(x)$ (Eq.~\eqref{3.12}) for some values of
parameters $a$ and~$b$\break (see the text for explanation)\break
Fig.~1A: $a=500.00000,\ b=-0.100$; \
Fig.~1B: $a=59.00000,\ b=-1.000$;\break
Fig.~1C: $a=11.50000,\ b=1.000$;\
Fig.~1D: $a=5.00000,\ b=0.000$;\break
Fig.~1E: $a=5.00000,\ b=-0.001$;\
Fig.~1F: $a=4.00000,\ b=0.100$;\break
Fig.~1G: $a=3.17300,\ b=0.000$;\
Fig.~1H: $a=3.00000,\ b=-0.260$.}

\obraz1
\grub0.2pt
\MT   0.000  800.000
\LT1400.000  800.000
\MT 160.000  800.000
\LT 160.000  810.000
\MT 220.000  800.000
\LT 220.000  810.000
\MT 280.000  800.000
\LT 280.000  810.000
\MT 340.000  800.000
\LT 340.000  810.000
\MT 400.000  800.000
\LT 400.000  810.000
\cput(400.000,730.000,5)
\MT 460.000  800.000
\LT 460.000  810.000
\MT 520.000  800.000
\LT 520.000  810.000
\MT 580.000  800.000
\LT 580.000  810.000
\MT 640.000  800.000
\LT 640.000  810.000
\MT 700.000  800.000
\LT 700.000  810.000
\cput(700.000,730.000,10)
\MT 760.000  800.000
\LT 760.000  810.000
\MT 820.000  800.000
\LT 820.000  810.000
\MT 880.000  800.000
\LT 880.000  810.000
\MT 940.000  800.000
\LT 940.000  810.000
\MT1000.000  800.000
\LT1000.000  810.000
\cput(1000.000,730.000,15)
\MT1060.000  800.000
\LT1060.000  810.000
\MT1120.000  800.000
\LT1120.000  810.000
\MT1180.000  800.000
\LT1180.000  810.000
\MT1240.000  800.000
\LT1240.000  810.000
\MT1300.000  800.000
\LT1300.000  810.000
\cput(1300.000,730.000,20)
\MT 100.000    0.000
\LT 100.000  850.000
\MT  92.000   50.000
\LT 108.000   50.000
\MT  92.000  100.000
\LT 108.000  100.000
\MT  92.000  150.000
\LT 108.000  150.000
\MT  92.000  200.000
\LT 108.000  200.000
\MT  92.000  250.000
\LT 108.000  250.000
\MT  92.000  300.000
\LT 108.000  300.000
\MT  92.000  350.000
\LT 108.000  350.000
\MT  92.000  400.000
\LT 108.000  400.000
\MT  92.000  450.000
\LT 108.000  450.000
\MT  92.000  500.000
\LT 108.000  500.000
\MT  92.000  550.000
\LT 108.000  550.000
\MT  92.000  600.000
\LT 108.000  600.000
\MT  92.000  650.000
\LT 108.000  650.000
\MT  92.000  700.000
\LT 108.000  700.000
\MT  92.000  750.000
\LT 108.000  750.000
\MT  92.000  800.000
\LT 108.000  800.000
\MT  84.000   50.000
\LT 116.000   50.000
\lput(80.000,25.000, -150)
\MT  84.000  300.000
\LT 116.000  300.000
\lput(80.000,275.000, -100)
\MT  84.000  550.000
\LT 116.000  550.000
\lput(80.000,525.000,  -50)
\MT  84.000  800.000
\LT 116.000  800.000
\grub0.6pt
\MT  100.000  805.000
\LT  101.000  804.768
\LT  102.000  804.074
\LT  103.000  802.915
\LT  104.000  801.294
\LT  105.000  799.210
\LT  106.000  796.662
\LT  107.000  793.652
\LT  108.000  790.178
\LT  109.000  786.243
\LT  110.000  781.846
\LT  111.000  776.987
\LT  112.000  771.670
\LT  113.000  765.893
\LT  114.000  759.660
\LT  115.000  752.972
\LT  116.000  745.833
\LT  117.000  738.246
\LT  118.000  730.214
\LT  119.000  721.743
\LT  120.000  712.838
\LT  121.000  703.506
\LT  122.000  693.754
\LT  123.000  683.591
\LT  124.000  673.026
\LT  125.000  662.071
\LT  126.000  650.738
\LT  127.000  639.039
\LT  128.000  626.991
\LT  129.000  614.608
\LT  130.000  601.910
\LT  131.000  588.914
\LT  132.000  575.640
\LT  133.000  562.112
\LT  134.000  548.350
\LT  135.000  534.380
\LT  136.000  520.227
\LT  137.000  505.917
\LT  138.000  491.476
\LT  139.000  476.932
\LT  140.000  462.315
\LT  143.000  418.302
\LT  144.000  403.674
\LT  145.000  389.115
\LT  146.000  374.652
\LT  147.000  360.312
\LT  148.000  346.122
\LT  149.000  332.105
\LT  150.000  318.287
\LT  151.000  304.689
\LT  152.000  291.332
\LT  153.000  278.235
\LT  154.000  265.417
\LT  155.000  252.893
\LT  156.000  240.677
\LT  157.000  228.783
\LT  158.000  217.221
\LT  159.000  206.001
\LT  160.000  195.131
\LT  161.000  184.615
\LT  162.000  174.460
\LT  163.000  164.669
\LT  164.000  155.242
\LT  165.000  146.181
\LT  166.000  137.485
\LT  167.000  129.152
\LT  168.000  121.180
\LT  169.000  113.564
\LT  170.000  106.301
\LT  171.000   99.385
\LT  172.000   92.810
\LT  173.000   86.571
\LT  174.000   80.659
\LT  175.000   75.068
\LT  176.000   69.790
\LT  177.000   64.817
\LT  178.000   60.142
\LT  179.000   55.755
\LT  180.000   51.648
\LT  181.000   47.813
\LT  182.000   44.241
\LT  183.000   40.924
\LT  184.000   37.853
\LT  185.000   35.019
\LT  186.000   32.416
\LT  187.000   30.033
\LT  188.000   27.864
\LT  189.000   25.900
\LT  190.000   24.133
\LT  191.000   22.557
\LT  192.000   21.162
\LT  193.000   19.943
\LT  194.000   18.892
\LT  195.000   18.002
\LT  196.000   17.266
\LT  197.000   16.679
\LT  198.000   16.234
\LT  199.000   15.924
\LT  200.000   15.744
\LT  201.000   15.689
\LT  202.000   15.753
\LT  203.000   15.930
\LT  204.000   16.215
\LT  205.000   16.604
\LT  206.000   17.092
\LT  207.000   17.675
\LT  208.000   18.347
\LT  209.000   19.105
\LT  210.000   19.944
\LT  211.000   20.862
\LT  212.000   21.853
\LT  213.000   22.915
\LT  214.000   24.044
\LT  215.000   25.236
\LT  216.000   26.489
\LT  217.000   27.799
\LT  218.000   29.164
\LT  219.000   30.581
\LT  220.000   32.047
\LT  221.000   33.559
\LT  222.000   35.115
\LT  223.000   36.713
\LT  224.000   38.350
\LT  225.000   40.024
\LT  226.000   41.733
\LT  227.000   43.475
\LT  228.000   45.248
\LT  229.000   47.051
\LT  230.000   48.881
\LT  231.000   50.738
\LT  232.000   52.618
\LT  233.000   54.521
\LT  235.000   58.389
\LT  237.000   62.332
\LT  239.000   66.338
\LT  241.000   70.399
\LT  243.000   74.505
\LT  246.000   80.735
\LT  249.000   87.027
\LT  254.000   97.599
\LT  263.000  116.684
\LT  267.000  125.115
\LT  271.000  133.484
\LT  274.000  139.708
\LT  277.000  145.881
\LT  280.000  151.996
\LT  283.000  158.050
\LT  286.000  164.038
\LT  289.000  169.958
\LT  292.000  175.807
\LT  295.000  181.582
\LT  298.000  187.283
\LT  301.000  192.907
\LT  304.000  198.454
\LT  307.000  203.923
\LT  310.000  209.314
\LT  313.000  214.625
\LT  316.000  219.858
\LT  319.000  225.013
\LT  322.000  230.089
\LT  325.000  235.087
\LT  328.000  240.007
\LT  331.000  244.851
\LT  334.000  249.618
\LT  337.000  254.310
\LT  340.000  258.928
\LT  343.000  263.471
\LT  346.000  267.942
\LT  349.000  272.341
\LT  352.000  276.669
\LT  355.000  280.927
\LT  358.000  285.116
\LT  361.000  289.238
\LT  364.000  293.292
\LT  367.000  297.281
\LT  370.000  301.205
\LT  373.000  305.065
\LT  376.000  308.862
\LT  379.000  312.598
\LT  382.000  316.273
\LT  385.000  319.889
\LT  388.000  323.446
\LT  391.000  326.945
\LT  394.000  330.388
\LT  397.000  333.775
\LT  400.000  337.108
\LT  403.000  340.386
\LT  406.000  343.613
\LT  409.000  346.787
\LT  412.000  349.910
\LT  415.000  352.983
\LT  418.000  356.007
\LT  421.000  358.983
\LT  424.000  361.911
\LT  427.000  364.793
\LT  430.000  367.629
\LT  434.000  371.340
\LT  438.000  374.972
\LT  442.000  378.529
\LT  446.000  382.011
\LT  450.000  385.420
\LT  454.000  388.757
\LT  458.000  392.025
\LT  462.000  395.225
\LT  466.000  398.358
\LT  470.000  401.426
\LT  474.000  404.431
\LT  478.000  407.373
\LT  482.000  410.254
\LT  486.000  413.076
\LT  490.000  415.840
\LT  494.000  418.546
\LT  498.000  421.197
\LT  502.000  423.793
\LT  507.000  426.963
\LT  512.000  430.052
\LT  517.000  433.062
\LT  522.000  435.994
\LT  527.000  438.851
\LT  532.000  441.635
\LT  537.000  444.347
\LT  542.000  446.989
\LT  547.000  449.564
\LT  552.000  452.072
\LT  557.000  454.515
\LT  562.000  456.895
\LT  567.000  459.213
\LT  572.000  461.470
\LT  577.000  463.669
\LT  582.000  465.810
\LT  587.000  467.895
\LT  593.000  470.324
\LT  599.000  472.675
\LT  605.000  474.952
\LT  611.000  477.155
\LT  617.000  479.287
\LT  623.000  481.350
\LT  629.000  483.344
\LT  635.000  485.272
\LT  641.000  487.136
\LT  647.000  488.936
\LT  653.000  490.675
\LT  660.000  492.628
\LT  667.000  494.501
\LT  674.000  496.297
\LT  681.000  498.017
\LT  688.000  499.663
\LT  695.000  501.238
\LT  702.000  502.742
\LT  709.000  504.178
\LT  716.000  505.548
\LT  723.000  506.852
\LT  730.000  508.093
\LT  737.000  509.272
\LT  745.000  510.545
\LT  753.000  511.740
\LT  761.000  512.860
\LT  769.000  513.907
\LT  777.000  514.882
\LT  785.000  515.787
\LT  793.000  516.623
\LT  801.000  517.393
\LT  809.000  518.097
\LT  818.000  518.813
\LT  827.000  519.449
\LT  836.000  520.009
\LT  845.000  520.493
\LT  854.000  520.904
\LT  863.000  521.243
\LT  872.000  521.512
\LT  881.000  521.711
\LT  890.000  521.844
\LT  899.000  521.910
\LT  909.000  521.908
\LT  919.000  521.828
\LT  929.000  521.671
\LT  939.000  521.439
\LT  949.000  521.134
\LT  959.000  520.758
\LT  969.000  520.311
\LT  979.000  519.794
\LT  989.000  519.210
\LT 1000.000  518.491
\LT 1011.000  517.693
\LT 1022.000  516.817
\LT 1033.000  515.865
\LT 1044.000  514.839
\LT 1055.000  513.739
\LT 1066.000  512.567
\LT 1077.000  511.325
\LT 1088.000  510.013
\LT 1099.000  508.632
\LT 1111.000  507.049
\LT 1123.000  505.386
\LT 1135.000  503.646
\LT 1147.000  501.830
\LT 1159.000  499.938
\LT 1171.000  497.972
\LT 1183.000  495.933
\LT 1195.000  493.821
\LT 1207.000  491.639
\LT 1219.000  489.386
\LT 1231.000  487.064
\LT 1244.000  484.472
\LT 1257.000  481.800
\LT 1270.000  479.050
\LT 1283.000  476.222
\LT 1296.000  473.319
\LT 1300.000  472.410
\koniec  500.00000  -0.100
\obraz2
\grub0.2pt
\MT   0.000  800.000
\LT1400.000  800.000
\MT 160.000  800.000
\LT 160.000  810.000
\MT 220.000  800.000
\LT 220.000  810.000
\MT 280.000  800.000
\LT 280.000  810.000
\MT 340.000  800.000
\LT 340.000  810.000
\MT 400.000  800.000
\LT 400.000  810.000
\cput(400.000,730.000,5)
\MT 460.000  800.000
\LT 460.000  810.000
\MT 520.000  800.000
\LT 520.000  810.000
\MT 580.000  800.000
\LT 580.000  810.000
\MT 640.000  800.000
\LT 640.000  810.000
\MT 700.000  800.000
\LT 700.000  810.000
\cput(700.000,730.000,10)
\MT 760.000  800.000
\LT 760.000  810.000
\MT 820.000  800.000
\LT 820.000  810.000
\MT 880.000  800.000
\LT 880.000  810.000
\MT 940.000  800.000
\LT 940.000  810.000
\MT1000.000  800.000
\LT1000.000  810.000
\cput(1000.000,730.000,15)
\MT1060.000  800.000
\LT1060.000  810.000
\MT1120.000  800.000
\LT1120.000  810.000
\MT1180.000  800.000
\LT1180.000  810.000
\MT1240.000  800.000
\LT1240.000  810.000
\MT1300.000  800.000
\LT1300.000  810.000
\cput(1300.000,730.000,20)
\MT 100.000    0.000
\LT 100.000  850.000
\MT  92.000   50.000
\LT 108.000   50.000
\MT  92.000   87.500
\LT 108.000   87.500
\MT  92.000  125.000
\LT 108.000  125.000
\MT  92.000  162.500
\LT 108.000  162.500
\MT  92.000  200.000
\LT 108.000  200.000
\MT  92.000  237.500
\LT 108.000  237.500
\MT  92.000  275.000
\LT 108.000  275.000
\MT  92.000  312.500
\LT 108.000  312.500
\MT  92.000  350.000
\LT 108.000  350.000
\MT  92.000  387.500
\LT 108.000  387.500
\MT  92.000  425.000
\LT 108.000  425.000
\MT  92.000  462.500
\LT 108.000  462.500
\MT  92.000  500.000
\LT 108.000  500.000
\MT  92.000  537.500
\LT 108.000  537.500
\MT  92.000  575.000
\LT 108.000  575.000
\MT  92.000  612.500
\LT 108.000  612.500
\MT  92.000  650.000
\LT 108.000  650.000
\MT  92.000  687.500
\LT 108.000  687.500
\MT  92.000  725.000
\LT 108.000  725.000
\MT  92.000  762.500
\LT 108.000  762.500
\MT  92.000  800.000
\LT 108.000  800.000
\MT  84.000   50.000
\LT 116.000   50.000
\lput(80.000,31.250, -400)
\MT  84.000  237.500
\LT 116.000  237.500
\lput(80.000,218.750, -300)
\MT  84.000  425.000
\LT 116.000  425.000
\lput(80.000,406.250, -200)
\MT  84.000  612.500
\LT 116.000  612.500
\lput(80.000,593.750, -100)
\MT  84.000  800.000
\LT 116.000  800.000
\grub0.6pt
\MT  100.000  801.875
\LT  101.000  801.864
\LT  102.000  801.832
\LT  103.000  801.778
\LT  104.000  801.703
\LT  105.000  801.606
\LT  106.000  801.488
\LT  107.000  801.348
\LT  108.000  801.186
\LT  109.000  801.003
\LT  110.000  800.799
\LT  111.000  800.573
\LT  112.000  800.326
\LT  113.000  800.058
\LT  114.000  799.768
\LT  115.000  799.457
\LT  116.000  799.125
\LT  117.000  798.772
\LT  118.000  798.399
\LT  119.000  798.005
\LT  121.000  797.157
\LT  123.000  796.230
\LT  125.000  795.229
\LT  127.000  794.156
\LT  129.000  793.017
\LT  131.000  791.819
\LT  133.000  790.567
\LT  136.000  788.607
\LT  139.000  786.575
\LT  147.000  781.061
\LT  150.000  779.051
\LT  152.000  777.753
\LT  154.000  776.498
\LT  156.000  775.290
\LT  158.000  774.134
\LT  160.000  773.035
\LT  162.000  771.995
\LT  164.000  771.015
\LT  166.000  770.095
\LT  168.000  769.236
\LT  170.000  768.436
\LT  172.000  767.692
\LT  174.000  767.004
\LT  176.000  766.369
\LT  178.000  765.784
\LT  180.000  765.245
\LT  182.000  764.751
\LT  185.000  764.085
\LT  188.000  763.501
\LT  191.000  762.990
\LT  194.000  762.542
\LT  197.000  762.149
\LT  201.000  761.698
\LT  205.000  761.314
\LT  210.000  760.909
\LT  216.000  760.500
\LT  225.000  759.983
\LT  244.000  758.970
\LT  253.000  758.430
\LT  261.000  757.888
\LT  269.000  757.276
\LT  276.000  756.675
\LT  283.000  756.011
\LT  290.000  755.281
\LT  297.000  754.483
\LT  304.000  753.616
\LT  311.000  752.680
\LT  318.000  751.675
\LT  325.000  750.600
\LT  332.000  749.457
\LT  339.000  748.244
\LT  346.000  746.963
\LT  353.000  745.615
\LT  360.000  744.199
\LT  367.000  742.717
\LT  374.000  741.169
\LT  381.000  739.555
\LT  388.000  737.876
\LT  395.000  736.133
\LT  402.000  734.326
\LT  409.000  732.456
\LT  416.000  730.523
\LT  423.000  728.529
\LT  430.000  726.472
\LT  438.000  724.046
\LT  446.000  721.541
\LT  454.000  718.958
\LT  462.000  716.296
\LT  470.000  713.556
\LT  478.000  710.739
\LT  486.000  707.846
\LT  494.000  704.876
\LT  502.000  701.831
\LT  510.000  698.710
\LT  518.000  695.515
\LT  526.000  692.244
\LT  534.000  688.900
\LT  542.000  685.482
\LT  550.000  681.990
\LT  558.000  678.424
\LT  566.000  674.786
\LT  574.000  671.075
\LT  582.000  667.292
\LT  590.000  663.436
\LT  598.000  659.508
\LT  606.000  655.508
\LT  614.000  651.437
\LT  622.000  647.294
\LT  630.000  643.080
\LT  638.000  638.795
\LT  646.000  634.440
\LT  654.000  630.013
\LT  662.000  625.516
\LT  670.000  620.948
\LT  678.000  616.310
\LT  686.000  611.602
\LT  694.000  606.823
\LT  702.000  601.975
\LT  710.000  597.057
\LT  718.000  592.070
\LT  726.000  587.012
\LT  734.000  581.885
\LT  742.000  576.689
\LT  750.000  571.423
\LT  758.000  566.088
\LT  766.000  560.684
\LT  774.000  555.211
\LT  782.000  549.669
\LT  790.000  544.058
\LT  798.000  538.378
\LT  806.000  532.629
\LT  814.000  526.812
\LT  822.000  520.926
\LT  830.000  514.971
\LT  838.000  508.948
\LT  846.000  502.856
\LT  854.000  496.696
\LT  862.000  490.467
\LT  870.000  484.170
\LT  878.000  477.805
\LT  886.000  471.371
\LT  894.000  464.869
\LT  902.000  458.300
\LT  910.000  451.662
\LT  918.000  444.955
\LT  926.000  438.181
\LT  934.000  431.339
\LT  942.000  424.429
\LT  950.000  417.451
\LT  958.000  410.405
\LT  966.000  403.291
\LT  974.000  396.109
\LT  982.000  388.859
\LT  990.000  381.542
\LT  998.000  374.157
\LT 1006.000  366.704
\LT 1014.000  359.184
\LT 1022.000  351.595
\LT 1030.000  343.939
\LT 1038.000  336.216
\LT 1046.000  328.425
\LT 1054.000  320.566
\LT 1062.000  312.640
\LT 1070.000  304.646
\LT 1078.000  296.584
\LT 1086.000  288.455
\LT 1094.000  280.259
\LT 1102.000  271.995
\LT 1110.000  263.664
\LT 1118.000  255.265
\LT 1126.000  246.799
\LT 1134.000  238.265
\LT 1142.000  229.664
\LT 1150.000  220.996
\LT 1158.000  212.260
\LT 1166.000  203.457
\LT 1174.000  194.587
\LT 1182.000  185.649
\LT 1190.000  176.644
\LT 1198.000  167.572
\LT 1206.000  158.433
\LT 1214.000  149.226
\LT 1222.000  139.952
\LT 1230.000  130.611
\LT 1238.000  121.202
\LT 1246.000  111.726
\LT 1254.000  102.183
\LT 1262.000   92.573
\LT 1270.000   82.896
\LT 1278.000   73.151
\LT 1286.000   63.340
\LT 1294.000   53.461
\LT 1300.000   46.008
\koniec  59.00000  -1.000 
\obraz3
\grub0.2pt
\MT   0.000   60.000
\LT1400.000   60.000
\MT 160.000   60.000
\LT 160.000   70.000
\MT 220.000   60.000
\LT 220.000   70.000
\MT 280.000   60.000
\LT 280.000   70.000
\MT 340.000   60.000
\LT 340.000   70.000
\cput(340.000,-10.000,2)
\MT 400.000   60.000
\LT 400.000   70.000
\MT 460.000   60.000
\LT 460.000   70.000
\MT 520.000   60.000
\LT 520.000   70.000
\MT 580.000   60.000
\LT 580.000   70.000
\cput(580.000,-10.000,4)
\MT 640.000   60.000
\LT 640.000   70.000
\MT 700.000   60.000
\LT 700.000   70.000
\MT 760.000   60.000
\LT 760.000   70.000
\MT 820.000   60.000
\LT 820.000   70.000
\cput(820.000,-10.000,6)
\MT 880.000   60.000
\LT 880.000   70.000
\MT 940.000   60.000
\LT 940.000   70.000
\MT1000.000   60.000
\LT1000.000   70.000
\MT1060.000   60.000
\LT1060.000   70.000
\cput(1060.000,-10.000,8)
\MT1120.000   60.000
\LT1120.000   70.000
\MT1180.000   60.000
\LT1180.000   70.000
\MT1240.000   60.000
\LT1240.000   70.000
\MT1300.000   60.000
\LT1300.000   70.000
\cput(1300.000,-10.000,10)
\MT 100.000    0.000
\LT 100.000  850.000
\MT  92.000   60.000
\LT 108.000   60.000
\MT  92.000   97.000
\LT 108.000   97.000
\MT  92.000  134.000
\LT 108.000  134.000
\MT  92.000  171.000
\LT 108.000  171.000
\MT  92.000  208.000
\LT 108.000  208.000
\MT  92.000  245.000
\LT 108.000  245.000
\MT  92.000  282.000
\LT 108.000  282.000
\MT  92.000  319.000
\LT 108.000  319.000
\MT  92.000  356.000
\LT 108.000  356.000
\MT  92.000  393.000
\LT 108.000  393.000
\MT  92.000  430.000
\LT 108.000  430.000
\MT  92.000  467.000
\LT 108.000  467.000
\MT  92.000  504.000
\LT 108.000  504.000
\MT  92.000  541.000
\LT 108.000  541.000
\MT  92.000  578.000
\LT 108.000  578.000
\MT  92.000  615.000
\LT 108.000  615.000
\MT  92.000  652.000
\LT 108.000  652.000
\MT  92.000  689.000
\LT 108.000  689.000
\MT  92.000  726.000
\LT 108.000  726.000
\MT  92.000  763.000
\LT 108.000  763.000
\MT  92.000  800.000
\LT 108.000  800.000
\MT  84.000   60.000
\LT 116.000   60.000
\MT  84.000  208.000
\LT 116.000  208.000
\lput(80.000,189.500,   20)
\MT  84.000  356.000
\LT 116.000  356.000
\lput(80.000,337.500,   40)
\MT  84.000  504.000
\LT 116.000  504.000
\lput(80.000,485.500,   60)
\MT  84.000  652.000
\LT 116.000  652.000
\lput(80.000,633.500,   80)
\MT  84.000  800.000
\LT 116.000  800.000
\lput(80.000,781.500,  100)
\grub0.6pt
\MT  100.000   67.400
\LT  105.000   67.364
\LT  110.000   67.254
\LT  115.000   67.072
\LT  120.000   66.818
\LT  125.000   66.491
\LT  130.000   66.093
\LT  135.000   65.624
\LT  140.000   65.087
\LT  145.000   64.485
\LT  150.000   63.822
\LT  156.000   62.956
\LT  163.000   61.864
\LT  173.000   60.208
\LT  185.000   58.208
\LT  191.000   57.262
\LT  196.000   56.527
\LT  201.000   55.854
\LT  206.000   55.256
\LT  210.000   54.839
\LT  214.000   54.479
\LT  218.000   54.182
\LT  222.000   53.948
\LT  226.000   53.780
\LT  230.000   53.678
\LT  234.000   53.643
\LT  238.000   53.673
\LT  242.000   53.767
\LT  246.000   53.924
\LT  250.000   54.142
\LT  254.000   54.420
\LT  258.000   54.754
\LT  263.000   55.248
\LT  268.000   55.824
\LT  273.000   56.475
\LT  278.000   57.199
\LT  283.000   57.990
\LT  288.000   58.846
\LT  293.000   59.761
\LT  299.000   60.934
\LT  305.000   62.182
\LT  311.000   63.502
\LT  317.000   64.889
\LT  323.000   66.338
\LT  330.000   68.104
\LT  337.000   69.945
\LT  344.000   71.858
\LT  351.000   73.839
\LT  358.000   75.885
\LT  366.000   78.298
\LT  374.000   80.789
\LT  382.000   83.355
\LT  390.000   85.993
\LT  398.000   88.701
\LT  406.000   91.477
\LT  414.000   94.321
\LT  422.000   97.230
\LT  430.000  100.205
\LT  439.000  103.628
\LT  448.000  107.131
\LT  457.000  110.713
\LT  466.000  114.375
\LT  475.000  118.115
\LT  484.000  121.933
\LT  493.000  125.828
\LT  502.000  129.802
\LT  511.000  133.853
\LT  520.000  137.981
\LT  529.000  142.187
\LT  538.000  146.470
\LT  547.000  150.831
\LT  556.000  155.269
\LT  565.000  159.785
\LT  574.000  164.379
\LT  583.000  169.050
\LT  592.000  173.799
\LT  601.000  178.626
\LT  610.000  183.530
\LT  619.000  188.514
\LT  628.000  193.575
\LT  637.000  198.715
\LT  646.000  203.933
\LT  655.000  209.230
\LT  664.000  214.606
\LT  673.000  220.060
\LT  682.000  225.594
\LT  691.000  231.207
\LT  700.000  236.898
\LT  709.000  242.670
\LT  718.000  248.520
\LT  727.000  254.451
\LT  736.000  260.460
\LT  745.000  266.550
\LT  754.000  272.719
\LT  763.000  278.969
\LT  772.000  285.298
\LT  781.000  291.708
\LT  790.000  298.197
\LT  799.000  304.767
\LT  808.000  311.418
\LT  817.000  318.148
\LT  826.000  324.960
\LT  835.000  331.851
\LT  844.000  338.824
\LT  853.000  345.877
\LT  862.000  353.011
\LT  871.000  360.226
\LT  880.000  367.522
\LT  888.000  374.075
\LT  896.000  380.692
\LT  904.000  387.374
\LT  912.000  394.119
\LT  920.000  400.929
\LT  928.000  407.802
\LT  936.000  414.740
\LT  944.000  421.743
\LT  952.000  428.809
\LT  960.000  435.940
\LT  968.000  443.135
\LT  976.000  450.395
\LT  984.000  457.719
\LT  992.000  465.107
\LT 1000.000  472.560
\LT 1008.000  480.077
\LT 1016.000  487.659
\LT 1024.000  495.306
\LT 1032.000  503.017
\LT 1040.000  510.792
\LT 1048.000  518.633
\LT 1056.000  526.538
\LT 1064.000  534.507
\LT 1072.000  542.542
\LT 1080.000  550.641
\LT 1088.000  558.804
\LT 1096.000  567.033
\LT 1104.000  575.326
\LT 1112.000  583.685
\LT 1120.000  592.108
\LT 1128.000  600.595
\LT 1136.000  609.148
\LT 1144.000  617.766
\LT 1152.000  626.448
\LT 1160.000  635.196
\LT 1168.000  644.008
\LT 1176.000  652.885
\LT 1184.000  661.827
\LT 1192.000  670.835
\LT 1200.000  679.907
\LT 1208.000  689.044
\LT 1216.000  698.246
\LT 1224.000  707.513
\LT 1232.000  716.846
\LT 1240.000  726.243
\LT 1248.000  735.706
\LT 1256.000  745.233
\LT 1264.000  754.826
\LT 1272.000  764.484
\LT 1280.000  774.207
\LT 1288.000  783.995
\LT 1296.000  793.848
\LT 1300.000  798.799
\koniec   11.50000   1.000
\obraz4
\grub0.2pt
\MT   0.000  300.000
\LT1400.000  300.000
\MT 140.000  300.000
\LT 140.000  310.000
\MT 180.000  300.000
\LT 180.000  310.000
\MT 220.000  300.000
\LT 220.000  310.000
\MT 260.000  300.000
\LT 260.000  310.000
\MT 300.000  300.000
\LT 300.000  310.000
\cput(300.000,230.000,5)
\MT 340.000  300.000
\LT 340.000  310.000
\MT 380.000  300.000
\LT 380.000  310.000
\MT 420.000  300.000
\LT 420.000  310.000
\MT 460.000  300.000
\LT 460.000  310.000
\MT 500.000  300.000
\LT 500.000  310.000
\cput(500.000,230.000,10)
\MT 540.000  300.000
\LT 540.000  310.000
\MT 580.000  300.000
\LT 580.000  310.000
\MT 620.000  300.000
\LT 620.000  310.000
\MT 660.000  300.000
\LT 660.000  310.000
\MT 700.000  300.000
\LT 700.000  310.000
\cput(700.000,230.000,15)
\MT 740.000  300.000
\LT 740.000  310.000
\MT 780.000  300.000
\LT 780.000  310.000
\MT 820.000  300.000
\LT 820.000  310.000
\MT 860.000  300.000
\LT 860.000  310.000
\MT 900.000  300.000
\LT 900.000  310.000
\cput(900.000,230.000,20)
\MT 940.000  300.000
\LT 940.000  310.000
\MT 980.000  300.000
\LT 980.000  310.000
\MT1020.000  300.000
\LT1020.000  310.000
\MT1060.000  300.000
\LT1060.000  310.000
\MT1100.000  300.000
\LT1100.000  310.000
\cput(1100.000,230.000,25)
\MT1140.000  300.000
\LT1140.000  310.000
\MT1180.000  300.000
\LT1180.000  310.000
\MT1220.000  300.000
\LT1220.000  310.000
\MT1260.000  300.000
\LT1260.000  310.000
\MT1300.000  300.000
\LT1300.000  310.000
\cput(1300.000,230.000,30)
\MT 100.000    0.000
\LT 100.000  850.000
\MT  92.000   50.000
\LT 108.000   50.000
\MT  92.000  100.000
\LT 108.000  100.000
\MT  92.000  150.000
\LT 108.000  150.000
\MT  92.000  200.000
\LT 108.000  200.000
\MT  92.000  250.000
\LT 108.000  250.000
\MT  92.000  300.000
\LT 108.000  300.000
\MT  92.000  350.000
\LT 108.000  350.000
\MT  92.000  400.000
\LT 108.000  400.000
\MT  92.000  450.000
\LT 108.000  450.000
\MT  92.000  500.000
\LT 108.000  500.000
\MT  92.000  550.000
\LT 108.000  550.000
\MT  92.000  600.000
\LT 108.000  600.000
\MT  92.000  650.000
\LT 108.000  650.000
\MT  92.000  700.000
\LT 108.000  700.000
\MT  92.000  750.000
\LT 108.000  750.000
\MT  92.000  800.000
\LT 108.000  800.000
\MT  84.000   50.000
\LT 116.000   50.000
\lput(80.000,25.000, -0.5)
\MT  84.000  300.000
\LT 116.000  300.000
\MT  84.000  550.000
\LT 116.000  550.000
\lput(80.000,525.000,  0.5)
\MT  84.000  800.000
\LT 116.000  800.000
\lput(80.000,775.000,  1.0)
\grub0.6pt
\MT  100.000  800.000
\LT  101.000  799.479
\LT  102.000  797.917
\LT  103.000  795.313
\LT  104.000  791.667
\LT  105.000  786.981
\LT  106.000  781.254
\LT  107.000  774.489
\LT  108.000  766.690
\LT  109.000  757.859
\LT  110.000  748.004
\LT  111.000  737.133
\LT  112.000  725.259
\LT  113.000  712.397
\LT  114.000  698.567
\LT  115.000  683.793
\LT  116.000  668.106
\LT  117.000  651.541
\LT  118.000  634.140
\LT  119.000  615.953
\LT  120.000  597.035
\LT  121.000  577.448
\LT  122.000  557.263
\LT  123.000  536.555
\LT  124.000  515.407
\LT  125.000  493.906
\LT  126.000  472.144
\LT  127.000  450.215
\LT  128.000  428.216
\LT  129.000  406.244
\LT  130.000  384.396
\LT  131.000  362.765
\LT  132.000  341.442
\LT  133.000  320.510
\LT  134.000  300.050
\LT  135.000  280.133
\LT  136.000  260.822
\LT  137.000  242.173
\LT  138.000  224.234
\LT  139.000  207.043
\LT  140.000  190.631
\LT  141.000  175.018
\LT  142.000  160.220
\LT  143.000  146.243
\LT  144.000  133.090
\LT  145.000  120.754
\LT  146.000  109.225
\LT  147.000   98.490
\LT  148.000   88.530
\LT  149.000   79.325
\LT  150.000   70.849
\LT  151.000   63.079
\LT  152.000   55.987
\LT  153.000   49.544
\LT  154.000   43.724
\LT  155.000   38.496
\LT  156.000   33.833
\LT  157.000   29.705
\LT  158.000   26.085
\LT  159.000   22.945
\LT  160.000   20.258
\LT  161.000   18.000
\LT  162.000   16.144
\LT  163.000   14.667
\LT  164.000   13.546
\LT  165.000   12.759
\LT  166.000   12.285
\LT  167.000   12.104
\LT  168.000   12.198
\LT  169.000   12.547
\LT  170.000   13.136
\LT  171.000   13.947
\LT  172.000   14.967
\LT  173.000   16.180
\LT  174.000   17.573
\LT  175.000   19.133
\LT  176.000   20.849
\LT  177.000   22.708
\LT  178.000   24.701
\LT  179.000   26.817
\LT  180.000   29.047
\LT  181.000   31.382
\LT  182.000   33.814
\LT  183.000   36.335
\LT  184.000   38.938
\LT  185.000   41.616
\LT  186.000   44.362
\LT  187.000   47.172
\LT  188.000   50.038
\LT  189.000   52.956
\LT  190.000   55.921
\LT  191.000   58.928
\LT  192.000   61.973
\LT  193.000   65.051
\LT  194.000   68.160
\LT  195.000   71.295
\LT  196.000   74.454
\LT  198.000   80.828
\LT  200.000   87.260
\LT  203.000   96.975
\LT  209.000  116.453
\LT  212.000  126.134
\LT  214.000  132.545
\LT  216.000  138.913
\LT  218.000  145.232
\LT  220.000  151.496
\LT  222.000  157.701
\LT  224.000  163.843
\LT  226.000  169.919
\LT  228.000  175.927
\LT  230.000  181.864
\LT  232.000  187.728
\LT  234.000  193.518
\LT  236.000  199.234
\LT  238.000  204.874
\LT  240.000  210.439
\LT  242.000  215.927
\LT  244.000  221.338
\LT  246.000  226.674
\LT  248.000  231.934
\LT  250.000  237.118
\LT  252.000  242.227
\LT  254.000  247.262
\LT  256.000  252.223
\LT  258.000  257.112
\LT  260.000  261.928
\LT  262.000  266.673
\LT  264.000  271.347
\LT  266.000  275.952
\LT  268.000  280.489
\LT  270.000  284.958
\LT  272.000  289.361
\LT  274.000  293.699
\LT  276.000  297.972
\LT  278.000  302.182
\LT  280.000  306.330
\LT  282.000  310.416
\LT  284.000  314.442
\LT  286.000  318.409
\LT  288.000  322.318
\LT  290.000  326.170
\LT  292.000  329.966
\LT  294.000  333.706
\LT  296.000  337.393
\LT  298.000  341.026
\LT  300.000  344.608
\LT  302.000  348.138
\LT  304.000  351.618
\LT  306.000  355.048
\LT  308.000  358.430
\LT  310.000  361.764
\LT  312.000  365.052
\LT  314.000  368.294
\LT  316.000  371.491
\LT  318.000  374.644
\LT  320.000  377.754
\LT  322.000  380.820
\LT  324.000  383.846
\LT  326.000  386.830
\LT  329.000  391.231
\LT  332.000  395.544
\LT  335.000  399.772
\LT  338.000  403.916
\LT  341.000  407.979
\LT  344.000  411.963
\LT  347.000  415.871
\LT  350.000  419.703
\LT  353.000  423.463
\LT  356.000  427.153
\LT  359.000  430.773
\LT  362.000  434.326
\LT  365.000  437.814
\LT  368.000  441.238
\LT  371.000  444.600
\LT  374.000  447.902
\LT  377.000  451.145
\LT  380.000  454.330
\LT  383.000  457.460
\LT  386.000  460.535
\LT  389.000  463.556
\LT  392.000  466.526
\LT  395.000  469.445
\LT  398.000  472.315
\LT  401.000  475.137
\LT  404.000  477.911
\LT  407.000  480.640
\LT  411.000  484.209
\LT  415.000  487.701
\LT  419.000  491.118
\LT  423.000  494.462
\LT  427.000  497.736
\LT  431.000  500.943
\LT  435.000  504.083
\LT  439.000  507.159
\LT  443.000  510.173
\LT  447.000  513.127
\LT  451.000  516.022
\LT  455.000  518.860
\LT  459.000  521.642
\LT  464.000  525.046
\LT  469.000  528.368
\LT  474.000  531.612
\LT  479.000  534.780
\LT  484.000  537.876
\LT  489.000  540.901
\LT  494.000  543.858
\LT  499.000  546.749
\LT  504.000  549.576
\LT  509.000  552.342
\LT  514.000  555.047
\LT  519.000  557.695
\LT  525.000  560.799
\LT  531.000  563.825
\LT  537.000  566.776
\LT  543.000  569.655
\LT  549.000  572.464
\LT  555.000  575.207
\LT  561.000  577.884
\LT  567.000  580.499
\LT  574.000  583.474
\LT  581.000  586.370
\LT  588.000  589.190
\LT  595.000  591.937
\LT  602.000  594.614
\LT  609.000  597.223
\LT  616.000  599.767
\LT  623.000  602.249
\LT  631.000  605.011
\LT  639.000  607.698
\LT  647.000  610.312
\LT  655.000  612.856
\LT  663.000  615.333
\LT  672.000  618.044
\LT  681.000  620.676
\LT  690.000  623.233
\LT  699.000  625.719
\LT  708.000  628.136
\LT  718.000  630.745
\LT  728.000  633.276
\LT  738.000  635.733
\LT  748.000  638.118
\LT  759.000  640.664
\LT  770.000  643.131
\LT  781.000  645.524
\LT  792.000  647.844
\LT  804.000  650.298
\LT  816.000  652.674
\LT  828.000  654.976
\LT  840.000  657.207
\LT  853.000  659.548
\LT  866.000  661.814
\LT  879.000  664.009
\LT  893.000  666.295
\LT  907.000  668.506
\LT  922.000  670.796
\LT  937.000  673.007
\LT  952.000  675.144
\LT  968.000  677.346
\LT  984.000  679.472
\LT 1001.000  681.651
\LT 1018.000  683.753
\LT 1036.000  685.899
\LT 1054.000  687.967
\LT 1073.000  690.071
\LT 1092.000  692.097
\LT 1112.000  694.151
\LT 1132.000  696.128
\LT 1153.000  698.126
\LT 1174.000  700.049
\LT 1196.000  701.987
\LT 1219.000  703.934
\LT 1242.000  705.806
\LT 1266.000  707.683
\LT 1291.000  709.560
\LT 1300.000  710.218
\koniec    5.00000   0.000
\obraz5
\grub0.2pt
\MT   0.000  300.000
\LT1400.000  300.000
\MT 140.000  300.000
\LT 140.000  310.000
\MT 180.000  300.000
\LT 180.000  310.000
\MT 220.000  300.000
\LT 220.000  310.000
\MT 260.000  300.000
\LT 260.000  310.000
\MT 300.000  300.000
\LT 300.000  310.000
\cput(300.000,230.000,5)
\MT 340.000  300.000
\LT 340.000  310.000
\MT 380.000  300.000
\LT 380.000  310.000
\MT 420.000  300.000
\LT 420.000  310.000
\MT 460.000  300.000
\LT 460.000  310.000
\MT 500.000  300.000
\LT 500.000  310.000
\cput(500.000,230.000,10)
\MT 540.000  300.000
\LT 540.000  310.000
\MT 580.000  300.000
\LT 580.000  310.000
\MT 620.000  300.000
\LT 620.000  310.000
\MT 660.000  300.000
\LT 660.000  310.000
\MT 700.000  300.000
\LT 700.000  310.000
\cput(700.000,230.000,15)
\MT 740.000  300.000
\LT 740.000  310.000
\MT 780.000  300.000
\LT 780.000  310.000
\MT 820.000  300.000
\LT 820.000  310.000
\MT 860.000  300.000
\LT 860.000  310.000
\MT 900.000  300.000
\LT 900.000  310.000
\cput(900.000,230.000,20)
\MT 940.000  300.000
\LT 940.000  310.000
\MT 980.000  300.000
\LT 980.000  310.000
\MT1020.000  300.000
\LT1020.000  310.000
\MT1060.000  300.000
\LT1060.000  310.000
\MT1100.000  300.000
\LT1100.000  310.000
\cput(1100.000,230.000,25)
\MT1140.000  300.000
\LT1140.000  310.000
\MT1180.000  300.000
\LT1180.000  310.000
\MT1220.000  300.000
\LT1220.000  310.000
\MT1260.000  300.000
\LT1260.000  310.000
\MT1300.000  300.000
\LT1300.000  310.000
\cput(1300.000,230.000,30)
\MT 100.000    0.000
\LT 100.000  850.000
\MT  92.000   50.000
\LT 108.000   50.000
\MT  92.000  100.000
\LT 108.000  100.000
\MT  92.000  150.000
\LT 108.000  150.000
\MT  92.000  200.000
\LT 108.000  200.000
\MT  92.000  250.000
\LT 108.000  250.000
\MT  92.000  300.000
\LT 108.000  300.000
\MT  92.000  350.000
\LT 108.000  350.000
\MT  92.000  400.000
\LT 108.000  400.000
\MT  92.000  450.000
\LT 108.000  450.000
\MT  92.000  500.000
\LT 108.000  500.000
\MT  92.000  550.000
\LT 108.000  550.000
\MT  92.000  600.000
\LT 108.000  600.000
\MT  92.000  650.000
\LT 108.000  650.000
\MT  92.000  700.000
\LT 108.000  700.000
\MT  92.000  750.000
\LT 108.000  750.000
\MT  92.000  800.000
\LT 108.000  800.000
\MT  84.000   50.000
\LT 116.000   50.000
\lput(80.000,25.000, -0.5)
\MT  84.000  300.000
\LT 116.000  300.000
\MT  84.000  550.000
\LT 116.000  550.000
\lput(80.000,525.000,  0.5)
\MT  84.000  800.000
\LT 116.000  800.000
\lput(80.000,775.000,  1.0)
\grub0.6pt
\MT  100.000  800.000
\LT  101.000  799.479
\LT  102.000  797.915
\LT  103.000  795.310
\LT  104.000  791.662
\LT  105.000  786.973
\LT  106.000  781.243
\LT  107.000  774.474
\LT  108.000  766.670
\LT  109.000  757.833
\LT  110.000  747.972
\LT  111.000  737.095
\LT  112.000  725.214
\LT  113.000  712.344
\LT  114.000  698.506
\LT  115.000  683.723
\LT  116.000  668.026
\LT  117.000  651.451
\LT  118.000  634.039
\LT  119.000  615.840
\LT  120.000  596.910
\LT  121.000  577.310
\LT  122.000  557.112
\LT  123.000  536.390
\LT  124.000  515.227
\LT  125.000  493.711
\LT  126.000  471.932
\LT  127.000  449.987
\LT  128.000  427.971
\LT  129.000  405.981
\LT  130.000  384.115
\LT  131.000  362.465
\LT  132.000  341.122
\LT  133.000  320.170
\LT  134.000  299.689
\LT  135.000  279.750
\LT  136.000  260.417
\LT  137.000  241.746
\LT  138.000  223.783
\LT  139.000  206.568
\LT  140.000  190.131
\LT  141.000  174.493
\LT  142.000  159.669
\LT  143.000  145.666
\LT  144.000  132.485
\LT  145.000  120.121
\LT  146.000  108.564
\LT  147.000   97.800
\LT  148.000   87.810
\LT  149.000   78.574
\LT  150.000   70.068
\LT  151.000   62.266
\LT  152.000   55.142
\LT  153.000   48.667
\LT  154.000   42.813
\LT  155.000   37.551
\LT  156.000   32.853
\LT  157.000   28.689
\LT  158.000   25.033
\LT  159.000   21.857
\LT  160.000   19.133
\LT  161.000   16.837
\LT  162.000   14.943
\LT  163.000   13.427
\LT  164.000   12.266
\LT  165.000   11.439
\LT  166.000   10.924
\LT  167.000   10.702
\LT  168.000   10.753
\LT  169.000   11.059
\LT  170.000   11.604
\LT  171.000   12.372
\LT  172.000   13.347
\LT  173.000   14.515
\LT  174.000   15.862
\LT  175.000   17.375
\LT  176.000   19.044
\LT  177.000   20.855
\LT  178.000   22.799
\LT  179.000   24.866
\LT  180.000   27.047
\LT  181.000   29.331
\LT  182.000   31.713
\LT  183.000   34.182
\LT  184.000   36.733
\LT  185.000   39.358
\LT  186.000   42.051
\LT  187.000   44.806
\LT  188.000   47.618
\LT  189.000   50.481
\LT  190.000   53.389
\LT  191.000   56.340
\LT  192.000   59.328
\LT  193.000   62.348
\LT  194.000   65.399
\LT  195.000   68.475
\LT  197.000   74.692
\LT  199.000   80.975
\LT  202.000   90.479
\LT  208.000  109.568
\LT  211.000  119.065
\LT  213.000  125.354
\LT  215.000  131.602
\LT  217.000  137.801
\LT  219.000  143.946
\LT  221.000  150.031
\LT  223.000  156.052
\LT  225.000  162.007
\LT  227.000  167.891
\LT  229.000  173.704
\LT  231.000  179.442
\LT  233.000  185.105
\LT  235.000  190.690
\LT  237.000  196.198
\LT  239.000  201.628
\LT  241.000  206.979
\LT  243.000  212.252
\LT  245.000  217.445
\LT  247.000  222.561
\LT  249.000  227.597
\LT  251.000  232.557
\LT  253.000  237.439
\LT  255.000  242.244
\LT  257.000  246.974
\LT  259.000  251.628
\LT  261.000  256.209
\LT  263.000  260.716
\LT  265.000  265.151
\LT  267.000  269.514
\LT  269.000  273.807
\LT  271.000  278.030
\LT  273.000  282.185
\LT  275.000  286.273
\LT  277.000  290.295
\LT  279.000  294.251
\LT  281.000  298.143
\LT  283.000  301.971
\LT  285.000  305.738
\LT  287.000  309.443
\LT  289.000  313.088
\LT  291.000  316.674
\LT  293.000  320.203
\LT  295.000  323.674
\LT  297.000  327.088
\LT  299.000  330.448
\LT  301.000  333.754
\LT  303.000  337.006
\LT  305.000  340.206
\LT  307.000  343.355
\LT  309.000  346.453
\LT  311.000  349.501
\LT  313.000  352.501
\LT  315.000  355.453
\LT  317.000  358.358
\LT  319.000  361.216
\LT  321.000  364.029
\LT  323.000  366.798
\LT  325.000  369.523
\LT  327.000  372.204
\LT  329.000  374.843
\LT  331.000  377.441
\LT  333.000  379.997
\LT  336.000  383.757
\LT  339.000  387.429
\LT  342.000  391.014
\LT  345.000  394.516
\LT  348.000  397.936
\LT  351.000  401.277
\LT  354.000  404.540
\LT  357.000  407.727
\LT  360.000  410.840
\LT  363.000  413.881
\LT  366.000  416.851
\LT  369.000  419.753
\LT  372.000  422.587
\LT  375.000  425.357
\LT  378.000  428.062
\LT  381.000  430.704
\LT  384.000  433.286
\LT  387.000  435.807
\LT  390.000  438.271
\LT  393.000  440.677
\LT  396.000  443.027
\LT  399.000  445.323
\LT  402.000  447.565
\LT  405.000  449.756
\LT  408.000  451.895
\LT  411.000  453.984
\LT  414.000  456.023
\LT  417.000  458.015
\LT  420.000  459.960
\LT  423.000  461.859
\LT  427.000  464.321
\LT  431.000  466.705
\LT  435.000  469.012
\LT  439.000  471.246
\LT  443.000  473.408
\LT  447.000  475.499
\LT  451.000  477.521
\LT  455.000  479.477
\LT  459.000  481.367
\LT  463.000  483.194
\LT  467.000  484.958
\LT  471.000  486.662
\LT  475.000  488.306
\LT  479.000  489.893
\LT  483.000  491.422
\LT  487.000  492.896
\LT  491.000  494.316
\LT  496.000  496.017
\LT  501.000  497.637
\LT  506.000  499.178
\LT  511.000  500.643
\LT  516.000  502.033
\LT  521.000  503.351
\LT  526.000  504.597
\LT  531.000  505.775
\LT  536.000  506.884
\LT  541.000  507.928
\LT  546.000  508.907
\LT  551.000  509.823
\LT  556.000  510.677
\LT  561.000  511.471
\LT  566.000  512.206
\LT  571.000  512.884
\LT  576.000  513.504
\LT  582.000  514.176
\LT  588.000  514.770
\LT  594.000  515.287
\LT  600.000  515.731
\LT  606.000  516.101
\LT  612.000  516.401
\LT  618.000  516.631
\LT  624.000  516.793
\LT  630.000  516.889
\LT  636.000  516.919
\LT  642.000  516.885
\LT  648.000  516.788
\LT  654.000  516.630
\LT  661.000  516.370
\LT  668.000  516.029
\LT  675.000  515.609
\LT  682.000  515.112
\LT  689.000  514.540
\LT  696.000  513.893
\LT  703.000  513.174
\LT  710.000  512.383
\LT  717.000  511.522
\LT  724.000  510.593
\LT  731.000  509.595
\LT  738.000  508.532
\LT  745.000  507.402
\LT  752.000  506.208
\LT  759.000  504.951
\LT  766.000  503.632
\LT  774.000  502.049
\LT  782.000  500.386
\LT  790.000  498.646
\LT  798.000  496.830
\LT  806.000  494.938
\LT  814.000  492.972
\LT  822.000  490.933
\LT  830.000  488.821
\LT  838.000  486.639
\LT  846.000  484.386
\LT  854.000  482.064
\LT  862.000  479.674
\LT  870.000  477.216
\LT  878.000  474.691
\LT  886.000  472.100
\LT  894.000  469.444
\LT  902.000  466.724
\LT  911.000  463.587
\LT  920.000  460.370
\LT  929.000  457.074
\LT  938.000  453.701
\LT  947.000  450.249
\LT  956.000  446.722
\LT  965.000  443.119
\LT  974.000  439.440
\LT  983.000  435.688
\LT  992.000  431.862
\LT 1001.000  427.963
\LT 1010.000  423.992
\LT 1019.000  419.949
\LT 1028.000  415.835
\LT 1037.000  411.651
\LT 1046.000  407.396
\LT 1055.000  403.072
\LT 1064.000  398.679
\LT 1073.000  394.218
\LT 1082.000  389.689
\LT 1091.000  385.092
\LT 1100.000  380.428
\LT 1109.000  375.697
\LT 1118.000  370.900
\LT 1127.000  366.038
\LT 1136.000  361.109
\LT 1145.000  356.116
\LT 1155.000  350.492
\LT 1165.000  344.788
\LT 1175.000  339.006
\LT 1185.000  333.144
\LT 1195.000  327.205
\LT 1205.000  321.188
\LT 1215.000  315.093
\LT 1225.000  308.922
\LT 1235.000  302.674
\LT 1245.000  296.350
\LT 1255.000  289.949
\LT 1265.000  283.474
\LT 1275.000  276.923
\LT 1285.000  270.297
\LT 1295.000  263.596
\LT 1300.000  260.218
\koniec    5.00000  -0.001
\obraz6
\grub0.2pt
\MT   0.000   60.000
\LT1400.000   60.000
\MT 160.000   60.000
\LT 160.000   70.000
\MT 220.000   60.000
\LT 220.000   70.000
\MT 280.000   60.000
\LT 280.000   70.000
\MT 340.000   60.000
\LT 340.000   70.000
\cput(340.000,-10.000,2)
\MT 400.000   60.000
\LT 400.000   70.000
\MT 460.000   60.000
\LT 460.000   70.000
\MT 520.000   60.000
\LT 520.000   70.000
\MT 580.000   60.000
\LT 580.000   70.000
\cput(580.000,-10.000,4)
\MT 640.000   60.000
\LT 640.000   70.000
\MT 700.000   60.000
\LT 700.000   70.000
\MT 760.000   60.000
\LT 760.000   70.000
\MT 820.000   60.000
\LT 820.000   70.000
\cput(820.000,-10.000,6)
\MT 880.000   60.000
\LT 880.000   70.000
\MT 940.000   60.000
\LT 940.000   70.000
\MT1000.000   60.000
\LT1000.000   70.000
\MT1060.000   60.000
\LT1060.000   70.000
\cput(1060.000,-10.000,8)
\MT1120.000   60.000
\LT1120.000   70.000
\MT1180.000   60.000
\LT1180.000   70.000
\MT1240.000   60.000
\LT1240.000   70.000
\MT1300.000   60.000
\LT1300.000   70.000
\cput(1300.000,-10.000,10)
\MT 100.000    0.000
\LT 100.000  850.000
\MT  92.000   60.000
\LT 108.000   60.000
\MT  92.000   97.000
\LT 108.000   97.000
\MT  92.000  134.000
\LT 108.000  134.000
\MT  92.000  171.000
\LT 108.000  171.000
\MT  92.000  208.000
\LT 108.000  208.000
\MT  92.000  245.000
\LT 108.000  245.000
\MT  92.000  282.000
\LT 108.000  282.000
\MT  92.000  319.000
\LT 108.000  319.000
\MT  92.000  356.000
\LT 108.000  356.000
\MT  92.000  393.000
\LT 108.000  393.000
\MT  92.000  430.000
\LT 108.000  430.000
\MT  92.000  467.000
\LT 108.000  467.000
\MT  92.000  504.000
\LT 108.000  504.000
\MT  92.000  541.000
\LT 108.000  541.000
\MT  92.000  578.000
\LT 108.000  578.000
\MT  92.000  615.000
\LT 108.000  615.000
\MT  92.000  652.000
\LT 108.000  652.000
\MT  92.000  689.000
\LT 108.000  689.000
\MT  92.000  726.000
\LT 108.000  726.000
\MT  92.000  763.000
\LT 108.000  763.000
\MT  92.000  800.000
\LT 108.000  800.000
\MT  84.000   60.000
\LT 116.000   60.000
\MT  84.000  208.000
\LT 116.000  208.000
\lput(80.000,189.500,  2.0)
\MT  84.000  356.000
\LT 116.000  356.000
\lput(80.000,337.500,  4.0)
\MT  84.000  504.000
\LT 116.000  504.000
\lput(80.000,485.500,  6.0)
\MT  84.000  652.000
\LT 116.000  652.000
\lput(80.000,633.500,  8.0)
\MT  84.000  800.000
\LT 116.000  800.000
\lput(80.000,781.500, 10.0)
\grub0.6pt
\MT  100.000  134.000
\LT  102.000  133.975
\LT  104.000  133.899
\LT  106.000  133.772
\LT  108.000  133.594
\LT  110.000  133.366
\LT  112.000  133.087
\LT  114.000  132.758
\LT  116.000  132.378
\LT  118.000  131.947
\LT  120.000  131.466
\LT  122.000  130.934
\LT  124.000  130.352
\LT  126.000  129.720
\LT  128.000  129.038
\LT  130.000  128.306
\LT  132.000  127.525
\LT  134.000  126.695
\LT  136.000  125.817
\LT  138.000  124.890
\LT  140.000  123.917
\LT  142.000  122.897
\LT  144.000  121.831
\LT  146.000  120.721
\LT  148.000  119.568
\LT  150.000  118.372
\LT  153.000  116.503
\LT  156.000  114.548
\LT  159.000  112.514
\LT  162.000  110.407
\LT  165.000  108.235
\LT  168.000  106.008
\LT  172.000  102.968
\LT  177.000   99.090
\LT  187.000   91.269
\LT  191.000   88.191
\LT  194.000   85.926
\LT  197.000   83.710
\LT  200.000   81.553
\LT  203.000   79.464
\LT  206.000   77.451
\LT  209.000   75.522
\LT  211.000   74.285
\LT  213.000   73.090
\LT  215.000   71.937
\LT  217.000   70.829
\LT  219.000   69.765
\LT  221.000   68.748
\LT  223.000   67.777
\LT  225.000   66.853
\LT  227.000   65.976
\LT  229.000   65.147
\LT  231.000   64.365
\LT  233.000   63.630
\LT  235.000   62.943
\LT  237.000   62.302
\LT  239.000   61.707
\LT  241.000   61.158
\LT  243.000   60.654
\LT  245.000   60.193
\LT  247.000   59.777
\LT  249.000   59.402
\LT  251.000   59.070
\LT  253.000   58.778
\LT  256.000   58.415
\LT  259.000   58.138
\LT  262.000   57.943
\LT  265.000   57.829
\LT  268.000   57.790
\LT  271.000   57.824
\LT  274.000   57.927
\LT  277.000   58.096
\LT  280.000   58.329
\LT  283.000   58.621
\LT  286.000   58.970
\LT  289.000   59.373
\LT  292.000   59.828
\LT  295.000   60.331
\LT  299.000   61.074
\LT  303.000   61.894
\LT  307.000   62.785
\LT  311.000   63.743
\LT  315.000   64.764
\LT  319.000   65.842
\LT  324.000   67.267
\LT  329.000   68.769
\LT  334.000   70.343
\LT  339.000   71.984
\LT  344.000   73.685
\LT  350.000   75.802
\LT  356.000   77.993
\LT  362.000   80.253
\LT  368.000   82.576
\LT  375.000   85.360
\LT  382.000   88.217
\LT  389.000   91.141
\LT  397.000   94.560
\LT  405.000   98.054
\LT  413.000  101.620
\LT  421.000  105.252
\LT  430.000  109.413
\LT  439.000  113.651
\LT  448.000  117.961
\LT  457.000  122.342
\LT  466.000  126.790
\LT  475.000  131.305
\LT  485.000  136.399
\LT  495.000  141.571
\LT  505.000  146.823
\LT  515.000  152.152
\LT  525.000  157.559
\LT  535.000  163.044
\LT  545.000  168.607
\LT  555.000  174.247
\LT  565.000  179.966
\LT  575.000  185.764
\LT  585.000  191.641
\LT  595.000  197.597
\LT  604.000  203.026
\LT  613.000  208.520
\LT  622.000  214.080
\LT  631.000  219.707
\LT  640.000  225.399
\LT  649.000  231.159
\LT  658.000  236.986
\LT  667.000  242.881
\LT  676.000  248.844
\LT  685.000  254.875
\LT  694.000  260.976
\LT  703.000  267.146
\LT  712.000  273.386
\LT  721.000  279.695
\LT  730.000  286.075
\LT  739.000  292.526
\LT  748.000  299.049
\LT  757.000  305.642
\LT  766.000  312.308
\LT  775.000  319.045
\LT  784.000  325.855
\LT  793.000  332.738
\LT  802.000  339.694
\LT  811.000  346.723
\LT  820.000  353.825
\LT  829.000  361.002
\LT  838.000  368.252
\LT  847.000  375.577
\LT  856.000  382.977
\LT  865.000  390.451
\LT  874.000  398.000
\LT  883.000  405.624
\LT  892.000  413.324
\LT  901.000  421.100
\LT  910.000  428.951
\LT  919.000  436.879
\LT  928.000  444.882
\LT  937.000  452.962
\LT  946.000  461.119
\LT  955.000  469.352
\LT  964.000  477.662
\LT  973.000  486.050
\LT  982.000  494.514
\LT  991.000  503.056
\LT 1000.000  511.675
\LT 1009.000  520.372
\LT 1018.000  529.147
\LT 1027.000  538.000
\LT 1036.000  546.931
\LT 1045.000  555.939
\LT 1054.000  565.026
\LT 1063.000  574.192
\LT 1072.000  583.436
\LT 1081.000  592.759
\LT 1090.000  602.160
\LT 1099.000  611.640
\LT 1108.000  621.199
\LT 1117.000  630.837
\LT 1126.000  640.554
\LT 1135.000  650.351
\LT 1144.000  660.227
\LT 1153.000  670.182
\LT 1162.000  680.216
\LT 1171.000  690.330
\LT 1180.000  700.524
\LT 1189.000  710.797
\LT 1198.000  721.150
\LT 1207.000  731.583
\LT 1216.000  742.096
\LT 1225.000  752.689
\LT 1234.000  763.362
\LT 1243.000  774.115
\LT 1252.000  784.949
\LT 1261.000  795.862
\LT 1270.000  806.856
\LT 1279.000  817.930
\LT 1288.000  829.085
\LT 1297.000  840.320
\LT 1300.000  844.083
\koniec    4.00000   0.100
\obraz7
\grub0.2pt
\MT   0.000   60.000
\LT1400.000   60.000
\MT 160.000   60.000
\LT 160.000   70.000
\MT 220.000   60.000
\LT 220.000   70.000
\MT 280.000   60.000
\LT 280.000   70.000
\MT 340.000   60.000
\LT 340.000   70.000
\cput(340.000,-10.000,2)
\MT 400.000   60.000
\LT 400.000   70.000
\MT 460.000   60.000
\LT 460.000   70.000
\MT 520.000   60.000
\LT 520.000   70.000
\MT 580.000   60.000
\LT 580.000   70.000
\cput(580.000,-10.000,4)
\MT 640.000   60.000
\LT 640.000   70.000
\MT 700.000   60.000
\LT 700.000   70.000
\MT 760.000   60.000
\LT 760.000   70.000
\MT 820.000   60.000
\LT 820.000   70.000
\cput(820.000,-10.000,6)
\MT 880.000   60.000
\LT 880.000   70.000
\MT 940.000   60.000
\LT 940.000   70.000
\MT1000.000   60.000
\LT1000.000   70.000
\MT1060.000   60.000
\LT1060.000   70.000
\cput(1060.000,-10.000,8)
\MT1120.000   60.000
\LT1120.000   70.000
\MT1180.000   60.000
\LT1180.000   70.000
\MT1240.000   60.000
\LT1240.000   70.000
\MT1300.000   60.000
\LT1300.000   70.000
\cput(1300.000,-10.000,10)
\MT 100.000    0.000
\LT 100.000  850.000
\MT  92.000   60.000
\LT 108.000   60.000
\MT  92.000   97.000
\LT 108.000   97.000
\MT  92.000  134.000
\LT 108.000  134.000
\MT  92.000  171.000
\LT 108.000  171.000
\MT  92.000  208.000
\LT 108.000  208.000
\MT  92.000  245.000
\LT 108.000  245.000
\MT  92.000  282.000
\LT 108.000  282.000
\MT  92.000  319.000
\LT 108.000  319.000
\MT  92.000  356.000
\LT 108.000  356.000
\MT  92.000  393.000
\LT 108.000  393.000
\MT  92.000  430.000
\LT 108.000  430.000
\MT  92.000  467.000
\LT 108.000  467.000
\MT  92.000  504.000
\LT 108.000  504.000
\MT  92.000  541.000
\LT 108.000  541.000
\MT  92.000  578.000
\LT 108.000  578.000
\MT  92.000  615.000
\LT 108.000  615.000
\MT  92.000  652.000
\LT 108.000  652.000
\MT  92.000  689.000
\LT 108.000  689.000
\MT  92.000  726.000
\LT 108.000  726.000
\MT  92.000  763.000
\LT 108.000  763.000
\MT  92.000  800.000
\LT 108.000  800.000
\MT  84.000   60.000
\LT 116.000   60.000
\MT  84.000  208.000
\LT 116.000  208.000
\lput(80.000,189.500, 0.20)
\MT  84.000  356.000
\LT 116.000  356.000
\lput(80.000,337.500, 0.40)
\MT  84.000  504.000
\LT 116.000  504.000
\lput(80.000,485.500, 0.60)
\MT  84.000  652.000
\LT 116.000  652.000
\lput(80.000,633.500, 0.80)
\MT  84.000  800.000
\LT 116.000  800.000
\lput(80.000,781.500, 1.00)
\grub0.6pt
\MT  100.000  800.000
\LT  101.000  799.946
\LT  102.000  799.783
\LT  103.000  799.511
\LT  104.000  799.130
\LT  105.000  798.641
\LT  106.000  798.043
\LT  107.000  797.337
\LT  108.000  796.521
\LT  109.000  795.598
\LT  110.000  794.565
\LT  111.000  793.424
\LT  112.000  792.174
\LT  113.000  790.815
\LT  114.000  789.348
\LT  115.000  787.772
\LT  116.000  786.088
\LT  117.000  784.295
\LT  118.000  782.394
\LT  119.000  780.384
\LT  120.000  778.266
\LT  121.000  776.040
\LT  122.000  773.706
\LT  123.000  771.264
\LT  124.000  768.715
\LT  125.000  766.057
\LT  126.000  763.292
\LT  127.000  760.421
\LT  128.000  757.442
\LT  129.000  754.356
\LT  130.000  751.165
\LT  131.000  747.867
\LT  132.000  744.463
\LT  133.000  740.955
\LT  134.000  737.342
\LT  135.000  733.624
\LT  136.000  729.803
\LT  137.000  725.878
\LT  138.000  721.851
\LT  139.000  717.723
\LT  140.000  713.493
\LT  141.000  709.163
\LT  142.000  704.733
\LT  143.000  700.205
\LT  144.000  695.580
\LT  145.000  690.858
\LT  146.000  686.040
\LT  147.000  681.129
\LT  148.000  676.124
\LT  149.000  671.028
\LT  150.000  665.842
\LT  151.000  660.566
\LT  152.000  655.204
\LT  153.000  649.756
\LT  154.000  644.223
\LT  155.000  638.609
\LT  156.000  632.915
\LT  157.000  627.142
\LT  158.000  621.292
\LT  159.000  615.369
\LT  160.000  609.373
\LT  161.000  603.308
\LT  162.000  597.175
\LT  163.000  590.978
\LT  164.000  584.717
\LT  165.000  578.397
\LT  166.000  572.019
\LT  167.000  565.587
\LT  168.000  559.103
\LT  169.000  552.571
\LT  170.000  545.992
\LT  171.000  539.370
\LT  172.000  532.708
\LT  173.000  526.009
\LT  174.000  519.277
\LT  175.000  512.514
\LT  176.000  505.724
\LT  177.000  498.910
\LT  179.000  485.222
\LT  182.000  464.594
\LT  186.000  437.051
\LT  188.000  423.324
\LT  190.000  409.661
\LT  191.000  402.862
\LT  192.000  396.089
\LT  193.000  389.346
\LT  194.000  382.634
\LT  195.000  375.957
\LT  196.000  369.318
\LT  197.000  362.721
\LT  198.000  356.167
\LT  199.000  349.659
\LT  200.000  343.201
\LT  201.000  336.795
\LT  202.000  330.443
\LT  203.000  324.148
\LT  204.000  317.911
\LT  205.000  311.736
\LT  206.000  305.625
\LT  207.000  299.578
\LT  208.000  293.600
\LT  209.000  287.690
\LT  210.000  281.851
\LT  211.000  276.085
\LT  212.000  270.393
\LT  213.000  264.776
\LT  214.000  259.236
\LT  215.000  253.775
\LT  216.000  248.392
\LT  217.000  243.090
\LT  218.000  237.870
\LT  219.000  232.731
\LT  220.000  227.675
\LT  221.000  222.703
\LT  222.000  217.815
\LT  223.000  213.012
\LT  224.000  208.294
\LT  225.000  203.661
\LT  226.000  199.113
\LT  227.000  194.652
\LT  228.000  190.276
\LT  229.000  185.987
\LT  230.000  181.783
\LT  231.000  177.665
\LT  232.000  173.633
\LT  233.000  169.686
\LT  234.000  165.824
\LT  235.000  162.046
\LT  236.000  158.354
\LT  237.000  154.745
\LT  238.000  151.219
\LT  239.000  147.776
\LT  240.000  144.415
\LT  241.000  141.136
\LT  242.000  137.938
\LT  243.000  134.821
\LT  244.000  131.782
\LT  245.000  128.822
\LT  246.000  125.941
\LT  247.000  123.136
\LT  248.000  120.408
\LT  249.000  117.754
\LT  250.000  115.176
\LT  251.000  112.671
\LT  252.000  110.239
\LT  253.000  107.878
\LT  254.000  105.588
\LT  255.000  103.368
\LT  256.000  101.217
\LT  257.000   99.133
\LT  258.000   97.117
\LT  259.000   95.166
\LT  260.000   93.280
\LT  261.000   91.458
\LT  262.000   89.699
\LT  263.000   88.002
\LT  264.000   86.366
\LT  265.000   84.790
\LT  266.000   83.272
\LT  267.000   81.812
\LT  268.000   80.409
\LT  269.000   79.063
\LT  270.000   77.771
\LT  271.000   76.533
\LT  272.000   75.348
\LT  273.000   74.214
\LT  274.000   73.132
\LT  275.000   72.101
\LT  276.000   71.118
\LT  277.000   70.183
\LT  278.000   69.296
\LT  279.000   68.456
\LT  280.000   67.660
\LT  281.000   66.910
\LT  282.000   66.203
\LT  283.000   65.539
\LT  284.000   64.917
\LT  285.000   64.337
\LT  286.000   63.796
\LT  287.000   63.295
\LT  288.000   62.833
\LT  289.000   62.409
\LT  290.000   62.022
\LT  291.000   61.671
\LT  292.000   61.356
\LT  293.000   61.076
\LT  294.000   60.830
\LT  295.000   60.617
\LT  296.000   60.437
\LT  297.000   60.289
\LT  298.000   60.172
\LT  299.000   60.086
\LT  300.000   60.029
\LT  301.000   60.002
\LT  302.000   60.004
\LT  303.000   60.033
\LT  304.000   60.090
\LT  305.000   60.173
\LT  306.000   60.283
\LT  307.000   60.418
\LT  308.000   60.578
\LT  309.000   60.762
\LT  310.000   60.971
\LT  311.000   61.202
\LT  312.000   61.457
\LT  313.000   61.733
\LT  314.000   62.031
\LT  315.000   62.351
\LT  316.000   62.691
\LT  318.000   63.431
\LT  320.000   64.248
\LT  322.000   65.138
\LT  324.000   66.098
\LT  326.000   67.125
\LT  328.000   68.215
\LT  330.000   69.365
\LT  332.000   70.571
\LT  334.000   71.833
\LT  336.000   73.145
\LT  338.000   74.507
\LT  340.000   75.914
\LT  342.000   77.366
\LT  344.000   78.859
\LT  347.000   81.172
\LT  350.000   83.566
\LT  353.000   86.035
\LT  356.000   88.573
\LT  359.000   91.172
\LT  362.000   93.829
\LT  365.000   96.538
\LT  369.000  100.222
\LT  373.000  103.979
\LT  377.000  107.799
\LT  382.000  112.650
\LT  387.000  117.570
\LT  393.000  123.546
\LT  401.000  131.600
\LT  414.000  144.798
\LT  429.000  160.033
\LT  438.000  169.111
\LT  446.000  177.113
\LT  453.000  184.051
\LT  460.000  190.920
\LT  467.000  197.713
\LT  473.000  203.472
\LT  479.000  209.168
\LT  485.000  214.799
\LT  491.000  220.364
\LT  497.000  225.860
\LT  503.000  231.287
\LT  509.000  236.644
\LT  515.000  241.929
\LT  521.000  247.143
\LT  527.000  252.286
\LT  533.000  257.357
\LT  539.000  262.356
\LT  545.000  267.284
\LT  551.000  272.141
\LT  557.000  276.928
\LT  563.000  281.646
\LT  569.000  286.294
\LT  575.000  290.873
\LT  581.000  295.386
\LT  587.000  299.831
\LT  593.000  304.210
\LT  599.000  308.525
\LT  605.000  312.775
\LT  611.000  316.962
\LT  617.000  321.087
\LT  623.000  325.151
\LT  630.000  329.816
\LT  637.000  334.400
\LT  644.000  338.906
\LT  651.000  343.334
\LT  658.000  347.686
\LT  665.000  351.964
\LT  672.000  356.169
\LT  679.000  360.303
\LT  686.000  364.367
\LT  693.000  368.363
\LT  700.000  372.292
\LT  707.000  376.155
\LT  714.000  379.955
\LT  721.000  383.691
\LT  729.000  387.887
\LT  737.000  392.004
\LT  745.000  396.045
\LT  753.000  400.012
\LT  761.000  403.906
\LT  769.000  407.729
\LT  777.000  411.483
\LT  785.000  415.170
\LT  793.000  418.791
\LT  802.000  422.788
\LT  811.000  426.706
\LT  820.000  430.547
\LT  829.000  434.313
\LT  838.000  438.007
\LT  847.000  441.630
\LT  856.000  445.184
\LT  865.000  448.671
\LT  875.000  452.469
\LT  885.000  456.188
\LT  895.000  459.832
\LT  905.000  463.402
\LT  915.000  466.900
\LT  925.000  470.328
\LT  935.000  473.689
\LT  946.000  477.309
\LT  957.000  480.853
\LT  968.000  484.322
\LT  979.000  487.718
\LT  990.000  491.043
\LT 1002.000  494.593
\LT 1014.000  498.065
\LT 1026.000  501.460
\LT 1038.000  504.782
\LT 1050.000  508.031
\LT 1063.000  511.474
\LT 1076.000  514.838
\LT 1089.000  518.126
\LT 1102.000  521.341
\LT 1115.000  524.484
\LT 1129.000  527.792
\LT 1143.000  531.023
\LT 1157.000  534.180
\LT 1171.000  537.264
\LT 1186.000  540.491
\LT 1201.000  543.642
\LT 1216.000  546.718
\LT 1231.000  549.722
\LT 1247.000  552.849
\LT 1263.000  555.901
\LT 1279.000  558.879
\LT 1296.000  561.965
\LT 1300.000  562.680
\koniec    3.17300   0.000
\obraz8
\grub0.2pt
\MT   0.000  800.000
\LT1400.000  800.000
\MT 160.000  800.000
\LT 160.000  810.000
\MT 220.000  800.000
\LT 220.000  810.000
\MT 280.000  800.000
\LT 280.000  810.000
\MT 340.000  800.000
\LT 340.000  810.000
\MT 400.000  800.000
\LT 400.000  810.000
\cput(400.000,730.000,5)
\MT 460.000  800.000
\LT 460.000  810.000
\MT 520.000  800.000
\LT 520.000  810.000
\MT 580.000  800.000
\LT 580.000  810.000
\MT 640.000  800.000
\LT 640.000  810.000
\MT 700.000  800.000
\LT 700.000  810.000
\cput(700.000,730.000,10)
\MT 760.000  800.000
\LT 760.000  810.000
\MT 820.000  800.000
\LT 820.000  810.000
\MT 880.000  800.000
\LT 880.000  810.000
\MT 940.000  800.000
\LT 940.000  810.000
\MT1000.000  800.000
\LT1000.000  810.000
\cput(1000.000,730.000,15)
\MT1060.000  800.000
\LT1060.000  810.000
\MT1120.000  800.000
\LT1120.000  810.000
\MT1180.000  800.000
\LT1180.000  810.000
\MT1240.000  800.000
\LT1240.000  810.000
\MT1300.000  800.000
\LT1300.000  810.000
\cput(1300.000,730.000,20)
\MT 100.000    0.000
\LT 100.000  850.000
\MT  92.000   50.000
\LT 108.000   50.000
\MT  92.000   87.500
\LT 108.000   87.500
\MT  92.000  125.000
\LT 108.000  125.000
\MT  92.000  162.500
\LT 108.000  162.500
\MT  92.000  200.000
\LT 108.000  200.000
\MT  92.000  237.500
\LT 108.000  237.500
\MT  92.000  275.000
\LT 108.000  275.000
\MT  92.000  312.500
\LT 108.000  312.500
\MT  92.000  350.000
\LT 108.000  350.000
\MT  92.000  387.500
\LT 108.000  387.500
\MT  92.000  425.000
\LT 108.000  425.000
\MT  92.000  462.500
\LT 108.000  462.500
\MT  92.000  500.000
\LT 108.000  500.000
\MT  92.000  537.500
\LT 108.000  537.500
\MT  92.000  575.000
\LT 108.000  575.000
\MT  92.000  612.500
\LT 108.000  612.500
\MT  92.000  650.000
\LT 108.000  650.000
\MT  92.000  687.500
\LT 108.000  687.500
\MT  92.000  725.000
\LT 108.000  725.000
\MT  92.000  762.500
\LT 108.000  762.500
\MT  92.000  800.000
\LT 108.000  800.000
\MT  84.000   50.000
\LT 116.000   50.000
\lput(80.000,31.250, -100)
\MT  84.000  200.000
\LT 116.000  200.000
\lput(80.000,181.250,  -80)
\MT  84.000  350.000
\LT 116.000  350.000
\lput(80.000,331.250,  -60)
\MT  84.000  500.000
\LT 116.000  500.000
\lput(80.000,481.250,  -40)
\MT  84.000  650.000
\LT 116.000  650.000
\lput(80.000,631.250,  -20)
\MT  84.000  800.000
\LT 116.000  800.000
\grub0.6pt
\MT  100.000  807.500
\LT  103.000  807.476
\LT  106.000  807.406
\LT  109.000  807.287
\LT  112.000  807.122
\LT  115.000  806.910
\LT  118.000  806.652
\LT  122.000  806.237
\LT  126.000  805.746
\LT  130.000  805.186
\LT  135.000  804.402
\LT  141.000  803.375
\LT  154.000  801.068
\LT  160.000  800.066
\LT  166.000  799.138
\LT  172.000  798.289
\LT  178.000  797.508
\LT  186.000  796.550
\LT  198.000  795.211
\LT  215.000  793.335
\LT  225.000  792.171
\LT  234.000  791.059
\LT  243.000  789.875
\LT  251.000  788.754
\LT  259.000  787.567
\LT  267.000  786.309
\LT  275.000  784.981
\LT  283.000  783.579
\LT  291.000  782.105
\LT  299.000  780.556
\LT  307.000  778.934
\LT  315.000  777.237
\LT  323.000  775.466
\LT  331.000  773.621
\LT  339.000  771.702
\LT  347.000  769.709
\LT  355.000  767.643
\LT  363.000  765.502
\LT  371.000  763.289
\LT  379.000  761.002
\LT  387.000  758.642
\LT  395.000  756.209
\LT  403.000  753.703
\LT  411.000  751.125
\LT  419.000  748.475
\LT  427.000  745.752
\LT  435.000  742.957
\LT  443.000  740.090
\LT  451.000  737.151
\LT  459.000  734.141
\LT  467.000  731.059
\LT  475.000  727.905
\LT  483.000  724.681
\LT  491.000  721.384
\LT  499.000  718.017
\LT  507.000  714.578
\LT  515.000  711.069
\LT  523.000  707.488
\LT  531.000  703.836
\LT  539.000  700.114
\LT  547.000  696.321
\LT  555.000  692.457
\LT  563.000  688.523
\LT  571.000  684.518
\LT  579.000  680.442
\LT  587.000  676.296
\LT  595.000  672.079
\LT  603.000  667.793
\LT  611.000  663.435
\LT  619.000  659.008
\LT  627.000  654.510
\LT  635.000  649.942
\LT  643.000  645.303
\LT  651.000  640.595
\LT  659.000  635.816
\LT  667.000  630.967
\LT  675.000  626.049
\LT  683.000  621.060
\LT  691.000  616.001
\LT  699.000  610.872
\LT  707.000  605.673
\LT  715.000  600.404
\LT  723.000  595.065
\LT  731.000  589.657
\LT  739.000  584.178
\LT  747.000  578.629
\LT  755.000  573.011
\LT  763.000  567.323
\LT  771.000  561.565
\LT  779.000  555.737
\LT  787.000  549.839
\LT  795.000  543.872
\LT  803.000  537.834
\LT  811.000  531.727
\LT  819.000  525.551
\LT  827.000  519.304
\LT  835.000  512.988
\LT  843.000  506.602
\LT  851.000  500.146
\LT  859.000  493.621
\LT  867.000  487.026
\LT  875.000  480.362
\LT  883.000  473.627
\LT  891.000  466.823
\LT  899.000  459.950
\LT  907.000  453.006
\LT  915.000  445.994
\LT  923.000  438.911
\LT  931.000  431.759
\LT  939.000  424.537
\LT  947.000  417.246
\LT  955.000  409.885
\LT  963.000  402.455
\LT  971.000  394.955
\LT  979.000  387.385
\LT  987.000  379.746
\LT  995.000  372.037
\LT 1003.000  364.259
\LT 1011.000  356.411
\LT 1019.000  348.494
\LT 1027.000  340.507
\LT 1035.000  332.450
\LT 1043.000  324.324
\LT 1051.000  316.129
\LT 1059.000  307.864
\LT 1067.000  299.529
\LT 1075.000  291.125
\LT 1083.000  282.652
\LT 1091.000  274.109
\LT 1099.000  265.496
\LT 1107.000  256.814
\LT 1115.000  248.063
\LT 1123.000  239.242
\LT 1131.000  230.351
\LT 1139.000  221.391
\LT 1147.000  212.362
\LT 1155.000  203.263
\LT 1163.000  194.095
\LT 1171.000  184.857
\LT 1179.000  175.549
\LT 1187.000  166.173
\LT 1195.000  156.726
\LT 1203.000  147.211
\LT 1211.000  137.625
\LT 1219.000  127.971
\LT 1227.000  118.247
\LT 1235.000  108.453
\LT 1243.000   98.590
\LT 1251.000   88.658
\LT 1259.000   78.656
\LT 1267.000   68.585
\LT 1275.000   58.444
\LT 1283.000   48.234
\LT 1291.000   37.954
\LT 1299.000   27.605
\LT 1300.000   26.307
\koniec    3.00000  -0.260
\eject

\def\opis{Plots of the \f\ $f(x)$ (Eq.~\eqref{3.12}) for some values of
parameters $a$ and~$b$\break (see the text for explanation)\break
Fig.~\the\nd A: $a=3.00000,\ b=-0.200$; \
Fig.~\the\nd B: $a=3.00000,\ b=-0.010$;\break
Fig.~\the\nd C: $a=3.00000,\ b=0.000$;\
Fig.~\the\nd D: $a=3.00000,\ b=0.001$;\break
Fig.~\the\nd E: $a=3.00000,\ b=0.010$;\
Fig.~\the\nd F: $a=3.00000,\ b=0.100$;\break
Fig.~\the\nd G: $a=3.00000,\ b=0.200$;\
Fig.~\the\nd H: $a=3.00000,\ b=1.000$.}

\obraz9
\grub0.2pt
\MT   0.000  800.000
\LT1400.000  800.000
\MT 160.000  800.000
\LT 160.000  810.000
\MT 220.000  800.000
\LT 220.000  810.000
\MT 280.000  800.000
\LT 280.000  810.000
\MT 340.000  800.000
\LT 340.000  810.000
\cput(340.000,730.000,2)
\MT 400.000  800.000
\LT 400.000  810.000
\MT 460.000  800.000
\LT 460.000  810.000
\MT 520.000  800.000
\LT 520.000  810.000
\MT 580.000  800.000
\LT 580.000  810.000
\cput(580.000,730.000,4)
\MT 640.000  800.000
\LT 640.000  810.000
\MT 700.000  800.000
\LT 700.000  810.000
\MT 760.000  800.000
\LT 760.000  810.000
\MT 820.000  800.000
\LT 820.000  810.000
\cput(820.000,730.000,6)
\MT 880.000  800.000
\LT 880.000  810.000
\MT 940.000  800.000
\LT 940.000  810.000
\MT1000.000  800.000
\LT1000.000  810.000
\MT1060.000  800.000
\LT1060.000  810.000
\cput(1060.000,730.000,8)
\MT1120.000  800.000
\LT1120.000  810.000
\MT1180.000  800.000
\LT1180.000  810.000
\MT1240.000  800.000
\LT1240.000  810.000
\MT1300.000  800.000
\LT1300.000  810.000
\cput(1300.000,730.000,10)
\MT 100.000    0.000
\LT 100.000  850.000
\MT  92.000   50.000
\LT 108.000   50.000
\MT  92.000   87.500
\LT 108.000   87.500
\MT  92.000  125.000
\LT 108.000  125.000
\MT  92.000  162.500
\LT 108.000  162.500
\MT  92.000  200.000
\LT 108.000  200.000
\MT  92.000  237.500
\LT 108.000  237.500
\MT  92.000  275.000
\LT 108.000  275.000
\MT  92.000  312.500
\LT 108.000  312.500
\MT  92.000  350.000
\LT 108.000  350.000
\MT  92.000  387.500
\LT 108.000  387.500
\MT  92.000  425.000
\LT 108.000  425.000
\MT  92.000  462.500
\LT 108.000  462.500
\MT  92.000  500.000
\LT 108.000  500.000
\MT  92.000  537.500
\LT 108.000  537.500
\MT  92.000  575.000
\LT 108.000  575.000
\MT  92.000  612.500
\LT 108.000  612.500
\MT  92.000  650.000
\LT 108.000  650.000
\MT  92.000  687.500
\LT 108.000  687.500
\MT  92.000  725.000
\LT 108.000  725.000
\MT  92.000  762.500
\LT 108.000  762.500
\MT  92.000  800.000
\LT 108.000  800.000
\MT  84.000   50.000
\LT 116.000   50.000
\lput(80.000,31.250,  -20)
\MT  84.000  237.500
\LT 116.000  237.500
\lput(80.000,218.750,  -15)
\MT  84.000  425.000
\LT 116.000  425.000
\lput(80.000,406.250,  -10)
\MT  84.000  612.500
\LT 116.000  612.500
\lput(80.000,593.750,   -5)
\MT  84.000  800.000
\LT 116.000  800.000
\grub0.6pt
\MT  100.000  837.500
\LT  103.000  837.472
\LT  106.000  837.388
\LT  109.000  837.247
\LT  112.000  837.050
\LT  115.000  836.797
\LT  118.000  836.488
\LT  121.000  836.122
\LT  124.000  835.701
\LT  127.000  835.224
\LT  130.000  834.691
\LT  133.000  834.104
\LT  136.000  833.462
\LT  139.000  832.766
\LT  142.000  832.017
\LT  145.000  831.216
\LT  148.000  830.365
\LT  151.000  829.465
\LT  154.000  828.518
\LT  158.000  827.186
\LT  162.000  825.780
\LT  166.000  824.308
\LT  170.000  822.778
\LT  175.000  820.796
\LT  181.000  818.342
\LT  199.000  810.818
\LT  205.000  808.364
\LT  210.000  806.372
\LT  215.000  804.441
\LT  220.000  802.578
\LT  225.000  800.790
\LT  230.000  799.077
\LT  235.000  797.442
\LT  240.000  795.881
\LT  245.000  794.392
\LT  250.000  792.969
\LT  255.000  791.609
\LT  261.000  790.051
\LT  267.000  788.564
\LT  274.000  786.905
\LT  282.000  785.090
\LT  291.000  783.124
\LT  304.000  780.374
\LT  330.000  774.941
\LT  343.000  772.157
\LT  354.000  769.730
\LT  364.000  767.452
\LT  373.000  765.335
\LT  382.000  763.148
\LT  391.000  760.889
\LT  400.000  758.552
\LT  408.000  756.407
\LT  416.000  754.197
\LT  424.000  751.920
\LT  432.000  749.574
\LT  440.000  747.160
\LT  448.000  744.676
\LT  456.000  742.121
\LT  464.000  739.495
\LT  472.000  736.798
\LT  480.000  734.028
\LT  488.000  731.187
\LT  496.000  728.273
\LT  504.000  725.286
\LT  512.000  722.227
\LT  520.000  719.095
\LT  528.000  715.890
\LT  536.000  712.612
\LT  544.000  709.262
\LT  552.000  705.839
\LT  560.000  702.343
\LT  568.000  698.775
\LT  576.000  695.134
\LT  584.000  691.421
\LT  592.000  687.636
\LT  600.000  683.778
\LT  608.000  679.848
\LT  616.000  675.846
\LT  624.000  671.773
\LT  632.000  667.627
\LT  640.000  663.410
\LT  648.000  659.121
\LT  656.000  654.761
\LT  664.000  650.329
\LT  672.000  645.827
\LT  680.000  641.253
\LT  688.000  636.608
\LT  696.000  631.892
\LT  704.000  627.105
\LT  712.000  622.248
\LT  720.000  617.320
\LT  728.000  612.321
\LT  736.000  607.252
\LT  744.000  602.113
\LT  752.000  596.904
\LT  760.000  591.624
\LT  768.000  586.274
\LT  776.000  580.854
\LT  784.000  575.365
\LT  792.000  569.805
\LT  800.000  564.176
\LT  808.000  558.477
\LT  816.000  552.709
\LT  824.000  546.871
\LT  832.000  540.963
\LT  840.000  534.986
\LT  848.000  528.940
\LT  856.000  522.825
\LT  864.000  516.640
\LT  872.000  510.386
\LT  880.000  504.063
\LT  888.000  497.671
\LT  896.000  491.211
\LT  904.000  484.681
\LT  912.000  478.082
\LT  920.000  471.414
\LT  928.000  464.678
\LT  936.000  457.873
\LT  944.000  450.999
\LT  952.000  444.057
\LT  960.000  437.046
\LT  968.000  429.967
\LT  976.000  422.819
\LT  984.000  415.602
\LT  992.000  408.317
\LT 1000.000  400.964
\LT 1008.000  393.542
\LT 1016.000  386.053
\LT 1024.000  378.494
\LT 1032.000  370.868
\LT 1040.000  363.173
\LT 1048.000  355.410
\LT 1056.000  347.579
\LT 1064.000  339.680
\LT 1072.000  331.713
\LT 1080.000  323.678
\LT 1088.000  315.574
\LT 1096.000  307.403
\LT 1104.000  299.164
\LT 1112.000  290.856
\LT 1120.000  282.481
\LT 1128.000  274.038
\LT 1136.000  265.527
\LT 1144.000  256.949
\LT 1152.000  248.302
\LT 1160.000  239.587
\LT 1168.000  230.805
\LT 1176.000  221.955
\LT 1184.000  213.038
\LT 1192.000  204.052
\LT 1200.000  194.999
\LT 1208.000  185.878
\LT 1216.000  176.690
\LT 1224.000  167.433
\LT 1232.000  158.110
\LT 1240.000  148.718
\LT 1248.000  139.259
\LT 1256.000  129.733
\LT 1264.000  120.139
\LT 1272.000  110.477
\LT 1280.000  100.748
\LT 1288.000   90.951
\LT 1296.000   81.087
\LT 1300.000   76.129
\koniec    3.00000  -0.200
\obraz10
\grub0.2pt
\MT   0.000  612.000
\LT1400.000  612.000
\MT 160.000  612.000
\LT 160.000  622.000
\MT 220.000  612.000
\LT 220.000  622.000
\MT 280.000  612.000
\LT 280.000  622.000
\MT 340.000  612.000
\LT 340.000  622.000
\MT 400.000  612.000
\LT 400.000  622.000
\cput(400.000,542.000,5)
\MT 460.000  612.000
\LT 460.000  622.000
\MT 520.000  612.000
\LT 520.000  622.000
\MT 580.000  612.000
\LT 580.000  622.000
\MT 640.000  612.000
\LT 640.000  622.000
\MT 700.000  612.000
\LT 700.000  622.000
\cput(700.000,542.000,10)
\MT 760.000  612.000
\LT 760.000  622.000
\MT 820.000  612.000
\LT 820.000  622.000
\MT 880.000  612.000
\LT 880.000  622.000
\MT 940.000  612.000
\LT 940.000  622.000
\MT1000.000  612.000
\LT1000.000  622.000
\cput(1000.000,542.000,15)
\MT1060.000  612.000
\LT1060.000  622.000
\MT1120.000  612.000
\LT1120.000  622.000
\MT1180.000  612.000
\LT1180.000  622.000
\MT1240.000  612.000
\LT1240.000  622.000
\MT1300.000  612.000
\LT1300.000  622.000
\cput(1300.000,542.000,20)
\MT 100.000    0.000
\LT 100.000  850.000
\MT  92.000   50.000
\LT 108.000   50.000
\MT  92.000   87.500
\LT 108.000   87.500
\MT  92.000  125.000
\LT 108.000  125.000
\MT  92.000  162.500
\LT 108.000  162.500
\MT  92.000  200.000
\LT 108.000  200.000
\MT  92.000  237.500
\LT 108.000  237.500
\MT  92.000  275.000
\LT 108.000  275.000
\MT  92.000  312.500
\LT 108.000  312.500
\MT  92.000  350.000
\LT 108.000  350.000
\MT  92.000  387.500
\LT 108.000  387.500
\MT  92.000  425.000
\LT 108.000  425.000
\MT  92.000  462.500
\LT 108.000  462.500
\MT  92.000  500.000
\LT 108.000  500.000
\MT  92.000  537.500
\LT 108.000  537.500
\MT  92.000  575.000
\LT 108.000  575.000
\MT  92.000  612.500
\LT 108.000  612.500
\MT  92.000  650.000
\LT 108.000  650.000
\MT  92.000  687.500
\LT 108.000  687.500
\MT  92.000  725.000
\LT 108.000  725.000
\MT  92.000  762.500
\LT 108.000  762.500
\MT  92.000  800.000
\LT 108.000  800.000
\MT  84.000   50.000
\LT 116.000   50.000
\lput(80.000,31.250, -3.0)
\MT  84.000  237.500
\LT 116.000  237.500
\lput(80.000,218.750, -2.0)
\MT  84.000  425.000
\LT 116.000  425.000
\lput(80.000,406.250, -1.0)
\MT  84.000  612.500
\LT 116.000  612.500
\MT  84.000  800.000
\LT 116.000  800.000
\lput(80.000,781.250,  1.0)
\grub0.6pt
\MT  100.000  800.000
\LT  101.000  799.947
\LT  102.000  799.790
\LT  103.000  799.527
\LT  104.000  799.158
\LT  105.000  798.685
\LT  106.000  798.106
\LT  107.000  797.423
\LT  108.000  796.634
\LT  109.000  795.740
\LT  110.000  794.741
\LT  111.000  793.638
\LT  112.000  792.430
\LT  113.000  791.118
\LT  114.000  789.703
\LT  115.000  788.184
\LT  116.000  786.562
\LT  117.000  784.839
\LT  118.000  783.015
\LT  119.000  781.090
\LT  120.000  779.068
\LT  121.000  776.948
\LT  122.000  774.733
\LT  123.000  772.424
\LT  124.000  770.024
\LT  125.000  767.535
\LT  126.000  764.960
\LT  127.000  762.302
\LT  128.000  759.564
\LT  129.000  756.750
\LT  130.000  753.864
\LT  131.000  750.910
\LT  132.000  747.893
\LT  133.000  744.817
\LT  134.000  741.688
\LT  135.000  738.511
\LT  136.000  735.292
\LT  137.000  732.036
\LT  138.000  728.750
\LT  140.000  722.112
\LT  144.000  708.754
\LT  145.000  705.434
\LT  146.000  702.136
\LT  147.000  698.864
\LT  148.000  695.624
\LT  149.000  692.423
\LT  150.000  689.266
\LT  151.000  686.157
\LT  152.000  683.101
\LT  153.000  680.103
\LT  154.000  677.166
\LT  155.000  674.295
\LT  156.000  671.492
\LT  157.000  668.761
\LT  158.000  666.103
\LT  159.000  663.521
\LT  160.000  661.017
\LT  161.000  658.592
\LT  162.000  656.247
\LT  163.000  653.982
\LT  164.000  651.799
\LT  165.000  649.697
\LT  166.000  647.676
\LT  167.000  645.736
\LT  168.000  643.877
\LT  169.000  642.096
\LT  170.000  640.394
\LT  171.000  638.769
\LT  172.000  637.219
\LT  173.000  635.744
\LT  174.000  634.342
\LT  175.000  633.011
\LT  176.000  631.750
\LT  177.000  630.556
\LT  178.000  629.428
\LT  179.000  628.364
\LT  180.000  627.362
\LT  181.000  626.421
\LT  182.000  625.537
\LT  183.000  624.710
\LT  184.000  623.937
\LT  185.000  623.217
\LT  186.000  622.548
\LT  187.000  621.927
\LT  188.000  621.353
\LT  189.000  620.825
\LT  190.000  620.339
\LT  191.000  619.896
\LT  192.000  619.493
\LT  193.000  619.128
\LT  194.000  618.800
\LT  195.000  618.507
\LT  196.000  618.248
\LT  197.000  618.021
\LT  198.000  617.826
\LT  199.000  617.660
\LT  200.000  617.522
\LT  201.000  617.411
\LT  202.000  617.326
\LT  203.000  617.265
\LT  204.000  617.228
\LT  205.000  617.213
\LT  206.000  617.220
\LT  208.000  617.293
\LT  210.000  617.438
\LT  212.000  617.651
\LT  214.000  617.922
\LT  216.000  618.247
\LT  218.000  618.620
\LT  221.000  619.258
\LT  224.000  619.976
\LT  227.000  620.760
\LT  230.000  621.599
\LT  234.000  622.784
\LT  239.000  624.342
\LT  247.000  626.932
\LT  259.000  630.845
\LT  266.000  633.062
\LT  272.000  634.893
\LT  277.000  636.360
\LT  282.000  637.767
\LT  287.000  639.109
\LT  292.000  640.384
\LT  297.000  641.588
\LT  302.000  642.720
\LT  307.000  643.780
\LT  312.000  644.765
\LT  317.000  645.678
\LT  322.000  646.516
\LT  327.000  647.282
\LT  332.000  647.975
\LT  337.000  648.595
\LT  342.000  649.145
\LT  347.000  649.625
\LT  352.000  650.035
\LT  357.000  650.377
\LT  362.000  650.652
\LT  367.000  650.861
\LT  372.000  651.005
\LT  377.000  651.085
\LT  382.000  651.103
\LT  387.000  651.058
\LT  392.000  650.953
\LT  397.000  650.789
\LT  402.000  650.565
\LT  407.000  650.284
\LT  412.000  649.947
\LT  418.000  649.468
\LT  424.000  648.911
\LT  430.000  648.276
\LT  436.000  647.565
\LT  442.000  646.780
\LT  448.000  645.922
\LT  454.000  644.993
\LT  460.000  643.993
\LT  466.000  642.923
\LT  472.000  641.785
\LT  478.000  640.580
\LT  484.000  639.309
\LT  490.000  637.973
\LT  496.000  636.573
\LT  502.000  635.110
\LT  508.000  633.584
\LT  514.000  631.997
\LT  520.000  630.349
\LT  527.000  628.351
\LT  534.000  626.273
\LT  541.000  624.115
\LT  548.000  621.880
\LT  555.000  619.567
\LT  562.000  617.178
\LT  569.000  614.713
\LT  576.000  612.175
\LT  583.000  609.563
\LT  590.000  606.878
\LT  597.000  604.121
\LT  604.000  601.293
\LT  611.000  598.394
\LT  618.000  595.426
\LT  625.000  592.389
\LT  632.000  589.282
\LT  639.000  586.108
\LT  646.000  582.867
\LT  653.000  579.558
\LT  660.000  576.183
\LT  667.000  572.742
\LT  674.000  569.236
\LT  681.000  565.665
\LT  688.000  562.029
\LT  695.000  558.329
\LT  702.000  554.565
\LT  709.000  550.738
\LT  716.000  546.848
\LT  723.000  542.895
\LT  730.000  538.880
\LT  737.000  534.803
\LT  744.000  530.665
\LT  752.000  525.860
\LT  760.000  520.975
\LT  768.000  516.012
\LT  776.000  510.969
\LT  784.000  505.848
\LT  792.000  500.649
\LT  800.000  495.372
\LT  808.000  490.017
\LT  816.000  484.586
\LT  824.000  479.077
\LT  832.000  473.493
\LT  840.000  467.832
\LT  848.000  462.095
\LT  856.000  456.282
\LT  864.000  450.394
\LT  872.000  444.431
\LT  880.000  438.393
\LT  888.000  432.281
\LT  896.000  426.094
\LT  904.000  419.832
\LT  912.000  413.497
\LT  920.000  407.088
\LT  928.000  400.605
\LT  936.000  394.048
\LT  944.000  387.419
\LT  952.000  380.716
\LT  960.000  373.940
\LT  968.000  367.092
\LT  976.000  360.171
\LT  984.000  353.177
\LT  992.000  346.111
\LT 1000.000  338.973
\LT 1008.000  331.763
\LT 1016.000  324.481
\LT 1024.000  317.127
\LT 1032.000  309.701
\LT 1040.000  302.204
\LT 1048.000  294.635
\LT 1056.000  286.995
\LT 1064.000  279.284
\LT 1072.000  271.502
\LT 1080.000  263.648
\LT 1088.000  255.724
\LT 1096.000  247.729
\LT 1104.000  239.663
\LT 1112.000  231.527
\LT 1120.000  223.320
\LT 1128.000  215.042
\LT 1136.000  206.694
\LT 1144.000  198.276
\LT 1152.000  189.788
\LT 1160.000  181.229
\LT 1168.000  172.600
\LT 1176.000  163.902
\LT 1184.000  155.133
\LT 1192.000  146.295
\LT 1200.000  137.386
\LT 1208.000  128.408
\LT 1216.000  119.361
\LT 1224.000  110.243
\LT 1232.000  101.056
\LT 1240.000   91.800
\LT 1248.000   82.474
\LT 1256.000   73.079
\LT 1264.000   63.614
\LT 1272.000   54.081
\LT 1280.000   44.477
\LT 1288.000   34.805
\LT 1296.000   25.064
\LT 1300.000   20.167
\koniec    3.00000  -0.010
\obraz11
\grub0.2pt
\MT   0.000   60.000
\LT1400.000   60.000
\MT 160.000   60.000
\LT 160.000   70.000
\MT 220.000   60.000
\LT 220.000   70.000
\MT 280.000   60.000
\LT 280.000   70.000
\MT 340.000   60.000
\LT 340.000   70.000
\cput(340.000,-10.000,2)
\MT 400.000   60.000
\LT 400.000   70.000
\MT 460.000   60.000
\LT 460.000   70.000
\MT 520.000   60.000
\LT 520.000   70.000
\MT 580.000   60.000
\LT 580.000   70.000
\cput(580.000,-10.000,4)
\MT 640.000   60.000
\LT 640.000   70.000
\MT 700.000   60.000
\LT 700.000   70.000
\MT 760.000   60.000
\LT 760.000   70.000
\MT 820.000   60.000
\LT 820.000   70.000
\cput(820.000,-10.000,6)
\MT 880.000   60.000
\LT 880.000   70.000
\MT 940.000   60.000
\LT 940.000   70.000
\MT1000.000   60.000
\LT1000.000   70.000
\MT1060.000   60.000
\LT1060.000   70.000
\cput(1060.000,-10.000,8)
\MT1120.000   60.000
\LT1120.000   70.000
\MT1180.000   60.000
\LT1180.000   70.000
\MT1240.000   60.000
\LT1240.000   70.000
\MT1300.000   60.000
\LT1300.000   70.000
\cput(1300.000,-10.000,10)
\MT 100.000    0.000
\LT 100.000  850.000
\MT  92.000   60.000
\LT 108.000   60.000
\MT  92.000   97.000
\LT 108.000   97.000
\MT  92.000  134.000
\LT 108.000  134.000
\MT  92.000  171.000
\LT 108.000  171.000
\MT  92.000  208.000
\LT 108.000  208.000
\MT  92.000  245.000
\LT 108.000  245.000
\MT  92.000  282.000
\LT 108.000  282.000
\MT  92.000  319.000
\LT 108.000  319.000
\MT  92.000  356.000
\LT 108.000  356.000
\MT  92.000  393.000
\LT 108.000  393.000
\MT  92.000  430.000
\LT 108.000  430.000
\MT  92.000  467.000
\LT 108.000  467.000
\MT  92.000  504.000
\LT 108.000  504.000
\MT  92.000  541.000
\LT 108.000  541.000
\MT  92.000  578.000
\LT 108.000  578.000
\MT  92.000  615.000
\LT 108.000  615.000
\MT  92.000  652.000
\LT 108.000  652.000
\MT  92.000  689.000
\LT 108.000  689.000
\MT  92.000  726.000
\LT 108.000  726.000
\MT  92.000  763.000
\LT 108.000  763.000
\MT  92.000  800.000
\LT 108.000  800.000
\MT  84.000   60.000
\LT 116.000   60.000
\MT  84.000  208.000
\LT 116.000  208.000
\lput(80.000,189.500, 0.20)
\MT  84.000  356.000
\LT 116.000  356.000
\lput(80.000,337.500, 0.40)
\MT  84.000  504.000
\LT 116.000  504.000
\lput(80.000,485.500, 0.60)
\MT  84.000  652.000
\LT 116.000  652.000
\lput(80.000,633.500, 0.80)
\MT  84.000  800.000
\LT 116.000  800.000
\lput(80.000,781.500, 1.00)
\grub0.6pt
\MT  100.000  800.000
\LT  101.000  799.949
\LT  102.000  799.794
\LT  103.000  799.538
\LT  104.000  799.178
\LT  105.000  798.715
\LT  106.000  798.150
\LT  107.000  797.482
\LT  108.000  796.711
\LT  109.000  795.838
\LT  110.000  794.861
\LT  111.000  793.782
\LT  112.000  792.600
\LT  113.000  791.316
\LT  114.000  789.929
\LT  115.000  788.439
\LT  116.000  786.846
\LT  117.000  785.151
\LT  118.000  783.354
\LT  119.000  781.454
\LT  120.000  779.451
\LT  121.000  777.347
\LT  122.000  775.140
\LT  123.000  772.831
\LT  124.000  770.420
\LT  125.000  767.908
\LT  126.000  765.294
\LT  127.000  762.579
\LT  128.000  759.762
\LT  129.000  756.845
\LT  130.000  753.827
\LT  131.000  750.709
\LT  132.000  747.491
\LT  133.000  744.174
\LT  134.000  740.758
\LT  135.000  737.243
\LT  136.000  733.630
\LT  137.000  729.920
\LT  138.000  726.112
\LT  139.000  722.209
\LT  140.000  718.209
\LT  141.000  714.115
\LT  142.000  709.927
\LT  143.000  705.646
\LT  144.000  701.273
\LT  145.000  696.809
\LT  146.000  692.254
\LT  147.000  687.610
\LT  148.000  682.878
\LT  149.000  678.060
\LT  150.000  673.156
\LT  151.000  668.169
\LT  152.000  663.098
\LT  153.000  657.947
\LT  154.000  652.717
\LT  155.000  647.409
\LT  156.000  642.025
\LT  157.000  636.566
\LT  158.000  631.036
\LT  159.000  625.436
\LT  160.000  619.767
\LT  161.000  614.032
\LT  162.000  608.234
\LT  163.000  602.374
\LT  164.000  596.455
\LT  165.000  590.479
\LT  166.000  584.450
\LT  167.000  578.368
\LT  168.000  572.238
\LT  169.000  566.061
\LT  170.000  559.841
\LT  171.000  553.580
\LT  172.000  547.282
\LT  173.000  540.948
\LT  174.000  534.583
\LT  175.000  528.189
\LT  176.000  521.769
\LT  178.000  508.864
\LT  180.000  495.893
\LT  186.000  456.840
\LT  188.000  443.861
\LT  190.000  430.944
\LT  191.000  424.515
\LT  192.000  418.112
\LT  193.000  411.735
\LT  194.000  405.390
\LT  195.000  399.077
\LT  196.000  392.800
\LT  197.000  386.562
\LT  198.000  380.366
\LT  199.000  374.213
\LT  200.000  368.107
\LT  201.000  362.050
\LT  202.000  356.044
\LT  203.000  350.092
\LT  204.000  344.196
\LT  205.000  338.358
\LT  206.000  332.579
\LT  207.000  326.863
\LT  208.000  321.210
\LT  209.000  315.622
\LT  210.000  310.102
\LT  211.000  304.650
\LT  212.000  299.268
\LT  213.000  293.958
\LT  214.000  288.720
\LT  215.000  283.556
\LT  216.000  278.468
\LT  217.000  273.455
\LT  218.000  268.519
\LT  219.000  263.660
\LT  220.000  258.880
\LT  221.000  254.179
\LT  222.000  249.557
\LT  223.000  245.016
\LT  224.000  240.555
\LT  225.000  236.175
\LT  226.000  231.875
\LT  227.000  227.657
\LT  228.000  223.520
\LT  229.000  219.464
\LT  230.000  215.490
\LT  231.000  211.596
\LT  232.000  207.784
\LT  233.000  204.052
\LT  234.000  200.401
\LT  235.000  196.829
\LT  236.000  193.338
\LT  237.000  189.926
\LT  238.000  186.592
\LT  239.000  183.337
\LT  240.000  180.160
\LT  241.000  177.059
\LT  242.000  174.036
\LT  243.000  171.088
\LT  244.000  168.215
\LT  245.000  165.417
\LT  246.000  162.692
\LT  247.000  160.040
\LT  248.000  157.461
\LT  249.000  154.952
\LT  250.000  152.514
\LT  251.000  150.146
\LT  252.000  147.846
\LT  253.000  145.614
\LT  254.000  143.449
\LT  255.000  141.350
\LT  256.000  139.316
\LT  257.000  137.346
\LT  258.000  135.440
\LT  259.000  133.595
\LT  260.000  131.812
\LT  261.000  130.090
\LT  262.000  128.427
\LT  263.000  126.822
\LT  264.000  125.275
\LT  265.000  123.785
\LT  266.000  122.350
\LT  267.000  120.970
\LT  268.000  119.643
\LT  269.000  118.370
\LT  270.000  117.148
\LT  271.000  115.978
\LT  272.000  114.857
\LT  273.000  113.786
\LT  274.000  112.763
\LT  275.000  111.788
\LT  276.000  110.858
\LT  277.000  109.975
\LT  278.000  109.136
\LT  279.000  108.341
\LT  280.000  107.589
\LT  281.000  106.880
\LT  282.000  106.212
\LT  283.000  105.584
\LT  284.000  104.996
\LT  285.000  104.447
\LT  286.000  103.936
\LT  287.000  103.462
\LT  288.000  103.026
\LT  289.000  102.624
\LT  290.000  102.259
\LT  291.000  101.927
\LT  292.000  101.629
\LT  293.000  101.364
\LT  294.000  101.131
\LT  295.000  100.930
\LT  296.000  100.760
\LT  297.000  100.620
\LT  298.000  100.509
\LT  299.000  100.428
\LT  300.000  100.374
\LT  301.000  100.349
\LT  302.000  100.350
\LT  303.000  100.378
\LT  304.000  100.432
\LT  305.000  100.510
\LT  306.000  100.614
\LT  307.000  100.742
\LT  308.000  100.893
\LT  309.000  101.068
\LT  310.000  101.264
\LT  311.000  101.483
\LT  312.000  101.724
\LT  313.000  101.985
\LT  314.000  102.267
\LT  316.000  102.891
\LT  318.000  103.590
\LT  320.000  104.363
\LT  322.000  105.205
\LT  324.000  106.113
\LT  326.000  107.083
\LT  328.000  108.113
\LT  330.000  109.201
\LT  332.000  110.342
\LT  334.000  111.534
\LT  336.000  112.775
\LT  338.000  114.062
\LT  340.000  115.393
\LT  342.000  116.766
\LT  345.000  118.898
\LT  348.000  121.111
\LT  351.000  123.399
\LT  354.000  125.755
\LT  357.000  128.174
\LT  360.000  130.651
\LT  363.000  133.180
\LT  367.000  136.625
\LT  371.000  140.144
\LT  375.000  143.727
\LT  380.000  148.283
\LT  385.000  152.910
\LT  391.000  158.537
\LT  398.000  165.178
\LT  408.000  174.753
\LT  428.000  193.968
\LT  438.000  203.509
\LT  446.000  211.075
\LT  453.000  217.634
\LT  460.000  224.128
\LT  467.000  230.551
\LT  474.000  236.898
\LT  480.000  242.273
\LT  486.000  247.587
\LT  492.000  252.838
\LT  498.000  258.024
\LT  504.000  263.144
\LT  510.000  268.197
\LT  516.000  273.183
\LT  522.000  278.102
\LT  528.000  282.952
\LT  534.000  287.735
\LT  540.000  292.451
\LT  546.000  297.099
\LT  552.000  301.680
\LT  558.000  306.195
\LT  564.000  310.645
\LT  570.000  315.028
\LT  576.000  319.348
\LT  582.000  323.603
\LT  588.000  327.796
\LT  594.000  331.926
\LT  600.000  335.995
\LT  607.000  340.666
\LT  614.000  345.256
\LT  621.000  349.767
\LT  628.000  354.199
\LT  635.000  358.555
\LT  642.000  362.836
\LT  649.000  367.044
\LT  656.000  371.179
\LT  663.000  375.244
\LT  670.000  379.239
\LT  677.000  383.167
\LT  684.000  387.028
\LT  691.000  390.824
\LT  698.000  394.557
\LT  705.000  398.227
\LT  713.000  402.347
\LT  721.000  406.390
\LT  729.000  410.356
\LT  737.000  414.249
\LT  745.000  418.070
\LT  753.000  421.820
\LT  761.000  425.502
\LT  769.000  429.116
\LT  777.000  432.666
\LT  786.000  436.583
\LT  795.000  440.422
\LT  804.000  444.184
\LT  813.000  447.872
\LT  822.000  451.488
\LT  831.000  455.034
\LT  840.000  458.511
\LT  849.000  461.922
\LT  859.000  465.635
\LT  869.000  469.272
\LT  879.000  472.833
\LT  889.000  476.320
\LT  899.000  479.737
\LT  909.000  483.085
\LT  920.000  486.690
\LT  931.000  490.217
\LT  942.000  493.667
\LT  953.000  497.044
\LT  964.000  500.349
\LT  975.000  503.584
\LT  987.000  507.037
\LT  999.000  510.413
\LT 1011.000  513.713
\LT 1023.000  516.941
\LT 1035.000  520.099
\LT 1048.000  523.443
\LT 1061.000  526.709
\LT 1074.000  529.901
\LT 1087.000  533.021
\LT 1101.000  536.302
\LT 1115.000  539.506
\LT 1129.000  542.634
\LT 1143.000  545.688
\LT 1158.000  548.883
\LT 1173.000  552.000
\LT 1188.000  555.042
\LT 1203.000  558.011
\LT 1219.000  561.101
\LT 1235.000  564.114
\LT 1251.000  567.053
\LT 1268.000  570.097
\LT 1285.000  573.064
\LT 1300.000  575.620
\koniec    3.00000   0.000
\obraz12
\grub0.2pt
\MT   0.000   60.000
\LT1400.000   60.000
\MT 160.000   60.000
\LT 160.000   70.000
\MT 220.000   60.000
\LT 220.000   70.000
\MT 280.000   60.000
\LT 280.000   70.000
\MT 340.000   60.000
\LT 340.000   70.000
\cput(340.000,-10.000,2)
\MT 400.000   60.000
\LT 400.000   70.000
\MT 460.000   60.000
\LT 460.000   70.000
\MT 520.000   60.000
\LT 520.000   70.000
\MT 580.000   60.000
\LT 580.000   70.000
\cput(580.000,-10.000,4)
\MT 640.000   60.000
\LT 640.000   70.000
\MT 700.000   60.000
\LT 700.000   70.000
\MT 760.000   60.000
\LT 760.000   70.000
\MT 820.000   60.000
\LT 820.000   70.000
\cput(820.000,-10.000,6)
\MT 880.000   60.000
\LT 880.000   70.000
\MT 940.000   60.000
\LT 940.000   70.000
\MT1000.000   60.000
\LT1000.000   70.000
\MT1060.000   60.000
\LT1060.000   70.000
\cput(1060.000,-10.000,8)
\MT1120.000   60.000
\LT1120.000   70.000
\MT1180.000   60.000
\LT1180.000   70.000
\MT1240.000   60.000
\LT1240.000   70.000
\MT1300.000   60.000
\LT1300.000   70.000
\cput(1300.000,-10.000,10)
\MT 100.000    0.000
\LT 100.000  850.000
\MT  92.000   60.000
\LT 108.000   60.000
\MT  92.000   97.000
\LT 108.000   97.000
\MT  92.000  134.000
\LT 108.000  134.000
\MT  92.000  171.000
\LT 108.000  171.000
\MT  92.000  208.000
\LT 108.000  208.000
\MT  92.000  245.000
\LT 108.000  245.000
\MT  92.000  282.000
\LT 108.000  282.000
\MT  92.000  319.000
\LT 108.000  319.000
\MT  92.000  356.000
\LT 108.000  356.000
\MT  92.000  393.000
\LT 108.000  393.000
\MT  92.000  430.000
\LT 108.000  430.000
\MT  92.000  467.000
\LT 108.000  467.000
\MT  92.000  504.000
\LT 108.000  504.000
\MT  92.000  541.000
\LT 108.000  541.000
\MT  92.000  578.000
\LT 108.000  578.000
\MT  92.000  615.000
\LT 108.000  615.000
\MT  92.000  652.000
\LT 108.000  652.000
\MT  92.000  689.000
\LT 108.000  689.000
\MT  92.000  726.000
\LT 108.000  726.000
\MT  92.000  763.000
\LT 108.000  763.000
\MT  92.000  800.000
\LT 108.000  800.000
\MT  84.000   60.000
\LT 116.000   60.000
\MT  84.000  208.000
\LT 116.000  208.000
\lput(80.000,189.500, 0.20)
\MT  84.000  356.000
\LT 116.000  356.000
\lput(80.000,337.500, 0.40)
\MT  84.000  504.000
\LT 116.000  504.000
\lput(80.000,485.500, 0.60)
\MT  84.000  652.000
\LT 116.000  652.000
\lput(80.000,633.500, 0.80)
\MT  84.000  800.000
\LT 116.000  800.000
\lput(80.000,781.500, 1.00)
\grub0.6pt
\MT  100.000  800.000
\LT  101.000  799.949
\LT  102.000  799.795
\LT  103.000  799.538
\LT  104.000  799.179
\LT  105.000  798.717
\LT  106.000  798.152
\LT  107.000  797.484
\LT  108.000  796.714
\LT  109.000  795.842
\LT  110.000  794.866
\LT  111.000  793.788
\LT  112.000  792.608
\LT  113.000  791.324
\LT  114.000  789.939
\LT  115.000  788.450
\LT  116.000  786.859
\LT  117.000  785.166
\LT  118.000  783.370
\LT  119.000  781.472
\LT  120.000  779.472
\LT  121.000  777.369
\LT  122.000  775.165
\LT  123.000  772.858
\LT  124.000  770.450
\LT  125.000  767.940
\LT  126.000  765.329
\LT  127.000  762.616
\LT  128.000  759.802
\LT  129.000  756.888
\LT  130.000  753.873
\LT  131.000  750.759
\LT  132.000  747.544
\LT  133.000  744.230
\LT  134.000  740.817
\LT  135.000  737.306
\LT  136.000  733.697
\LT  137.000  729.990
\LT  138.000  726.186
\LT  139.000  722.287
\LT  140.000  718.292
\LT  141.000  714.202
\LT  142.000  710.018
\LT  143.000  705.741
\LT  144.000  701.373
\LT  145.000  696.913
\LT  146.000  692.363
\LT  147.000  687.723
\LT  148.000  682.997
\LT  149.000  678.183
\LT  150.000  673.285
\LT  151.000  668.302
\LT  152.000  663.237
\LT  153.000  658.092
\LT  154.000  652.867
\LT  155.000  647.564
\LT  156.000  642.186
\LT  157.000  636.733
\LT  158.000  631.209
\LT  159.000  625.614
\LT  160.000  619.952
\LT  161.000  614.223
\LT  162.000  608.431
\LT  163.000  602.578
\LT  164.000  596.665
\LT  165.000  590.696
\LT  166.000  584.673
\LT  167.000  578.599
\LT  168.000  572.475
\LT  169.000  566.306
\LT  170.000  560.093
\LT  171.000  553.839
\LT  172.000  547.548
\LT  173.000  541.222
\LT  174.000  534.864
\LT  175.000  528.478
\LT  176.000  522.065
\LT  178.000  509.176
\LT  180.000  496.222
\LT  186.000  457.220
\LT  188.000  444.259
\LT  190.000  431.360
\LT  191.000  424.941
\LT  192.000  418.547
\LT  193.000  412.180
\LT  194.000  405.844
\LT  195.000  399.541
\LT  196.000  393.274
\LT  197.000  387.046
\LT  198.000  380.859
\LT  199.000  374.717
\LT  200.000  368.621
\LT  201.000  362.574
\LT  202.000  356.579
\LT  203.000  350.638
\LT  204.000  344.752
\LT  205.000  338.924
\LT  206.000  333.157
\LT  207.000  327.451
\LT  208.000  321.809
\LT  209.000  316.233
\LT  210.000  310.724
\LT  211.000  305.283
\LT  212.000  299.913
\LT  213.000  294.614
\LT  214.000  289.388
\LT  215.000  284.236
\LT  216.000  279.159
\LT  217.000  274.158
\LT  218.000  269.234
\LT  219.000  264.388
\LT  220.000  259.620
\LT  221.000  254.931
\LT  222.000  250.322
\LT  223.000  245.793
\LT  224.000  241.345
\LT  225.000  236.977
\LT  226.000  232.691
\LT  227.000  228.486
\LT  228.000  224.362
\LT  229.000  220.319
\LT  230.000  216.358
\LT  231.000  212.478
\LT  232.000  208.679
\LT  233.000  204.961
\LT  234.000  201.323
\LT  235.000  197.766
\LT  236.000  194.288
\LT  237.000  190.890
\LT  238.000  187.571
\LT  239.000  184.330
\LT  240.000  181.167
\LT  241.000  178.081
\LT  242.000  175.072
\LT  243.000  172.139
\LT  244.000  169.281
\LT  245.000  166.497
\LT  246.000  163.787
\LT  247.000  161.151
\LT  248.000  158.586
\LT  249.000  156.093
\LT  250.000  153.670
\LT  251.000  151.318
\LT  252.000  149.033
\LT  253.000  146.817
\LT  254.000  144.668
\LT  255.000  142.585
\LT  256.000  140.567
\LT  257.000  138.613
\LT  258.000  136.723
\LT  259.000  134.895
\LT  260.000  133.128
\LT  261.000  131.422
\LT  262.000  129.775
\LT  263.000  128.188
\LT  264.000  126.657
\LT  265.000  125.184
\LT  266.000  123.766
\LT  267.000  122.403
\LT  268.000  121.094
\LT  269.000  119.838
\LT  270.000  118.634
\LT  271.000  117.481
\LT  272.000  116.378
\LT  273.000  115.324
\LT  274.000  114.319
\LT  275.000  113.361
\LT  276.000  112.450
\LT  277.000  111.585
\LT  278.000  110.764
\LT  279.000  109.988
\LT  280.000  109.254
\LT  281.000  108.563
\LT  282.000  107.914
\LT  283.000  107.305
\LT  284.000  106.736
\LT  285.000  106.206
\LT  286.000  105.714
\LT  287.000  105.259
\LT  288.000  104.842
\LT  289.000  104.460
\LT  290.000  104.114
\LT  291.000  103.802
\LT  292.000  103.523
\LT  293.000  103.278
\LT  294.000  103.065
\LT  295.000  102.884
\LT  296.000  102.734
\LT  297.000  102.614
\LT  298.000  102.524
\LT  299.000  102.463
\LT  300.000  102.430
\LT  301.000  102.425
\LT  302.000  102.447
\LT  303.000  102.496
\LT  304.000  102.570
\LT  305.000  102.670
\LT  306.000  102.795
\LT  307.000  102.944
\LT  308.000  103.116
\LT  309.000  103.312
\LT  310.000  103.531
\LT  311.000  103.771
\LT  312.000  104.033
\LT  313.000  104.317
\LT  314.000  104.621
\LT  316.000  105.288
\LT  318.000  106.033
\LT  320.000  106.850
\LT  322.000  107.737
\LT  324.000  108.691
\LT  326.000  109.708
\LT  328.000  110.785
\LT  330.000  111.919
\LT  332.000  113.108
\LT  334.000  114.348
\LT  336.000  115.637
\LT  338.000  116.973
\LT  340.000  118.353
\LT  342.000  119.775
\LT  345.000  121.982
\LT  348.000  124.271
\LT  351.000  126.636
\LT  354.000  129.071
\LT  357.000  131.569
\LT  360.000  134.125
\LT  363.000  136.734
\LT  366.000  139.392
\LT  370.000  143.004
\LT  374.000  146.684
\LT  378.000  150.423
\LT  383.000  155.167
\LT  388.000  159.977
\LT  394.000  165.817
\LT  402.000  173.686
\LT  414.000  185.588
\LT  430.000  201.479
\LT  440.000  211.347
\LT  448.000  219.178
\LT  456.000  226.938
\LT  463.000  233.662
\LT  470.000  240.316
\LT  477.000  246.897
\LT  484.000  253.401
\LT  490.000  258.911
\LT  496.000  264.361
\LT  502.000  269.749
\LT  508.000  275.074
\LT  514.000  280.336
\LT  520.000  285.535
\LT  526.000  290.669
\LT  532.000  295.739
\LT  538.000  300.745
\LT  544.000  305.688
\LT  550.000  310.567
\LT  556.000  315.383
\LT  562.000  320.137
\LT  568.000  324.830
\LT  574.000  329.461
\LT  581.000  334.788
\LT  588.000  340.034
\LT  595.000  345.200
\LT  602.000  350.288
\LT  609.000  355.300
\LT  616.000  360.236
\LT  623.000  365.097
\LT  630.000  369.887
\LT  637.000  374.605
\LT  644.000  379.254
\LT  651.000  383.834
\LT  658.000  388.348
\LT  665.000  392.797
\LT  672.000  397.182
\LT  679.000  401.504
\LT  687.000  406.370
\LT  695.000  411.158
\LT  703.000  415.870
\LT  711.000  420.509
\LT  719.000  425.076
\LT  727.000  429.574
\LT  735.000  434.004
\LT  743.000  438.368
\LT  752.000  443.201
\LT  761.000  447.955
\LT  770.000  452.632
\LT  779.000  457.236
\LT  788.000  461.767
\LT  797.000  466.229
\LT  806.000  470.624
\LT  816.000  475.430
\LT  826.000  480.158
\LT  836.000  484.811
\LT  846.000  489.391
\LT  856.000  493.900
\LT  866.000  498.342
\LT  877.000  503.151
\LT  888.000  507.884
\LT  899.000  512.544
\LT  910.000  517.132
\LT  922.000  522.060
\LT  934.000  526.909
\LT  946.000  531.683
\LT  958.000  536.385
\LT  971.000  541.401
\LT  984.000  546.340
\LT  997.000  551.204
\LT 1010.000  555.996
\LT 1024.000  561.082
\LT 1038.000  566.092
\LT 1052.000  571.030
\LT 1067.000  576.245
\LT 1082.000  581.385
\LT 1098.000  586.789
\LT 1114.000  592.117
\LT 1131.000  597.699
\LT 1148.000  603.203
\LT 1166.000  608.951
\LT 1184.000  614.623
\LT 1203.000  620.531
\LT 1222.000  626.364
\LT 1242.000  632.428
\LT 1263.000  638.717
\LT 1285.000  645.225
\LT 1300.000  649.620
\koniec    3.00000   0.001
\obraz13
\grub0.2pt
\MT   0.000   60.000
\LT1400.000   60.000
\MT 160.000   60.000
\LT 160.000   70.000
\MT 220.000   60.000
\LT 220.000   70.000
\MT 280.000   60.000
\LT 280.000   70.000
\MT 340.000   60.000
\LT 340.000   70.000
\cput(340.000,-10.000,2)
\MT 400.000   60.000
\LT 400.000   70.000
\MT 460.000   60.000
\LT 460.000   70.000
\MT 520.000   60.000
\LT 520.000   70.000
\MT 580.000   60.000
\LT 580.000   70.000
\cput(580.000,-10.000,4)
\MT 640.000   60.000
\LT 640.000   70.000
\MT 700.000   60.000
\LT 700.000   70.000
\MT 760.000   60.000
\LT 760.000   70.000
\MT 820.000   60.000
\LT 820.000   70.000
\cput(820.000,-10.000,6)
\MT 880.000   60.000
\LT 880.000   70.000
\MT 940.000   60.000
\LT 940.000   70.000
\MT1000.000   60.000
\LT1000.000   70.000
\MT1060.000   60.000
\LT1060.000   70.000
\cput(1060.000,-10.000,8)
\MT1120.000   60.000
\LT1120.000   70.000
\MT1180.000   60.000
\LT1180.000   70.000
\MT1240.000   60.000
\LT1240.000   70.000
\MT1300.000   60.000
\LT1300.000   70.000
\cput(1300.000,-10.000,10)
\MT 100.000    0.000
\LT 100.000  850.000
\MT  92.000   60.000
\LT 108.000   60.000
\MT  92.000  109.333
\LT 108.000  109.333
\MT  92.000  158.667
\LT 108.000  158.667
\MT  92.000  208.000
\LT 108.000  208.000
\MT  92.000  257.333
\LT 108.000  257.333
\MT  92.000  306.667
\LT 108.000  306.667
\MT  92.000  356.000
\LT 108.000  356.000
\MT  92.000  405.333
\LT 108.000  405.333
\MT  92.000  454.667
\LT 108.000  454.667
\MT  92.000  504.000
\LT 108.000  504.000
\MT  92.000  553.333
\LT 108.000  553.333
\MT  92.000  602.667
\LT 108.000  602.667
\MT  92.000  652.000
\LT 108.000  652.000
\MT  92.000  701.333
\LT 108.000  701.333
\MT  92.000  750.667
\LT 108.000  750.667
\MT  92.000  800.000
\LT 108.000  800.000
\MT  84.000   60.000
\LT 116.000   60.000
\MT  84.000  306.667
\LT 116.000  306.667
\lput(80.000,282.000,  0.5)
\MT  84.000  553.333
\LT 116.000  553.333
\lput(80.000,528.667,  1.0)
\MT  84.000  800.000
\LT 116.000  800.000
\lput(80.000,775.333,  1.5)
\grub0.6pt
\MT  100.000  553.333
\LT  101.000  553.299
\LT  102.000  553.198
\LT  103.000  553.028
\LT  104.000  552.791
\LT  105.000  552.485
\LT  106.000  552.112
\LT  107.000  551.671
\LT  108.000  551.163
\LT  109.000  550.586
\LT  110.000  549.942
\LT  111.000  549.230
\LT  112.000  548.450
\LT  113.000  547.602
\LT  114.000  546.686
\LT  115.000  545.703
\LT  116.000  544.652
\LT  117.000  543.533
\LT  118.000  542.347
\LT  119.000  541.093
\LT  120.000  539.771
\LT  121.000  538.382
\LT  122.000  536.926
\LT  123.000  535.402
\LT  124.000  533.811
\LT  125.000  532.153
\LT  126.000  530.428
\LT  127.000  528.635
\LT  128.000  526.777
\LT  129.000  524.851
\LT  130.000  522.860
\LT  131.000  520.802
\LT  132.000  518.678
\LT  133.000  516.489
\LT  134.000  514.235
\LT  135.000  511.915
\LT  136.000  509.531
\LT  137.000  507.082
\LT  138.000  504.570
\LT  139.000  501.993
\LT  140.000  499.354
\LT  141.000  496.653
\LT  142.000  493.889
\LT  143.000  491.064
\LT  144.000  488.179
\LT  145.000  485.233
\LT  146.000  482.227
\LT  147.000  479.163
\LT  148.000  476.041
\LT  149.000  472.862
\LT  150.000  469.627
\LT  151.000  466.337
\LT  152.000  462.992
\LT  153.000  459.594
\LT  154.000  456.144
\LT  155.000  452.642
\LT  156.000  449.091
\LT  157.000  445.491
\LT  158.000  441.843
\LT  159.000  438.150
\LT  160.000  434.411
\LT  161.000  430.630
\LT  162.000  426.806
\LT  163.000  422.942
\LT  164.000  419.040
\LT  165.000  415.100
\LT  166.000  411.125
\LT  167.000  407.117
\LT  168.000  403.076
\LT  169.000  399.005
\LT  170.000  394.906
\LT  171.000  390.780
\LT  172.000  386.630
\LT  173.000  382.458
\LT  175.000  374.053
\LT  177.000  365.582
\LT  180.000  352.788
\LT  186.000  327.094
\LT  188.000  318.561
\LT  190.000  310.071
\LT  192.000  301.641
\LT  193.000  297.453
\LT  194.000  293.287
\LT  195.000  289.143
\LT  196.000  285.024
\LT  197.000  280.932
\LT  198.000  276.867
\LT  199.000  272.833
\LT  200.000  268.831
\LT  201.000  264.862
\LT  202.000  260.927
\LT  203.000  257.029
\LT  204.000  253.170
\LT  205.000  249.349
\LT  206.000  245.569
\LT  207.000  241.831
\LT  208.000  238.136
\LT  209.000  234.485
\LT  210.000  230.880
\LT  211.000  227.321
\LT  212.000  223.810
\LT  213.000  220.346
\LT  214.000  216.932
\LT  215.000  213.568
\LT  216.000  210.255
\LT  217.000  206.993
\LT  218.000  203.783
\LT  219.000  200.625
\LT  220.000  197.520
\LT  221.000  194.469
\LT  222.000  191.471
\LT  223.000  188.527
\LT  224.000  185.638
\LT  225.000  182.803
\LT  226.000  180.022
\LT  227.000  177.297
\LT  228.000  174.626
\LT  229.000  172.011
\LT  230.000  169.450
\LT  231.000  166.943
\LT  232.000  164.492
\LT  233.000  162.095
\LT  234.000  159.752
\LT  235.000  157.463
\LT  236.000  155.228
\LT  237.000  153.047
\LT  238.000  150.919
\LT  239.000  148.844
\LT  240.000  146.821
\LT  241.000  144.851
\LT  242.000  142.932
\LT  243.000  141.064
\LT  244.000  139.247
\LT  245.000  137.481
\LT  246.000  135.764
\LT  247.000  134.097
\LT  248.000  132.478
\LT  249.000  130.907
\LT  250.000  129.384
\LT  251.000  127.909
\LT  252.000  126.479
\LT  253.000  125.096
\LT  254.000  123.758
\LT  255.000  122.464
\LT  256.000  121.215
\LT  257.000  120.009
\LT  258.000  118.846
\LT  259.000  117.725
\LT  260.000  116.645
\LT  261.000  115.607
\LT  262.000  114.609
\LT  263.000  113.650
\LT  264.000  112.731
\LT  265.000  111.850
\LT  266.000  111.007
\LT  267.000  110.201
\LT  268.000  109.432
\LT  269.000  108.698
\LT  270.000  108.000
\LT  271.000  107.336
\LT  272.000  106.707
\LT  273.000  106.111
\LT  274.000  105.548
\LT  275.000  105.017
\LT  276.000  104.518
\LT  277.000  104.050
\LT  278.000  103.612
\LT  279.000  103.204
\LT  280.000  102.826
\LT  281.000  102.477
\LT  282.000  102.156
\LT  283.000  101.862
\LT  284.000  101.596
\LT  285.000  101.356
\LT  286.000  101.143
\LT  287.000  100.955
\LT  288.000  100.792
\LT  289.000  100.654
\LT  290.000  100.540
\LT  291.000  100.449
\LT  292.000  100.382
\LT  293.000  100.337
\LT  294.000  100.315
\LT  295.000  100.314
\LT  296.000  100.334
\LT  297.000  100.376
\LT  299.000  100.519
\LT  301.000  100.740
\LT  303.000  101.036
\LT  305.000  101.404
\LT  307.000  101.841
\LT  309.000  102.343
\LT  311.000  102.908
\LT  313.000  103.533
\LT  315.000  104.216
\LT  317.000  104.953
\LT  319.000  105.743
\LT  321.000  106.583
\LT  323.000  107.470
\LT  325.000  108.404
\LT  327.000  109.381
\LT  329.000  110.399
\LT  332.000  112.001
\LT  335.000  113.686
\LT  338.000  115.447
\LT  341.000  117.281
\LT  344.000  119.182
\LT  347.000  121.144
\LT  350.000  123.164
\LT  353.000  125.237
\LT  356.000  127.360
\LT  360.000  130.260
\LT  364.000  133.233
\LT  368.000  136.272
\LT  372.000  139.369
\LT  377.000  143.313
\LT  382.000  147.328
\LT  388.000  152.226
\LT  394.000  157.196
\LT  401.000  163.068
\LT  409.000  169.854
\LT  419.000  178.416
\LT  434.000  191.353
\LT  460.000  213.819
\LT  474.000  225.852
\LT  487.000  236.954
\LT  499.000  247.129
\LT  511.000  257.226
\LT  522.000  266.411
\LT  533.000  275.528
\LT  544.000  284.575
\LT  556.000  294.369
\LT  568.000  304.086
\LT  580.000  313.728
\LT  593.000  324.095
\LT  606.000  334.385
\LT  620.000  345.388
\LT  634.000  356.317
\LT  649.000  367.954
\LT  666.000  381.061
\LT  684.000  394.862
\LT  705.000  410.882
\LT  731.000  430.631
\LT  807.000  488.192
\LT  834.000  508.707
\LT  857.000  526.255
\LT  878.000  542.352
\LT  897.000  556.991
\LT  915.000  570.933
\LT  932.000  584.173
\LT  949.000  597.491
\LT  965.000  610.100
\LT  981.000  622.787
\LT  996.000  634.756
\LT 1011.000  646.800
\LT 1026.000  658.923
\LT 1040.000  670.311
\LT 1054.000  681.772
\LT 1068.000  693.309
\LT 1082.000  704.923
\LT 1095.000  715.779
\LT 1108.000  726.705
\LT 1121.000  737.703
\LT 1134.000  748.773
\LT 1147.000  759.919
\LT 1160.000  771.139
\LT 1173.000  782.437
\LT 1186.000  793.813
\LT 1199.000  805.268
\LT 1211.000  815.913
\LT 1223.000  826.627
\LT 1235.000  837.412
\LT 1247.000  848.268
\LT 1259.000  859.195
\LT 1271.000  870.194
\LT 1283.000  881.267
\LT 1295.000  892.413
\LT 1300.000  897.080
\koniec    3.00000   0.010
\obraz14
\grub0.2pt
\MT   0.000   60.000
\LT1400.000   60.000
\MT 160.000   60.000
\LT 160.000   70.000
\MT 220.000   60.000
\LT 220.000   70.000
\MT 280.000   60.000
\LT 280.000   70.000
\MT 340.000   60.000
\LT 340.000   70.000
\cput(340.000,-10.000,2)
\MT 400.000   60.000
\LT 400.000   70.000
\MT 460.000   60.000
\LT 460.000   70.000
\MT 520.000   60.000
\LT 520.000   70.000
\MT 580.000   60.000
\LT 580.000   70.000
\cput(580.000,-10.000,4)
\MT 640.000   60.000
\LT 640.000   70.000
\MT 700.000   60.000
\LT 700.000   70.000
\MT 760.000   60.000
\LT 760.000   70.000
\MT 820.000   60.000
\LT 820.000   70.000
\cput(820.000,-10.000,6)
\MT 880.000   60.000
\LT 880.000   70.000
\MT 940.000   60.000
\LT 940.000   70.000
\MT1000.000   60.000
\LT1000.000   70.000
\MT1060.000   60.000
\LT1060.000   70.000
\cput(1060.000,-10.000,8)
\MT1120.000   60.000
\LT1120.000   70.000
\MT1180.000   60.000
\LT1180.000   70.000
\MT1240.000   60.000
\LT1240.000   70.000
\MT1300.000   60.000
\LT1300.000   70.000
\cput(1300.000,-10.000,10)
\MT 100.000    0.000
\LT 100.000  850.000
\MT  92.000   60.000
\LT 108.000   60.000
\MT  92.000   97.000
\LT 108.000   97.000
\MT  92.000  134.000
\LT 108.000  134.000
\MT  92.000  171.000
\LT 108.000  171.000
\MT  92.000  208.000
\LT 108.000  208.000
\MT  92.000  245.000
\LT 108.000  245.000
\MT  92.000  282.000
\LT 108.000  282.000
\MT  92.000  319.000
\LT 108.000  319.000
\MT  92.000  356.000
\LT 108.000  356.000
\MT  92.000  393.000
\LT 108.000  393.000
\MT  92.000  430.000
\LT 108.000  430.000
\MT  92.000  467.000
\LT 108.000  467.000
\MT  92.000  504.000
\LT 108.000  504.000
\MT  92.000  541.000
\LT 108.000  541.000
\MT  92.000  578.000
\LT 108.000  578.000
\MT  92.000  615.000
\LT 108.000  615.000
\MT  92.000  652.000
\LT 108.000  652.000
\MT  92.000  689.000
\LT 108.000  689.000
\MT  92.000  726.000
\LT 108.000  726.000
\MT  92.000  763.000
\LT 108.000  763.000
\MT  92.000  800.000
\LT 108.000  800.000
\MT  84.000   60.000
\LT 116.000   60.000
\MT  84.000  208.000
\LT 116.000  208.000
\lput(80.000,189.500,  2.0)
\MT  84.000  356.000
\LT 116.000  356.000
\lput(80.000,337.500,  4.0)
\MT  84.000  504.000
\LT 116.000  504.000
\lput(80.000,485.500,  6.0)
\MT  84.000  652.000
\LT 116.000  652.000
\lput(80.000,633.500,  8.0)
\MT  84.000  800.000
\LT 116.000  800.000
\lput(80.000,781.500, 10.0)
\grub0.6pt
\MT  100.000  134.000
\LT  103.000  133.958
\LT  106.000  133.834
\LT  109.000  133.625
\LT  112.000  133.334
\LT  115.000  132.959
\LT  118.000  132.502
\LT  121.000  131.961
\LT  124.000  131.338
\LT  127.000  130.632
\LT  130.000  129.845
\LT  133.000  128.977
\LT  136.000  128.029
\LT  139.000  127.002
\LT  142.000  125.899
\LT  145.000  124.721
\LT  148.000  123.472
\LT  151.000  122.153
\LT  154.000  120.770
\LT  157.000  119.326
\LT  160.000  117.827
\LT  163.000  116.277
\LT  167.000  114.144
\LT  172.000  111.392
\LT  179.000  107.446
\LT  187.000  102.924
\LT  192.000  100.161
\LT  196.000   98.016
\LT  199.000   96.458
\LT  202.000   94.951
\LT  205.000   93.501
\LT  208.000   92.115
\LT  211.000   90.797
\LT  214.000   89.551
\LT  217.000   88.380
\LT  220.000   87.288
\LT  223.000   86.276
\LT  226.000   85.346
\LT  229.000   84.498
\LT  232.000   83.732
\LT  235.000   83.049
\LT  238.000   82.446
\LT  241.000   81.923
\LT  244.000   81.478
\LT  247.000   81.109
\LT  250.000   80.814
\LT  253.000   80.591
\LT  256.000   80.438
\LT  259.000   80.351
\LT  262.000   80.329
\LT  265.000   80.369
\LT  268.000   80.468
\LT  271.000   80.624
\LT  274.000   80.835
\LT  277.000   81.097
\LT  280.000   81.409
\LT  283.000   81.768
\LT  287.000   82.316
\LT  291.000   82.940
\LT  295.000   83.634
\LT  299.000   84.393
\LT  303.000   85.215
\LT  307.000   86.094
\LT  312.000   87.269
\LT  317.000   88.522
\LT  322.000   89.847
\LT  327.000   91.239
\LT  332.000   92.694
\LT  338.000   94.515
\LT  344.000   96.413
\LT  350.000   98.381
\LT  356.000  100.414
\LT  363.000  102.863
\LT  370.000  105.388
\LT  377.000  107.984
\LT  384.000  110.646
\LT  392.000  113.764
\LT  400.000  116.959
\LT  408.000  120.225
\LT  416.000  123.559
\LT  424.000  126.959
\LT  433.000  130.859
\LT  442.000  134.837
\LT  451.000  138.888
\LT  460.000  143.013
\LT  469.000  147.209
\LT  478.000  151.475
\LT  487.000  155.811
\LT  496.000  160.216
\LT  505.000  164.690
\LT  514.000  169.231
\LT  523.000  173.841
\LT  532.000  178.519
\LT  541.000  183.265
\LT  550.000  188.079
\LT  559.000  192.961
\LT  568.000  197.911
\LT  577.000  202.931
\LT  586.000  208.019
\LT  595.000  213.176
\LT  604.000  218.403
\LT  613.000  223.700
\LT  622.000  229.067
\LT  631.000  234.504
\LT  640.000  240.012
\LT  649.000  245.591
\LT  658.000  251.241
\LT  667.000  256.963
\LT  676.000  262.757
\LT  685.000  268.623
\LT  694.000  274.562
\LT  703.000  280.573
\LT  712.000  286.658
\LT  721.000  292.816
\LT  730.000  299.047
\LT  739.000  305.353
\LT  748.000  311.732
\LT  757.000  318.187
\LT  766.000  324.715
\LT  775.000  331.319
\LT  784.000  337.998
\LT  793.000  344.752
\LT  802.000  351.582
\LT  811.000  358.487
\LT  820.000  365.469
\LT  829.000  372.527
\LT  838.000  379.661
\LT  847.000  386.872
\LT  856.000  394.159
\LT  865.000  401.523
\LT  874.000  408.965
\LT  883.000  416.483
\LT  892.000  424.079
\LT  901.000  431.753
\LT  910.000  439.504
\LT  919.000  447.333
\LT  928.000  455.240
\LT  937.000  463.225
\LT  946.000  471.289
\LT  955.000  479.431
\LT  964.000  487.651
\LT  973.000  495.950
\LT  982.000  504.327
\LT  991.000  512.784
\LT 1000.000  521.319
\LT 1009.000  529.933
\LT 1018.000  538.627
\LT 1027.000  547.400
\LT 1036.000  556.252
\LT 1045.000  565.183
\LT 1054.000  574.194
\LT 1063.000  583.285
\LT 1072.000  592.455
\LT 1081.000  601.706
\LT 1090.000  611.036
\LT 1099.000  620.445
\LT 1108.000  629.935
\LT 1117.000  639.505
\LT 1126.000  649.155
\LT 1135.000  658.886
\LT 1144.000  668.696
\LT 1153.000  678.587
\LT 1162.000  688.559
\LT 1171.000  698.611
\LT 1180.000  708.743
\LT 1189.000  718.956
\LT 1198.000  729.249
\LT 1207.000  739.624
\LT 1216.000  750.079
\LT 1225.000  760.615
\LT 1234.000  771.231
\LT 1243.000  781.929
\LT 1252.000  792.707
\LT 1260.000  802.356
\LT 1268.000  812.069
\LT 1276.000  821.846
\LT 1284.000  831.687
\LT 1292.000  841.593
\LT 1300.000  851.562
\koniec    3.00000   0.100
\obraz15
\grub0.2pt
\MT   0.000   60.000
\LT1400.000   60.000
\MT 160.000   60.000
\LT 160.000   70.000
\MT 220.000   60.000
\LT 220.000   70.000
\MT 280.000   60.000
\LT 280.000   70.000
\MT 340.000   60.000
\LT 340.000   70.000
\cput(340.000,-10.000,2)
\MT 400.000   60.000
\LT 400.000   70.000
\MT 460.000   60.000
\LT 460.000   70.000
\MT 520.000   60.000
\LT 520.000   70.000
\MT 580.000   60.000
\LT 580.000   70.000
\cput(580.000,-10.000,4)
\MT 640.000   60.000
\LT 640.000   70.000
\MT 700.000   60.000
\LT 700.000   70.000
\MT 760.000   60.000
\LT 760.000   70.000
\MT 820.000   60.000
\LT 820.000   70.000
\cput(820.000,-10.000,6)
\MT 880.000   60.000
\LT 880.000   70.000
\MT 940.000   60.000
\LT 940.000   70.000
\MT1000.000   60.000
\LT1000.000   70.000
\MT1060.000   60.000
\LT1060.000   70.000
\cput(1060.000,-10.000,8)
\MT1120.000   60.000
\LT1120.000   70.000
\MT1180.000   60.000
\LT1180.000   70.000
\MT1240.000   60.000
\LT1240.000   70.000
\MT1300.000   60.000
\LT1300.000   70.000
\cput(1300.000,-10.000,10)
\MT 100.000    0.000
\LT 100.000  850.000
\MT  92.000   60.000
\LT 108.000   60.000
\MT  92.000   97.000
\LT 108.000   97.000
\MT  92.000  134.000
\LT 108.000  134.000
\MT  92.000  171.000
\LT 108.000  171.000
\MT  92.000  208.000
\LT 108.000  208.000
\MT  92.000  245.000
\LT 108.000  245.000
\MT  92.000  282.000
\LT 108.000  282.000
\MT  92.000  319.000
\LT 108.000  319.000
\MT  92.000  356.000
\LT 108.000  356.000
\MT  92.000  393.000
\LT 108.000  393.000
\MT  92.000  430.000
\LT 108.000  430.000
\MT  92.000  467.000
\LT 108.000  467.000
\MT  92.000  504.000
\LT 108.000  504.000
\MT  92.000  541.000
\LT 108.000  541.000
\MT  92.000  578.000
\LT 108.000  578.000
\MT  92.000  615.000
\LT 108.000  615.000
\MT  92.000  652.000
\LT 108.000  652.000
\MT  92.000  689.000
\LT 108.000  689.000
\MT  92.000  726.000
\LT 108.000  726.000
\MT  92.000  763.000
\LT 108.000  763.000
\MT  92.000  800.000
\LT 108.000  800.000
\MT  84.000   60.000
\LT 116.000   60.000
\MT  84.000  245.000
\LT 116.000  245.000
\lput(80.000,226.500,    5)
\MT  84.000  430.000
\LT 116.000  430.000
\lput(80.000,411.500,   10)
\MT  84.000  615.000
\LT 116.000  615.000
\lput(80.000,596.500,   15)
\MT  84.000  800.000
\LT 116.000  800.000
\lput(80.000,781.500,   20)
\grub0.6pt
\MT  100.000   97.000
\LT  104.000   96.967
\LT  108.000   96.868
\LT  112.000   96.704
\LT  116.000   96.474
\LT  120.000   96.178
\LT  124.000   95.817
\LT  128.000   95.391
\LT  132.000   94.901
\LT  136.000   94.348
\LT  140.000   93.733
\LT  144.000   93.059
\LT  148.000   92.328
\LT  153.000   91.341
\LT  158.000   90.281
\LT  164.000   88.928
\LT  171.000   87.270
\LT  185.000   83.880
\LT  191.000   82.481
\LT  196.000   81.376
\LT  201.000   80.345
\LT  205.000   79.584
\LT  209.000   78.887
\LT  213.000   78.260
\LT  217.000   77.707
\LT  221.000   77.233
\LT  224.000   76.929
\LT  227.000   76.671
\LT  230.000   76.459
\LT  233.000   76.293
\LT  236.000   76.172
\LT  240.000   76.080
\LT  244.000   76.067
\LT  248.000   76.129
\LT  252.000   76.265
\LT  256.000   76.472
\LT  260.000   76.746
\LT  264.000   77.085
\LT  268.000   77.486
\LT  272.000   77.946
\LT  276.000   78.461
\LT  281.000   79.180
\LT  286.000   79.975
\LT  291.000   80.844
\LT  296.000   81.780
\LT  301.000   82.779
\LT  306.000   83.838
\LT  312.000   85.182
\LT  318.000   86.602
\LT  324.000   88.091
\LT  330.000   89.645
\LT  337.000   91.535
\LT  344.000   93.504
\LT  351.000   95.545
\LT  358.000   97.656
\LT  365.000   99.832
\LT  373.000  102.396
\LT  381.000  105.037
\LT  389.000  107.753
\LT  397.000  110.541
\LT  405.000  113.398
\LT  413.000  116.323
\LT  421.000  119.314
\LT  429.000  122.370
\LT  438.000  125.884
\LT  447.000  129.478
\LT  456.000  133.150
\LT  465.000  136.899
\LT  474.000  140.726
\LT  483.000  144.629
\LT  492.000  148.608
\LT  501.000  152.663
\LT  510.000  156.795
\LT  519.000  161.001
\LT  528.000  165.284
\LT  537.000  169.642
\LT  546.000  174.076
\LT  555.000  178.585
\LT  564.000  183.170
\LT  573.000  187.832
\LT  582.000  192.569
\LT  591.000  197.382
\LT  600.000  202.272
\LT  609.000  207.238
\LT  618.000  212.281
\LT  627.000  217.400
\LT  636.000  222.597
\LT  645.000  227.870
\LT  654.000  233.221
\LT  663.000  238.649
\LT  672.000  244.155
\LT  681.000  249.738
\LT  690.000  255.399
\LT  699.000  261.138
\LT  708.000  266.955
\LT  717.000  272.851
\LT  726.000  278.825
\LT  735.000  284.877
\LT  744.000  291.008
\LT  753.000  297.218
\LT  762.000  303.507
\LT  771.000  309.874
\LT  780.000  316.321
\LT  789.000  322.847
\LT  798.000  329.453
\LT  807.000  336.138
\LT  816.000  342.902
\LT  825.000  349.747
\LT  834.000  356.671
\LT  843.000  363.675
\LT  852.000  370.758
\LT  861.000  377.922
\LT  870.000  385.166
\LT  879.000  392.490
\LT  888.000  399.895
\LT  897.000  407.380
\LT  906.000  414.945
\LT  915.000  422.591
\LT  924.000  430.317
\LT  933.000  438.124
\LT  942.000  446.012
\LT  951.000  453.981
\LT  960.000  462.030
\LT  969.000  470.160
\LT  978.000  478.371
\LT  986.000  485.738
\LT  994.000  493.169
\LT 1002.000  500.664
\LT 1010.000  508.223
\LT 1018.000  515.847
\LT 1026.000  523.534
\LT 1034.000  531.286
\LT 1042.000  539.102
\LT 1050.000  546.982
\LT 1058.000  554.927
\LT 1066.000  562.936
\LT 1074.000  571.009
\LT 1082.000  579.147
\LT 1090.000  587.349
\LT 1098.000  595.616
\LT 1106.000  603.947
\LT 1114.000  612.342
\LT 1122.000  620.803
\LT 1130.000  629.327
\LT 1138.000  637.917
\LT 1146.000  646.571
\LT 1154.000  655.289
\LT 1162.000  664.073
\LT 1170.000  672.921
\LT 1178.000  681.833
\LT 1186.000  690.811
\LT 1194.000  699.853
\LT 1202.000  708.959
\LT 1210.000  718.131
\LT 1218.000  727.368
\LT 1226.000  736.669
\LT 1234.000  746.035
\LT 1242.000  755.466
\LT 1250.000  764.962
\LT 1258.000  774.522
\LT 1266.000  784.148
\LT 1274.000  793.838
\LT 1282.000  803.594
\LT 1290.000  813.414
\LT 1298.000  823.299
\LT 1300.000  825.781
\koniec    3.00000   0.200
\obraz16
\grub0.2pt
\MT   0.000   60.000
\LT1400.000   60.000
\MT 160.000   60.000
\LT 160.000   70.000
\MT 220.000   60.000
\LT 220.000   70.000
\MT 280.000   60.000
\LT 280.000   70.000
\MT 340.000   60.000
\LT 340.000   70.000
\cput(340.000,-10.000,2)
\MT 400.000   60.000
\LT 400.000   70.000
\MT 460.000   60.000
\LT 460.000   70.000
\MT 520.000   60.000
\LT 520.000   70.000
\MT 580.000   60.000
\LT 580.000   70.000
\cput(580.000,-10.000,4)
\MT 640.000   60.000
\LT 640.000   70.000
\MT 700.000   60.000
\LT 700.000   70.000
\MT 760.000   60.000
\LT 760.000   70.000
\MT 820.000   60.000
\LT 820.000   70.000
\cput(820.000,-10.000,6)
\MT 880.000   60.000
\LT 880.000   70.000
\MT 940.000   60.000
\LT 940.000   70.000
\MT1000.000   60.000
\LT1000.000   70.000
\MT1060.000   60.000
\LT1060.000   70.000
\cput(1060.000,-10.000,8)
\MT1120.000   60.000
\LT1120.000   70.000
\MT1180.000   60.000
\LT1180.000   70.000
\MT1240.000   60.000
\LT1240.000   70.000
\MT1300.000   60.000
\LT1300.000   70.000
\cput(1300.000,-10.000,10)
\MT 100.000    0.000
\LT 100.000  850.000
\MT  92.000   60.000
\LT 108.000   60.000
\MT  92.000   97.000
\LT 108.000   97.000
\MT  92.000  134.000
\LT 108.000  134.000
\MT  92.000  171.000
\LT 108.000  171.000
\MT  92.000  208.000
\LT 108.000  208.000
\MT  92.000  245.000
\LT 108.000  245.000
\MT  92.000  282.000
\LT 108.000  282.000
\MT  92.000  319.000
\LT 108.000  319.000
\MT  92.000  356.000
\LT 108.000  356.000
\MT  92.000  393.000
\LT 108.000  393.000
\MT  92.000  430.000
\LT 108.000  430.000
\MT  92.000  467.000
\LT 108.000  467.000
\MT  92.000  504.000
\LT 108.000  504.000
\MT  92.000  541.000
\LT 108.000  541.000
\MT  92.000  578.000
\LT 108.000  578.000
\MT  92.000  615.000
\LT 108.000  615.000
\MT  92.000  652.000
\LT 108.000  652.000
\MT  92.000  689.000
\LT 108.000  689.000
\MT  92.000  726.000
\LT 108.000  726.000
\MT  92.000  763.000
\LT 108.000  763.000
\MT  92.000  800.000
\LT 108.000  800.000
\MT  84.000   60.000
\LT 116.000   60.000
\MT  84.000  208.000
\LT 116.000  208.000
\lput(80.000,189.500,   20)
\MT  84.000  356.000
\LT 116.000  356.000
\lput(80.000,337.500,   40)
\MT  84.000  504.000
\LT 116.000  504.000
\lput(80.000,485.500,   60)
\MT  84.000  652.000
\LT 116.000  652.000
\lput(80.000,633.500,   80)
\MT  84.000  800.000
\LT 116.000  800.000
\lput(80.000,781.500,  100)
\grub0.6pt
\MT  100.000   67.400
\LT  150.000   67.416
\LT  165.000   67.476
\LT  176.000   67.586
\LT  185.000   67.746
\LT  193.000   67.962
\LT  200.000   68.220
\LT  207.000   68.552
\LT  213.000   68.901
\LT  219.000   69.314
\LT  225.000   69.791
\LT  231.000   70.335
\LT  237.000   70.944
\LT  243.000   71.619
\LT  249.000   72.358
\LT  255.000   73.160
\LT  262.000   74.171
\LT  269.000   75.261
\LT  276.000   76.427
\LT  283.000   77.665
\LT  290.000   78.974
\LT  297.000   80.350
\LT  304.000   81.790
\LT  311.000   83.294
\LT  319.000   85.086
\LT  327.000   86.956
\LT  335.000   88.901
\LT  343.000   90.919
\LT  351.000   93.010
\LT  359.000   95.170
\LT  367.000   97.401
\LT  375.000   99.700
\LT  383.000  102.067
\LT  391.000  104.502
\LT  399.000  107.004
\LT  407.000  109.571
\LT  415.000  112.205
\LT  423.000  114.905
\LT  431.000  117.671
\LT  439.000  120.501
\LT  447.000  123.397
\LT  455.000  126.358
\LT  463.000  129.384
\LT  471.000  132.474
\LT  479.000  135.629
\LT  487.000  138.849
\LT  495.000  142.134
\LT  503.000  145.483
\LT  511.000  148.897
\LT  519.000  152.375
\LT  527.000  155.918
\LT  535.000  159.526
\LT  543.000  163.198
\LT  551.000  166.935
\LT  559.000  170.736
\LT  567.000  174.602
\LT  575.000  178.533
\LT  583.000  182.528
\LT  591.000  186.588
\LT  599.000  190.712
\LT  607.000  194.901
\LT  615.000  199.155
\LT  623.000  203.474
\LT  631.000  207.857
\LT  639.000  212.306
\LT  647.000  216.819
\LT  655.000  221.397
\LT  663.000  226.039
\LT  671.000  230.747
\LT  679.000  235.519
\LT  687.000  240.357
\LT  695.000  245.259
\LT  703.000  250.226
\LT  711.000  255.259
\LT  719.000  260.356
\LT  727.000  265.518
\LT  735.000  270.746
\LT  743.000  276.038
\LT  751.000  281.396
\LT  759.000  286.818
\LT  767.000  292.306
\LT  775.000  297.858
\LT  783.000  303.476
\LT  791.000  309.159
\LT  799.000  314.908
\LT  807.000  320.721
\LT  815.000  326.600
\LT  823.000  332.543
\LT  831.000  338.553
\LT  839.000  344.627
\LT  847.000  350.766
\LT  855.000  356.971
\LT  863.000  363.241
\LT  871.000  369.577
\LT  879.000  375.977
\LT  887.000  382.443
\LT  895.000  388.974
\LT  903.000  395.571
\LT  911.000  402.233
\LT  919.000  408.960
\LT  927.000  415.753
\LT  935.000  422.611
\LT  943.000  429.534
\LT  951.000  436.523
\LT  959.000  443.577
\LT  967.000  450.697
\LT  975.000  457.882
\LT  983.000  465.132
\LT  991.000  472.448
\LT  999.000  479.830
\LT 1007.000  487.276
\LT 1015.000  494.789
\LT 1023.000  502.366
\LT 1031.000  510.009
\LT 1039.000  517.718
\LT 1047.000  525.492
\LT 1055.000  533.332
\LT 1063.000  541.237
\LT 1071.000  549.207
\LT 1079.000  557.243
\LT 1087.000  565.345
\LT 1095.000  573.512
\LT 1103.000  581.744
\LT 1111.000  590.043
\LT 1119.000  598.406
\LT 1127.000  606.835
\LT 1135.000  615.330
\LT 1143.000  623.890
\LT 1151.000  632.516
\LT 1159.000  641.208
\LT 1167.000  649.964
\LT 1175.000  658.787
\LT 1183.000  667.675
\LT 1191.000  676.629
\LT 1199.000  685.648
\LT 1207.000  694.733
\LT 1215.000  703.883
\LT 1223.000  713.099
\LT 1231.000  722.380
\LT 1239.000  731.727
\LT 1247.000  741.140
\LT 1255.000  750.618
\LT 1263.000  760.162
\LT 1271.000  769.772
\LT 1279.000  779.447
\LT 1287.000  789.188
\LT 1295.000  798.994
\LT 1300.000  805.156
\koniec    3.00000   1.000
\eject

\def\opis{Plots of the \f\ $f(x)$ (Eq.~\eqref{3.12}) for some values of
parameters $a$ and~$b$\break (see the text for explanation)\break
Fig.~\the\nd A: $a=2.00000,\ b=0.010$; \
Fig.~\the\nd B: $a=0.79126,\ b=0.001$;\break
Fig.~\the\nd C: $a=0.78126,\ b=-0.100$;\
Fig.~\the\nd D: $a=0.78126,\ b=-0.010$;\break
Fig.~\the\nd E: $a=0.78126,\ b=-0.001$;\
Fig.~\the\nd F: $a=0.78126,\ b=0.000$;\break
Fig.~\the\nd G: $a=0.78126,\ b=0.001$;\
Fig.~\the\nd H: $a=0.78126,\ b=0.010$.}

\obraz17
\grub0.2pt
\MT   0.000   60.000
\LT1400.000   60.000
\MT 160.000   60.000
\LT 160.000   70.000
\MT 220.000   60.000
\LT 220.000   70.000
\MT 280.000   60.000
\LT 280.000   70.000
\MT 340.000   60.000
\LT 340.000   70.000
\cput(340.000,-10.000,2)
\MT 400.000   60.000
\LT 400.000   70.000
\MT 460.000   60.000
\LT 460.000   70.000
\MT 520.000   60.000
\LT 520.000   70.000
\MT 580.000   60.000
\LT 580.000   70.000
\cput(580.000,-10.000,4)
\MT 640.000   60.000
\LT 640.000   70.000
\MT 700.000   60.000
\LT 700.000   70.000
\MT 760.000   60.000
\LT 760.000   70.000
\MT 820.000   60.000
\LT 820.000   70.000
\cput(820.000,-10.000,6)
\MT 880.000   60.000
\LT 880.000   70.000
\MT 940.000   60.000
\LT 940.000   70.000
\MT1000.000   60.000
\LT1000.000   70.000
\MT1060.000   60.000
\LT1060.000   70.000
\cput(1060.000,-10.000,8)
\MT1120.000   60.000
\LT1120.000   70.000
\MT1180.000   60.000
\LT1180.000   70.000
\MT1240.000   60.000
\LT1240.000   70.000
\MT1300.000   60.000
\LT1300.000   70.000
\cput(1300.000,-10.000,10)
\MT 100.000    0.000
\LT 100.000   70.000
\multi(100.000,70.000)(0.0000,4.0000){25}{\linia(0,0)(0.0000,2.0000)}
\MT 100.000  170.000
\LT 100.000  850.000
\MT  92.000   60.000
\LT 108.000   60.000
\MT  92.000   97.000
\LT 108.000   97.000
\MT  92.000  134.000
\LT 108.000  134.000
\MT  92.000  171.000
\LT 108.000  171.000
\MT  92.000  208.000
\LT 108.000  208.000
\MT  92.000  245.000
\LT 108.000  245.000
\MT  92.000  282.000
\LT 108.000  282.000
\MT  92.000  319.000
\LT 108.000  319.000
\MT  92.000  356.000
\LT 108.000  356.000
\MT  92.000  393.000
\LT 108.000  393.000
\MT  92.000  430.000
\LT 108.000  430.000
\MT  92.000  467.000
\LT 108.000  467.000
\MT  92.000  504.000
\LT 108.000  504.000
\MT  92.000  541.000
\LT 108.000  541.000
\MT  92.000  578.000
\LT 108.000  578.000
\MT  92.000  615.000
\LT 108.000  615.000
\MT  92.000  652.000
\LT 108.000  652.000
\MT  92.000  689.000
\LT 108.000  689.000
\MT  92.000  726.000
\LT 108.000  726.000
\MT  92.000  763.000
\LT 108.000  763.000
\MT  92.000  800.000
\LT 108.000  800.000
\MT  84.000   60.000
\LT 116.000   60.000
\MT  84.000  245.000
\LT 116.000  245.000
\lput(80.000,226.500,  0.5)
\MT  84.000  430.000
\LT 116.000  430.000
\lput(80.000,411.500,  1.0)
\MT  84.000  615.000
\LT 116.000  615.000
\lput(80.000,596.500,  1.5)
\MT  84.000  800.000
\LT 116.000  800.000
\lput(80.000,781.500,  2.0)
\grub0.6pt
\MT  100.000  430.000
\LT  101.000  429.983
\LT  102.000  429.933
\LT  103.000  429.848
\LT  104.000  429.730
\LT  105.000  429.578
\LT  106.000  429.393
\LT  107.000  429.173
\LT  108.000  428.920
\LT  109.000  428.633
\LT  110.000  428.313
\LT  111.000  427.958
\LT  112.000  427.570
\LT  113.000  427.149
\LT  114.000  426.693
\LT  115.000  426.204
\LT  116.000  425.681
\LT  117.000  425.125
\LT  118.000  424.534
\LT  119.000  423.911
\LT  120.000  423.253
\LT  121.000  422.562
\LT  122.000  421.838
\LT  123.000  421.080
\LT  124.000  420.288
\LT  125.000  419.463
\LT  126.000  418.605
\LT  127.000  417.714
\LT  128.000  416.789
\LT  129.000  415.831
\LT  130.000  414.840
\LT  131.000  413.817
\LT  132.000  412.760
\LT  133.000  411.671
\LT  134.000  410.550
\LT  135.000  409.396
\LT  136.000  408.210
\LT  137.000  406.992
\LT  138.000  405.742
\LT  139.000  404.460
\LT  140.000  403.148
\LT  141.000  401.804
\LT  142.000  400.429
\LT  143.000  399.024
\LT  144.000  397.588
\LT  145.000  396.123
\LT  146.000  394.628
\LT  147.000  393.104
\LT  148.000  391.551
\LT  149.000  389.970
\LT  150.000  388.361
\LT  151.000  386.724
\LT  152.000  385.061
\LT  153.000  383.371
\LT  154.000  381.655
\LT  155.000  379.913
\LT  156.000  378.147
\LT  157.000  376.357
\LT  158.000  374.543
\LT  159.000  372.706
\LT  160.000  370.847
\LT  161.000  368.967
\LT  163.000  365.144
\LT  165.000  361.245
\LT  167.000  357.276
\LT  169.000  353.244
\LT  171.000  349.155
\LT  173.000  345.019
\LT  176.000  338.740
\LT  180.000  330.275
\LT  187.000  315.393
\LT  190.000  309.062
\LT  193.000  302.801
\LT  195.000  298.678
\LT  197.000  294.605
\LT  199.000  290.589
\LT  201.000  286.638
\LT  203.000  282.757
\LT  205.000  278.952
\LT  206.000  277.080
\LT  207.000  275.229
\LT  208.000  273.400
\LT  209.000  271.594
\LT  210.000  269.810
\LT  211.000  268.049
\LT  212.000  266.313
\LT  213.000  264.600
\LT  214.000  262.913
\LT  215.000  261.250
\LT  216.000  259.613
\LT  217.000  258.002
\LT  218.000  256.417
\LT  219.000  254.859
\LT  220.000  253.327
\LT  221.000  251.822
\LT  222.000  250.344
\LT  223.000  248.893
\LT  224.000  247.469
\LT  225.000  246.073
\LT  226.000  244.704
\LT  227.000  243.363
\LT  228.000  242.050
\LT  229.000  240.764
\LT  230.000  239.506
\LT  231.000  238.275
\LT  232.000  237.072
\LT  233.000  235.896
\LT  234.000  234.747
\LT  235.000  233.626
\LT  236.000  232.532
\LT  237.000  231.464
\LT  238.000  230.424
\LT  239.000  229.410
\LT  240.000  228.423
\LT  241.000  227.461
\LT  242.000  226.526
\LT  243.000  225.617
\LT  244.000  224.733
\LT  245.000  223.874
\LT  246.000  223.041
\LT  247.000  222.232
\LT  248.000  221.448
\LT  249.000  220.688
\LT  250.000  219.953
\LT  251.000  219.241
\LT  252.000  218.552
\LT  253.000  217.886
\LT  254.000  217.243
\LT  255.000  216.623
\LT  256.000  216.025
\LT  257.000  215.449
\LT  258.000  214.894
\LT  259.000  214.361
\LT  260.000  213.849
\LT  261.000  213.357
\LT  262.000  212.886
\LT  264.000  212.002
\LT  266.000  211.197
\LT  268.000  210.466
\LT  270.000  209.808
\LT  272.000  209.221
\LT  274.000  208.700
\LT  276.000  208.245
\LT  278.000  207.853
\LT  280.000  207.521
\LT  282.000  207.248
\LT  284.000  207.031
\LT  286.000  206.868
\LT  288.000  206.757
\LT  290.000  206.695
\LT  292.000  206.682
\LT  294.000  206.714
\LT  296.000  206.791
\LT  298.000  206.910
\LT  301.000  207.164
\LT  304.000  207.504
\LT  307.000  207.924
\LT  310.000  208.419
\LT  313.000  208.986
\LT  316.000  209.618
\LT  319.000  210.313
\LT  322.000  211.065
\LT  325.000  211.871
\LT  328.000  212.728
\LT  331.000  213.632
\LT  335.000  214.906
\LT  339.000  216.251
\LT  343.000  217.661
\LT  347.000  219.131
\LT  352.000  221.043
\LT  357.000  223.029
\LT  362.000  225.082
\LT  368.000  227.621
\LT  374.000  230.232
\LT  381.000  233.356
\LT  388.000  236.550
\LT  396.000  240.270
\LT  405.000  244.527
\LT  416.000  249.806
\LT  430.000  256.609
\LT  451.000  266.911
\LT  497.000  289.551
\LT  524.000  302.768
\LT  550.000  315.418
\LT  580.000  329.931
\LT  622.000  350.148
\LT  677.000  376.600
\LT  708.000  391.577
\LT  733.000  403.725
\LT  756.000  414.975
\LT  777.000  425.320
\LT  797.000  435.247
\LT  816.000  444.753
\LT  834.000  453.830
\LT  852.000  462.984
\LT  869.000  471.704
\LT  886.000  480.500
\LT  902.000  488.851
\LT  918.000  497.275
\LT  934.000  505.774
\LT  950.000  514.352
\LT  965.000  522.467
\LT  980.000  530.655
\LT  995.000  538.918
\LT 1010.000  547.256
\LT 1025.000  555.672
\LT 1040.000  564.168
\LT 1054.000  572.169
\LT 1068.000  580.242
\LT 1082.000  588.388
\LT 1096.000  596.606
\LT 1110.000  604.899
\LT 1124.000  613.267
\LT 1138.000  621.712
\LT 1152.000  630.233
\LT 1166.000  638.832
\LT 1180.000  647.510
\LT 1194.000  656.266
\LT 1207.000  664.469
\LT 1220.000  672.741
\LT 1233.000  681.084
\LT 1246.000  689.497
\LT 1259.000  697.980
\LT 1272.000  706.535
\LT 1285.000  715.162
\LT 1298.000  723.862
\LT 1300.000  725.207
\koniec    2.00000   0.010
\obraz18
\grub0.2pt
\MT   0.000   60.000
\LT1400.000   60.000
\MT 160.000   60.000
\LT 160.000   70.000
\MT 220.000   60.000
\LT 220.000   70.000
\MT 280.000   60.000
\LT 280.000   70.000
\MT 340.000   60.000
\LT 340.000   70.000
\cput(340.000,-10.000,2)
\MT 400.000   60.000
\LT 400.000   70.000
\MT 460.000   60.000
\LT 460.000   70.000
\MT 520.000   60.000
\LT 520.000   70.000
\MT 580.000   60.000
\LT 580.000   70.000
\cput(580.000,-10.000,4)
\MT 640.000   60.000
\LT 640.000   70.000
\MT 700.000   60.000
\LT 700.000   70.000
\MT 760.000   60.000
\LT 760.000   70.000
\MT 820.000   60.000
\LT 820.000   70.000
\cput(820.000,-10.000,6)
\MT 880.000   60.000
\LT 880.000   70.000
\MT 940.000   60.000
\LT 940.000   70.000
\MT1000.000   60.000
\LT1000.000   70.000
\MT1060.000   60.000
\LT1060.000   70.000
\cput(1060.000,-10.000,8)
\MT1120.000   60.000
\LT1120.000   70.000
\MT1180.000   60.000
\LT1180.000   70.000
\MT1240.000   60.000
\LT1240.000   70.000
\MT1300.000   60.000
\LT1300.000   70.000
\cput(1300.000,-10.000,10)
\MT 100.000    0.000
\LT 100.000   70.000
\multi(100.000,70.000)(0.0000,4.0000){25}{\linia(0,0)(0.0000,2.0000)}
\MT 100.000  170.000
\LT 100.000  850.000
\MT  92.000  200.000
\LT 108.000  200.000
\MT  92.000  224.000
\LT 108.000  224.000
\MT  92.000  248.000
\LT 108.000  248.000
\MT  92.000  272.000
\LT 108.000  272.000
\MT  92.000  296.000
\LT 108.000  296.000
\MT  92.000  320.000
\LT 108.000  320.000
\MT  92.000  344.000
\LT 108.000  344.000
\MT  92.000  368.000
\LT 108.000  368.000
\MT  92.000  392.000
\LT 108.000  392.000
\MT  92.000  416.000
\LT 108.000  416.000
\MT  92.000  440.000
\LT 108.000  440.000
\MT  92.000  464.000
\LT 108.000  464.000
\MT  92.000  488.000
\LT 108.000  488.000
\MT  92.000  512.000
\LT 108.000  512.000
\MT  92.000  536.000
\LT 108.000  536.000
\MT  92.000  560.000
\LT 108.000  560.000
\MT  92.000  584.000
\LT 108.000  584.000
\MT  92.000  608.000
\LT 108.000  608.000
\MT  92.000  632.000
\LT 108.000  632.000
\MT  92.000  656.000
\LT 108.000  656.000
\MT  92.000  680.000
\LT 108.000  680.000
\MT  92.000  704.000
\LT 108.000  704.000
\MT  92.000  728.000
\LT 108.000  728.000
\MT  92.000  752.000
\LT 108.000  752.000
\MT  92.000  776.000
\LT 108.000  776.000
\MT  92.000  800.000
\LT 108.000  800.000
\MT  84.000  200.000
\LT 116.000  200.000
\lput(80.000,188.000, 0.75)
\MT  84.000  320.000
\LT 116.000  320.000
\lput(80.000,308.000, 0.80)
\MT  84.000  440.000
\LT 116.000  440.000
\lput(80.000,428.000, 0.85)
\MT  84.000  560.000
\LT 116.000  560.000
\lput(80.000,548.000, 0.90)
\MT  84.000  680.000
\LT 116.000  680.000
\lput(80.000,668.000, 0.95)
\MT  84.000  800.000
\LT 116.000  800.000
\lput(80.000,788.000, 1.00)
\grub0.6pt
\MT  100.000  800.000
\LT  101.000  799.956
\LT  102.000  799.825
\LT  103.000  799.606
\LT  104.000  799.299
\LT  105.000  798.905
\LT  106.000  798.423
\LT  107.000  797.854
\LT  108.000  797.197
\LT  109.000  796.453
\LT  110.000  795.621
\LT  111.000  794.701
\LT  112.000  793.694
\LT  113.000  792.600
\LT  114.000  791.417
\LT  115.000  790.148
\LT  116.000  788.791
\LT  117.000  787.346
\LT  118.000  785.814
\LT  119.000  784.195
\LT  120.000  782.489
\LT  121.000  780.695
\LT  122.000  778.815
\LT  123.000  776.847
\LT  124.000  774.793
\LT  125.000  772.652
\LT  126.000  770.424
\LT  127.000  768.111
\LT  128.000  765.711
\LT  129.000  763.225
\LT  130.000  760.653
\LT  131.000  757.996
\LT  132.000  755.254
\LT  133.000  752.427
\LT  134.000  749.516
\LT  135.000  746.521
\LT  136.000  743.442
\LT  137.000  740.280
\LT  138.000  737.036
\LT  139.000  733.709
\LT  140.000  730.302
\LT  141.000  726.813
\LT  142.000  723.245
\LT  143.000  719.597
\LT  144.000  715.870
\LT  145.000  712.066
\LT  146.000  708.185
\LT  147.000  704.228
\LT  148.000  700.196
\LT  149.000  696.091
\LT  150.000  691.912
\LT  151.000  687.663
\LT  152.000  683.343
\LT  153.000  678.954
\LT  154.000  674.498
\LT  155.000  669.975
\LT  156.000  665.388
\LT  157.000  660.738
\LT  158.000  656.026
\LT  159.000  651.255
\LT  160.000  646.426
\LT  161.000  641.540
\LT  162.000  636.601
\LT  163.000  631.609
\LT  164.000  626.567
\LT  165.000  621.477
\LT  166.000  616.341
\LT  167.000  611.161
\LT  168.000  605.939
\LT  169.000  600.678
\LT  170.000  595.381
\LT  171.000  590.049
\LT  172.000  584.685
\LT  173.000  579.291
\LT  174.000  573.870
\LT  175.000  568.426
\LT  177.000  557.473
\LT  179.000  546.455
\LT  182.000  529.852
\LT  186.000  507.688
\LT  188.000  496.644
\LT  190.000  485.653
\LT  191.000  480.185
\LT  192.000  474.737
\LT  193.000  469.314
\LT  194.000  463.916
\LT  195.000  458.548
\LT  196.000  453.211
\LT  197.000  447.907
\LT  198.000  442.639
\LT  199.000  437.409
\LT  200.000  432.219
\LT  201.000  427.071
\LT  202.000  421.967
\LT  203.000  416.910
\LT  204.000  411.901
\LT  205.000  406.941
\LT  206.000  402.033
\LT  207.000  397.179
\LT  208.000  392.379
\LT  209.000  387.636
\LT  210.000  382.950
\LT  211.000  378.323
\LT  212.000  373.757
\LT  213.000  369.251
\LT  214.000  364.809
\LT  215.000  360.430
\LT  216.000  356.115
\LT  217.000  351.866
\LT  218.000  347.683
\LT  219.000  343.566
\LT  220.000  339.517
\LT  221.000  335.536
\LT  222.000  331.623
\LT  223.000  327.779
\LT  224.000  324.004
\LT  225.000  320.299
\LT  226.000  316.663
\LT  227.000  313.097
\LT  228.000  309.600
\LT  229.000  306.174
\LT  230.000  302.817
\LT  231.000  299.530
\LT  232.000  296.313
\LT  233.000  293.164
\LT  234.000  290.086
\LT  235.000  287.075
\LT  236.000  284.134
\LT  237.000  281.261
\LT  238.000  278.455
\LT  239.000  275.717
\LT  240.000  273.045
\LT  241.000  270.440
\LT  242.000  267.901
\LT  243.000  265.426
\LT  244.000  263.017
\LT  245.000  260.671
\LT  246.000  258.389
\LT  247.000  256.170
\LT  248.000  254.012
\LT  249.000  251.916
\LT  250.000  249.880
\LT  251.000  247.904
\LT  252.000  245.988
\LT  253.000  244.129
\LT  254.000  242.328
\LT  255.000  240.584
\LT  256.000  238.896
\LT  257.000  237.263
\LT  258.000  235.685
\LT  259.000  234.160
\LT  260.000  232.688
\LT  261.000  231.268
\LT  262.000  229.899
\LT  263.000  228.581
\LT  264.000  227.312
\LT  265.000  226.092
\LT  266.000  224.920
\LT  267.000  223.795
\LT  268.000  222.716
\LT  269.000  221.683
\LT  270.000  220.694
\LT  271.000  219.750
\LT  272.000  218.849
\LT  273.000  217.990
\LT  274.000  217.172
\LT  275.000  216.396
\LT  276.000  215.660
\LT  277.000  214.963
\LT  278.000  214.304
\LT  279.000  213.684
\LT  280.000  213.101
\LT  281.000  212.554
\LT  282.000  212.043
\LT  283.000  211.567
\LT  284.000  211.125
\LT  285.000  210.717
\LT  286.000  210.341
\LT  287.000  209.999
\LT  288.000  209.687
\LT  289.000  209.407
\LT  290.000  209.157
\LT  291.000  208.937
\LT  292.000  208.746
\LT  293.000  208.584
\LT  294.000  208.449
\LT  295.000  208.342
\LT  296.000  208.261
\LT  297.000  208.207
\LT  298.000  208.178
\LT  299.000  208.175
\LT  300.000  208.196
\LT  301.000  208.240
\LT  302.000  208.309
\LT  303.000  208.400
\LT  304.000  208.514
\LT  305.000  208.649
\LT  306.000  208.807
\LT  307.000  208.985
\LT  308.000  209.183
\LT  310.000  209.640
\LT  312.000  210.174
\LT  314.000  210.781
\LT  316.000  211.457
\LT  318.000  212.201
\LT  320.000  213.007
\LT  322.000  213.875
\LT  324.000  214.800
\LT  326.000  215.780
\LT  328.000  216.813
\LT  330.000  217.896
\LT  332.000  219.026
\LT  334.000  220.201
\LT  336.000  221.419
\LT  339.000  223.323
\LT  342.000  225.311
\LT  345.000  227.378
\LT  348.000  229.518
\LT  351.000  231.724
\LT  354.000  233.993
\LT  357.000  236.318
\LT  360.000  238.695
\LT  364.000  241.938
\LT  368.000  245.256
\LT  372.000  248.640
\LT  376.000  252.083
\LT  381.000  256.458
\LT  386.000  260.899
\LT  392.000  266.301
\LT  399.000  272.679
\LT  408.000  280.964
\LT  424.000  295.806
\LT  440.000  310.642
\LT  451.000  320.770
\LT  460.000  328.990
\LT  469.000  337.136
\LT  477.000  344.307
\LT  485.000  351.408
\LT  493.000  358.433
\LT  501.000  365.382
\LT  508.000  371.397
\LT  515.000  377.350
\LT  522.000  383.240
\LT  529.000  389.068
\LT  536.000  394.834
\LT  543.000  400.538
\LT  550.000  406.179
\LT  558.000  412.552
\LT  566.000  418.846
\LT  574.000  425.063
\LT  582.000  431.203
\LT  590.000  437.269
\LT  598.000  443.262
\LT  606.000  449.185
\LT  614.000  455.037
\LT  622.000  460.823
\LT  630.000  466.543
\LT  639.000  472.902
\LT  648.000  479.182
\LT  657.000  485.388
\LT  666.000  491.521
\LT  675.000  497.584
\LT  684.000  503.580
\LT  694.000  510.165
\LT  704.000  516.674
\LT  714.000  523.110
\LT  724.000  529.475
\LT  735.000  536.400
\LT  746.000  543.248
\LT  757.000  550.022
\LT  768.000  556.727
\LT  780.000  563.968
\LT  792.000  571.134
\LT  805.000  578.820
\LT  818.000  586.431
\LT  832.000  594.547
\LT  846.000  602.589
\LT  861.000  611.127
\LT  877.000  620.153
\LT  893.000  629.104
\LT  910.000  638.539
\LT  928.000  648.454
\LT  948.000  659.392
\LT  970.000  671.340
\LT  994.000  684.293
\LT 1022.000  699.320
\LT 1056.000  717.478
\LT 1115.000  748.873
\LT 1170.000  778.146
\LT 1207.000  797.907
\LT 1238.000  814.536
\LT 1266.000  829.627
\LT 1293.000  844.255
\LT 1300.000  848.061
\koniec    0.79126   0.001
\obraz19
\grub0.2pt
\MT   0.000  800.000
\LT1400.000  800.000
\MT 160.000  800.000
\LT 160.000  810.000
\MT 220.000  800.000
\LT 220.000  810.000
\MT 280.000  800.000
\LT 280.000  810.000
\MT 340.000  800.000
\LT 340.000  810.000
\MT 400.000  800.000
\LT 400.000  810.000
\cput(400.000,730.000,5)
\MT 460.000  800.000
\LT 460.000  810.000
\MT 520.000  800.000
\LT 520.000  810.000
\MT 580.000  800.000
\LT 580.000  810.000
\MT 640.000  800.000
\LT 640.000  810.000
\MT 700.000  800.000
\LT 700.000  810.000
\cput(700.000,730.000,10)
\MT 760.000  800.000
\LT 760.000  810.000
\MT 820.000  800.000
\LT 820.000  810.000
\MT 880.000  800.000
\LT 880.000  810.000
\MT 940.000  800.000
\LT 940.000  810.000
\MT1000.000  800.000
\LT1000.000  810.000
\cput(1000.000,730.000,15)
\MT1060.000  800.000
\LT1060.000  810.000
\MT1120.000  800.000
\LT1120.000  810.000
\MT1180.000  800.000
\LT1180.000  810.000
\MT1240.000  800.000
\LT1240.000  810.000
\MT1300.000  800.000
\LT1300.000  810.000
\cput(1300.000,730.000,20)
\MT 100.000    0.000
\LT 100.000  850.000
\MT  92.000   50.000
\LT 108.000   50.000
\MT  92.000   87.500
\LT 108.000   87.500
\MT  92.000  125.000
\LT 108.000  125.000
\MT  92.000  162.500
\LT 108.000  162.500
\MT  92.000  200.000
\LT 108.000  200.000
\MT  92.000  237.500
\LT 108.000  237.500
\MT  92.000  275.000
\LT 108.000  275.000
\MT  92.000  312.500
\LT 108.000  312.500
\MT  92.000  350.000
\LT 108.000  350.000
\MT  92.000  387.500
\LT 108.000  387.500
\MT  92.000  425.000
\LT 108.000  425.000
\MT  92.000  462.500
\LT 108.000  462.500
\MT  92.000  500.000
\LT 108.000  500.000
\MT  92.000  537.500
\LT 108.000  537.500
\MT  92.000  575.000
\LT 108.000  575.000
\MT  92.000  612.500
\LT 108.000  612.500
\MT  92.000  650.000
\LT 108.000  650.000
\MT  92.000  687.500
\LT 108.000  687.500
\MT  92.000  725.000
\LT 108.000  725.000
\MT  92.000  762.500
\LT 108.000  762.500
\MT  92.000  800.000
\LT 108.000  800.000
\MT  84.000   50.000
\LT 116.000   50.000
\lput(80.000,31.250,  -40)
\MT  84.000  237.500
\LT 116.000  237.500
\lput(80.000,218.750,  -30)
\MT  84.000  425.000
\LT 116.000  425.000
\lput(80.000,406.250,  -20)
\MT  84.000  612.500
\LT 116.000  612.500
\lput(80.000,593.750,  -10)
\MT  84.000  800.000
\LT 116.000  800.000
\grub0.6pt
\MT  100.000  818.750
\LT  104.000  818.720
\LT  108.000  818.630
\LT  112.000  818.480
\LT  116.000  818.270
\LT  120.000  818.002
\LT  124.000  817.677
\LT  129.000  817.197
\LT  134.000  816.645
\LT  140.000  815.910
\LT  158.000  813.557
\LT  166.000  812.574
\LT  175.000  811.548
\LT  187.000  810.273
\LT  206.000  808.290
\LT  217.000  807.077
\LT  227.000  805.900
\LT  236.000  804.768
\LT  245.000  803.561
\LT  253.000  802.419
\LT  261.000  801.213
\LT  269.000  799.939
\LT  277.000  798.596
\LT  285.000  797.185
\LT  293.000  795.704
\LT  301.000  794.154
\LT  309.000  792.534
\LT  317.000  790.844
\LT  325.000  789.085
\LT  333.000  787.255
\LT  341.000  785.356
\LT  349.000  783.387
\LT  357.000  781.349
\LT  365.000  779.241
\LT  373.000  777.064
\LT  381.000  774.818
\LT  389.000  772.503
\LT  397.000  770.118
\LT  405.000  767.665
\LT  413.000  765.144
\LT  421.000  762.553
\LT  429.000  759.894
\LT  437.000  757.167
\LT  445.000  754.371
\LT  453.000  751.507
\LT  461.000  748.575
\LT  469.000  745.574
\LT  477.000  742.506
\LT  485.000  739.370
\LT  493.000  736.165
\LT  501.000  732.893
\LT  509.000  729.553
\LT  517.000  726.145
\LT  525.000  722.669
\LT  533.000  719.126
\LT  541.000  715.515
\LT  549.000  711.837
\LT  557.000  708.091
\LT  565.000  704.277
\LT  573.000  700.396
\LT  581.000  696.448
\LT  589.000  692.432
\LT  597.000  688.349
\LT  605.000  684.198
\LT  613.000  679.980
\LT  621.000  675.695
\LT  629.000  671.342
\LT  637.000  666.923
\LT  645.000  662.436
\LT  653.000  657.882
\LT  661.000  653.260
\LT  669.000  648.572
\LT  677.000  643.816
\LT  685.000  638.993
\LT  693.000  634.103
\LT  701.000  629.146
\LT  709.000  624.122
\LT  717.000  619.031
\LT  725.000  613.873
\LT  733.000  608.647
\LT  741.000  603.355
\LT  749.000  597.995
\LT  757.000  592.569
\LT  765.000  587.076
\LT  773.000  581.515
\LT  781.000  575.888
\LT  789.000  570.194
\LT  797.000  564.432
\LT  805.000  558.604
\LT  813.000  552.709
\LT  821.000  546.747
\LT  829.000  540.718
\LT  837.000  534.622
\LT  845.000  528.459
\LT  853.000  522.229
\LT  861.000  515.933
\LT  869.000  509.569
\LT  877.000  503.139
\LT  885.000  496.641
\LT  893.000  490.077
\LT  901.000  483.446
\LT  909.000  476.748
\LT  917.000  469.984
\LT  925.000  463.152
\LT  933.000  456.254
\LT  941.000  449.288
\LT  949.000  442.256
\LT  957.000  435.157
\LT  965.000  427.991
\LT  973.000  420.759
\LT  981.000  413.459
\LT  989.000  406.093
\LT  997.000  398.660
\LT 1005.000  391.160
\LT 1013.000  383.593
\LT 1021.000  375.960
\LT 1029.000  368.260
\LT 1037.000  360.493
\LT 1045.000  352.659
\LT 1053.000  344.758
\LT 1061.000  336.791
\LT 1069.000  328.757
\LT 1077.000  320.656
\LT 1085.000  312.488
\LT 1093.000  304.253
\LT 1101.000  295.952
\LT 1109.000  287.584
\LT 1117.000  279.149
\LT 1125.000  270.647
\LT 1133.000  262.079
\LT 1141.000  253.444
\LT 1149.000  244.742
\LT 1157.000  235.973
\LT 1165.000  227.138
\LT 1173.000  218.235
\LT 1181.000  209.267
\LT 1189.000  200.231
\LT 1197.000  191.128
\LT 1205.000  181.959
\LT 1213.000  172.723
\LT 1221.000  163.421
\LT 1229.000  154.051
\LT 1237.000  144.615
\LT 1245.000  135.112
\LT 1253.000  125.542
\LT 1261.000  115.906
\LT 1269.000  106.203
\LT 1277.000   96.433
\LT 1285.000   86.597
\LT 1293.000   76.693
\LT 1300.000   67.973
\koniec    0.78126  -0.100
\obraz20
\grub0.2pt
\MT   0.000  612.000
\LT1400.000  612.000
\MT 160.000  612.000
\LT 160.000  622.000
\MT 220.000  612.000
\LT 220.000  622.000
\MT 280.000  612.000
\LT 280.000  622.000
\MT 340.000  612.000
\LT 340.000  622.000
\MT 400.000  612.000
\LT 400.000  622.000
\cput(400.000,542.000,5)
\MT 460.000  612.000
\LT 460.000  622.000
\MT 520.000  612.000
\LT 520.000  622.000
\MT 580.000  612.000
\LT 580.000  622.000
\MT 640.000  612.000
\LT 640.000  622.000
\MT 700.000  612.000
\LT 700.000  622.000
\cput(700.000,542.000,10)
\MT 760.000  612.000
\LT 760.000  622.000
\MT 820.000  612.000
\LT 820.000  622.000
\MT 880.000  612.000
\LT 880.000  622.000
\MT 940.000  612.000
\LT 940.000  622.000
\MT1000.000  612.000
\LT1000.000  622.000
\cput(1000.000,542.000,15)
\MT1060.000  612.000
\LT1060.000  622.000
\MT1120.000  612.000
\LT1120.000  622.000
\MT1180.000  612.000
\LT1180.000  622.000
\MT1240.000  612.000
\LT1240.000  622.000
\MT1300.000  612.000
\LT1300.000  622.000
\cput(1300.000,542.000,20)
\MT 100.000    0.000
\LT 100.000  850.000
\MT  92.000   50.000
\LT 108.000   50.000
\MT  92.000   87.500
\LT 108.000   87.500
\MT  92.000  125.000
\LT 108.000  125.000
\MT  92.000  162.500
\LT 108.000  162.500
\MT  92.000  200.000
\LT 108.000  200.000
\MT  92.000  237.500
\LT 108.000  237.500
\MT  92.000  275.000
\LT 108.000  275.000
\MT  92.000  312.500
\LT 108.000  312.500
\MT  92.000  350.000
\LT 108.000  350.000
\MT  92.000  387.500
\LT 108.000  387.500
\MT  92.000  425.000
\LT 108.000  425.000
\MT  92.000  462.500
\LT 108.000  462.500
\MT  92.000  500.000
\LT 108.000  500.000
\MT  92.000  537.500
\LT 108.000  537.500
\MT  92.000  575.000
\LT 108.000  575.000
\MT  92.000  612.500
\LT 108.000  612.500
\MT  92.000  650.000
\LT 108.000  650.000
\MT  92.000  687.500
\LT 108.000  687.500
\MT  92.000  725.000
\LT 108.000  725.000
\MT  92.000  762.500
\LT 108.000  762.500
\MT  92.000  800.000
\LT 108.000  800.000
\MT  84.000   50.000
\LT 116.000   50.000
\lput(80.000,31.250, -3.0)
\MT  84.000  237.500
\LT 116.000  237.500
\lput(80.000,218.750, -2.0)
\MT  84.000  425.000
\LT 116.000  425.000
\lput(80.000,406.250, -1.0)
\MT  84.000  612.500
\LT 116.000  612.500
\MT  84.000  800.000
\LT 116.000  800.000
\lput(80.000,781.250,  1.0)
\grub0.6pt
\MT  100.000  800.000
\LT  101.000  799.986
\LT  102.000  799.944
\LT  103.000  799.873
\LT  104.000  799.775
\LT  105.000  799.648
\LT  106.000  799.493
\LT  107.000  799.310
\LT  108.000  799.099
\LT  109.000  798.859
\LT  110.000  798.592
\LT  111.000  798.297
\LT  112.000  797.973
\LT  113.000  797.622
\LT  114.000  797.243
\LT  115.000  796.836
\LT  116.000  796.402
\LT  117.000  795.940
\LT  118.000  795.452
\LT  119.000  794.937
\LT  120.000  794.395
\LT  121.000  793.827
\LT  122.000  793.233
\LT  123.000  792.615
\LT  124.000  791.972
\LT  125.000  791.305
\LT  126.000  790.614
\LT  127.000  789.902
\LT  128.000  789.168
\LT  130.000  787.639
\LT  132.000  786.036
\LT  134.000  784.369
\LT  136.000  782.649
\LT  139.000  779.997
\LT  146.000  773.699
\LT  149.000  771.060
\LT  151.000  769.351
\LT  153.000  767.694
\LT  155.000  766.099
\LT  157.000  764.571
\LT  159.000  763.117
\LT  161.000  761.741
\LT  162.000  761.083
\LT  163.000  760.445
\LT  164.000  759.828
\LT  165.000  759.231
\LT  166.000  758.654
\LT  167.000  758.098
\LT  169.000  757.045
\LT  171.000  756.070
\LT  173.000  755.172
\LT  175.000  754.346
\LT  177.000  753.590
\LT  179.000  752.899
\LT  181.000  752.269
\LT  183.000  751.697
\LT  185.000  751.179
\LT  187.000  750.711
\LT  189.000  750.288
\LT  192.000  749.732
\LT  195.000  749.259
\LT  198.000  748.859
\LT  201.000  748.521
\LT  204.000  748.236
\LT  208.000  747.926
\LT  212.000  747.681
\LT  217.000  747.442
\LT  224.000  747.195
\LT  236.000  746.885
\LT  249.000  746.548
\LT  258.000  746.249
\LT  266.000  745.911
\LT  273.000  745.552
\LT  280.000  745.126
\LT  287.000  744.631
\LT  294.000  744.063
\LT  301.000  743.421
\LT  308.000  742.705
\LT  315.000  741.913
\LT  322.000  741.046
\LT  329.000  740.103
\LT  336.000  739.086
\LT  343.000  737.995
\LT  350.000  736.831
\LT  357.000  735.593
\LT  364.000  734.284
\LT  371.000  732.903
\LT  378.000  731.452
\LT  385.000  729.931
\LT  392.000  728.342
\LT  399.000  726.684
\LT  406.000  724.959
\LT  413.000  723.166
\LT  420.000  721.308
\LT  427.000  719.384
\LT  434.000  717.396
\LT  441.000  715.343
\LT  448.000  713.226
\LT  455.000  711.046
\LT  462.000  708.804
\LT  469.000  706.499
\LT  476.000  704.133
\LT  484.000  701.353
\LT  492.000  698.494
\LT  500.000  695.556
\LT  508.000  692.540
\LT  516.000  689.447
\LT  524.000  686.275
\LT  532.000  683.028
\LT  540.000  679.703
\LT  548.000  676.303
\LT  556.000  672.827
\LT  564.000  669.276
\LT  572.000  665.651
\LT  580.000  661.951
\LT  588.000  658.176
\LT  596.000  654.328
\LT  604.000  650.406
\LT  612.000  646.411
\LT  620.000  642.343
\LT  628.000  638.202
\LT  636.000  633.989
\LT  644.000  629.703
\LT  652.000  625.345
\LT  660.000  620.916
\LT  668.000  616.414
\LT  676.000  611.841
\LT  684.000  607.197
\LT  692.000  602.481
\LT  700.000  597.694
\LT  708.000  592.837
\LT  716.000  587.908
\LT  724.000  582.910
\LT  732.000  577.840
\LT  740.000  572.701
\LT  748.000  567.491
\LT  756.000  562.211
\LT  764.000  556.860
\LT  772.000  551.441
\LT  780.000  545.951
\LT  788.000  540.391
\LT  796.000  534.762
\LT  804.000  529.064
\LT  812.000  523.296
\LT  820.000  517.458
\LT  828.000  511.552
\LT  836.000  505.576
\LT  844.000  499.531
\LT  852.000  493.417
\LT  860.000  487.235
\LT  868.000  480.983
\LT  876.000  474.662
\LT  884.000  468.273
\LT  892.000  461.815
\LT  900.000  455.288
\LT  908.000  448.692
\LT  916.000  442.028
\LT  924.000  435.296
\LT  932.000  428.495
\LT  940.000  421.626
\LT  948.000  414.688
\LT  956.000  407.682
\LT  964.000  400.607
\LT  972.000  393.465
\LT  980.000  386.254
\LT  988.000  378.975
\LT  996.000  371.628
\LT 1004.000  364.213
\LT 1012.000  356.730
\LT 1020.000  349.178
\LT 1028.000  341.559
\LT 1036.000  333.872
\LT 1044.000  326.117
\LT 1052.000  318.294
\LT 1060.000  310.403
\LT 1068.000  302.444
\LT 1076.000  294.418
\LT 1084.000  286.324
\LT 1092.000  278.162
\LT 1100.000  269.932
\LT 1108.000  261.634
\LT 1116.000  253.269
\LT 1124.000  244.836
\LT 1132.000  236.336
\LT 1140.000  227.767
\LT 1148.000  219.132
\LT 1156.000  210.428
\LT 1164.000  201.657
\LT 1172.000  192.819
\LT 1180.000  183.913
\LT 1188.000  174.939
\LT 1196.000  165.898
\LT 1204.000  156.790
\LT 1212.000  147.614
\LT 1220.000  138.371
\LT 1228.000  129.060
\LT 1236.000  119.682
\LT 1244.000  110.236
\LT 1252.000  100.723
\LT 1260.000   91.143
\LT 1268.000   81.495
\LT 1276.000   71.780
\LT 1284.000   61.998
\LT 1292.000   52.148
\LT 1300.000   42.231
\koniec    0.78126  -0.010
\obraz21
\grub0.2pt
\MT   0.000   60.000
\LT1400.000   60.000
\MT 160.000   60.000
\LT 160.000   70.000
\MT 220.000   60.000
\LT 220.000   70.000
\MT 280.000   60.000
\LT 280.000   70.000
\MT 340.000   60.000
\LT 340.000   70.000
\MT 400.000   60.000
\LT 400.000   70.000
\cput(400.000,-10.000,5)
\MT 460.000   60.000
\LT 460.000   70.000
\MT 520.000   60.000
\LT 520.000   70.000
\MT 580.000   60.000
\LT 580.000   70.000
\MT 640.000   60.000
\LT 640.000   70.000
\MT 700.000   60.000
\LT 700.000   70.000
\cput(700.000,-10.000,10)
\MT 760.000   60.000
\LT 760.000   70.000
\MT 820.000   60.000
\LT 820.000   70.000
\MT 880.000   60.000
\LT 880.000   70.000
\MT 940.000   60.000
\LT 940.000   70.000
\MT1000.000   60.000
\LT1000.000   70.000
\cput(1000.000,-10.000,15)
\MT1060.000   60.000
\LT1060.000   70.000
\MT1120.000   60.000
\LT1120.000   70.000
\MT1180.000   60.000
\LT1180.000   70.000
\MT1240.000   60.000
\LT1240.000   70.000
\MT1300.000   60.000
\LT1300.000   70.000
\cput(1300.000,-10.000,20)
\MT 100.000    0.000
\LT 100.000   70.000
\multi(100.000,70.000)(0.0000,4.0000){25}{\linia(0,0)(0.0000,2.0000)}
\MT 100.000  170.000
\LT 100.000  850.000
\MT  92.000  200.000
\LT 108.000  200.000
\MT  92.000  230.000
\LT 108.000  230.000
\MT  92.000  260.000
\LT 108.000  260.000
\MT  92.000  290.000
\LT 108.000  290.000
\MT  92.000  320.000
\LT 108.000  320.000
\MT  92.000  350.000
\LT 108.000  350.000
\MT  92.000  380.000
\LT 108.000  380.000
\MT  92.000  410.000
\LT 108.000  410.000
\MT  92.000  440.000
\LT 108.000  440.000
\MT  92.000  470.000
\LT 108.000  470.000
\MT  92.000  500.000
\LT 108.000  500.000
\MT  92.000  530.000
\LT 108.000  530.000
\MT  92.000  560.000
\LT 108.000  560.000
\MT  92.000  590.000
\LT 108.000  590.000
\MT  92.000  620.000
\LT 108.000  620.000
\MT  92.000  650.000
\LT 108.000  650.000
\MT  92.000  680.000
\LT 108.000  680.000
\MT  92.000  710.000
\LT 108.000  710.000
\MT  92.000  740.000
\LT 108.000  740.000
\MT  92.000  770.000
\LT 108.000  770.000
\MT  92.000  800.000
\LT 108.000  800.000
\MT  84.000  200.000
\LT 116.000  200.000
\lput(80.000,185.000, 0.60)
\MT  84.000  350.000
\LT 116.000  350.000
\lput(80.000,335.000, 0.70)
\MT  84.000  500.000
\LT 116.000  500.000
\lput(80.000,485.000, 0.80)
\MT  84.000  650.000
\LT 116.000  650.000
\lput(80.000,635.000, 0.90)
\MT  84.000  800.000
\LT 116.000  800.000
\lput(80.000,785.000, 1.00)
\grub0.6pt
\MT  100.000  800.000
\LT  101.000  799.891
\LT  102.000  799.564
\LT  103.000  799.020
\LT  104.000  798.257
\LT  105.000  797.277
\LT  106.000  796.079
\LT  107.000  794.663
\LT  108.000  793.030
\LT  109.000  791.179
\LT  110.000  789.111
\LT  111.000  786.826
\LT  112.000  784.326
\LT  113.000  781.609
\LT  114.000  778.678
\LT  115.000  775.533
\LT  116.000  772.175
\LT  117.000  768.607
\LT  118.000  764.830
\LT  119.000  760.846
\LT  120.000  756.658
\LT  121.000  752.269
\LT  122.000  747.683
\LT  123.000  742.903
\LT  124.000  737.934
\LT  125.000  732.781
\LT  126.000  727.451
\LT  127.000  721.949
\LT  128.000  716.282
\LT  129.000  710.457
\LT  130.000  704.484
\LT  131.000  698.370
\LT  132.000  692.126
\LT  133.000  685.762
\LT  134.000  679.288
\LT  135.000  672.715
\LT  136.000  666.055
\LT  137.000  659.322
\LT  138.000  652.526
\LT  139.000  645.682
\LT  140.000  638.802
\LT  143.000  618.083
\LT  144.000  611.195
\LT  145.000  604.339
\LT  146.000  597.528
\LT  147.000  590.773
\LT  148.000  584.088
\LT  149.000  577.484
\LT  150.000  570.971
\LT  151.000  564.562
\LT  152.000  558.264
\LT  153.000  552.088
\LT  154.000  546.042
\LT  155.000  540.133
\LT  156.000  534.368
\LT  157.000  528.753
\LT  158.000  523.293
\LT  159.000  517.992
\LT  160.000  512.854
\LT  161.000  507.883
\LT  162.000  503.079
\LT  163.000  498.445
\LT  164.000  493.982
\LT  165.000  489.689
\LT  166.000  485.567
\LT  167.000  481.614
\LT  168.000  477.829
\LT  169.000  474.211
\LT  170.000  470.758
\LT  171.000  467.467
\LT  172.000  464.334
\LT  173.000  461.359
\LT  174.000  458.536
\LT  175.000  455.863
\LT  176.000  453.335
\LT  177.000  450.951
\LT  178.000  448.704
\LT  179.000  446.593
\LT  180.000  444.612
\LT  181.000  442.757
\LT  182.000  441.026
\LT  183.000  439.413
\LT  184.000  437.914
\LT  185.000  436.527
\LT  186.000  435.246
\LT  187.000  434.069
\LT  188.000  432.990
\LT  189.000  432.007
\LT  190.000  431.116
\LT  191.000  430.314
\LT  192.000  429.596
\LT  193.000  428.959
\LT  194.000  428.401
\LT  195.000  427.917
\LT  196.000  427.505
\LT  197.000  427.162
\LT  198.000  426.885
\LT  199.000  426.670
\LT  200.000  426.516
\LT  201.000  426.420
\LT  202.000  426.378
\LT  203.000  426.389
\LT  204.000  426.450
\LT  205.000  426.559
\LT  206.000  426.713
\LT  207.000  426.911
\LT  208.000  427.151
\LT  209.000  427.430
\LT  210.000  427.747
\LT  211.000  428.099
\LT  212.000  428.485
\LT  213.000  428.904
\LT  214.000  429.353
\LT  215.000  429.832
\LT  216.000  430.338
\LT  217.000  430.870
\LT  218.000  431.427
\LT  219.000  432.008
\LT  220.000  432.611
\LT  222.000  433.879
\LT  224.000  435.222
\LT  226.000  436.633
\LT  228.000  438.102
\LT  230.000  439.623
\LT  232.000  441.191
\LT  235.000  443.615
\LT  238.000  446.110
\LT  241.000  448.662
\LT  245.000  452.128
\LT  251.000  457.408
\LT  262.000  467.130
\LT  267.000  471.495
\LT  272.000  475.797
\LT  276.000  479.182
\LT  280.000  482.510
\LT  284.000  485.775
\LT  288.000  488.973
\LT  292.000  492.102
\LT  296.000  495.158
\LT  300.000  498.141
\LT  304.000  501.047
\LT  308.000  503.878
\LT  312.000  506.632
\LT  316.000  509.309
\LT  320.000  511.909
\LT  324.000  514.433
\LT  328.000  516.881
\LT  332.000  519.253
\LT  336.000  521.551
\LT  340.000  523.775
\LT  344.000  525.927
\LT  348.000  528.007
\LT  352.000  530.016
\LT  356.000  531.956
\LT  360.000  533.827
\LT  364.000  535.631
\LT  368.000  537.369
\LT  372.000  539.043
\LT  376.000  540.652
\LT  380.000  542.199
\LT  384.000  543.684
\LT  388.000  545.109
\LT  392.000  546.474
\LT  396.000  547.782
\LT  400.000  549.032
\LT  404.000  550.227
\LT  408.000  551.366
\LT  412.000  552.451
\LT  417.000  553.734
\LT  422.000  554.935
\LT  427.000  556.058
\LT  432.000  557.103
\LT  437.000  558.073
\LT  442.000  558.969
\LT  447.000  559.793
\LT  452.000  560.546
\LT  457.000  561.229
\LT  462.000  561.845
\LT  467.000  562.395
\LT  472.000  562.880
\LT  477.000  563.301
\LT  482.000  563.659
\LT  487.000  563.957
\LT  492.000  564.194
\LT  497.000  564.373
\LT  502.000  564.493
\LT  507.000  564.557
\LT  513.000  564.560
\LT  519.000  564.485
\LT  525.000  564.333
\LT  531.000  564.106
\LT  537.000  563.804
\LT  543.000  563.431
\LT  549.000  562.986
\LT  555.000  562.471
\LT  561.000  561.889
\LT  567.000  561.238
\LT  573.000  560.522
\LT  579.000  559.741
\LT  585.000  558.896
\LT  591.000  557.988
\LT  597.000  557.018
\LT  603.000  555.988
\LT  610.000  554.710
\LT  617.000  553.352
\LT  624.000  551.915
\LT  631.000  550.400
\LT  638.000  548.809
\LT  645.000  547.143
\LT  652.000  545.402
\LT  659.000  543.589
\LT  666.000  541.704
\LT  673.000  539.748
\LT  680.000  537.721
\LT  687.000  535.626
\LT  694.000  533.463
\LT  701.000  531.232
\LT  708.000  528.934
\LT  715.000  526.571
\LT  722.000  524.142
\LT  729.000  521.650
\LT  736.000  519.093
\LT  743.000  516.474
\LT  750.000  513.792
\LT  757.000  511.048
\LT  765.000  507.838
\LT  773.000  504.549
\LT  781.000  501.181
\LT  789.000  497.736
\LT  797.000  494.214
\LT  805.000  490.616
\LT  813.000  486.943
\LT  821.000  483.195
\LT  829.000  479.374
\LT  837.000  475.478
\LT  845.000  471.510
\LT  853.000  467.470
\LT  861.000  463.358
\LT  869.000  459.175
\LT  877.000  454.921
\LT  885.000  450.597
\LT  893.000  446.203
\LT  901.000  441.740
\LT  909.000  437.208
\LT  917.000  432.608
\LT  925.000  427.940
\LT  933.000  423.204
\LT  941.000  418.400
\LT  949.000  413.530
\LT  957.000  408.594
\LT  965.000  403.591
\LT  973.000  398.522
\LT  981.000  393.387
\LT  989.000  388.188
\LT  997.000  382.923
\LT 1005.000  377.594
\LT 1013.000  372.200
\LT 1022.000  366.056
\LT 1031.000  359.830
\LT 1040.000  353.524
\LT 1049.000  347.138
\LT 1058.000  340.673
\LT 1067.000  334.127
\LT 1076.000  327.503
\LT 1085.000  320.800
\LT 1094.000  314.018
\LT 1103.000  307.158
\LT 1112.000  300.220
\LT 1121.000  293.204
\LT 1130.000  286.111
\LT 1139.000  278.941
\LT 1148.000  271.693
\LT 1157.000  264.369
\LT 1166.000  256.968
\LT 1175.000  249.490
\LT 1184.000  241.937
\LT 1193.000  234.308
\LT 1202.000  226.602
\LT 1211.000  218.822
\LT 1220.000  210.966
\LT 1229.000  203.034
\LT 1238.000  195.028
\LT 1247.000  186.947
\LT 1256.000  178.791
\LT 1265.000  170.561
\LT 1274.000  162.256
\LT 1283.000  153.877
\LT 1292.000  145.424
\LT 1300.000  137.848
\koniec    0.78126  -0.001
\obraz22
\grub0.2pt
\MT   0.000   60.000
\LT1400.000   60.000
\MT 160.000   60.000
\LT 160.000   70.000
\MT 220.000   60.000
\LT 220.000   70.000
\MT 280.000   60.000
\LT 280.000   70.000
\MT 340.000   60.000
\LT 340.000   70.000
\cput(340.000,-10.000,2)
\MT 400.000   60.000
\LT 400.000   70.000
\MT 460.000   60.000
\LT 460.000   70.000
\MT 520.000   60.000
\LT 520.000   70.000
\MT 580.000   60.000
\LT 580.000   70.000
\cput(580.000,-10.000,4)
\MT 640.000   60.000
\LT 640.000   70.000
\MT 700.000   60.000
\LT 700.000   70.000
\MT 760.000   60.000
\LT 760.000   70.000
\MT 820.000   60.000
\LT 820.000   70.000
\cput(820.000,-10.000,6)
\MT 880.000   60.000
\LT 880.000   70.000
\MT 940.000   60.000
\LT 940.000   70.000
\MT1000.000   60.000
\LT1000.000   70.000
\MT1060.000   60.000
\LT1060.000   70.000
\cput(1060.000,-10.000,8)
\MT1120.000   60.000
\LT1120.000   70.000
\MT1180.000   60.000
\LT1180.000   70.000
\MT1240.000   60.000
\LT1240.000   70.000
\MT1300.000   60.000
\LT1300.000   70.000
\cput(1300.000,-10.000,10)
\MT 100.000    0.000
\LT 100.000   70.000
\multi(100.000,70.000)(0.0000,4.0000){25}{\linia(0,0)(0.0000,2.0000)}
\MT 100.000  170.000
\LT 100.000  850.000
\MT  92.000  200.000
\LT 108.000  200.000
\MT  92.000  224.000
\LT 108.000  224.000
\MT  92.000  248.000
\LT 108.000  248.000
\MT  92.000  272.000
\LT 108.000  272.000
\MT  92.000  296.000
\LT 108.000  296.000
\MT  92.000  320.000
\LT 108.000  320.000
\MT  92.000  344.000
\LT 108.000  344.000
\MT  92.000  368.000
\LT 108.000  368.000
\MT  92.000  392.000
\LT 108.000  392.000
\MT  92.000  416.000
\LT 108.000  416.000
\MT  92.000  440.000
\LT 108.000  440.000
\MT  92.000  464.000
\LT 108.000  464.000
\MT  92.000  488.000
\LT 108.000  488.000
\MT  92.000  512.000
\LT 108.000  512.000
\MT  92.000  536.000
\LT 108.000  536.000
\MT  92.000  560.000
\LT 108.000  560.000
\MT  92.000  584.000
\LT 108.000  584.000
\MT  92.000  608.000
\LT 108.000  608.000
\MT  92.000  632.000
\LT 108.000  632.000
\MT  92.000  656.000
\LT 108.000  656.000
\MT  92.000  680.000
\LT 108.000  680.000
\MT  92.000  704.000
\LT 108.000  704.000
\MT  92.000  728.000
\LT 108.000  728.000
\MT  92.000  752.000
\LT 108.000  752.000
\MT  92.000  776.000
\LT 108.000  776.000
\MT  92.000  800.000
\LT 108.000  800.000
\MT  84.000  200.000
\LT 116.000  200.000
\lput(80.000,188.000, 0.75)
\MT  84.000  320.000
\LT 116.000  320.000
\lput(80.000,308.000, 0.80)
\MT  84.000  440.000
\LT 116.000  440.000
\lput(80.000,428.000, 0.85)
\MT  84.000  560.000
\LT 116.000  560.000
\lput(80.000,548.000, 0.90)
\MT  84.000  680.000
\LT 116.000  680.000
\lput(80.000,668.000, 0.95)
\MT  84.000  800.000
\LT 116.000  800.000
\lput(80.000,788.000, 1.00)
\grub0.6pt
\MT  100.000  800.000
\LT  101.000  799.957
\LT  102.000  799.826
\LT  103.000  799.609
\LT  104.000  799.306
\LT  105.000  798.915
\LT  106.000  798.437
\LT  107.000  797.873
\LT  108.000  797.222
\LT  109.000  796.484
\LT  110.000  795.660
\LT  111.000  794.748
\LT  112.000  793.750
\LT  113.000  792.665
\LT  114.000  791.494
\LT  115.000  790.235
\LT  116.000  788.890
\LT  117.000  787.459
\LT  118.000  785.940
\LT  119.000  784.336
\LT  120.000  782.644
\LT  121.000  780.867
\LT  122.000  779.003
\LT  123.000  777.053
\LT  124.000  775.017
\LT  125.000  772.895
\LT  126.000  770.687
\LT  127.000  768.394
\LT  128.000  766.015
\LT  129.000  763.551
\LT  130.000  761.002
\LT  131.000  758.369
\LT  132.000  755.651
\LT  133.000  752.849
\LT  134.000  749.964
\LT  135.000  746.995
\LT  136.000  743.944
\LT  137.000  740.810
\LT  138.000  737.594
\LT  139.000  734.297
\LT  140.000  730.919
\LT  141.000  727.461
\LT  142.000  723.924
\LT  143.000  720.308
\LT  144.000  716.615
\LT  145.000  712.844
\LT  146.000  708.997
\LT  147.000  705.075
\LT  148.000  701.078
\LT  149.000  697.009
\LT  150.000  692.867
\LT  151.000  688.654
\LT  152.000  684.372
\LT  153.000  680.021
\LT  154.000  675.604
\LT  155.000  671.120
\LT  156.000  666.573
\LT  157.000  661.963
\LT  158.000  657.292
\LT  159.000  652.562
\LT  160.000  647.774
\LT  161.000  642.931
\LT  162.000  638.033
\LT  163.000  633.084
\LT  164.000  628.085
\LT  165.000  623.038
\LT  166.000  617.945
\LT  167.000  612.808
\LT  168.000  607.631
\LT  169.000  602.414
\LT  170.000  597.160
\LT  171.000  591.872
\LT  172.000  586.553
\LT  173.000  581.203
\LT  174.000  575.827
\LT  175.000  570.427
\LT  177.000  559.563
\LT  179.000  548.633
\LT  182.000  532.160
\LT  186.000  510.166
\LT  188.000  499.203
\LT  190.000  488.293
\LT  191.000  482.864
\LT  192.000  477.455
\LT  193.000  472.070
\LT  194.000  466.710
\LT  195.000  461.378
\LT  196.000  456.077
\LT  197.000  450.808
\LT  198.000  445.575
\LT  199.000  440.378
\LT  200.000  435.221
\LT  201.000  430.105
\LT  202.000  425.033
\LT  203.000  420.006
\LT  204.000  415.026
\LT  205.000  410.094
\LT  206.000  405.214
\LT  207.000  400.386
\LT  208.000  395.611
\LT  209.000  390.892
\LT  210.000  386.229
\LT  211.000  381.625
\LT  212.000  377.079
\LT  213.000  372.594
\LT  214.000  368.170
\LT  215.000  363.809
\LT  216.000  359.511
\LT  217.000  355.277
\LT  218.000  351.108
\LT  219.000  347.004
\LT  220.000  342.967
\LT  221.000  338.997
\LT  222.000  335.093
\LT  223.000  331.257
\LT  224.000  327.490
\LT  225.000  323.790
\LT  226.000  320.159
\LT  227.000  316.596
\LT  228.000  313.102
\LT  229.000  309.676
\LT  230.000  306.319
\LT  231.000  303.031
\LT  232.000  299.811
\LT  233.000  296.659
\LT  234.000  293.575
\LT  235.000  290.559
\LT  236.000  287.610
\LT  237.000  284.728
\LT  238.000  281.912
\LT  239.000  279.163
\LT  240.000  276.479
\LT  241.000  273.861
\LT  242.000  271.307
\LT  243.000  268.817
\LT  244.000  266.391
\LT  245.000  264.028
\LT  246.000  261.726
\LT  247.000  259.487
\LT  248.000  257.308
\LT  249.000  255.189
\LT  250.000  253.130
\LT  251.000  251.130
\LT  252.000  249.187
\LT  253.000  247.302
\LT  254.000  245.474
\LT  255.000  243.701
\LT  256.000  241.983
\LT  257.000  240.319
\LT  258.000  238.709
\LT  259.000  237.151
\LT  260.000  235.645
\LT  261.000  234.190
\LT  262.000  232.786
\LT  263.000  231.430
\LT  264.000  230.124
\LT  265.000  228.865
\LT  266.000  227.653
\LT  267.000  226.487
\LT  268.000  225.367
\LT  269.000  224.292
\LT  270.000  223.260
\LT  271.000  222.271
\LT  272.000  221.325
\LT  273.000  220.420
\LT  274.000  219.556
\LT  275.000  218.732
\LT  276.000  217.947
\LT  277.000  217.201
\LT  278.000  216.493
\LT  279.000  215.821
\LT  280.000  215.186
\LT  281.000  214.587
\LT  282.000  214.022
\LT  283.000  213.492
\LT  284.000  212.996
\LT  285.000  212.532
\LT  286.000  212.101
\LT  287.000  211.701
\LT  288.000  211.332
\LT  289.000  210.993
\LT  290.000  210.684
\LT  291.000  210.404
\LT  292.000  210.152
\LT  293.000  209.928
\LT  294.000  209.732
\LT  295.000  209.562
\LT  296.000  209.418
\LT  297.000  209.300
\LT  298.000  209.206
\LT  299.000  209.137
\LT  300.000  209.092
\LT  301.000  209.071
\LT  302.000  209.072
\LT  303.000  209.095
\LT  304.000  209.141
\LT  305.000  209.207
\LT  306.000  209.295
\LT  308.000  209.531
\LT  310.000  209.844
\LT  312.000  210.232
\LT  314.000  210.691
\LT  316.000  211.218
\LT  318.000  211.809
\LT  320.000  212.461
\LT  322.000  213.172
\LT  324.000  213.939
\LT  326.000  214.759
\LT  328.000  215.629
\LT  330.000  216.547
\LT  332.000  217.511
\LT  334.000  218.518
\LT  337.000  220.105
\LT  340.000  221.778
\LT  343.000  223.529
\LT  346.000  225.353
\LT  349.000  227.244
\LT  352.000  229.196
\LT  355.000  231.205
\LT  358.000  233.265
\LT  362.000  236.084
\LT  366.000  238.976
\LT  370.000  241.933
\LT  375.000  245.708
\LT  380.000  249.556
\LT  386.000  254.252
\LT  393.000  259.814
\LT  402.000  267.053
\LT  430.000  289.759
\LT  440.000  297.803
\LT  449.000  304.971
\LT  457.000  311.272
\LT  465.000  317.497
\LT  472.000  322.876
\LT  479.000  328.188
\LT  486.000  333.429
\LT  493.000  338.598
\LT  500.000  343.692
\LT  507.000  348.709
\LT  514.000  353.650
\LT  521.000  358.513
\LT  528.000  363.299
\LT  535.000  368.006
\LT  542.000  372.636
\LT  549.000  377.189
\LT  556.000  381.665
\LT  563.000  386.065
\LT  570.000  390.390
\LT  577.000  394.641
\LT  584.000  398.819
\LT  591.000  402.925
\LT  598.000  406.959
\LT  605.000  410.924
\LT  612.000  414.820
\LT  619.000  418.649
\LT  626.000  422.411
\LT  633.000  426.109
\LT  640.000  429.742
\LT  647.000  433.314
\LT  655.000  437.320
\LT  663.000  441.248
\LT  671.000  445.100
\LT  679.000  448.878
\LT  687.000  452.582
\LT  695.000  456.216
\LT  703.000  459.780
\LT  711.000  463.276
\LT  719.000  466.707
\LT  728.000  470.489
\LT  737.000  474.193
\LT  746.000  477.819
\LT  755.000  481.370
\LT  764.000  484.848
\LT  773.000  488.255
\LT  782.000  491.594
\LT  791.000  494.865
\LT  801.000  498.424
\LT  811.000  501.904
\LT  821.000  505.308
\LT  831.000  508.639
\LT  841.000  511.899
\LT  851.000  515.090
\LT  862.000  518.522
\LT  873.000  521.876
\LT  884.000  525.153
\LT  895.000  528.357
\LT  906.000  531.489
\LT  917.000  534.553
\LT  929.000  537.819
\LT  941.000  541.007
\LT  953.000  544.121
\LT  965.000  547.164
\LT  978.000  550.381
\LT  991.000  553.520
\LT 1004.000  556.582
\LT 1017.000  559.572
\LT 1030.000  562.490
\LT 1044.000  565.556
\LT 1058.000  568.546
\LT 1072.000  571.462
\LT 1087.000  574.508
\LT 1102.000  577.475
\LT 1117.000  580.366
\LT 1132.000  583.185
\LT 1148.000  586.114
\LT 1164.000  588.966
\LT 1180.000  591.745
\LT 1197.000  594.619
\LT 1214.000  597.416
\LT 1231.000  600.139
\LT 1249.000  602.944
\LT 1267.000  605.673
\LT 1285.000  608.328
\LT 1300.000  610.487
\koniec    0.78126   0.000
\obraz23
\grub0.2pt
\MT   0.000   60.000
\LT1400.000   60.000
\MT 160.000   60.000
\LT 160.000   70.000
\MT 220.000   60.000
\LT 220.000   70.000
\MT 280.000   60.000
\LT 280.000   70.000
\MT 340.000   60.000
\LT 340.000   70.000
\cput(340.000,-10.000,2)
\MT 400.000   60.000
\LT 400.000   70.000
\MT 460.000   60.000
\LT 460.000   70.000
\MT 520.000   60.000
\LT 520.000   70.000
\MT 580.000   60.000
\LT 580.000   70.000
\cput(580.000,-10.000,4)
\MT 640.000   60.000
\LT 640.000   70.000
\MT 700.000   60.000
\LT 700.000   70.000
\MT 760.000   60.000
\LT 760.000   70.000
\MT 820.000   60.000
\LT 820.000   70.000
\cput(820.000,-10.000,6)
\MT 880.000   60.000
\LT 880.000   70.000
\MT 940.000   60.000
\LT 940.000   70.000
\MT1000.000   60.000
\LT1000.000   70.000
\MT1060.000   60.000
\LT1060.000   70.000
\cput(1060.000,-10.000,8)
\MT1120.000   60.000
\LT1120.000   70.000
\MT1180.000   60.000
\LT1180.000   70.000
\MT1240.000   60.000
\LT1240.000   70.000
\MT1300.000   60.000
\LT1300.000   70.000
\cput(1300.000,-10.000,10)
\MT 100.000    0.000
\LT 100.000   70.000
\multi(100.000,70.000)(0.0000,4.0000){25}{\linia(0,0)(0.0000,2.0000)}
\MT 100.000  170.000
\LT 100.000  850.000
\MT  92.000  200.000
\LT 108.000  200.000
\MT  92.000  224.000
\LT 108.000  224.000
\MT  92.000  248.000
\LT 108.000  248.000
\MT  92.000  272.000
\LT 108.000  272.000
\MT  92.000  296.000
\LT 108.000  296.000
\MT  92.000  320.000
\LT 108.000  320.000
\MT  92.000  344.000
\LT 108.000  344.000
\MT  92.000  368.000
\LT 108.000  368.000
\MT  92.000  392.000
\LT 108.000  392.000
\MT  92.000  416.000
\LT 108.000  416.000
\MT  92.000  440.000
\LT 108.000  440.000
\MT  92.000  464.000
\LT 108.000  464.000
\MT  92.000  488.000
\LT 108.000  488.000
\MT  92.000  512.000
\LT 108.000  512.000
\MT  92.000  536.000
\LT 108.000  536.000
\MT  92.000  560.000
\LT 108.000  560.000
\MT  92.000  584.000
\LT 108.000  584.000
\MT  92.000  608.000
\LT 108.000  608.000
\MT  92.000  632.000
\LT 108.000  632.000
\MT  92.000  656.000
\LT 108.000  656.000
\MT  92.000  680.000
\LT 108.000  680.000
\MT  92.000  704.000
\LT 108.000  704.000
\MT  92.000  728.000
\LT 108.000  728.000
\MT  92.000  752.000
\LT 108.000  752.000
\MT  92.000  776.000
\LT 108.000  776.000
\MT  92.000  800.000
\LT 108.000  800.000
\MT  84.000  200.000
\LT 116.000  200.000
\lput(80.000,188.000, 0.75)
\MT  84.000  320.000
\LT 116.000  320.000
\lput(80.000,308.000, 0.80)
\MT  84.000  440.000
\LT 116.000  440.000
\lput(80.000,428.000, 0.85)
\MT  84.000  560.000
\LT 116.000  560.000
\lput(80.000,548.000, 0.90)
\MT  84.000  680.000
\LT 116.000  680.000
\lput(80.000,668.000, 0.95)
\MT  84.000  800.000
\LT 116.000  800.000
\lput(80.000,788.000, 1.00)
\grub0.6pt
\MT  100.000  800.000
\LT  101.000  799.957
\LT  102.000  799.827
\LT  103.000  799.611
\LT  104.000  799.308
\LT  105.000  798.919
\LT  106.000  798.443
\LT  107.000  797.881
\LT  108.000  797.233
\LT  109.000  796.498
\LT  110.000  795.676
\LT  111.000  794.769
\LT  112.000  793.774
\LT  113.000  792.693
\LT  114.000  791.526
\LT  115.000  790.273
\LT  116.000  788.933
\LT  117.000  787.507
\LT  118.000  785.994
\LT  119.000  784.396
\LT  120.000  782.711
\LT  121.000  780.940
\LT  122.000  779.084
\LT  123.000  777.141
\LT  124.000  775.113
\LT  125.000  772.999
\LT  126.000  770.800
\LT  127.000  768.515
\LT  128.000  766.146
\LT  129.000  763.691
\LT  130.000  761.152
\LT  131.000  758.529
\LT  132.000  755.822
\LT  133.000  753.031
\LT  134.000  750.156
\LT  135.000  747.199
\LT  136.000  744.160
\LT  137.000  741.038
\LT  138.000  737.835
\LT  139.000  734.550
\LT  140.000  731.186
\LT  141.000  727.742
\LT  142.000  724.218
\LT  143.000  720.617
\LT  144.000  716.937
\LT  145.000  713.181
\LT  146.000  709.350
\LT  147.000  705.443
\LT  148.000  701.462
\LT  149.000  697.409
\LT  150.000  693.284
\LT  151.000  689.088
\LT  152.000  684.823
\LT  153.000  680.490
\LT  154.000  676.090
\LT  155.000  671.625
\LT  156.000  667.096
\LT  157.000  662.505
\LT  158.000  657.853
\LT  159.000  653.142
\LT  160.000  648.374
\LT  161.000  643.551
\LT  162.000  638.674
\LT  163.000  633.746
\LT  164.000  628.767
\LT  165.000  623.742
\LT  166.000  618.671
\LT  167.000  613.557
\LT  168.000  608.401
\LT  169.000  603.207
\LT  170.000  597.977
\LT  171.000  592.713
\LT  172.000  587.417
\LT  173.000  582.092
\LT  174.000  576.740
\LT  175.000  571.364
\LT  177.000  560.551
\LT  179.000  549.673
\LT  182.000  533.281
\LT  186.000  511.398
\LT  188.000  500.494
\LT  190.000  489.643
\LT  191.000  484.244
\LT  192.000  478.866
\LT  193.000  473.511
\LT  194.000  468.183
\LT  195.000  462.882
\LT  196.000  457.613
\LT  197.000  452.376
\LT  198.000  447.175
\LT  199.000  442.012
\LT  200.000  436.888
\LT  201.000  431.805
\LT  202.000  426.767
\LT  203.000  421.774
\LT  204.000  416.828
\LT  205.000  411.932
\LT  206.000  407.087
\LT  207.000  402.294
\LT  208.000  397.555
\LT  209.000  392.872
\LT  210.000  388.246
\LT  211.000  383.678
\LT  212.000  379.170
\LT  213.000  374.722
\LT  214.000  370.336
\LT  215.000  366.013
\LT  216.000  361.754
\LT  217.000  357.558
\LT  218.000  353.429
\LT  219.000  349.365
\LT  220.000  345.367
\LT  221.000  341.437
\LT  222.000  337.574
\LT  223.000  333.779
\LT  224.000  330.052
\LT  225.000  326.394
\LT  226.000  322.805
\LT  227.000  319.284
\LT  228.000  315.832
\LT  229.000  312.450
\LT  230.000  309.136
\LT  231.000  305.891
\LT  232.000  302.715
\LT  233.000  299.607
\LT  234.000  296.568
\LT  235.000  293.596
\LT  236.000  290.692
\LT  237.000  287.856
\LT  238.000  285.086
\LT  239.000  282.383
\LT  240.000  279.746
\LT  241.000  277.174
\LT  242.000  274.668
\LT  243.000  272.226
\LT  244.000  269.847
\LT  245.000  267.532
\LT  246.000  265.279
\LT  247.000  263.088
\LT  248.000  260.958
\LT  249.000  258.889
\LT  250.000  256.880
\LT  251.000  254.930
\LT  252.000  253.038
\LT  253.000  251.204
\LT  254.000  249.426
\LT  255.000  247.705
\LT  256.000  246.039
\LT  257.000  244.427
\LT  258.000  242.869
\LT  259.000  241.365
\LT  260.000  239.912
\LT  261.000  238.510
\LT  262.000  237.160
\LT  263.000  235.859
\LT  264.000  234.606
\LT  265.000  233.402
\LT  266.000  232.246
\LT  267.000  231.135
\LT  268.000  230.071
\LT  269.000  229.052
\LT  270.000  228.077
\LT  271.000  227.145
\LT  272.000  226.256
\LT  273.000  225.408
\LT  274.000  224.602
\LT  275.000  223.836
\LT  276.000  223.110
\LT  277.000  222.423
\LT  278.000  221.773
\LT  279.000  221.161
\LT  280.000  220.586
\LT  281.000  220.047
\LT  282.000  219.543
\LT  283.000  219.074
\LT  284.000  218.638
\LT  285.000  218.236
\LT  286.000  217.867
\LT  287.000  217.529
\LT  288.000  217.222
\LT  289.000  216.946
\LT  290.000  216.700
\LT  291.000  216.484
\LT  292.000  216.296
\LT  293.000  216.137
\LT  294.000  216.004
\LT  295.000  215.899
\LT  296.000  215.821
\LT  297.000  215.768
\LT  298.000  215.740
\LT  299.000  215.738
\LT  300.000  215.759
\LT  301.000  215.804
\LT  302.000  215.873
\LT  303.000  215.964
\LT  304.000  216.077
\LT  305.000  216.211
\LT  306.000  216.368
\LT  307.000  216.544
\LT  309.000  216.958
\LT  311.000  217.449
\LT  313.000  218.014
\LT  315.000  218.650
\LT  317.000  219.353
\LT  319.000  220.121
\LT  321.000  220.950
\LT  323.000  221.837
\LT  325.000  222.780
\LT  327.000  223.776
\LT  329.000  224.822
\LT  331.000  225.917
\LT  333.000  227.057
\LT  335.000  228.241
\LT  338.000  230.094
\LT  341.000  232.033
\LT  344.000  234.052
\LT  347.000  236.144
\LT  350.000  238.305
\LT  353.000  240.528
\LT  356.000  242.808
\LT  359.000  245.142
\LT  362.000  247.524
\LT  366.000  250.769
\LT  370.000  254.083
\LT  374.000  257.460
\LT  379.000  261.755
\LT  384.000  266.121
\LT  390.000  271.438
\LT  397.000  277.723
\LT  405.000  284.985
\LT  417.000  295.970
\LT  438.000  315.241
\LT  449.000  325.271
\LT  458.000  333.415
\LT  467.000  341.489
\LT  475.000  348.599
\LT  483.000  355.640
\LT  491.000  362.609
\LT  499.000  369.502
\LT  507.000  376.318
\LT  515.000  383.054
\LT  522.000  388.882
\LT  529.000  394.649
\LT  536.000  400.355
\LT  544.000  406.801
\LT  552.000  413.167
\LT  560.000  419.455
\LT  568.000  425.666
\LT  576.000  431.801
\LT  584.000  437.862
\LT  592.000  443.849
\LT  600.000  449.766
\LT  608.000  455.613
\LT  616.000  461.392
\LT  624.000  467.106
\LT  633.000  473.457
\LT  642.000  479.730
\LT  651.000  485.927
\LT  660.000  492.051
\LT  669.000  498.105
\LT  678.000  504.090
\LT  688.000  510.664
\LT  698.000  517.161
\LT  708.000  523.584
\LT  718.000  529.936
\LT  729.000  536.845
\LT  740.000  543.676
\LT  751.000  550.434
\LT  762.000  557.122
\LT  774.000  564.342
\LT  786.000  571.489
\LT  799.000  579.152
\LT  812.000  586.738
\LT  826.000  594.829
\LT  840.000  602.843
\LT  855.000  611.351
\LT  870.000  619.785
\LT  886.000  628.707
\LT  903.000  638.110
\LT  921.000  647.990
\LT  941.000  658.887
\LT  962.000  670.250
\LT  986.000  683.154
\LT 1013.000  697.588
\LT 1046.000  715.141
\LT 1097.000  742.163
\LT 1160.000  775.527
\LT 1198.000  795.719
\LT 1230.000  812.797
\LT 1259.000  828.350
\LT 1286.000  842.907
\LT 1300.000  850.487
\koniec    0.78126   0.001
\obraz24
\grub0.2pt
\MT   0.000   60.000
\LT1400.000   60.000
\MT 160.000   60.000
\LT 160.000   70.000
\MT 220.000   60.000
\LT 220.000   70.000
\MT 280.000   60.000
\LT 280.000   70.000
\MT 340.000   60.000
\LT 340.000   70.000
\cput(340.000,-10.000,2)
\MT 400.000   60.000
\LT 400.000   70.000
\MT 460.000   60.000
\LT 460.000   70.000
\MT 520.000   60.000
\LT 520.000   70.000
\MT 580.000   60.000
\LT 580.000   70.000
\cput(580.000,-10.000,4)
\MT 640.000   60.000
\LT 640.000   70.000
\MT 700.000   60.000
\LT 700.000   70.000
\MT 760.000   60.000
\LT 760.000   70.000
\MT 820.000   60.000
\LT 820.000   70.000
\cput(820.000,-10.000,6)
\MT 880.000   60.000
\LT 880.000   70.000
\MT 940.000   60.000
\LT 940.000   70.000
\MT1000.000   60.000
\LT1000.000   70.000
\MT1060.000   60.000
\LT1060.000   70.000
\cput(1060.000,-10.000,8)
\MT1120.000   60.000
\LT1120.000   70.000
\MT1180.000   60.000
\LT1180.000   70.000
\MT1240.000   60.000
\LT1240.000   70.000
\MT1300.000   60.000
\LT1300.000   70.000
\cput(1300.000,-10.000,10)
\MT 100.000    0.000
\LT 100.000   70.000
\multi(100.000,70.000)(0.0000,4.0000){25}{\linia(0,0)(0.0000,2.0000)}
\MT 100.000  170.000
\LT 100.000  850.000
\MT  92.000  200.000
\LT 108.000  200.000
\MT  92.000  225.000
\LT 108.000  225.000
\MT  92.000  250.000
\LT 108.000  250.000
\MT  92.000  275.000
\LT 108.000  275.000
\MT  92.000  300.000
\LT 108.000  300.000
\MT  92.000  325.000
\LT 108.000  325.000
\MT  92.000  350.000
\LT 108.000  350.000
\MT  92.000  375.000
\LT 108.000  375.000
\MT  92.000  400.000
\LT 108.000  400.000
\MT  92.000  425.000
\LT 108.000  425.000
\MT  92.000  450.000
\LT 108.000  450.000
\MT  92.000  475.000
\LT 108.000  475.000
\MT  92.000  500.000
\LT 108.000  500.000
\MT  92.000  525.000
\LT 108.000  525.000
\MT  92.000  550.000
\LT 108.000  550.000
\MT  92.000  575.000
\LT 108.000  575.000
\MT  92.000  600.000
\LT 108.000  600.000
\MT  92.000  625.000
\LT 108.000  625.000
\MT  92.000  650.000
\LT 108.000  650.000
\MT  92.000  675.000
\LT 108.000  675.000
\MT  92.000  700.000
\LT 108.000  700.000
\MT  92.000  725.000
\LT 108.000  725.000
\MT  92.000  750.000
\LT 108.000  750.000
\MT  92.000  775.000
\LT 108.000  775.000
\MT  92.000  800.000
\LT 108.000  800.000
\MT  84.000  200.000
\LT 116.000  200.000
\lput(80.000,187.500, 0.80)
\MT  84.000  300.000
\LT 116.000  300.000
\lput(80.000,287.500, 1.00)
\MT  84.000  400.000
\LT 116.000  400.000
\lput(80.000,387.500, 1.20)
\MT  84.000  500.000
\LT 116.000  500.000
\lput(80.000,487.500, 1.40)
\MT  84.000  600.000
\LT 116.000  600.000
\lput(80.000,587.500, 1.60)
\MT  84.000  700.000
\LT 116.000  700.000
\lput(80.000,687.500, 1.80)
\MT  84.000  800.000
\LT 116.000  800.000
\lput(80.000,787.500, 2.00)
\grub0.6pt
\MT  100.000  300.000
\LT  102.000  299.965
\LT  104.000  299.861
\LT  106.000  299.687
\LT  108.000  299.444
\LT  110.000  299.131
\LT  112.000  298.748
\LT  114.000  298.296
\LT  116.000  297.774
\LT  118.000  297.183
\LT  120.000  296.523
\LT  122.000  295.794
\LT  124.000  294.995
\LT  126.000  294.128
\LT  128.000  293.192
\LT  130.000  292.188
\LT  132.000  291.116
\LT  134.000  289.977
\LT  136.000  288.772
\LT  138.000  287.500
\LT  140.000  286.164
\LT  142.000  284.763
\LT  144.000  283.300
\LT  146.000  281.776
\LT  148.000  280.191
\LT  150.000  278.549
\LT  152.000  276.850
\LT  154.000  275.097
\LT  156.000  273.292
\LT  158.000  271.437
\LT  160.000  269.536
\LT  162.000  267.592
\LT  165.000  264.600
\LT  168.000  261.529
\LT  171.000  258.390
\LT  175.000  254.125
\LT  181.000  247.623
\LT  188.000  240.023
\LT  192.000  235.742
\LT  195.000  232.587
\LT  198.000  229.496
\LT  201.000  226.481
\LT  203.000  224.518
\LT  205.000  222.598
\LT  207.000  220.722
\LT  209.000  218.895
\LT  211.000  217.117
\LT  213.000  215.391
\LT  215.000  213.719
\LT  217.000  212.102
\LT  219.000  210.543
\LT  221.000  209.041
\LT  223.000  207.598
\LT  225.000  206.215
\LT  227.000  204.891
\LT  229.000  203.627
\LT  231.000  202.423
\LT  233.000  201.279
\LT  235.000  200.195
\LT  237.000  199.169
\LT  239.000  198.201
\LT  241.000  197.291
\LT  243.000  196.437
\LT  245.000  195.639
\LT  247.000  194.896
\LT  249.000  194.206
\LT  251.000  193.569
\LT  253.000  192.983
\LT  255.000  192.446
\LT  257.000  191.958
\LT  259.000  191.518
\LT  261.000  191.123
\LT  263.000  190.773
\LT  265.000  190.467
\LT  267.000  190.202
\LT  270.000  189.881
\LT  273.000  189.646
\LT  276.000  189.495
\LT  279.000  189.421
\LT  282.000  189.423
\LT  285.000  189.495
\LT  288.000  189.633
\LT  291.000  189.834
\LT  294.000  190.096
\LT  297.000  190.413
\LT  300.000  190.783
\LT  303.000  191.204
\LT  306.000  191.671
\LT  310.000  192.363
\LT  314.000  193.129
\LT  318.000  193.962
\LT  322.000  194.857
\LT  326.000  195.809
\LT  331.000  197.075
\LT  336.000  198.415
\LT  341.000  199.824
\LT  346.000  201.294
\LT  352.000  203.133
\LT  358.000  205.043
\LT  364.000  207.017
\LT  371.000  209.392
\LT  378.000  211.837
\LT  386.000  214.704
\LT  394.000  217.640
\LT  403.000  221.016
\LT  413.000  224.845
\LT  424.000  229.138
\LT  435.000  233.507
\LT  447.000  238.347
\LT  460.000  243.670
\LT  473.000  249.067
\LT  487.000  254.956
\LT  501.000  260.920
\LT  515.000  266.957
\LT  529.000  273.065
\LT  543.000  279.244
\LT  557.000  285.496
\LT  571.000  291.820
\LT  585.000  298.219
\LT  599.000  304.694
\LT  613.000  311.247
\LT  626.000  317.404
\LT  639.000  323.631
\LT  652.000  329.931
\LT  665.000  336.304
\LT  678.000  342.753
\LT  691.000  349.280
\LT  704.000  355.885
\LT  716.000  362.053
\LT  728.000  368.291
\LT  740.000  374.599
\LT  752.000  380.979
\LT  764.000  387.432
\LT  776.000  393.959
\LT  788.000  400.560
\LT  800.000  407.237
\LT  812.000  413.990
\LT  824.000  420.821
\LT  836.000  427.730
\LT  848.000  434.718
\LT  859.000  441.194
\LT  870.000  447.737
\LT  881.000  454.348
\LT  892.000  461.027
\LT  903.000  467.776
\LT  914.000  474.594
\LT  925.000  481.482
\LT  936.000  488.441
\LT  947.000  495.470
\LT  958.000  502.570
\LT  969.000  509.742
\LT  980.000  516.987
\LT  991.000  524.303
\LT 1002.000  531.692
\LT 1013.000  539.154
\LT 1024.000  546.690
\LT 1035.000  554.299
\LT 1046.000  561.982
\LT 1057.000  569.740
\LT 1068.000  577.572
\LT 1079.000  585.478
\LT 1090.000  593.460
\LT 1101.000  601.517
\LT 1112.000  609.649
\LT 1123.000  617.858
\LT 1134.000  626.142
\LT 1145.000  634.503
\LT 1156.000  642.939
\LT 1167.000  651.453
\LT 1178.000  660.043
\LT 1189.000  668.711
\LT 1200.000  677.455
\LT 1211.000  686.277
\LT 1222.000  695.176
\LT 1233.000  704.153
\LT 1244.000  713.208
\LT 1255.000  722.341
\LT 1266.000  731.552
\LT 1277.000  740.841
\LT 1288.000  750.209
\LT 1299.000  759.655
\LT 1300.000  760.518
\koniec    0.78126   0.010
\eject

\def\opis{Plots of the \f\ $f(x)$ (Eq.~\eqref{3.12}) for some values of
parameters $a$ and~$b$\break (see the text for explanation)\break
Fig.~\the\nd A: $a=0.78126,\ b=0.100$; \
Fig.~\the\nd B: $a=0.10000,\ b=-0.100$;\break
Fig.~\the\nd C: $a=0.10000,\ b=-0.010$;\
Fig.~\the\nd D: $a=0.10000,\ b=-0.001$;\break
Fig.~\the\nd E: $a=0.10000,\ b=0.000$;\
Fig.~\the\nd F: $a=0.10000,\ b=0.001$;\break
Fig.~\the\nd G: $a=0.10000,\ b=0.010$;\
Fig.~\the\nd H: $a=0.10000,\ b=0.100$.}

\obraz25
\grub0.2pt
\MT   0.000   60.000
\LT1400.000   60.000
\MT 160.000   60.000
\LT 160.000   70.000
\MT 220.000   60.000
\LT 220.000   70.000
\MT 280.000   60.000
\LT 280.000   70.000
\MT 340.000   60.000
\LT 340.000   70.000
\cput(340.000,-10.000,2)
\MT 400.000   60.000
\LT 400.000   70.000
\MT 460.000   60.000
\LT 460.000   70.000
\MT 520.000   60.000
\LT 520.000   70.000
\MT 580.000   60.000
\LT 580.000   70.000
\cput(580.000,-10.000,4)
\MT 640.000   60.000
\LT 640.000   70.000
\MT 700.000   60.000
\LT 700.000   70.000
\MT 760.000   60.000
\LT 760.000   70.000
\MT 820.000   60.000
\LT 820.000   70.000
\cput(820.000,-10.000,6)
\MT 880.000   60.000
\LT 880.000   70.000
\MT 940.000   60.000
\LT 940.000   70.000
\MT1000.000   60.000
\LT1000.000   70.000
\MT1060.000   60.000
\LT1060.000   70.000
\cput(1060.000,-10.000,8)
\MT1120.000   60.000
\LT1120.000   70.000
\MT1180.000   60.000
\LT1180.000   70.000
\MT1240.000   60.000
\LT1240.000   70.000
\MT1300.000   60.000
\LT1300.000   70.000
\cput(1300.000,-10.000,10)
\MT 100.000    0.000
\LT 100.000   70.000
\multi(100.000,70.000)(0.0000,4.0000){25}{\linia(0,0)(0.0000,2.0000)}
\MT 100.000  170.000
\LT 100.000  850.000
\MT  92.000   60.000
\LT 108.000   60.000
\MT  92.000   97.000
\LT 108.000   97.000
\MT  92.000  134.000
\LT 108.000  134.000
\MT  92.000  171.000
\LT 108.000  171.000
\MT  92.000  208.000
\LT 108.000  208.000
\MT  92.000  245.000
\LT 108.000  245.000
\MT  92.000  282.000
\LT 108.000  282.000
\MT  92.000  319.000
\LT 108.000  319.000
\MT  92.000  356.000
\LT 108.000  356.000
\MT  92.000  393.000
\LT 108.000  393.000
\MT  92.000  430.000
\LT 108.000  430.000
\MT  92.000  467.000
\LT 108.000  467.000
\MT  92.000  504.000
\LT 108.000  504.000
\MT  92.000  541.000
\LT 108.000  541.000
\MT  92.000  578.000
\LT 108.000  578.000
\MT  92.000  615.000
\LT 108.000  615.000
\MT  92.000  652.000
\LT 108.000  652.000
\MT  92.000  689.000
\LT 108.000  689.000
\MT  92.000  726.000
\LT 108.000  726.000
\MT  92.000  763.000
\LT 108.000  763.000
\MT  92.000  800.000
\LT 108.000  800.000
\MT  84.000   60.000
\LT 116.000   60.000
\MT  84.000  208.000
\LT 116.000  208.000
\lput(80.000,189.500,  2.0)
\MT  84.000  356.000
\LT 116.000  356.000
\lput(80.000,337.500,  4.0)
\MT  84.000  504.000
\LT 116.000  504.000
\lput(80.000,485.500,  6.0)
\MT  84.000  652.000
\LT 116.000  652.000
\lput(80.000,633.500,  8.0)
\MT  84.000  800.000
\LT 116.000  800.000
\lput(80.000,781.500, 10.0)
\grub0.6pt
\MT  100.000  134.000
\LT  107.000  133.960
\LT  114.000  133.838
\LT  121.000  133.637
\LT  128.000  133.355
\LT  135.000  132.995
\LT  142.000  132.561
\LT  149.000  132.058
\LT  157.000  131.413
\LT  169.000  130.354
\LT  182.000  129.197
\LT  189.000  128.628
\LT  195.000  128.197
\LT  201.000  127.837
\LT  206.000  127.601
\LT  211.000  127.432
\LT  216.000  127.333
\LT  221.000  127.310
\LT  226.000  127.363
\LT  231.000  127.496
\LT  236.000  127.706
\LT  241.000  127.994
\LT  246.000  128.357
\LT  251.000  128.794
\LT  256.000  129.300
\LT  261.000  129.875
\LT  266.000  130.513
\LT  271.000  131.213
\LT  276.000  131.972
\LT  282.000  132.954
\LT  288.000  134.012
\LT  294.000  135.141
\LT  300.000  136.336
\LT  306.000  137.594
\LT  313.000  139.137
\LT  320.000  140.756
\LT  327.000  142.448
\LT  334.000  144.209
\LT  341.000  146.036
\LT  348.000  147.927
\LT  356.000  150.161
\LT  364.000  152.473
\LT  372.000  154.859
\LT  380.000  157.317
\LT  388.000  159.846
\LT  396.000  162.443
\LT  404.000  165.109
\LT  412.000  167.841
\LT  420.000  170.640
\LT  428.000  173.504
\LT  436.000  176.433
\LT  444.000  179.426
\LT  453.000  182.869
\LT  462.000  186.393
\LT  471.000  189.997
\LT  480.000  193.681
\LT  489.000  197.445
\LT  498.000  201.288
\LT  507.000  205.210
\LT  516.000  209.212
\LT  525.000  213.293
\LT  534.000  217.454
\LT  543.000  221.693
\LT  552.000  226.012
\LT  561.000  230.411
\LT  570.000  234.888
\LT  579.000  239.446
\LT  588.000  244.082
\LT  597.000  248.799
\LT  606.000  253.595
\LT  615.000  258.471
\LT  624.000  263.426
\LT  633.000  268.462
\LT  642.000  273.577
\LT  651.000  278.773
\LT  660.000  284.049
\LT  669.000  289.405
\LT  678.000  294.841
\LT  687.000  300.358
\LT  696.000  305.955
\LT  705.000  311.633
\LT  714.000  317.392
\LT  723.000  323.231
\LT  732.000  329.151
\LT  741.000  335.151
\LT  750.000  341.233
\LT  758.000  346.707
\LT  766.000  352.245
\LT  774.000  357.847
\LT  782.000  363.513
\LT  790.000  369.243
\LT  798.000  375.037
\LT  806.000  380.896
\LT  814.000  386.819
\LT  822.000  392.806
\LT  830.000  398.858
\LT  838.000  404.973
\LT  846.000  411.154
\LT  854.000  417.398
\LT  862.000  423.708
\LT  870.000  430.081
\LT  878.000  436.519
\LT  886.000  443.022
\LT  894.000  449.590
\LT  902.000  456.221
\LT  910.000  462.918
\LT  918.000  469.679
\LT  926.000  476.505
\LT  934.000  483.396
\LT  942.000  490.351
\LT  950.000  497.371
\LT  958.000  504.456
\LT  966.000  511.606
\LT  974.000  518.821
\LT  982.000  526.100
\LT  990.000  533.444
\LT  998.000  540.853
\LT 1006.000  548.327
\LT 1014.000  555.866
\LT 1022.000  563.470
\LT 1030.000  571.139
\LT 1038.000  578.873
\LT 1046.000  586.672
\LT 1054.000  594.536
\LT 1062.000  602.465
\LT 1070.000  610.459
\LT 1078.000  618.518
\LT 1086.000  626.642
\LT 1094.000  634.831
\LT 1102.000  643.085
\LT 1110.000  651.405
\LT 1118.000  659.789
\LT 1126.000  668.239
\LT 1134.000  676.754
\LT 1142.000  685.334
\LT 1150.000  693.979
\LT 1158.000  702.689
\LT 1166.000  711.465
\LT 1174.000  720.305
\LT 1182.000  729.211
\LT 1190.000  738.183
\LT 1198.000  747.219
\LT 1206.000  756.321
\LT 1214.000  765.488
\LT 1222.000  774.720
\LT 1230.000  784.017
\LT 1238.000  793.380
\LT 1246.000  802.808
\LT 1254.000  812.302
\LT 1262.000  821.861
\LT 1270.000  831.485
\LT 1278.000  841.174
\LT 1286.000  850.929
\LT 1294.000  860.749
\LT 1300.000  868.157
\koniec    0.78126   0.100
\obraz26
\grub0.2pt
\MT   0.000  800.000
\LT1400.000  800.000
\MT 160.000  800.000
\LT 160.000  810.000
\MT 220.000  800.000
\LT 220.000  810.000
\MT 280.000  800.000
\LT 280.000  810.000
\MT 340.000  800.000
\LT 340.000  810.000
\MT 400.000  800.000
\LT 400.000  810.000
\cput(400.000,730.000,5)
\MT 460.000  800.000
\LT 460.000  810.000
\MT 520.000  800.000
\LT 520.000  810.000
\MT 580.000  800.000
\LT 580.000  810.000
\MT 640.000  800.000
\LT 640.000  810.000
\MT 700.000  800.000
\LT 700.000  810.000
\cput(700.000,730.000,10)
\MT 760.000  800.000
\LT 760.000  810.000
\MT 820.000  800.000
\LT 820.000  810.000
\MT 880.000  800.000
\LT 880.000  810.000
\MT 940.000  800.000
\LT 940.000  810.000
\MT1000.000  800.000
\LT1000.000  810.000
\cput(1000.000,730.000,15)
\MT1060.000  800.000
\LT1060.000  810.000
\MT1120.000  800.000
\LT1120.000  810.000
\MT1180.000  800.000
\LT1180.000  810.000
\MT1240.000  800.000
\LT1240.000  810.000
\MT1300.000  800.000
\LT1300.000  810.000
\cput(1300.000,730.000,20)
\MT 100.000    0.000
\LT 100.000  850.000
\MT  92.000   50.000
\LT 108.000   50.000
\MT  92.000   87.500
\LT 108.000   87.500
\MT  92.000  125.000
\LT 108.000  125.000
\MT  92.000  162.500
\LT 108.000  162.500
\MT  92.000  200.000
\LT 108.000  200.000
\MT  92.000  237.500
\LT 108.000  237.500
\MT  92.000  275.000
\LT 108.000  275.000
\MT  92.000  312.500
\LT 108.000  312.500
\MT  92.000  350.000
\LT 108.000  350.000
\MT  92.000  387.500
\LT 108.000  387.500
\MT  92.000  425.000
\LT 108.000  425.000
\MT  92.000  462.500
\LT 108.000  462.500
\MT  92.000  500.000
\LT 108.000  500.000
\MT  92.000  537.500
\LT 108.000  537.500
\MT  92.000  575.000
\LT 108.000  575.000
\MT  92.000  612.500
\LT 108.000  612.500
\MT  92.000  650.000
\LT 108.000  650.000
\MT  92.000  687.500
\LT 108.000  687.500
\MT  92.000  725.000
\LT 108.000  725.000
\MT  92.000  762.500
\LT 108.000  762.500
\MT  92.000  800.000
\LT 108.000  800.000
\MT  84.000   50.000
\LT 116.000   50.000
\lput(80.000,31.250,  -40)
\MT  84.000  237.500
\LT 116.000  237.500
\lput(80.000,218.750,  -30)
\MT  84.000  425.000
\LT 116.000  425.000
\lput(80.000,406.250,  -20)
\MT  84.000  612.500
\LT 116.000  612.500
\lput(80.000,593.750,  -10)
\MT  84.000  800.000
\LT 116.000  800.000
\grub0.6pt
\MT  100.000  818.750
\LT  107.000  818.716
\LT  114.000  818.614
\LT  121.000  818.444
\LT  128.000  818.208
\LT  136.000  817.862
\LT  144.000  817.441
\LT  153.000  816.892
\LT  163.000  816.203
\LT  173.000  815.436
\LT  182.000  814.678
\LT  191.000  813.851
\LT  200.000  812.951
\LT  209.000  811.974
\LT  218.000  810.917
\LT  226.000  809.910
\LT  234.000  808.838
\LT  242.000  807.701
\LT  250.000  806.497
\LT  258.000  805.226
\LT  266.000  803.889
\LT  274.000  802.485
\LT  282.000  801.015
\LT  290.000  799.477
\LT  298.000  797.872
\LT  306.000  796.200
\LT  314.000  794.461
\LT  322.000  792.655
\LT  330.000  790.782
\LT  338.000  788.842
\LT  346.000  786.835
\LT  354.000  784.761
\LT  362.000  782.619
\LT  370.000  780.411
\LT  378.000  778.136
\LT  386.000  775.793
\LT  394.000  773.384
\LT  402.000  770.908
\LT  410.000  768.365
\LT  418.000  765.755
\LT  426.000  763.078
\LT  434.000  760.334
\LT  442.000  757.524
\LT  450.000  754.646
\LT  458.000  751.702
\LT  466.000  748.690
\LT  474.000  745.612
\LT  482.000  742.467
\LT  490.000  739.255
\LT  498.000  735.977
\LT  506.000  732.631
\LT  514.000  729.219
\LT  522.000  725.740
\LT  530.000  722.194
\LT  538.000  718.581
\LT  546.000  714.902
\LT  554.000  711.155
\LT  562.000  707.342
\LT  570.000  703.463
\LT  578.000  699.516
\LT  586.000  695.503
\LT  594.000  691.423
\LT  602.000  687.276
\LT  610.000  683.062
\LT  618.000  678.782
\LT  626.000  674.435
\LT  634.000  670.021
\LT  642.000  665.540
\LT  650.000  660.993
\LT  658.000  656.379
\LT  666.000  651.698
\LT  674.000  646.951
\LT  682.000  642.136
\LT  690.000  637.256
\LT  698.000  632.308
\LT  706.000  627.293
\LT  714.000  622.212
\LT  722.000  617.064
\LT  730.000  611.850
\LT  738.000  606.569
\LT  746.000  601.221
\LT  754.000  595.806
\LT  762.000  590.325
\LT  770.000  584.776
\LT  778.000  579.162
\LT  786.000  573.480
\LT  794.000  567.732
\LT  802.000  561.917
\LT  810.000  556.035
\LT  818.000  550.087
\LT  826.000  544.072
\LT  834.000  537.990
\LT  842.000  531.842
\LT  850.000  525.627
\LT  858.000  519.345
\LT  866.000  512.996
\LT  874.000  506.581
\LT  882.000  500.099
\LT  890.000  493.551
\LT  898.000  486.935
\LT  906.000  480.253
\LT  914.000  473.505
\LT  922.000  466.689
\LT  930.000  459.807
\LT  938.000  452.858
\LT  946.000  445.843
\LT  954.000  438.761
\LT  962.000  431.612
\LT  970.000  424.397
\LT  978.000  417.114
\LT  986.000  409.765
\LT  994.000  402.350
\LT 1002.000  394.868
\LT 1010.000  387.319
\LT 1018.000  379.703
\LT 1026.000  372.021
\LT 1034.000  364.272
\LT 1042.000  356.456
\LT 1050.000  348.574
\LT 1058.000  340.625
\LT 1066.000  332.609
\LT 1074.000  324.527
\LT 1082.000  316.378
\LT 1090.000  308.162
\LT 1098.000  299.879
\LT 1106.000  291.530
\LT 1114.000  283.115
\LT 1122.000  274.632
\LT 1130.000  266.083
\LT 1138.000  257.467
\LT 1146.000  248.785
\LT 1154.000  240.035
\LT 1162.000  231.220
\LT 1170.000  222.337
\LT 1178.000  213.388
\LT 1186.000  204.372
\LT 1194.000  195.289
\LT 1202.000  186.140
\LT 1210.000  176.924
\LT 1218.000  167.642
\LT 1226.000  158.292
\LT 1234.000  148.876
\LT 1242.000  139.394
\LT 1250.000  129.844
\LT 1258.000  120.228
\LT 1266.000  110.546
\LT 1274.000  100.796
\LT 1282.000   90.980
\LT 1290.000   81.098
\LT 1298.000   71.148
\LT 1300.000   68.651
\koniec    0.10000  -0.100
\obraz27
\grub0.2pt
\MT   0.000  612.000
\LT1400.000  612.000
\MT 160.000  612.000
\LT 160.000  622.000
\MT 220.000  612.000
\LT 220.000  622.000
\MT 280.000  612.000
\LT 280.000  622.000
\MT 340.000  612.000
\LT 340.000  622.000
\MT 400.000  612.000
\LT 400.000  622.000
\cput(400.000,542.000,5)
\MT 460.000  612.000
\LT 460.000  622.000
\MT 520.000  612.000
\LT 520.000  622.000
\MT 580.000  612.000
\LT 580.000  622.000
\MT 640.000  612.000
\LT 640.000  622.000
\MT 700.000  612.000
\LT 700.000  622.000
\cput(700.000,542.000,10)
\MT 760.000  612.000
\LT 760.000  622.000
\MT 820.000  612.000
\LT 820.000  622.000
\MT 880.000  612.000
\LT 880.000  622.000
\MT 940.000  612.000
\LT 940.000  622.000
\MT1000.000  612.000
\LT1000.000  622.000
\cput(1000.000,542.000,15)
\MT1060.000  612.000
\LT1060.000  622.000
\MT1120.000  612.000
\LT1120.000  622.000
\MT1180.000  612.000
\LT1180.000  622.000
\MT1240.000  612.000
\LT1240.000  622.000
\MT1300.000  612.000
\LT1300.000  622.000
\cput(1300.000,542.000,20)
\MT 100.000    0.000
\LT 100.000  850.000
\MT  92.000   50.000
\LT 108.000   50.000
\MT  92.000   87.500
\LT 108.000   87.500
\MT  92.000  125.000
\LT 108.000  125.000
\MT  92.000  162.500
\LT 108.000  162.500
\MT  92.000  200.000
\LT 108.000  200.000
\MT  92.000  237.500
\LT 108.000  237.500
\MT  92.000  275.000
\LT 108.000  275.000
\MT  92.000  312.500
\LT 108.000  312.500
\MT  92.000  350.000
\LT 108.000  350.000
\MT  92.000  387.500
\LT 108.000  387.500
\MT  92.000  425.000
\LT 108.000  425.000
\MT  92.000  462.500
\LT 108.000  462.500
\MT  92.000  500.000
\LT 108.000  500.000
\MT  92.000  537.500
\LT 108.000  537.500
\MT  92.000  575.000
\LT 108.000  575.000
\MT  92.000  612.500
\LT 108.000  612.500
\MT  92.000  650.000
\LT 108.000  650.000
\MT  92.000  687.500
\LT 108.000  687.500
\MT  92.000  725.000
\LT 108.000  725.000
\MT  92.000  762.500
\LT 108.000  762.500
\MT  92.000  800.000
\LT 108.000  800.000
\MT  84.000   50.000
\LT 116.000   50.000
\lput(80.000,31.250, -3.0)
\MT  84.000  237.500
\LT 116.000  237.500
\lput(80.000,218.750, -2.0)
\MT  84.000  425.000
\LT 116.000  425.000
\lput(80.000,406.250, -1.0)
\MT  84.000  612.500
\LT 116.000  612.500
\MT  84.000  800.000
\LT 116.000  800.000
\lput(80.000,781.250,  1.0)
\grub0.6pt
\MT  100.000  800.000
\LT  104.000  799.964
\LT  108.000  799.856
\LT  112.000  799.675
\LT  116.000  799.423
\LT  120.000  799.101
\LT  124.000  798.711
\LT  128.000  798.257
\LT  133.000  797.612
\LT  139.000  796.749
\LT  156.000  794.138
\LT  163.000  793.134
\LT  169.000  792.340
\LT  176.000  791.484
\LT  185.000  790.470
\LT  212.000  787.606
\LT  223.000  786.374
\LT  232.000  785.300
\LT  241.000  784.157
\LT  250.000  782.936
\LT  258.000  781.782
\LT  266.000  780.562
\LT  274.000  779.273
\LT  282.000  777.915
\LT  290.000  776.487
\LT  298.000  774.989
\LT  306.000  773.420
\LT  314.000  771.781
\LT  322.000  770.071
\LT  330.000  768.290
\LT  338.000  766.438
\LT  346.000  764.516
\LT  354.000  762.524
\LT  362.000  760.461
\LT  370.000  758.329
\LT  378.000  756.126
\LT  386.000  753.854
\LT  394.000  751.512
\LT  402.000  749.100
\LT  410.000  746.619
\LT  418.000  744.069
\LT  426.000  741.450
\LT  434.000  738.762
\LT  442.000  736.005
\LT  450.000  733.179
\LT  458.000  730.284
\LT  466.000  727.321
\LT  474.000  724.289
\LT  482.000  721.189
\LT  490.000  718.021
\LT  498.000  714.784
\LT  506.000  711.480
\LT  514.000  708.107
\LT  522.000  704.666
\LT  530.000  701.157
\LT  538.000  697.580
\LT  546.000  693.936
\LT  554.000  690.223
\LT  562.000  686.443
\LT  570.000  682.595
\LT  578.000  678.679
\LT  586.000  674.696
\LT  594.000  670.645
\LT  602.000  666.527
\LT  610.000  662.341
\LT  618.000  658.087
\LT  626.000  653.766
\LT  634.000  649.378
\LT  642.000  644.922
\LT  650.000  640.399
\LT  658.000  635.809
\LT  666.000  631.151
\LT  674.000  626.426
\LT  682.000  621.634
\LT  690.000  616.774
\LT  698.000  611.847
\LT  706.000  606.853
\LT  714.000  601.792
\LT  722.000  596.663
\LT  730.000  591.468
\LT  738.000  586.205
\LT  746.000  580.875
\LT  754.000  575.478
\LT  762.000  570.014
\LT  770.000  564.483
\LT  778.000  558.885
\LT  786.000  553.220
\LT  794.000  547.488
\LT  802.000  541.688
\LT  810.000  535.822
\LT  818.000  529.889
\LT  826.000  523.888
\LT  834.000  517.821
\LT  842.000  511.686
\LT  850.000  505.485
\LT  858.000  499.217
\LT  866.000  492.882
\LT  874.000  486.480
\LT  882.000  480.010
\LT  890.000  473.474
\LT  898.000  466.871
\LT  906.000  460.201
\LT  914.000  453.465
\LT  922.000  446.661
\LT  930.000  439.790
\LT  938.000  432.853
\LT  946.000  425.849
\LT  954.000  418.777
\LT  962.000  411.639
\LT  970.000  404.434
\LT  978.000  397.162
\LT  986.000  389.824
\LT  994.000  382.418
\LT 1002.000  374.946
\LT 1010.000  367.406
\LT 1018.000  359.800
\LT 1026.000  352.127
\LT 1034.000  344.387
\LT 1042.000  336.581
\LT 1050.000  328.707
\LT 1058.000  320.767
\LT 1066.000  312.760
\LT 1074.000  304.686
\LT 1082.000  296.545
\LT 1090.000  288.338
\LT 1098.000  280.064
\LT 1106.000  271.723
\LT 1114.000  263.315
\LT 1122.000  254.840
\LT 1130.000  246.298
\LT 1138.000  237.690
\LT 1146.000  229.015
\LT 1154.000  220.273
\LT 1162.000  211.464
\LT 1170.000  202.589
\LT 1178.000  193.647
\LT 1186.000  184.638
\LT 1194.000  175.562
\LT 1202.000  166.420
\LT 1210.000  157.210
\LT 1218.000  147.934
\LT 1226.000  138.591
\LT 1234.000  129.182
\LT 1242.000  119.705
\LT 1250.000  110.162
\LT 1258.000  100.553
\LT 1266.000   90.876
\LT 1274.000   81.133
\LT 1282.000   71.322
\LT 1290.000   61.446
\LT 1298.000   51.502
\LT 1300.000   49.006
\koniec    0.10000  -0.010
\obraz28
\grub0.2pt
\MT   0.000   60.000
\LT1400.000   60.000
\MT 160.000   60.000
\LT 160.000   70.000
\MT 220.000   60.000
\LT 220.000   70.000
\MT 280.000   60.000
\LT 280.000   70.000
\MT 340.000   60.000
\LT 340.000   70.000
\cput(340.000,-10.000,2)
\MT 400.000   60.000
\LT 400.000   70.000
\MT 460.000   60.000
\LT 460.000   70.000
\MT 520.000   60.000
\LT 520.000   70.000
\MT 580.000   60.000
\LT 580.000   70.000
\cput(580.000,-10.000,4)
\MT 640.000   60.000
\LT 640.000   70.000
\MT 700.000   60.000
\LT 700.000   70.000
\MT 760.000   60.000
\LT 760.000   70.000
\MT 820.000   60.000
\LT 820.000   70.000
\cput(820.000,-10.000,6)
\MT 880.000   60.000
\LT 880.000   70.000
\MT 940.000   60.000
\LT 940.000   70.000
\MT1000.000   60.000
\LT1000.000   70.000
\MT1060.000   60.000
\LT1060.000   70.000
\cput(1060.000,-10.000,8)
\MT1120.000   60.000
\LT1120.000   70.000
\MT1180.000   60.000
\LT1180.000   70.000
\MT1240.000   60.000
\LT1240.000   70.000
\MT1300.000   60.000
\LT1300.000   70.000
\cput(1300.000,-10.000,10)
\MT 100.000    0.000
\LT 100.000   70.000
\multi(100.000,70.000)(0.0000,4.0000){25}{\linia(0,0)(0.0000,2.0000)}
\MT 100.000  170.000
\LT 100.000  850.000
\MT  92.000  200.000
\LT 108.000  200.000
\MT  92.000  225.000
\LT 108.000  225.000
\MT  92.000  250.000
\LT 108.000  250.000
\MT  92.000  275.000
\LT 108.000  275.000
\MT  92.000  300.000
\LT 108.000  300.000
\MT  92.000  325.000
\LT 108.000  325.000
\MT  92.000  350.000
\LT 108.000  350.000
\MT  92.000  375.000
\LT 108.000  375.000
\MT  92.000  400.000
\LT 108.000  400.000
\MT  92.000  425.000
\LT 108.000  425.000
\MT  92.000  450.000
\LT 108.000  450.000
\MT  92.000  475.000
\LT 108.000  475.000
\MT  92.000  500.000
\LT 108.000  500.000
\MT  92.000  525.000
\LT 108.000  525.000
\MT  92.000  550.000
\LT 108.000  550.000
\MT  92.000  575.000
\LT 108.000  575.000
\MT  92.000  600.000
\LT 108.000  600.000
\MT  92.000  625.000
\LT 108.000  625.000
\MT  92.000  650.000
\LT 108.000  650.000
\MT  92.000  675.000
\LT 108.000  675.000
\MT  92.000  700.000
\LT 108.000  700.000
\MT  92.000  725.000
\LT 108.000  725.000
\MT  92.000  750.000
\LT 108.000  750.000
\MT  92.000  775.000
\LT 108.000  775.000
\MT  92.000  800.000
\LT 108.000  800.000
\MT  84.000  200.000
\LT 116.000  200.000
\lput(80.000,187.500,0.880)
\MT  84.000  300.000
\LT 116.000  300.000
\lput(80.000,287.500,0.900)
\MT  84.000  400.000
\LT 116.000  400.000
\lput(80.000,387.500,0.920)
\MT  84.000  500.000
\LT 116.000  500.000
\lput(80.000,487.500,0.940)
\MT  84.000  600.000
\LT 116.000  600.000
\lput(80.000,587.500,0.960)
\MT  84.000  700.000
\LT 116.000  700.000
\lput(80.000,687.500,0.980)
\MT  84.000  800.000
\LT 116.000  800.000
\lput(80.000,787.500,1.000)
\grub0.6pt
\MT  100.000  800.000
\LT  101.000  799.988
\LT  102.000  799.952
\LT  103.000  799.893
\LT  104.000  799.809
\LT  105.000  799.702
\LT  106.000  799.571
\LT  107.000  799.416
\LT  108.000  799.237
\LT  109.000  799.034
\LT  110.000  798.808
\LT  111.000  798.558
\LT  112.000  798.283
\LT  113.000  797.985
\LT  114.000  797.664
\LT  115.000  797.318
\LT  116.000  796.949
\LT  117.000  796.555
\LT  118.000  796.138
\LT  119.000  795.698
\LT  120.000  795.233
\LT  121.000  794.745
\LT  122.000  794.233
\LT  123.000  793.697
\LT  124.000  793.138
\LT  125.000  792.555
\LT  126.000  791.949
\LT  127.000  791.319
\LT  128.000  790.665
\LT  129.000  789.988
\LT  130.000  789.288
\LT  131.000  788.565
\LT  132.000  787.818
\LT  133.000  787.048
\LT  134.000  786.256
\LT  135.000  785.440
\LT  136.000  784.602
\LT  137.000  783.741
\LT  138.000  782.857
\LT  139.000  781.951
\LT  140.000  781.023
\LT  141.000  780.073
\LT  142.000  779.101
\LT  143.000  778.107
\LT  144.000  777.092
\LT  145.000  776.056
\LT  146.000  774.998
\LT  147.000  773.920
\LT  148.000  772.821
\LT  150.000  770.564
\LT  152.000  768.227
\LT  154.000  765.816
\LT  156.000  763.331
\LT  158.000  760.777
\LT  160.000  758.157
\LT  162.000  755.475
\LT  164.000  752.734
\LT  166.000  749.940
\LT  168.000  747.097
\LT  170.000  744.209
\LT  173.000  739.805
\LT  176.000  735.330
\LT  180.000  729.285
\LT  189.000  715.581
\LT  193.000  709.550
\LT  196.000  705.088
\LT  199.000  700.699
\LT  201.000  697.821
\LT  203.000  694.986
\LT  205.000  692.198
\LT  207.000  689.462
\LT  209.000  686.781
\LT  211.000  684.157
\LT  213.000  681.593
\LT  215.000  679.092
\LT  217.000  676.656
\LT  219.000  674.286
\LT  221.000  671.984
\LT  223.000  669.750
\LT  225.000  667.587
\LT  227.000  665.494
\LT  229.000  663.471
\LT  231.000  661.518
\LT  233.000  659.635
\LT  235.000  657.823
\LT  237.000  656.079
\LT  239.000  654.403
\LT  241.000  652.795
\LT  243.000  651.253
\LT  245.000  649.775
\LT  247.000  648.362
\LT  249.000  647.010
\LT  251.000  645.719
\LT  253.000  644.488
\LT  255.000  643.313
\LT  257.000  642.195
\LT  259.000  641.131
\LT  261.000  640.119
\LT  263.000  639.158
\LT  265.000  638.246
\LT  267.000  637.382
\LT  269.000  636.563
\LT  271.000  635.788
\LT  273.000  635.055
\LT  276.000  634.032
\LT  279.000  633.096
\LT  282.000  632.240
\LT  285.000  631.460
\LT  288.000  630.752
\LT  291.000  630.109
\LT  294.000  629.529
\LT  297.000  629.007
\LT  300.000  628.538
\LT  303.000  628.119
\LT  307.000  627.631
\LT  311.000  627.218
\LT  315.000  626.871
\LT  319.000  626.583
\LT  324.000  626.297
\LT  329.000  626.082
\LT  335.000  625.903
\LT  341.000  625.796
\LT  348.000  625.741
\LT  357.000  625.754
\LT  369.000  625.860
\LT  394.000  626.153
\LT  406.000  626.234
\LT  417.000  626.236
\LT  427.000  626.163
\LT  436.000  626.027
\LT  445.000  625.819
\LT  454.000  625.534
\LT  462.000  625.213
\LT  470.000  624.826
\LT  478.000  624.371
\LT  486.000  623.848
\LT  494.000  623.255
\LT  502.000  622.591
\LT  510.000  621.856
\LT  518.000  621.050
\LT  526.000  620.173
\LT  534.000  619.224
\LT  542.000  618.203
\LT  550.000  617.111
\LT  558.000  615.948
\LT  566.000  614.714
\LT  574.000  613.410
\LT  582.000  612.035
\LT  590.000  610.591
\LT  598.000  609.078
\LT  606.000  607.496
\LT  614.000  605.845
\LT  622.000  604.127
\LT  630.000  602.342
\LT  638.000  600.489
\LT  646.000  598.571
\LT  654.000  596.586
\LT  662.000  594.536
\LT  670.000  592.421
\LT  678.000  590.242
\LT  687.000  587.714
\LT  696.000  585.106
\LT  705.000  582.419
\LT  714.000  579.652
\LT  723.000  576.807
\LT  732.000  573.884
\LT  741.000  570.884
\LT  750.000  567.808
\LT  759.000  564.656
\LT  768.000  561.428
\LT  777.000  558.125
\LT  786.000  554.748
\LT  795.000  551.297
\LT  804.000  547.772
\LT  813.000  544.175
\LT  822.000  540.505
\LT  831.000  536.763
\LT  840.000  532.949
\LT  849.000  529.064
\LT  858.000  525.108
\LT  867.000  521.082
\LT  876.000  516.986
\LT  885.000  512.820
\LT  894.000  508.585
\LT  903.000  504.280
\LT  912.000  499.907
\LT  921.000  495.466
\LT  930.000  490.956
\LT  939.000  486.379
\LT  948.000  481.734
\LT  957.000  477.022
\LT  966.000  472.243
\LT  975.000  467.398
\LT  984.000  462.486
\LT  993.000  457.508
\LT 1002.000  452.464
\LT 1011.000  447.354
\LT 1020.000  442.179
\LT 1029.000  436.939
\LT 1038.000  431.633
\LT 1048.000  425.662
\LT 1058.000  419.612
\LT 1068.000  413.482
\LT 1078.000  407.272
\LT 1088.000  400.984
\LT 1098.000  394.617
\LT 1108.000  388.172
\LT 1118.000  381.648
\LT 1128.000  375.046
\LT 1138.000  368.366
\LT 1148.000  361.609
\LT 1158.000  354.774
\LT 1168.000  347.862
\LT 1178.000  340.873
\LT 1188.000  333.807
\LT 1198.000  326.664
\LT 1208.000  319.445
\LT 1218.000  312.149
\LT 1228.000  304.778
\LT 1238.000  297.330
\LT 1248.000  289.806
\LT 1258.000  282.206
\LT 1268.000  274.531
\LT 1278.000  266.780
\LT 1288.000  258.954
\LT 1298.000  251.053
\LT 1300.000  249.464
\koniec    0.10000  -0.001
\obraz29
\grub0.2pt
\MT   0.000   60.000
\LT1400.000   60.000
\MT 160.000   60.000
\LT 160.000   70.000
\MT 220.000   60.000
\LT 220.000   70.000
\MT 280.000   60.000
\LT 280.000   70.000
\MT 340.000   60.000
\LT 340.000   70.000
\cput(340.000,-10.000,2)
\MT 400.000   60.000
\LT 400.000   70.000
\MT 460.000   60.000
\LT 460.000   70.000
\MT 520.000   60.000
\LT 520.000   70.000
\MT 580.000   60.000
\LT 580.000   70.000
\cput(580.000,-10.000,4)
\MT 640.000   60.000
\LT 640.000   70.000
\MT 700.000   60.000
\LT 700.000   70.000
\MT 760.000   60.000
\LT 760.000   70.000
\MT 820.000   60.000
\LT 820.000   70.000
\cput(820.000,-10.000,6)
\MT 880.000   60.000
\LT 880.000   70.000
\MT 940.000   60.000
\LT 940.000   70.000
\MT1000.000   60.000
\LT1000.000   70.000
\MT1060.000   60.000
\LT1060.000   70.000
\cput(1060.000,-10.000,8)
\MT1120.000   60.000
\LT1120.000   70.000
\MT1180.000   60.000
\LT1180.000   70.000
\MT1240.000   60.000
\LT1240.000   70.000
\MT1300.000   60.000
\LT1300.000   70.000
\cput(1300.000,-10.000,10)
\MT 100.000    0.000
\LT 100.000   70.000
\multi(100.000,70.000)(0.0000,4.0000){25}{\linia(0,0)(0.0000,2.0000)}
\MT 100.000  170.000
\LT 100.000  850.000
\MT  92.000  200.000
\LT 108.000  200.000
\MT  92.000  240.000
\LT 108.000  240.000
\MT  92.000  280.000
\LT 108.000  280.000
\MT  92.000  320.000
\LT 108.000  320.000
\MT  92.000  360.000
\LT 108.000  360.000
\MT  92.000  400.000
\LT 108.000  400.000
\MT  92.000  440.000
\LT 108.000  440.000
\MT  92.000  480.000
\LT 108.000  480.000
\MT  92.000  520.000
\LT 108.000  520.000
\MT  92.000  560.000
\LT 108.000  560.000
\MT  92.000  600.000
\LT 108.000  600.000
\MT  92.000  640.000
\LT 108.000  640.000
\MT  92.000  680.000
\LT 108.000  680.000
\MT  92.000  720.000
\LT 108.000  720.000
\MT  92.000  760.000
\LT 108.000  760.000
\MT  92.000  800.000
\LT 108.000  800.000
\MT  84.000  200.000
\LT 116.000  200.000
\lput(80.000,180.000,0.970)
\MT  84.000  400.000
\LT 116.000  400.000
\lput(80.000,380.000,0.980)
\MT  84.000  600.000
\LT 116.000  600.000
\lput(80.000,580.000,0.990)
\MT  84.000  800.000
\LT 116.000  800.000
\lput(80.000,780.000,1.000)
\grub0.6pt
\MT  100.000  800.000
\LT  101.000  799.954
\LT  102.000  799.815
\LT  103.000  799.583
\LT  104.000  799.259
\LT  105.000  798.843
\LT  106.000  798.333
\LT  107.000  797.731
\LT  108.000  797.037
\LT  109.000  796.250
\LT  110.000  795.370
\LT  111.000  794.398
\LT  112.000  793.334
\LT  113.000  792.176
\LT  114.000  790.927
\LT  115.000  789.584
\LT  116.000  788.150
\LT  117.000  786.623
\LT  118.000  785.003
\LT  119.000  783.292
\LT  120.000  781.488
\LT  121.000  779.592
\LT  122.000  777.603
\LT  123.000  775.523
\LT  124.000  773.352
\LT  125.000  771.088
\LT  126.000  768.733
\LT  127.000  766.287
\LT  128.000  763.750
\LT  129.000  761.122
\LT  130.000  758.403
\LT  131.000  755.594
\LT  132.000  752.695
\LT  133.000  749.706
\LT  134.000  746.629
\LT  135.000  743.462
\LT  136.000  740.207
\LT  137.000  736.864
\LT  138.000  733.434
\LT  139.000  729.918
\LT  140.000  726.315
\LT  141.000  722.627
\LT  142.000  718.854
\LT  143.000  714.997
\LT  144.000  711.057
\LT  145.000  707.035
\LT  146.000  702.931
\LT  147.000  698.748
\LT  148.000  694.485
\LT  149.000  690.144
\LT  150.000  685.726
\LT  151.000  681.233
\LT  152.000  676.665
\LT  153.000  672.025
\LT  154.000  667.312
\LT  155.000  662.530
\LT  156.000  657.680
\LT  157.000  652.762
\LT  158.000  647.780
\LT  159.000  642.735
\LT  160.000  637.628
\LT  161.000  632.461
\LT  162.000  627.238
\LT  163.000  621.959
\LT  164.000  616.626
\LT  165.000  611.243
\LT  166.000  605.810
\LT  167.000  600.332
\LT  168.000  594.809
\LT  169.000  589.244
\LT  170.000  583.640
\LT  171.000  578.000
\LT  172.000  572.326
\LT  173.000  566.620
\LT  174.000  560.885
\LT  175.000  555.125
\LT  176.000  549.341
\LT  178.000  537.715
\LT  180.000  526.029
\LT  186.000  490.847
\LT  188.000  479.154
\LT  190.000  467.517
\LT  191.000  461.726
\LT  192.000  455.956
\LT  193.000  450.212
\LT  194.000  444.495
\LT  195.000  438.808
\LT  196.000  433.153
\LT  197.000  427.533
\LT  198.000  421.951
\LT  199.000  416.408
\LT  200.000  410.907
\LT  201.000  405.451
\LT  202.000  400.040
\LT  203.000  394.678
\LT  204.000  389.366
\LT  205.000  384.106
\LT  206.000  378.900
\LT  207.000  373.750
\LT  208.000  368.657
\LT  209.000  363.624
\LT  210.000  358.650
\LT  211.000  353.739
\LT  212.000  348.890
\LT  213.000  344.106
\LT  214.000  339.388
\LT  215.000  334.735
\LT  216.000  330.151
\LT  217.000  325.635
\LT  218.000  321.188
\LT  219.000  316.811
\LT  220.000  312.505
\LT  221.000  308.269
\LT  222.000  304.106
\LT  223.000  300.014
\LT  224.000  295.995
\LT  225.000  292.049
\LT  226.000  288.176
\LT  227.000  284.376
\LT  228.000  280.649
\LT  229.000  276.995
\LT  230.000  273.414
\LT  231.000  269.906
\LT  232.000  266.472
\LT  233.000  263.110
\LT  234.000  259.820
\LT  235.000  256.603
\LT  236.000  253.457
\LT  237.000  250.383
\LT  238.000  247.380
\LT  239.000  244.448
\LT  240.000  241.585
\LT  241.000  238.792
\LT  242.000  236.068
\LT  243.000  233.412
\LT  244.000  230.824
\LT  245.000  228.303
\LT  246.000  225.849
\LT  247.000  223.460
\LT  248.000  221.136
\LT  249.000  218.876
\LT  250.000  216.679
\LT  251.000  214.546
\LT  252.000  212.474
\LT  253.000  210.463
\LT  254.000  208.513
\LT  255.000  206.622
\LT  256.000  204.789
\LT  257.000  203.015
\LT  258.000  201.297
\LT  259.000  199.636
\LT  260.000  198.029
\LT  261.000  196.477
\LT  262.000  194.979
\LT  263.000  193.533
\LT  264.000  192.140
\LT  265.000  190.797
\LT  266.000  189.504
\LT  267.000  188.261
\LT  268.000  187.066
\LT  269.000  185.919
\LT  270.000  184.818
\LT  271.000  183.764
\LT  272.000  182.754
\LT  273.000  181.789
\LT  274.000  180.868
\LT  275.000  179.989
\LT  276.000  179.152
\LT  277.000  178.356
\LT  278.000  177.600
\LT  279.000  176.884
\LT  280.000  176.207
\LT  281.000  175.567
\LT  282.000  174.965
\LT  283.000  174.400
\LT  284.000  173.870
\LT  285.000  173.375
\LT  286.000  172.915
\LT  287.000  172.489
\LT  288.000  172.095
\LT  289.000  171.734
\LT  290.000  171.404
\LT  291.000  171.105
\LT  292.000  170.837
\LT  293.000  170.598
\LT  294.000  170.389
\LT  295.000  170.207
\LT  296.000  170.054
\LT  297.000  169.928
\LT  298.000  169.828
\LT  299.000  169.755
\LT  300.000  169.707
\LT  301.000  169.684
\LT  302.000  169.685
\LT  303.000  169.710
\LT  304.000  169.758
\LT  305.000  169.829
\LT  306.000  169.923
\LT  307.000  170.038
\LT  308.000  170.174
\LT  309.000  170.331
\LT  311.000  170.706
\LT  313.000  171.158
\LT  315.000  171.684
\LT  317.000  172.280
\LT  319.000  172.944
\LT  321.000  173.671
\LT  323.000  174.460
\LT  325.000  175.306
\LT  327.000  176.208
\LT  329.000  177.162
\LT  331.000  178.166
\LT  333.000  179.218
\LT  335.000  180.314
\LT  337.000  181.453
\LT  340.000  183.237
\LT  343.000  185.105
\LT  346.000  187.051
\LT  349.000  189.068
\LT  352.000  191.150
\LT  355.000  193.293
\LT  358.000  195.490
\LT  361.000  197.737
\LT  365.000  200.804
\LT  369.000  203.942
\LT  373.000  207.142
\LT  378.000  211.217
\LT  383.000  215.362
\LT  389.000  220.409
\LT  396.000  226.373
\LT  406.000  234.984
\LT  428.000  254.026
\LT  438.000  262.621
\LT  447.000  270.284
\LT  455.000  277.024
\LT  462.000  282.857
\LT  469.000  288.624
\LT  476.000  294.321
\LT  483.000  299.944
\LT  490.000  305.491
\LT  497.000  310.959
\LT  504.000  316.346
\LT  510.000  320.898
\LT  516.000  325.390
\LT  522.000  329.821
\LT  528.000  334.191
\LT  534.000  338.500
\LT  540.000  342.749
\LT  546.000  346.936
\LT  552.000  351.064
\LT  559.000  355.803
\LT  566.000  360.462
\LT  573.000  365.042
\LT  580.000  369.542
\LT  587.000  373.965
\LT  594.000  378.312
\LT  601.000  382.583
\LT  608.000  386.781
\LT  615.000  390.906
\LT  622.000  394.959
\LT  629.000  398.942
\LT  636.000  402.857
\LT  643.000  406.704
\LT  650.000  410.485
\LT  657.000  414.202
\LT  664.000  417.855
\LT  671.000  421.445
\LT  679.000  425.474
\LT  687.000  429.426
\LT  695.000  433.301
\LT  703.000  437.103
\LT  711.000  440.833
\LT  719.000  444.492
\LT  727.000  448.082
\LT  735.000  451.606
\LT  743.000  455.064
\LT  752.000  458.878
\LT  761.000  462.614
\LT  770.000  466.274
\LT  779.000  469.859
\LT  788.000  473.372
\LT  797.000  476.815
\LT  806.000  480.189
\LT  815.000  483.497
\LT  825.000  487.097
\LT  835.000  490.619
\LT  845.000  494.066
\LT  855.000  497.441
\LT  865.000  500.744
\LT  875.000  503.979
\LT  886.000  507.461
\LT  897.000  510.864
\LT  908.000  514.192
\LT  919.000  517.446
\LT  930.000  520.630
\LT  941.000  523.745
\LT  953.000  527.066
\LT  965.000  530.311
\LT  977.000  533.482
\LT  989.000  536.581
\LT 1002.000  539.860
\LT 1015.000  543.061
\LT 1028.000  546.185
\LT 1041.000  549.236
\LT 1054.000  552.216
\LT 1068.000  555.349
\LT 1082.000  558.405
\LT 1096.000  561.387
\LT 1110.000  564.298
\LT 1125.000  567.340
\LT 1140.000  570.306
\LT 1155.000  573.199
\LT 1171.000  576.206
\LT 1187.000  579.136
\LT 1203.000  581.992
\LT 1220.000  584.947
\LT 1237.000  587.824
\LT 1254.000  590.627
\LT 1272.000  593.515
\LT 1290.000  596.326
\LT 1300.000  597.855
\koniec    0.10000   0.000
\obraz30
\grub0.2pt
\MT   0.000   60.000
\LT1400.000   60.000
\MT 160.000   60.000
\LT 160.000   70.000
\MT 220.000   60.000
\LT 220.000   70.000
\MT 280.000   60.000
\LT 280.000   70.000
\MT 340.000   60.000
\LT 340.000   70.000
\cput(340.000,-10.000,2)
\MT 400.000   60.000
\LT 400.000   70.000
\MT 460.000   60.000
\LT 460.000   70.000
\MT 520.000   60.000
\LT 520.000   70.000
\MT 580.000   60.000
\LT 580.000   70.000
\cput(580.000,-10.000,4)
\MT 640.000   60.000
\LT 640.000   70.000
\MT 700.000   60.000
\LT 700.000   70.000
\MT 760.000   60.000
\LT 760.000   70.000
\MT 820.000   60.000
\LT 820.000   70.000
\cput(820.000,-10.000,6)
\MT 880.000   60.000
\LT 880.000   70.000
\MT 940.000   60.000
\LT 940.000   70.000
\MT1000.000   60.000
\LT1000.000   70.000
\MT1060.000   60.000
\LT1060.000   70.000
\cput(1060.000,-10.000,8)
\MT1120.000   60.000
\LT1120.000   70.000
\MT1180.000   60.000
\LT1180.000   70.000
\MT1240.000   60.000
\LT1240.000   70.000
\MT1300.000   60.000
\LT1300.000   70.000
\cput(1300.000,-10.000,10)
\MT 100.000    0.000
\LT 100.000   70.000
\multi(100.000,70.000)(0.0000,4.0000){25}{\linia(0,0)(0.0000,2.0000)}
\MT 100.000  170.000
\LT 100.000  850.000
\MT  92.000  200.000
\LT 108.000  200.000
\MT  92.000  230.000
\LT 108.000  230.000
\MT  92.000  260.000
\LT 108.000  260.000
\MT  92.000  290.000
\LT 108.000  290.000
\MT  92.000  320.000
\LT 108.000  320.000
\MT  92.000  350.000
\LT 108.000  350.000
\MT  92.000  380.000
\LT 108.000  380.000
\MT  92.000  410.000
\LT 108.000  410.000
\MT  92.000  440.000
\LT 108.000  440.000
\MT  92.000  470.000
\LT 108.000  470.000
\MT  92.000  500.000
\LT 108.000  500.000
\MT  92.000  530.000
\LT 108.000  530.000
\MT  92.000  560.000
\LT 108.000  560.000
\MT  92.000  590.000
\LT 108.000  590.000
\MT  92.000  620.000
\LT 108.000  620.000
\MT  92.000  650.000
\LT 108.000  650.000
\MT  92.000  680.000
\LT 108.000  680.000
\MT  92.000  710.000
\LT 108.000  710.000
\MT  92.000  740.000
\LT 108.000  740.000
\MT  92.000  770.000
\LT 108.000  770.000
\MT  92.000  800.000
\LT 108.000  800.000
\MT  84.000  200.000
\LT 116.000  200.000
\lput(80.000,185.000,0.980)
\MT  84.000  320.000
\LT 116.000  320.000
\lput(80.000,305.000,1.000)
\MT  84.000  440.000
\LT 116.000  440.000
\lput(80.000,425.000,1.020)
\MT  84.000  560.000
\LT 116.000  560.000
\lput(80.000,545.000,1.040)
\MT  84.000  680.000
\LT 116.000  680.000
\lput(80.000,665.000,1.060)
\MT  84.000  800.000
\LT 116.000  800.000
\lput(80.000,785.000,1.080)
\grub0.6pt
\MT  100.000  320.000
\LT  101.000  319.987
\LT  102.000  319.946
\LT  103.000  319.879
\LT  104.000  319.784
\LT  105.000  319.663
\LT  106.000  319.515
\LT  107.000  319.340
\LT  108.000  319.138
\LT  109.000  318.909
\LT  110.000  318.653
\LT  111.000  318.370
\LT  112.000  318.060
\LT  113.000  317.723
\LT  114.000  317.360
\LT  115.000  316.969
\LT  116.000  316.552
\LT  117.000  316.107
\LT  118.000  315.636
\LT  119.000  315.138
\LT  120.000  314.613
\LT  121.000  314.061
\LT  122.000  313.483
\LT  123.000  312.877
\LT  124.000  312.245
\LT  125.000  311.587
\LT  126.000  310.902
\LT  127.000  310.190
\LT  128.000  309.452
\LT  129.000  308.687
\LT  130.000  307.896
\LT  131.000  307.079
\LT  132.000  306.235
\LT  133.000  305.366
\LT  134.000  304.470
\LT  135.000  303.549
\LT  136.000  302.602
\LT  137.000  301.630
\LT  138.000  300.632
\LT  139.000  299.609
\LT  140.000  298.561
\LT  141.000  297.488
\LT  142.000  296.391
\LT  143.000  295.269
\LT  144.000  294.124
\LT  145.000  292.954
\LT  146.000  291.761
\LT  147.000  290.545
\LT  148.000  289.305
\LT  149.000  288.044
\LT  150.000  286.760
\LT  151.000  285.454
\LT  152.000  284.126
\LT  153.000  282.778
\LT  154.000  281.409
\LT  156.000  278.611
\LT  158.000  275.736
\LT  160.000  272.788
\LT  162.000  269.773
\LT  164.000  266.695
\LT  166.000  263.558
\LT  168.000  260.369
\LT  170.000  257.134
\LT  173.000  252.206
\LT  176.000  247.209
\LT  181.000  238.785
\LT  187.000  228.652
\LT  190.000  223.630
\LT  193.000  218.667
\LT  195.000  215.403
\LT  197.000  212.180
\LT  199.000  209.006
\LT  201.000  205.886
\LT  203.000  202.824
\LT  205.000  199.826
\LT  207.000  196.895
\LT  209.000  194.038
\LT  211.000  191.255
\LT  212.000  189.894
\LT  213.000  188.552
\LT  214.000  187.231
\LT  215.000  185.931
\LT  216.000  184.652
\LT  217.000  183.394
\LT  218.000  182.158
\LT  219.000  180.944
\LT  220.000  179.751
\LT  221.000  178.581
\LT  222.000  177.433
\LT  223.000  176.308
\LT  224.000  175.205
\LT  225.000  174.125
\LT  226.000  173.068
\LT  227.000  172.033
\LT  228.000  171.021
\LT  229.000  170.032
\LT  230.000  169.066
\LT  231.000  168.122
\LT  232.000  167.202
\LT  233.000  166.303
\LT  234.000  165.428
\LT  235.000  164.575
\LT  236.000  163.744
\LT  237.000  162.935
\LT  238.000  162.149
\LT  239.000  161.385
\LT  240.000  160.642
\LT  241.000  159.921
\LT  242.000  159.222
\LT  243.000  158.544
\LT  244.000  157.887
\LT  245.000  157.251
\LT  246.000  156.636
\LT  247.000  156.042
\LT  248.000  155.467
\LT  250.000  154.379
\LT  252.000  153.369
\LT  254.000  152.435
\LT  256.000  151.577
\LT  258.000  150.791
\LT  260.000  150.075
\LT  262.000  149.429
\LT  264.000  148.849
\LT  266.000  148.333
\LT  268.000  147.880
\LT  270.000  147.487
\LT  272.000  147.153
\LT  274.000  146.875
\LT  276.000  146.652
\LT  278.000  146.482
\LT  280.000  146.362
\LT  282.000  146.291
\LT  284.000  146.268
\LT  286.000  146.290
\LT  288.000  146.355
\LT  290.000  146.463
\LT  293.000  146.700
\LT  296.000  147.023
\LT  299.000  147.427
\LT  302.000  147.907
\LT  305.000  148.459
\LT  308.000  149.079
\LT  311.000  149.762
\LT  314.000  150.505
\LT  317.000  151.304
\LT  320.000  152.157
\LT  323.000  153.058
\LT  327.000  154.333
\LT  331.000  155.684
\LT  335.000  157.105
\LT  339.000  158.590
\LT  343.000  160.135
\LT  348.000  162.143
\LT  353.000  164.228
\LT  358.000  166.382
\LT  363.000  168.599
\LT  369.000  171.333
\LT  375.000  174.139
\LT  382.000  177.493
\LT  389.000  180.923
\LT  397.000  184.923
\LT  405.000  188.996
\LT  414.000  193.655
\LT  424.000  198.911
\LT  435.000  204.775
\LT  447.000  211.256
\LT  460.000  218.359
\LT  473.000  225.537
\LT  487.000  233.341
\LT  502.000  241.779
\LT  517.000  250.294
\LT  532.000  258.881
\LT  547.000  267.542
\LT  562.000  276.278
\LT  577.000  285.091
\LT  591.000  293.388
\LT  605.000  301.758
\LT  619.000  310.203
\LT  632.000  318.115
\LT  645.000  326.098
\LT  658.000  334.153
\LT  671.000  342.284
\LT  684.000  350.493
\LT  696.000  358.141
\LT  708.000  365.859
\LT  720.000  373.650
\LT  732.000  381.514
\LT  744.000  389.454
\LT  756.000  397.471
\LT  768.000  405.567
\LT  779.000  413.058
\LT  790.000  420.618
\LT  801.000  428.247
\LT  812.000  435.947
\LT  823.000  443.719
\LT  834.000  451.563
\LT  845.000  459.480
\LT  856.000  467.472
\LT  867.000  475.539
\LT  878.000  483.683
\LT  889.000  491.903
\LT  900.000  500.200
\LT  911.000  508.576
\LT  922.000  517.031
\LT  933.000  525.566
\LT  943.000  533.395
\LT  953.000  541.290
\LT  963.000  549.253
\LT  973.000  557.284
\LT  983.000  565.383
\LT  993.000  573.550
\LT 1003.000  581.787
\LT 1013.000  590.092
\LT 1023.000  598.468
\LT 1033.000  606.914
\LT 1043.000  615.430
\LT 1053.000  624.017
\LT 1063.000  632.675
\LT 1073.000  641.405
\LT 1083.000  650.207
\LT 1093.000  659.080
\LT 1103.000  668.026
\LT 1113.000  677.044
\LT 1123.000  686.135
\LT 1133.000  695.300
\LT 1143.000  704.538
\LT 1153.000  713.849
\LT 1163.000  723.234
\LT 1173.000  732.693
\LT 1183.000  742.227
\LT 1193.000  751.835
\LT 1203.000  761.518
\LT 1213.000  771.276
\LT 1223.000  781.109
\LT 1233.000  791.017
\LT 1243.000  801.000
\LT 1253.000  811.060
\LT 1263.000  821.195
\LT 1273.000  831.406
\LT 1283.000  841.693
\LT 1293.000  852.057
\LT 1300.000  859.357
\koniec    0.10000   0.001
\obraz31
\grub0.2pt
\MT   0.000   60.000
\LT1400.000   60.000
\MT 160.000   60.000
\LT 160.000   70.000
\MT 220.000   60.000
\LT 220.000   70.000
\MT 280.000   60.000
\LT 280.000   70.000
\MT 340.000   60.000
\LT 340.000   70.000
\cput(340.000,-10.000,2)
\MT 400.000   60.000
\LT 400.000   70.000
\MT 460.000   60.000
\LT 460.000   70.000
\MT 520.000   60.000
\LT 520.000   70.000
\MT 580.000   60.000
\LT 580.000   70.000
\cput(580.000,-10.000,4)
\MT 640.000   60.000
\LT 640.000   70.000
\MT 700.000   60.000
\LT 700.000   70.000
\MT 760.000   60.000
\LT 760.000   70.000
\MT 820.000   60.000
\LT 820.000   70.000
\cput(820.000,-10.000,6)
\MT 880.000   60.000
\LT 880.000   70.000
\MT 940.000   60.000
\LT 940.000   70.000
\MT1000.000   60.000
\LT1000.000   70.000
\MT1060.000   60.000
\LT1060.000   70.000
\cput(1060.000,-10.000,8)
\MT1120.000   60.000
\LT1120.000   70.000
\MT1180.000   60.000
\LT1180.000   70.000
\MT1240.000   60.000
\LT1240.000   70.000
\MT1300.000   60.000
\LT1300.000   70.000
\cput(1300.000,-10.000,10)
\MT 100.000    0.000
\LT 100.000   70.000
\multi(100.000,70.000)(0.0000,4.0000){25}{\linia(0,0)(0.0000,2.0000)}
\MT 100.000  170.000
\LT 100.000  850.000
\MT  92.000  200.000
\LT 108.000  200.000
\MT  92.000  230.000
\LT 108.000  230.000
\MT  92.000  260.000
\LT 108.000  260.000
\MT  92.000  290.000
\LT 108.000  290.000
\MT  92.000  320.000
\LT 108.000  320.000
\MT  92.000  350.000
\LT 108.000  350.000
\MT  92.000  380.000
\LT 108.000  380.000
\MT  92.000  410.000
\LT 108.000  410.000
\MT  92.000  440.000
\LT 108.000  440.000
\MT  92.000  470.000
\LT 108.000  470.000
\MT  92.000  500.000
\LT 108.000  500.000
\MT  92.000  530.000
\LT 108.000  530.000
\MT  92.000  560.000
\LT 108.000  560.000
\MT  92.000  590.000
\LT 108.000  590.000
\MT  92.000  620.000
\LT 108.000  620.000
\MT  92.000  650.000
\LT 108.000  650.000
\MT  92.000  680.000
\LT 108.000  680.000
\MT  92.000  710.000
\LT 108.000  710.000
\MT  92.000  740.000
\LT 108.000  740.000
\MT  92.000  770.000
\LT 108.000  770.000
\MT  92.000  800.000
\LT 108.000  800.000
\MT  84.000  200.000
\LT 116.000  200.000
\lput(80.000,185.000, 1.00)
\MT  84.000  320.000
\LT 116.000  320.000
\lput(80.000,305.000, 1.20)
\MT  84.000  440.000
\LT 116.000  440.000
\lput(80.000,425.000, 1.40)
\MT  84.000  560.000
\LT 116.000  560.000
\lput(80.000,545.000, 1.60)
\MT  84.000  680.000
\LT 116.000  680.000
\lput(80.000,665.000, 1.80)
\MT  84.000  800.000
\LT 116.000  800.000
\lput(80.000,785.000, 2.00)
\grub0.6pt
\MT  100.000  200.000
\LT  106.000  199.965
\LT  112.000  199.860
\LT  118.000  199.685
\LT  124.000  199.441
\LT  130.000  199.127
\LT  136.000  198.746
\LT  142.000  198.301
\LT  149.000  197.705
\LT  156.000  197.037
\LT  165.000  196.098
\LT  184.000  194.017
\LT  192.000  193.205
\LT  198.000  192.660
\LT  204.000  192.188
\LT  209.000  191.859
\LT  214.000  191.597
\LT  219.000  191.405
\LT  224.000  191.287
\LT  229.000  191.244
\LT  234.000  191.276
\LT  239.000  191.384
\LT  244.000  191.565
\LT  249.000  191.817
\LT  254.000  192.137
\LT  259.000  192.523
\LT  264.000  192.971
\LT  269.000  193.478
\LT  275.000  194.160
\LT  281.000  194.917
\LT  287.000  195.745
\LT  293.000  196.638
\LT  300.000  197.758
\LT  307.000  198.955
\LT  314.000  200.224
\LT  321.000  201.561
\LT  329.000  203.165
\LT  337.000  204.847
\LT  345.000  206.602
\LT  353.000  208.426
\LT  361.000  210.316
\LT  370.000  212.517
\LT  379.000  214.795
\LT  388.000  217.147
\LT  397.000  219.571
\LT  406.000  222.065
\LT  415.000  224.627
\LT  424.000  227.257
\LT  433.000  229.954
\LT  442.000  232.716
\LT  451.000  235.544
\LT  461.000  238.761
\LT  471.000  242.058
\LT  481.000  245.434
\LT  491.000  248.889
\LT  501.000  252.422
\LT  511.000  256.033
\LT  521.000  259.723
\LT  531.000  263.491
\LT  541.000  267.337
\LT  551.000  271.262
\LT  561.000  275.265
\LT  571.000  279.346
\LT  581.000  283.506
\LT  591.000  287.744
\LT  601.000  292.061
\LT  611.000  296.457
\LT  621.000  300.932
\LT  631.000  305.486
\LT  641.000  310.119
\LT  651.000  314.831
\LT  661.000  319.623
\LT  671.000  324.494
\LT  681.000  329.444
\LT  691.000  334.475
\LT  701.000  339.585
\LT  711.000  344.775
\LT  720.000  349.515
\LT  729.000  354.320
\LT  738.000  359.189
\LT  747.000  364.123
\LT  756.000  369.123
\LT  765.000  374.188
\LT  774.000  379.318
\LT  783.000  384.513
\LT  792.000  389.774
\LT  801.000  395.100
\LT  810.000  400.492
\LT  819.000  405.949
\LT  828.000  411.472
\LT  837.000  417.060
\LT  846.000  422.714
\LT  855.000  428.434
\LT  864.000  434.219
\LT  873.000  440.071
\LT  882.000  445.988
\LT  891.000  451.971
\LT  900.000  458.020
\LT  909.000  464.135
\LT  918.000  470.316
\LT  927.000  476.563
\LT  936.000  482.877
\LT  945.000  489.256
\LT  954.000  495.702
\LT  963.000  502.214
\LT  972.000  508.792
\LT  981.000  515.436
\LT  990.000  522.147
\LT  999.000  528.924
\LT 1008.000  535.767
\LT 1017.000  542.677
\LT 1026.000  549.653
\LT 1035.000  556.696
\LT 1044.000  563.805
\LT 1053.000  570.980
\LT 1062.000  578.222
\LT 1071.000  585.531
\LT 1080.000  592.906
\LT 1089.000  600.348
\LT 1098.000  607.856
\LT 1107.000  615.431
\LT 1116.000  623.072
\LT 1125.000  630.781
\LT 1134.000  638.556
\LT 1143.000  646.397
\LT 1152.000  654.305
\LT 1161.000  662.280
\LT 1170.000  670.322
\LT 1179.000  678.431
\LT 1188.000  686.606
\LT 1197.000  694.848
\LT 1206.000  703.157
\LT 1215.000  711.533
\LT 1224.000  719.976
\LT 1233.000  728.485
\LT 1242.000  737.061
\LT 1251.000  745.705
\LT 1260.000  754.415
\LT 1269.000  763.192
\LT 1278.000  772.035
\LT 1287.000  780.946
\LT 1296.000  789.924
\LT 1300.000  793.936
\koniec    0.10000   0.010
\obraz32
\grub0.2pt
\MT   0.000   60.000
\LT1400.000   60.000
\MT 160.000   60.000
\LT 160.000   70.000
\MT 220.000   60.000
\LT 220.000   70.000
\MT 280.000   60.000
\LT 280.000   70.000
\MT 340.000   60.000
\LT 340.000   70.000
\cput(340.000,-10.000,2)
\MT 400.000   60.000
\LT 400.000   70.000
\MT 460.000   60.000
\LT 460.000   70.000
\MT 520.000   60.000
\LT 520.000   70.000
\MT 580.000   60.000
\LT 580.000   70.000
\cput(580.000,-10.000,4)
\MT 640.000   60.000
\LT 640.000   70.000
\MT 700.000   60.000
\LT 700.000   70.000
\MT 760.000   60.000
\LT 760.000   70.000
\MT 820.000   60.000
\LT 820.000   70.000
\cput(820.000,-10.000,6)
\MT 880.000   60.000
\LT 880.000   70.000
\MT 940.000   60.000
\LT 940.000   70.000
\MT1000.000   60.000
\LT1000.000   70.000
\MT1060.000   60.000
\LT1060.000   70.000
\cput(1060.000,-10.000,8)
\MT1120.000   60.000
\LT1120.000   70.000
\MT1180.000   60.000
\LT1180.000   70.000
\MT1240.000   60.000
\LT1240.000   70.000
\MT1300.000   60.000
\LT1300.000   70.000
\cput(1300.000,-10.000,10)
\MT 100.000    0.000
\LT 100.000   70.000
\multi(100.000,70.000)(0.0000,4.0000){25}{\linia(0,0)(0.0000,2.0000)}
\MT 100.000  170.000
\LT 100.000  850.000
\MT  92.000   60.000
\LT 108.000   60.000
\MT  92.000   97.000
\LT 108.000   97.000
\MT  92.000  134.000
\LT 108.000  134.000
\MT  92.000  171.000
\LT 108.000  171.000
\MT  92.000  208.000
\LT 108.000  208.000
\MT  92.000  245.000
\LT 108.000  245.000
\MT  92.000  282.000
\LT 108.000  282.000
\MT  92.000  319.000
\LT 108.000  319.000
\MT  92.000  356.000
\LT 108.000  356.000
\MT  92.000  393.000
\LT 108.000  393.000
\MT  92.000  430.000
\LT 108.000  430.000
\MT  92.000  467.000
\LT 108.000  467.000
\MT  92.000  504.000
\LT 108.000  504.000
\MT  92.000  541.000
\LT 108.000  541.000
\MT  92.000  578.000
\LT 108.000  578.000
\MT  92.000  615.000
\LT 108.000  615.000
\MT  92.000  652.000
\LT 108.000  652.000
\MT  92.000  689.000
\LT 108.000  689.000
\MT  92.000  726.000
\LT 108.000  726.000
\MT  92.000  763.000
\LT 108.000  763.000
\MT  92.000  800.000
\LT 108.000  800.000
\MT  84.000   60.000
\LT 116.000   60.000
\MT  84.000  208.000
\LT 116.000  208.000
\lput(80.000,189.500,  2.0)
\MT  84.000  356.000
\LT 116.000  356.000
\lput(80.000,337.500,  4.0)
\MT  84.000  504.000
\LT 116.000  504.000
\lput(80.000,485.500,  6.0)
\MT  84.000  652.000
\LT 116.000  652.000
\lput(80.000,633.500,  8.0)
\MT  84.000  800.000
\LT 116.000  800.000
\lput(80.000,781.500, 10.0)
\grub0.6pt
\MT  100.000  134.000
\LT  110.000  134.034
\LT  120.000  134.137
\LT  130.000  134.309
\LT  140.000  134.550
\LT  150.000  134.862
\LT  160.000  135.249
\LT  169.000  135.667
\LT  178.000  136.156
\LT  186.000  136.657
\LT  194.000  137.225
\LT  202.000  137.867
\LT  210.000  138.585
\LT  217.000  139.279
\LT  224.000  140.037
\LT  231.000  140.858
\LT  238.000  141.742
\LT  245.000  142.689
\LT  252.000  143.699
\LT  260.000  144.928
\LT  268.000  146.236
\LT  276.000  147.621
\LT  284.000  149.082
\LT  292.000  150.616
\LT  300.000  152.223
\LT  308.000  153.903
\LT  316.000  155.652
\LT  324.000  157.472
\LT  332.000  159.361
\LT  340.000  161.318
\LT  348.000  163.343
\LT  356.000  165.436
\LT  364.000  167.596
\LT  372.000  169.823
\LT  380.000  172.116
\LT  388.000  174.476
\LT  396.000  176.902
\LT  404.000  179.395
\LT  412.000  181.953
\LT  420.000  184.577
\LT  428.000  187.266
\LT  436.000  190.021
\LT  444.000  192.842
\LT  452.000  195.729
\LT  460.000  198.680
\LT  468.000  201.698
\LT  476.000  204.781
\LT  484.000  207.929
\LT  492.000  211.142
\LT  500.000  214.421
\LT  508.000  217.766
\LT  516.000  221.175
\LT  524.000  224.651
\LT  532.000  228.191
\LT  540.000  231.797
\LT  548.000  235.468
\LT  556.000  239.205
\LT  564.000  243.007
\LT  572.000  246.874
\LT  580.000  250.807
\LT  588.000  254.806
\LT  596.000  258.869
\LT  604.000  262.998
\LT  612.000  267.193
\LT  620.000  271.453
\LT  628.000  275.778
\LT  636.000  280.169
\LT  644.000  284.625
\LT  652.000  289.147
\LT  660.000  293.734
\LT  668.000  298.387
\LT  676.000  303.105
\LT  684.000  307.888
\LT  692.000  312.737
\LT  700.000  317.652
\LT  708.000  322.632
\LT  716.000  327.678
\LT  724.000  332.789
\LT  732.000  337.966
\LT  740.000  343.208
\LT  748.000  348.516
\LT  756.000  353.889
\LT  764.000  359.328
\LT  772.000  364.832
\LT  780.000  370.402
\LT  788.000  376.038
\LT  796.000  381.739
\LT  804.000  387.506
\LT  812.000  393.338
\LT  820.000  399.236
\LT  828.000  405.199
\LT  836.000  411.228
\LT  844.000  417.323
\LT  852.000  423.483
\LT  860.000  429.709
\LT  868.000  436.000
\LT  876.000  442.357
\LT  884.000  448.780
\LT  892.000  455.269
\LT  900.000  461.822
\LT  908.000  468.442
\LT  916.000  475.127
\LT  924.000  481.878
\LT  932.000  488.695
\LT  940.000  495.577
\LT  948.000  502.525
\LT  956.000  509.538
\LT  964.000  516.617
\LT  972.000  523.762
\LT  980.000  530.972
\LT  988.000  538.248
\LT  996.000  545.590
\LT 1004.000  552.998
\LT 1012.000  560.471
\LT 1020.000  568.009
\LT 1028.000  575.614
\LT 1036.000  583.284
\LT 1044.000  591.020
\LT 1052.000  598.821
\LT 1060.000  606.688
\LT 1068.000  614.621
\LT 1076.000  622.620
\LT 1084.000  630.684
\LT 1092.000  638.814
\LT 1100.000  647.009
\LT 1108.000  655.270
\LT 1116.000  663.597
\LT 1124.000  671.990
\LT 1132.000  680.448
\LT 1140.000  688.972
\LT 1148.000  697.562
\LT 1156.000  706.218
\LT 1164.000  714.939
\LT 1172.000  723.726
\LT 1180.000  732.578
\LT 1188.000  741.496
\LT 1196.000  750.480
\LT 1204.000  759.530
\LT 1212.000  768.645
\LT 1220.000  777.827
\LT 1228.000  787.073
\LT 1236.000  796.386
\LT 1244.000  805.764
\LT 1252.000  815.208
\LT 1260.000  824.718
\LT 1268.000  834.293
\LT 1276.000  843.934
\LT 1284.000  853.641
\LT 1292.000  863.414
\LT 1300.000  873.252
\koniec    0.10000   0.100
\eject

\def\obraz#1 {\global\nr=#1
\ifodd\nr \vtop to850\jedn \bgroup \vfill \bps \fi
\hfil \vbox \bgroup \hsize=\hs
\leftline \bgroup \hskip-0.7\hs
\nd=\nr \global\divide\nd by8 \nf=\nd \global\multiply\nf by8
\ng=\nr \advance\ng by-\nf
\lput(-150,770,{\rm\ifcase\ng D\or A\or B\or C\or D\or E\or F\or G\fi})}
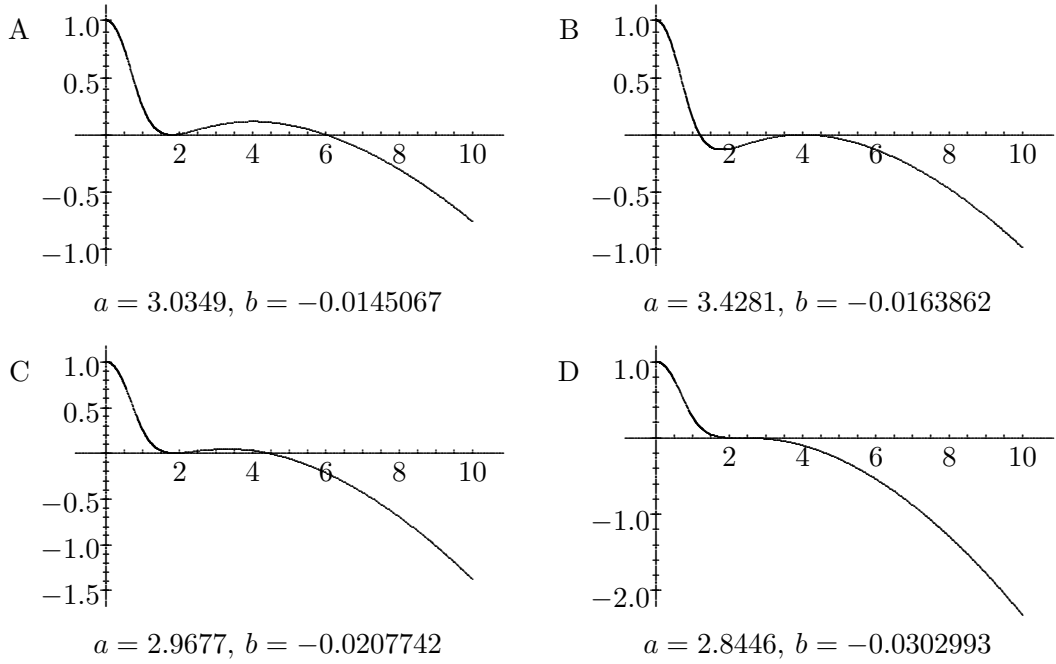
\def\opis{Plots of the \f\ $f(x)$ (Eq.~\eqref{3.12} for some interesting
values of $a$ and~$b$ $(b<0)$, see the text for an explanation)\break
Fig.~\the\nd A: $a=3.0349,\ b=-0.0145067$;\
Fig.~\the\nd B: $a=3.4281,\ b=-0.0163862$;\break
Fig.~\the\nd C: $a=2.9677,\ b=-0.0207742$;
Fig.~\the\nd D: $a=2.8446,\ b=-0.0302993$.}

\vbox{%
\obraz33
\grub0.2pt
\MT   0.000  425.000
\LT1400.000  425.000
\MT 160.000  425.000
\LT 160.000  435.000
\MT 220.000  425.000
\LT 220.000  435.000
\MT 280.000  425.000
\LT 280.000  435.000
\MT 340.000  425.000
\LT 340.000  435.000
\cput(340.000,365.000,2)
\MT 400.000  425.000
\LT 400.000  435.000
\MT 460.000  425.000
\LT 460.000  435.000
\MT 520.000  425.000
\LT 520.000  435.000
\MT 580.000  425.000
\LT 580.000  435.000
\cput(580.000,365.000,4)
\MT 640.000  425.000
\LT 640.000  435.000
\MT 700.000  425.000
\LT 700.000  435.000
\MT 760.000  425.000
\LT 760.000  435.000
\MT 820.000  425.000
\LT 820.000  435.000
\cput(820.000,365.000,6)
\MT 880.000  425.000
\LT 880.000  435.000
\MT 940.000  425.000
\LT 940.000  435.000
\MT1000.000  425.000
\LT1000.000  435.000
\MT1060.000  425.000
\LT1060.000  435.000
\cput(1060.000,365.000,8)
\MT1120.000  425.000
\LT1120.000  435.000
\MT1180.000  425.000
\LT1180.000  435.000
\MT1240.000  425.000
\LT1240.000  435.000
\MT1300.000  425.000
\LT1300.000  435.000
\cput(1300.000,365.000,10)
\MT 100.000    0.000
\LT 100.000  850.000
\MT  92.000   50.000
\LT 108.000   50.000
\MT  92.000   87.500
\LT 108.000   87.500
\MT  92.000  125.000
\LT 108.000  125.000
\MT  92.000  162.500
\LT 108.000  162.500
\MT  92.000  200.000
\LT 108.000  200.000
\MT  92.000  237.500
\LT 108.000  237.500
\MT  92.000  275.000
\LT 108.000  275.000
\MT  92.000  312.500
\LT 108.000  312.500
\MT  92.000  350.000
\LT 108.000  350.000
\MT  92.000  387.500
\LT 108.000  387.500
\MT  92.000  425.000
\LT 108.000  425.000
\MT  92.000  462.500
\LT 108.000  462.500
\MT  92.000  500.000
\LT 108.000  500.000
\MT  92.000  537.500
\LT 108.000  537.500
\MT  92.000  575.000
\LT 108.000  575.000
\MT  92.000  612.500
\LT 108.000  612.500
\MT  92.000  650.000
\LT 108.000  650.000
\MT  92.000  687.500
\LT 108.000  687.500
\MT  92.000  725.000
\LT 108.000  725.000
\MT  92.000  762.500
\LT 108.000  762.500
\MT  92.000  800.000
\LT 108.000  800.000
\MT  84.000   50.000
\LT 116.000   50.000
\lput(80.000,31.250, -1.0)
\MT  84.000  237.500
\LT 116.000  237.500
\lput(80.000,218.750, -0.5)
\MT  84.000  425.000
\LT 116.000  425.000
\MT  84.000  612.500
\LT 116.000  612.500
\lput(80.000,593.750,  0.5)
\MT  84.000  800.000
\LT 116.000  800.000
\lput(80.000,781.250,  1.0)
\grub0.6pt
\MT  100.000  800.000
\LT  101.000  799.973
\LT  102.000  799.893
\LT  103.000  799.760
\LT  104.000  799.572
\LT  105.000  799.332
\LT  106.000  799.038
\LT  107.000  798.691
\LT  108.000  798.290
\LT  109.000  797.836
\LT  110.000  797.328
\LT  111.000  796.767
\LT  112.000  796.152
\LT  113.000  795.484
\LT  114.000  794.763
\LT  115.000  793.988
\LT  116.000  793.160
\LT  117.000  792.279
\LT  118.000  791.344
\LT  119.000  790.356
\LT  120.000  789.315
\LT  121.000  788.220
\LT  122.000  787.073
\LT  123.000  785.872
\LT  124.000  784.618
\LT  125.000  783.312
\LT  126.000  781.953
\LT  127.000  780.541
\LT  128.000  779.076
\LT  129.000  777.559
\LT  130.000  775.990
\LT  131.000  774.368
\LT  132.000  772.695
\LT  133.000  770.970
\LT  134.000  769.193
\LT  135.000  767.365
\LT  136.000  765.486
\LT  137.000  763.556
\LT  138.000  761.576
\LT  139.000  759.546
\LT  140.000  757.466
\LT  141.000  755.336
\LT  142.000  753.158
\LT  143.000  750.931
\LT  144.000  748.656
\LT  145.000  746.334
\LT  146.000  743.965
\LT  147.000  741.549
\LT  148.000  739.087
\LT  149.000  736.581
\LT  150.000  734.029
\LT  151.000  731.434
\LT  152.000  728.796
\LT  153.000  726.116
\LT  154.000  723.394
\LT  155.000  720.632
\LT  156.000  717.829
\LT  157.000  714.989
\LT  158.000  712.110
\LT  159.000  709.195
\LT  160.000  706.244
\LT  161.000  703.258
\LT  162.000  700.239
\LT  163.000  697.188
\LT  164.000  694.106
\LT  165.000  690.993
\LT  166.000  687.853
\LT  167.000  684.685
\LT  168.000  681.491
\LT  169.000  678.273
\LT  170.000  675.032
\LT  171.000  671.769
\LT  173.000  665.184
\LT  175.000  658.531
\LT  177.000  651.822
\LT  180.000  641.682
\LT  187.000  617.890
\LT  190.000  607.744
\LT  192.000  601.028
\LT  194.000  594.366
\LT  196.000  587.768
\LT  197.000  584.497
\LT  198.000  581.247
\LT  199.000  578.018
\LT  200.000  574.813
\LT  201.000  571.632
\LT  202.000  568.477
\LT  203.000  565.348
\LT  204.000  562.247
\LT  205.000  559.175
\LT  206.000  556.133
\LT  207.000  553.122
\LT  208.000  550.143
\LT  209.000  547.196
\LT  210.000  544.283
\LT  211.000  541.405
\LT  212.000  538.562
\LT  213.000  535.754
\LT  214.000  532.984
\LT  215.000  530.250
\LT  216.000  527.554
\LT  217.000  524.896
\LT  218.000  522.277
\LT  219.000  519.697
\LT  220.000  517.156
\LT  221.000  514.655
\LT  222.000  512.194
\LT  223.000  509.773
\LT  224.000  507.393
\LT  225.000  505.053
\LT  226.000  502.754
\LT  227.000  500.496
\LT  228.000  498.279
\LT  229.000  496.103
\LT  230.000  493.967
\LT  231.000  491.873
\LT  232.000  489.819
\LT  233.000  487.806
\LT  234.000  485.833
\LT  235.000  483.901
\LT  236.000  482.008
\LT  237.000  480.156
\LT  238.000  478.343
\LT  239.000  476.570
\LT  240.000  474.835
\LT  241.000  473.140
\LT  242.000  471.483
\LT  243.000  469.864
\LT  244.000  468.283
\LT  245.000  466.739
\LT  246.000  465.232
\LT  247.000  463.762
\LT  248.000  462.328
\LT  249.000  460.930
\LT  250.000  459.567
\LT  251.000  458.240
\LT  252.000  456.946
\LT  253.000  455.687
\LT  254.000  454.461
\LT  255.000  453.268
\LT  256.000  452.108
\LT  257.000  450.980
\LT  258.000  449.883
\LT  259.000  448.818
\LT  260.000  447.784
\LT  261.000  446.779
\LT  262.000  445.805
\LT  263.000  444.859
\LT  264.000  443.943
\LT  265.000  443.054
\LT  266.000  442.194
\LT  267.000  441.360
\LT  268.000  440.554
\LT  269.000  439.774
\LT  270.000  439.019
\LT  271.000  438.291
\LT  272.000  437.587
\LT  273.000  436.907
\LT  274.000  436.252
\LT  275.000  435.620
\LT  276.000  435.011
\LT  277.000  434.424
\LT  278.000  433.860
\LT  279.000  433.318
\LT  280.000  432.797
\LT  281.000  432.297
\LT  282.000  431.817
\LT  284.000  430.917
\LT  286.000  430.094
\LT  288.000  429.345
\LT  290.000  428.666
\LT  292.000  428.055
\LT  294.000  427.508
\LT  296.000  427.023
\LT  298.000  426.597
\LT  300.000  426.227
\LT  302.000  425.911
\LT  304.000  425.646
\LT  306.000  425.430
\LT  308.000  425.260
\LT  310.000  425.134
\LT  312.000  425.051
\LT  315.000  425.000
\LT  318.000  425.033
\LT  321.000  425.143
\LT  324.000  425.324
\LT  327.000  425.571
\LT  330.000  425.878
\LT  333.000  426.241
\LT  336.000  426.655
\LT  340.000  427.278
\LT  344.000  427.973
\LT  348.000  428.734
\LT  353.000  429.762
\LT  358.000  430.863
\LT  364.000  432.264
\LT  371.000  433.981
\LT  379.000  436.021
\LT  394.000  439.956
\LT  409.000  443.890
\LT  419.000  446.447
\LT  428.000  448.676
\LT  436.000  450.585
\LT  443.000  452.193
\LT  450.000  453.736
\LT  457.000  455.210
\LT  464.000  456.612
\LT  471.000  457.939
\LT  478.000  459.189
\LT  485.000  460.359
\LT  491.000  461.297
\LT  497.000  462.175
\LT  503.000  462.992
\LT  509.000  463.747
\LT  515.000  464.442
\LT  521.000  465.074
\LT  527.000  465.644
\LT  533.000  466.153
\LT  539.000  466.599
\LT  545.000  466.984
\LT  551.000  467.308
\LT  557.000  467.570
\LT  563.000  467.771
\LT  569.000  467.911
\LT  575.000  467.991
\LT  582.000  468.009
\LT  589.000  467.946
\LT  596.000  467.802
\LT  603.000  467.579
\LT  610.000  467.276
\LT  617.000  466.896
\LT  624.000  466.438
\LT  631.000  465.904
\LT  638.000  465.293
\LT  645.000  464.608
\LT  652.000  463.848
\LT  659.000  463.014
\LT  666.000  462.108
\LT  673.000  461.129
\LT  680.000  460.079
\LT  687.000  458.958
\LT  694.000  457.767
\LT  701.000  456.506
\LT  708.000  455.177
\LT  715.000  453.780
\LT  722.000  452.316
\LT  729.000  450.785
\LT  736.000  449.188
\LT  743.000  447.525
\LT  750.000  445.798
\LT  757.000  444.007
\LT  764.000  442.152
\LT  771.000  440.234
\LT  778.000  438.253
\LT  785.000  436.210
\LT  793.000  433.801
\LT  801.000  431.312
\LT  809.000  428.744
\LT  817.000  426.098
\LT  825.000  423.376
\LT  833.000  420.576
\LT  841.000  417.701
\LT  849.000  414.750
\LT  857.000  411.725
\LT  865.000  408.626
\LT  873.000  405.454
\LT  881.000  402.208
\LT  889.000  398.891
\LT  897.000  395.502
\LT  905.000  392.041
\LT  913.000  388.511
\LT  921.000  384.910
\LT  929.000  381.239
\LT  937.000  377.499
\LT  945.000  373.691
\LT  953.000  369.814
\LT  961.000  365.870
\LT  969.000  361.859
\LT  977.000  357.780
\LT  985.000  353.635
\LT  993.000  349.424
\LT 1001.000  345.147
\LT 1009.000  340.805
\LT 1017.000  336.397
\LT 1025.000  331.926
\LT 1033.000  327.389
\LT 1042.000  322.210
\LT 1051.000  316.950
\LT 1060.000  311.609
\LT 1069.000  306.190
\LT 1078.000  300.691
\LT 1087.000  295.113
\LT 1096.000  289.457
\LT 1105.000  283.723
\LT 1114.000  277.911
\LT 1123.000  272.022
\LT 1132.000  266.056
\LT 1141.000  260.014
\LT 1150.000  253.896
\LT 1159.000  247.701
\LT 1168.000  241.431
\LT 1177.000  235.086
\LT 1186.000  228.666
\LT 1195.000  222.171
\LT 1204.000  215.602
\LT 1213.000  208.958
\LT 1222.000  202.241
\LT 1231.000  195.450
\LT 1240.000  188.586
\LT 1249.000  181.649
\LT 1258.000  174.639
\LT 1267.000  167.557
\LT 1276.000  160.402
\LT 1285.000  153.175
\LT 1294.000  145.876
\LT 1300.000  140.970
\koniec     3.0349 -0.0145067
\obraz34
\grub0.2pt
\MT   0.000  425.000
\LT1400.000  425.000
\MT 160.000  425.000
\LT 160.000  435.000
\MT 220.000  425.000
\LT 220.000  435.000
\MT 280.000  425.000
\LT 280.000  435.000
\MT 340.000  425.000
\LT 340.000  435.000
\cput(340.000,365.000,2)
\MT 400.000  425.000
\LT 400.000  435.000
\MT 460.000  425.000
\LT 460.000  435.000
\MT 520.000  425.000
\LT 520.000  435.000
\MT 580.000  425.000
\LT 580.000  435.000
\cput(580.000,365.000,4)
\MT 640.000  425.000
\LT 640.000  435.000
\MT 700.000  425.000
\LT 700.000  435.000
\MT 760.000  425.000
\LT 760.000  435.000
\MT 820.000  425.000
\LT 820.000  435.000
\cput(820.000,365.000,6)
\MT 880.000  425.000
\LT 880.000  435.000
\MT 940.000  425.000
\LT 940.000  435.000
\MT1000.000  425.000
\LT1000.000  435.000
\MT1060.000  425.000
\LT1060.000  435.000
\cput(1060.000,365.000,8)
\MT1120.000  425.000
\LT1120.000  435.000
\MT1180.000  425.000
\LT1180.000  435.000
\MT1240.000  425.000
\LT1240.000  435.000
\MT1300.000  425.000
\LT1300.000  435.000
\cput(1300.000,365.000,10)
\MT 100.000    0.000
\LT 100.000  850.000
\MT  92.000   50.000
\LT 108.000   50.000
\MT  92.000   87.500
\LT 108.000   87.500
\MT  92.000  125.000
\LT 108.000  125.000
\MT  92.000  162.500
\LT 108.000  162.500
\MT  92.000  200.000
\LT 108.000  200.000
\MT  92.000  237.500
\LT 108.000  237.500
\MT  92.000  275.000
\LT 108.000  275.000
\MT  92.000  312.500
\LT 108.000  312.500
\MT  92.000  350.000
\LT 108.000  350.000
\MT  92.000  387.500
\LT 108.000  387.500
\MT  92.000  425.000
\LT 108.000  425.000
\MT  92.000  462.500
\LT 108.000  462.500
\MT  92.000  500.000
\LT 108.000  500.000
\MT  92.000  537.500
\LT 108.000  537.500
\MT  92.000  575.000
\LT 108.000  575.000
\MT  92.000  612.500
\LT 108.000  612.500
\MT  92.000  650.000
\LT 108.000  650.000
\MT  92.000  687.500
\LT 108.000  687.500
\MT  92.000  725.000
\LT 108.000  725.000
\MT  92.000  762.500
\LT 108.000  762.500
\MT  92.000  800.000
\LT 108.000  800.000
\MT  84.000   50.000
\LT 116.000   50.000
\lput(80.000,31.250, -1.0)
\MT  84.000  237.500
\LT 116.000  237.500
\lput(80.000,218.750, -0.5)
\MT  84.000  425.000
\LT 116.000  425.000
\MT  84.000  612.500
\LT 116.000  612.500
\lput(80.000,593.750,  0.5)
\MT  84.000  800.000
\LT 116.000  800.000
\lput(80.000,781.250,  1.0)
\grub0.6pt
\MT  100.000  800.000
\LT  101.000  799.970
\LT  102.000  799.879
\LT  103.000  799.728
\LT  104.000  799.517
\LT  105.000  799.245
\LT  106.000  798.913
\LT  107.000  798.521
\LT  108.000  798.068
\LT  109.000  797.555
\LT  110.000  796.982
\LT  111.000  796.348
\LT  112.000  795.654
\LT  113.000  794.899
\LT  114.000  794.084
\LT  115.000  793.209
\LT  116.000  792.274
\LT  117.000  791.278
\LT  118.000  790.222
\LT  119.000  789.106
\LT  120.000  787.930
\LT  121.000  786.694
\LT  122.000  785.398
\LT  123.000  784.042
\LT  124.000  782.626
\LT  125.000  781.150
\LT  126.000  779.614
\LT  127.000  778.019
\LT  128.000  776.365
\LT  129.000  774.651
\LT  130.000  772.879
\LT  131.000  771.047
\LT  132.000  769.157
\LT  133.000  767.208
\LT  134.000  765.202
\LT  135.000  763.137
\LT  136.000  761.014
\LT  137.000  758.835
\LT  138.000  756.598
\LT  139.000  754.305
\LT  140.000  751.955
\LT  141.000  749.550
\LT  142.000  747.089
\LT  143.000  744.574
\LT  144.000  742.004
\LT  145.000  739.381
\LT  146.000  736.705
\LT  147.000  733.976
\LT  148.000  731.196
\LT  149.000  728.364
\LT  150.000  725.482
\LT  151.000  722.551
\LT  152.000  719.571
\LT  153.000  716.543
\LT  154.000  713.469
\LT  155.000  710.349
\LT  156.000  707.183
\LT  157.000  703.975
\LT  158.000  700.723
\LT  159.000  697.430
\LT  160.000  694.097
\LT  161.000  690.724
\LT  162.000  687.314
\LT  163.000  683.868
\LT  164.000  680.386
\LT  165.000  676.871
\LT  166.000  673.323
\LT  167.000  669.745
\LT  168.000  666.137
\LT  169.000  662.502
\LT  170.000  658.841
\LT  171.000  655.155
\LT  172.000  651.447
\LT  174.000  643.969
\LT  176.000  636.421
\LT  178.000  628.816
\LT  181.000  617.337
\LT  187.000  594.296
\LT  190.000  582.835
\LT  192.000  575.249
\LT  194.000  567.724
\LT  195.000  563.988
\LT  196.000  560.271
\LT  197.000  556.577
\LT  198.000  552.905
\LT  199.000  549.259
\LT  200.000  545.638
\LT  201.000  542.045
\LT  202.000  538.480
\LT  203.000  534.946
\LT  204.000  531.444
\LT  205.000  527.974
\LT  206.000  524.538
\LT  207.000  521.136
\LT  208.000  517.771
\LT  209.000  514.443
\LT  210.000  511.153
\LT  211.000  507.902
\LT  212.000  504.690
\LT  213.000  501.519
\LT  214.000  498.389
\LT  215.000  495.301
\LT  216.000  492.256
\LT  217.000  489.254
\LT  218.000  486.295
\LT  219.000  483.381
\LT  220.000  480.511
\LT  221.000  477.686
\LT  222.000  474.906
\LT  223.000  472.171
\LT  224.000  469.483
\LT  225.000  466.840
\LT  226.000  464.243
\LT  227.000  461.693
\LT  228.000  459.188
\LT  229.000  456.730
\LT  230.000  454.318
\LT  231.000  451.952
\LT  232.000  449.632
\LT  233.000  447.358
\LT  234.000  445.130
\LT  235.000  442.947
\LT  236.000  440.809
\LT  237.000  438.717
\LT  238.000  436.669
\LT  239.000  434.666
\LT  240.000  432.707
\LT  241.000  430.792
\LT  242.000  428.920
\LT  243.000  427.092
\LT  244.000  425.306
\LT  245.000  423.562
\LT  246.000  421.860
\LT  247.000  420.200
\LT  248.000  418.580
\LT  249.000  417.001
\LT  250.000  415.461
\LT  251.000  413.961
\LT  252.000  412.500
\LT  253.000  411.078
\LT  254.000  409.693
\LT  255.000  408.346
\LT  256.000  407.035
\LT  257.000  405.761
\LT  258.000  404.523
\LT  259.000  403.319
\LT  260.000  402.151
\LT  261.000  401.016
\LT  262.000  399.915
\LT  263.000  398.847
\LT  264.000  397.812
\LT  265.000  396.809
\LT  266.000  395.837
\LT  267.000  394.895
\LT  268.000  393.984
\LT  269.000  393.103
\LT  270.000  392.251
\LT  271.000  391.428
\LT  272.000  390.633
\LT  273.000  389.865
\LT  274.000  389.125
\LT  275.000  388.411
\LT  276.000  387.723
\LT  277.000  387.061
\LT  278.000  386.423
\LT  279.000  385.811
\LT  280.000  385.222
\LT  281.000  384.657
\LT  282.000  384.115
\LT  283.000  383.596
\LT  284.000  383.099
\LT  285.000  382.624
\LT  286.000  382.170
\LT  287.000  381.736
\LT  289.000  380.930
\LT  291.000  380.202
\LT  293.000  379.548
\LT  295.000  378.966
\LT  297.000  378.452
\LT  299.000  378.002
\LT  301.000  377.615
\LT  303.000  377.287
\LT  305.000  377.016
\LT  307.000  376.798
\LT  309.000  376.632
\LT  311.000  376.514
\LT  313.000  376.443
\LT  315.000  376.416
\LT  317.000  376.430
\LT  320.000  376.526
\LT  323.000  376.705
\LT  326.000  376.960
\LT  329.000  377.284
\LT  332.000  377.674
\LT  335.000  378.122
\LT  338.000  378.625
\LT  341.000  379.177
\LT  345.000  379.982
\LT  349.000  380.858
\LT  353.000  381.794
\LT  358.000  383.038
\LT  363.000  384.351
\LT  369.000  385.997
\LT  376.000  387.992
\LT  386.000  390.928
\LT  407.000  397.166
\LT  417.000  400.070
\LT  425.000  402.329
\LT  433.000  404.516
\LT  440.000  406.361
\LT  447.000  408.136
\LT  454.000  409.836
\LT  460.000  411.229
\LT  466.000  412.560
\LT  472.000  413.829
\LT  478.000  415.033
\LT  484.000  416.172
\LT  490.000  417.243
\LT  496.000  418.246
\LT  502.000  419.180
\LT  508.000  420.045
\LT  514.000  420.841
\LT  520.000  421.567
\LT  526.000  422.223
\LT  532.000  422.809
\LT  538.000  423.325
\LT  544.000  423.771
\LT  550.000  424.148
\LT  556.000  424.456
\LT  562.000  424.694
\LT  568.000  424.864
\LT  574.000  424.966
\LT  580.000  425.000
\LT  586.000  424.967
\LT  592.000  424.866
\LT  598.000  424.700
\LT  604.000  424.467
\LT  610.000  424.169
\LT  616.000  423.806
\LT  622.000  423.379
\LT  628.000  422.888
\LT  634.000  422.333
\LT  640.000  421.716
\LT  646.000  421.037
\LT  652.000  420.296
\LT  658.000  419.494
\LT  664.000  418.631
\LT  671.000  417.549
\LT  678.000  416.386
\LT  685.000  415.142
\LT  692.000  413.819
\LT  699.000  412.418
\LT  706.000  410.939
\LT  713.000  409.383
\LT  720.000  407.750
\LT  727.000  406.043
\LT  734.000  404.260
\LT  741.000  402.403
\LT  748.000  400.473
\LT  755.000  398.470
\LT  762.000  396.395
\LT  769.000  394.249
\LT  776.000  392.032
\LT  783.000  389.744
\LT  790.000  387.387
\LT  797.000  384.962
\LT  804.000  382.468
\LT  811.000  379.906
\LT  818.000  377.276
\LT  825.000  374.580
\LT  832.000  371.818
\LT  839.000  368.990
\LT  846.000  366.097
\LT  853.000  363.139
\LT  860.000  360.117
\LT  867.000  357.032
\LT  874.000  353.883
\LT  881.000  350.671
\LT  888.000  347.396
\LT  895.000  344.060
\LT  902.000  340.662
\LT  910.000  336.703
\LT  918.000  332.665
\LT  926.000  328.549
\LT  934.000  324.354
\LT  942.000  320.081
\LT  950.000  315.731
\LT  958.000  311.304
\LT  966.000  306.801
\LT  974.000  302.223
\LT  982.000  297.569
\LT  990.000  292.840
\LT  998.000  288.037
\LT 1006.000  283.160
\LT 1014.000  278.209
\LT 1022.000  273.185
\LT 1030.000  268.088
\LT 1038.000  262.919
\LT 1046.000  257.678
\LT 1054.000  252.365
\LT 1062.000  246.981
\LT 1070.000  241.526
\LT 1078.000  236.000
\LT 1086.000  230.405
\LT 1094.000  224.739
\LT 1102.000  219.003
\LT 1110.000  213.198
\LT 1118.000  207.324
\LT 1126.000  201.381
\LT 1134.000  195.370
\LT 1142.000  189.290
\LT 1150.000  183.143
\LT 1158.000  176.927
\LT 1166.000  170.644
\LT 1174.000  164.295
\LT 1182.000  157.878
\LT 1190.000  151.394
\LT 1198.000  144.844
\LT 1206.000  138.227
\LT 1214.000  131.544
\LT 1222.000  124.796
\LT 1230.000  117.982
\LT 1238.000  111.102
\LT 1246.000  104.157
\LT 1254.000   97.147
\LT 1262.000   90.072
\LT 1270.000   82.933
\LT 1278.000   75.729
\LT 1286.000   68.460
\LT 1294.000   61.128
\LT 1300.000   55.586
\koniec     3.4281 -0.0163862
\vskip30pt
\obraz35
\grub0.2pt
\MT   0.000  500.000
\LT1400.000  500.000
\MT 160.000  500.000
\LT 160.000  510.000
\MT 220.000  500.000
\LT 220.000  510.000
\MT 280.000  500.000
\LT 280.000  510.000
\MT 340.000  500.000
\LT 340.000  510.000
\cput(340.000,440.000,2)
\MT 400.000  500.000
\LT 400.000  510.000
\MT 460.000  500.000
\LT 460.000  510.000
\MT 520.000  500.000
\LT 520.000  510.000
\MT 580.000  500.000
\LT 580.000  510.000
\cput(580.000,440.000,4)
\MT 640.000  500.000
\LT 640.000  510.000
\MT 700.000  500.000
\LT 700.000  510.000
\MT 760.000  500.000
\LT 760.000  510.000
\MT 820.000  500.000
\LT 820.000  510.000
\cput(820.000,440.000,6)
\MT 880.000  500.000
\LT 880.000  510.000
\MT 940.000  500.000
\LT 940.000  510.000
\MT1000.000  500.000
\LT1000.000  510.000
\MT1060.000  500.000
\LT1060.000  510.000
\cput(1060.000,440.000,8)
\MT1120.000  500.000
\LT1120.000  510.000
\MT1180.000  500.000
\LT1180.000  510.000
\MT1240.000  500.000
\LT1240.000  510.000
\MT1300.000  500.000
\LT1300.000  510.000
\cput(1300.000,440.000,10)
\MT 100.000    0.000
\LT 100.000  850.000
\MT  92.000   50.000
\LT 108.000   50.000
\MT  92.000   80.000
\LT 108.000   80.000
\MT  92.000  110.000
\LT 108.000  110.000
\MT  92.000  140.000
\LT 108.000  140.000
\MT  92.000  170.000
\LT 108.000  170.000
\MT  92.000  200.000
\LT 108.000  200.000
\MT  92.000  230.000
\LT 108.000  230.000
\MT  92.000  260.000
\LT 108.000  260.000
\MT  92.000  290.000
\LT 108.000  290.000
\MT  92.000  320.000
\LT 108.000  320.000
\MT  92.000  350.000
\LT 108.000  350.000
\MT  92.000  380.000
\LT 108.000  380.000
\MT  92.000  410.000
\LT 108.000  410.000
\MT  92.000  440.000
\LT 108.000  440.000
\MT  92.000  470.000
\LT 108.000  470.000
\MT  92.000  500.000
\LT 108.000  500.000
\MT  92.000  530.000
\LT 108.000  530.000
\MT  92.000  560.000
\LT 108.000  560.000
\MT  92.000  590.000
\LT 108.000  590.000
\MT  92.000  620.000
\LT 108.000  620.000
\MT  92.000  650.000
\LT 108.000  650.000
\MT  92.000  680.000
\LT 108.000  680.000
\MT  92.000  710.000
\LT 108.000  710.000
\MT  92.000  740.000
\LT 108.000  740.000
\MT  92.000  770.000
\LT 108.000  770.000
\MT  92.000  800.000
\LT 108.000  800.000
\MT  84.000   50.000
\LT 116.000   50.000
\lput(80.000,35.000, -1.5)
\MT  84.000  200.000
\LT 116.000  200.000
\lput(80.000,185.000, -1.0)
\MT  84.000  350.000
\LT 116.000  350.000
\lput(80.000,335.000, -0.5)
\MT  84.000  500.000
\LT 116.000  500.000
\MT  84.000  650.000
\LT 116.000  650.000
\lput(80.000,635.000,  0.5)
\MT  84.000  800.000
\LT 116.000  800.000
\lput(80.000,785.000,  1.0)
\grub0.6pt
\MT  100.000  800.000
\LT  101.000  799.979
\LT  102.000  799.916
\LT  103.000  799.811
\LT  104.000  799.663
\LT  105.000  799.474
\LT  106.000  799.242
\LT  107.000  798.969
\LT  108.000  798.653
\LT  109.000  798.296
\LT  110.000  797.896
\LT  111.000  797.454
\LT  112.000  796.970
\LT  113.000  796.444
\LT  114.000  795.876
\LT  115.000  795.266
\LT  116.000  794.614
\LT  117.000  793.920
\LT  118.000  793.184
\LT  119.000  792.406
\LT  120.000  791.586
\LT  121.000  790.724
\LT  122.000  789.820
\LT  123.000  788.875
\LT  124.000  787.888
\LT  125.000  786.859
\LT  126.000  785.789
\LT  127.000  784.677
\LT  128.000  783.523
\LT  129.000  782.329
\LT  130.000  781.093
\LT  131.000  779.816
\LT  132.000  778.498
\LT  133.000  777.140
\LT  134.000  775.741
\LT  135.000  774.301
\LT  136.000  772.822
\LT  137.000  771.302
\LT  138.000  769.743
\LT  139.000  768.144
\LT  140.000  766.506
\LT  141.000  764.829
\LT  142.000  763.113
\LT  143.000  761.360
\LT  144.000  759.568
\LT  145.000  757.739
\LT  146.000  755.873
\LT  147.000  753.970
\LT  148.000  752.032
\LT  149.000  750.057
\LT  150.000  748.048
\LT  151.000  746.004
\LT  152.000  743.926
\LT  153.000  741.814
\LT  154.000  739.671
\LT  155.000  737.495
\LT  156.000  735.287
\LT  157.000  733.049
\LT  158.000  730.782
\LT  159.000  728.485
\LT  160.000  726.160
\LT  161.000  723.808
\LT  162.000  721.429
\LT  163.000  719.025
\LT  164.000  716.596
\LT  165.000  714.144
\LT  166.000  711.669
\LT  167.000  709.172
\LT  169.000  704.119
\LT  171.000  698.992
\LT  173.000  693.802
\LT  175.000  688.556
\LT  178.000  680.608
\LT  182.000  669.911
\LT  188.000  653.820
\LT  191.000  645.829
\LT  193.000  640.545
\LT  195.000  635.305
\LT  197.000  630.120
\LT  199.000  624.998
\LT  201.000  619.947
\LT  202.000  617.450
\LT  203.000  614.975
\LT  204.000  612.520
\LT  205.000  610.088
\LT  206.000  607.680
\LT  207.000  605.295
\LT  208.000  602.935
\LT  209.000  600.600
\LT  210.000  598.291
\LT  211.000  596.009
\LT  212.000  593.754
\LT  213.000  591.527
\LT  214.000  589.328
\LT  215.000  587.158
\LT  216.000  585.017
\LT  217.000  582.906
\LT  218.000  580.825
\LT  219.000  578.774
\LT  220.000  576.753
\LT  221.000  574.764
\LT  222.000  572.805
\LT  223.000  570.878
\LT  224.000  568.982
\LT  225.000  567.117
\LT  226.000  565.284
\LT  227.000  563.483
\LT  228.000  561.714
\LT  229.000  559.976
\LT  230.000  558.270
\LT  231.000  556.595
\LT  232.000  554.953
\LT  233.000  553.341
\LT  234.000  551.761
\LT  235.000  550.213
\LT  236.000  548.695
\LT  237.000  547.209
\LT  238.000  545.753
\LT  239.000  544.327
\LT  240.000  542.932
\LT  241.000  541.567
\LT  242.000  540.232
\LT  243.000  538.927
\LT  244.000  537.650
\LT  245.000  536.403
\LT  246.000  535.184
\LT  247.000  533.994
\LT  248.000  532.832
\LT  249.000  531.697
\LT  250.000  530.590
\LT  251.000  529.510
\LT  252.000  528.457
\LT  253.000  527.429
\LT  254.000  526.428
\LT  255.000  525.453
\LT  256.000  524.502
\LT  257.000  523.577
\LT  258.000  522.676
\LT  259.000  521.799
\LT  260.000  520.946
\LT  261.000  520.116
\LT  262.000  519.310
\LT  263.000  518.525
\LT  264.000  517.763
\LT  265.000  517.023
\LT  266.000  516.305
\LT  267.000  515.607
\LT  268.000  514.930
\LT  270.000  513.637
\LT  272.000  512.422
\LT  274.000  511.283
\LT  276.000  510.216
\LT  278.000  509.219
\LT  280.000  508.288
\LT  282.000  507.423
\LT  284.000  506.618
\LT  286.000  505.873
\LT  288.000  505.184
\LT  290.000  504.549
\LT  292.000  503.966
\LT  294.000  503.432
\LT  296.000  502.946
\LT  298.000  502.504
\LT  300.000  502.106
\LT  303.000  501.584
\LT  306.000  501.147
\LT  309.000  500.791
\LT  312.000  500.507
\LT  315.000  500.292
\LT  318.000  500.139
\LT  321.000  500.044
\LT  324.000  500.003
\LT  327.000  500.010
\LT  331.000  500.089
\LT  335.000  500.238
\LT  339.000  500.450
\LT  344.000  500.790
\LT  349.000  501.203
\LT  355.000  501.775
\LT  362.000  502.525
\LT  370.000  503.461
\LT  383.000  505.079
\LT  401.000  507.344
\LT  411.000  508.543
\LT  420.000  509.556
\LT  428.000  510.391
\LT  436.000  511.156
\LT  443.000  511.761
\LT  450.000  512.302
\LT  457.000  512.776
\LT  464.000  513.180
\LT  471.000  513.512
\LT  478.000  513.770
\LT  485.000  513.952
\LT  492.000  514.057
\LT  499.000  514.085
\LT  506.000  514.034
\LT  513.000  513.904
\LT  520.000  513.695
\LT  527.000  513.406
\LT  534.000  513.038
\LT  541.000  512.591
\LT  548.000  512.065
\LT  555.000  511.460
\LT  562.000  510.776
\LT  569.000  510.014
\LT  576.000  509.175
\LT  583.000  508.258
\LT  590.000  507.265
\LT  597.000  506.195
\LT  604.000  505.050
\LT  611.000  503.829
\LT  618.000  502.534
\LT  625.000  501.166
\LT  632.000  499.723
\LT  639.000  498.208
\LT  646.000  496.621
\LT  653.000  494.963
\LT  660.000  493.233
\LT  667.000  491.433
\LT  674.000  489.563
\LT  681.000  487.623
\LT  688.000  485.615
\LT  695.000  483.539
\LT  702.000  481.395
\LT  709.000  479.185
\LT  716.000  476.907
\LT  723.000  474.564
\LT  730.000  472.155
\LT  737.000  469.681
\LT  744.000  467.142
\LT  751.000  464.539
\LT  758.000  461.873
\LT  765.000  459.144
\LT  772.000  456.351
\LT  779.000  453.497
\LT  786.000  450.581
\LT  794.000  447.173
\LT  802.000  443.685
\LT  810.000  440.118
\LT  818.000  436.473
\LT  826.000  432.749
\LT  834.000  428.949
\LT  842.000  425.071
\LT  850.000  421.117
\LT  858.000  417.088
\LT  866.000  412.982
\LT  874.000  408.802
\LT  882.000  404.548
\LT  890.000  400.220
\LT  898.000  395.818
\LT  906.000  391.342
\LT  914.000  386.794
\LT  922.000  382.174
\LT  930.000  377.482
\LT  938.000  372.718
\LT  946.000  367.883
\LT  954.000  362.977
\LT  962.000  358.000
\LT  970.000  352.954
\LT  978.000  347.837
\LT  986.000  342.651
\LT  994.000  337.395
\LT 1002.000  332.071
\LT 1010.000  326.678
\LT 1018.000  321.217
\LT 1026.000  315.687
\LT 1034.000  310.090
\LT 1042.000  304.425
\LT 1050.000  298.693
\LT 1058.000  292.893
\LT 1066.000  287.027
\LT 1074.000  281.094
\LT 1082.000  275.095
\LT 1090.000  269.030
\LT 1098.000  262.899
\LT 1106.000  256.702
\LT 1114.000  250.440
\LT 1122.000  244.112
\LT 1130.000  237.719
\LT 1138.000  231.261
\LT 1146.000  224.739
\LT 1154.000  218.152
\LT 1162.000  211.500
\LT 1170.000  204.785
\LT 1179.000  197.153
\LT 1188.000  189.440
\LT 1197.000  181.647
\LT 1206.000  173.774
\LT 1215.000  165.820
\LT 1224.000  157.787
\LT 1233.000  149.673
\LT 1242.000  141.481
\LT 1251.000  133.208
\LT 1260.000  124.857
\LT 1269.000  116.427
\LT 1278.000  107.917
\LT 1287.000   99.330
\LT 1296.000   90.663
\LT 1300.000   86.786
\koniec     2.9677 -0.0207742
\obraz40
\grub0.2pt
\MT   0.000  550.000
\LT1400.000  550.000
\MT 160.000  550.000
\LT 160.000  560.000
\MT 220.000  550.000
\LT 220.000  560.000
\MT 280.000  550.000
\LT 280.000  560.000
\MT 340.000  550.000
\LT 340.000  560.000
\cput(340.000,490.000,2)
\MT 400.000  550.000
\LT 400.000  560.000
\MT 460.000  550.000
\LT 460.000  560.000
\MT 520.000  550.000
\LT 520.000  560.000
\MT 580.000  550.000
\LT 580.000  560.000
\cput(580.000,490.000,4)
\MT 640.000  550.000
\LT 640.000  560.000
\MT 700.000  550.000
\LT 700.000  560.000
\MT 760.000  550.000
\LT 760.000  560.000
\MT 820.000  550.000
\LT 820.000  560.000
\cput(820.000,490.000,6)
\MT 880.000  550.000
\LT 880.000  560.000
\MT 940.000  550.000
\LT 940.000  560.000
\MT1000.000  550.000
\LT1000.000  560.000
\MT1060.000  550.000
\LT1060.000  560.000
\cput(1060.000,490.000,8)
\MT1120.000  550.000
\LT1120.000  560.000
\MT1180.000  550.000
\LT1180.000  560.000
\MT1240.000  550.000
\LT1240.000  560.000
\MT1300.000  550.000
\LT1300.000  560.000
\cput(1300.000,490.000,10)
\MT 100.000    0.000
\LT 100.000  850.000
\MT  92.000   50.000
\LT 108.000   50.000
\MT  92.000  100.000
\LT 108.000  100.000
\MT  92.000  150.000
\LT 108.000  150.000
\MT  92.000  200.000
\LT 108.000  200.000
\MT  92.000  250.000
\LT 108.000  250.000
\MT  92.000  300.000
\LT 108.000  300.000
\MT  92.000  350.000
\LT 108.000  350.000
\MT  92.000  400.000
\LT 108.000  400.000
\MT  92.000  450.000
\LT 108.000  450.000
\MT  92.000  500.000
\LT 108.000  500.000
\MT  92.000  550.000
\LT 108.000  550.000
\MT  92.000  600.000
\LT 108.000  600.000
\MT  92.000  650.000
\LT 108.000  650.000
\MT  92.000  700.000
\LT 108.000  700.000
\MT  92.000  750.000
\LT 108.000  750.000
\MT  92.000  800.000
\LT 108.000  800.000
\MT  84.000   50.000
\LT 116.000   50.000
\lput(80.000,25.000, -2.0)
\MT  84.000  300.000
\LT 116.000  300.000
\lput(80.000,275.000, -1.0)
\MT  84.000  550.000
\LT 116.000  550.000
\MT  84.000  800.000
\LT 116.000  800.000
\lput(80.000,775.000,  1.0)
\grub0.6pt
\MT  100.000  800.000
\LT  101.000  799.983
\LT  102.000  799.932
\LT  103.000  799.847
\LT  104.000  799.728
\LT  105.000  799.575
\LT  106.000  799.388
\LT  107.000  799.168
\LT  108.000  798.913
\LT  109.000  798.624
\LT  110.000  798.301
\LT  111.000  797.945
\LT  112.000  797.554
\LT  113.000  797.129
\LT  114.000  796.671
\LT  115.000  796.178
\LT  116.000  795.652
\LT  117.000  795.091
\LT  118.000  794.497
\LT  119.000  793.869
\LT  120.000  793.207
\LT  121.000  792.511
\LT  122.000  791.782
\LT  123.000  791.018
\LT  124.000  790.222
\LT  125.000  789.391
\LT  126.000  788.527
\LT  127.000  787.629
\LT  128.000  786.698
\LT  129.000  785.733
\LT  130.000  784.736
\LT  131.000  783.705
\LT  132.000  782.641
\LT  133.000  781.544
\LT  134.000  780.414
\LT  135.000  779.252
\LT  136.000  778.057
\LT  137.000  776.830
\LT  138.000  775.571
\LT  139.000  774.280
\LT  140.000  772.958
\LT  141.000  771.604
\LT  142.000  770.218
\LT  143.000  768.802
\LT  144.000  767.356
\LT  145.000  765.879
\LT  146.000  764.372
\LT  147.000  762.835
\LT  148.000  761.269
\LT  149.000  759.675
\LT  150.000  758.052
\LT  151.000  756.401
\LT  152.000  754.723
\LT  153.000  753.018
\LT  154.000  751.286
\LT  155.000  749.528
\LT  156.000  747.745
\LT  157.000  745.937
\LT  158.000  744.105
\LT  159.000  742.249
\LT  160.000  740.371
\LT  161.000  738.470
\LT  162.000  736.548
\LT  164.000  732.642
\LT  166.000  728.660
\LT  168.000  724.607
\LT  170.000  720.490
\LT  172.000  716.318
\LT  174.000  712.096
\LT  177.000  705.689
\LT  180.000  699.216
\LT  189.000  679.677
\LT  192.000  673.214
\LT  195.000  666.822
\LT  197.000  662.611
\LT  199.000  658.449
\LT  201.000  654.342
\LT  203.000  650.297
\LT  205.000  646.319
\LT  207.000  642.414
\LT  209.000  638.586
\LT  210.000  636.702
\LT  211.000  634.839
\LT  212.000  632.998
\LT  213.000  631.179
\LT  214.000  629.381
\LT  215.000  627.607
\LT  216.000  625.855
\LT  217.000  624.127
\LT  218.000  622.422
\LT  219.000  620.741
\LT  220.000  619.084
\LT  221.000  617.451
\LT  222.000  615.843
\LT  223.000  614.259
\LT  224.000  612.700
\LT  225.000  611.166
\LT  226.000  609.657
\LT  227.000  608.173
\LT  228.000  606.713
\LT  229.000  605.279
\LT  230.000  603.869
\LT  231.000  602.485
\LT  232.000  601.125
\LT  233.000  599.790
\LT  234.000  598.480
\LT  235.000  597.195
\LT  236.000  595.934
\LT  237.000  594.697
\LT  238.000  593.485
\LT  239.000  592.296
\LT  240.000  591.131
\LT  241.000  589.991
\LT  242.000  588.873
\LT  243.000  587.779
\LT  244.000  586.708
\LT  245.000  585.659
\LT  246.000  584.633
\LT  247.000  583.630
\LT  248.000  582.648
\LT  249.000  581.688
\LT  250.000  580.750
\LT  251.000  579.833
\LT  252.000  578.937
\LT  253.000  578.062
\LT  254.000  577.207
\LT  256.000  575.557
\LT  258.000  573.984
\LT  260.000  572.488
\LT  262.000  571.065
\LT  264.000  569.712
\LT  266.000  568.428
\LT  268.000  567.209
\LT  270.000  566.055
\LT  272.000  564.961
\LT  274.000  563.926
\LT  276.000  562.948
\LT  278.000  562.023
\LT  280.000  561.151
\LT  282.000  560.329
\LT  284.000  559.555
\LT  286.000  558.826
\LT  288.000  558.141
\LT  290.000  557.497
\LT  293.000  556.606
\LT  296.000  555.799
\LT  299.000  555.069
\LT  302.000  554.412
\LT  305.000  553.821
\LT  308.000  553.291
\LT  311.000  552.819
\LT  314.000  552.400
\LT  317.000  552.028
\LT  321.000  551.601
\LT  325.000  551.244
\LT  329.000  550.949
\LT  334.000  550.655
\LT  339.000  550.432
\LT  345.000  550.242
\LT  352.000  550.103
\LT  360.000  550.023
\LT  373.000  549.990
\LT  391.000  549.970
\LT  402.000  549.891
\LT  411.000  549.755
\LT  419.000  549.569
\LT  427.000  549.310
\LT  434.000  549.020
\LT  441.000  548.665
\LT  448.000  548.243
\LT  455.000  547.750
\LT  462.000  547.184
\LT  469.000  546.543
\LT  476.000  545.825
\LT  483.000  545.030
\LT  490.000  544.156
\LT  497.000  543.202
\LT  504.000  542.168
\LT  511.000  541.054
\LT  518.000  539.858
\LT  525.000  538.582
\LT  532.000  537.224
\LT  539.000  535.786
\LT  546.000  534.266
\LT  553.000  532.666
\LT  560.000  530.985
\LT  567.000  529.224
\LT  574.000  527.384
\LT  581.000  525.464
\LT  588.000  523.464
\LT  595.000  521.386
\LT  602.000  519.230
\LT  609.000  516.996
\LT  616.000  514.685
\LT  623.000  512.297
\LT  630.000  509.833
\LT  637.000  507.292
\LT  644.000  504.676
\LT  651.000  501.985
\LT  658.000  499.220
\LT  665.000  496.381
\LT  672.000  493.468
\LT  679.000  490.481
\LT  686.000  487.422
\LT  693.000  484.291
\LT  700.000  481.088
\LT  707.000  477.814
\LT  714.000  474.469
\LT  721.000  471.053
\LT  728.000  467.567
\LT  735.000  464.012
\LT  742.000  460.387
\LT  749.000  456.693
\LT  756.000  452.930
\LT  763.000  449.099
\LT  770.000  445.201
\LT  777.000  441.234
\LT  784.000  437.201
\LT  791.000  433.101
\LT  798.000  428.934
\LT  805.000  424.701
\LT  812.000  420.402
\LT  819.000  416.038
\LT  826.000  411.608
\LT  833.000  407.114
\LT  840.000  402.554
\LT  847.000  397.931
\LT  854.000  393.243
\LT  861.000  388.491
\LT  868.000  383.675
\LT  875.000  378.796
\LT  882.000  373.854
\LT  889.000  368.849
\LT  896.000  363.781
\LT  903.000  358.651
\LT  910.000  353.458
\LT  917.000  348.204
\LT  924.000  342.887
\LT  931.000  337.509
\LT  938.000  332.070
\LT  946.000  325.778
\LT  954.000  319.407
\LT  962.000  312.956
\LT  970.000  306.427
\LT  978.000  299.818
\LT  986.000  293.131
\LT  994.000  286.365
\LT 1002.000  279.521
\LT 1010.000  272.599
\LT 1018.000  265.600
\LT 1026.000  258.523
\LT 1034.000  251.369
\LT 1042.000  244.137
\LT 1050.000  236.829
\LT 1058.000  229.444
\LT 1066.000  221.983
\LT 1074.000  214.446
\LT 1082.000  206.832
\LT 1090.000  199.142
\LT 1098.000  191.377
\LT 1106.000  183.536
\LT 1114.000  175.620
\LT 1122.000  167.628
\LT 1130.000  159.562
\LT 1138.000  151.420
\LT 1146.000  143.204
\LT 1154.000  134.913
\LT 1162.000  126.547
\LT 1170.000  118.107
\LT 1178.000  109.593
\LT 1186.000  101.005
\LT 1194.000   92.343
\LT 1202.000   83.607
\LT 1210.000   74.797
\LT 1218.000   65.913
\LT 1226.000   56.957
\LT 1234.000   47.926
\LT 1242.000   38.823
\LT 1250.000   29.646
\LT 1258.000   20.396
\LT 1266.000   11.073
\LT 1274.000    1.678
\LT 1282.000   -7.791
\LT 1290.000  -17.332
\LT 1298.000  -26.945
\LT 1300.000  -29.360
\koniec     2.8446 -0.0302993 }

\noindent
Eq. \eqref{3.14} has one \so\ for $b\ge0$. However in the case of $b_0<b<0$ it
has two \so s, one \so\ for $b=b_0$ and no \so s for $b<b_0$. Thus we can write
$a_{\rm crt}=a_{\rm crt}(b)$, $x_{\rm crt}(b)$ only for $b\ge0$ (and
$b=b_0$). In the remaining case $b_0<b<0$ we rewrite these \e s as
\bea{3.16}
b&=&a_{\rm crt}\,\frac{x^3_{\rm crt}+g(x_{\rm crt})(1+x^4_{\rm crt})}
{2x^3_{\rm crt}(x^4_{\rm crt}+1)}\\
a_{\rm crt}&=&-\frac{2x_{\rm crt}(x^4_{\rm crt}+1)}{x^3_{\rm crt}
+3g(x_{\rm crt})(x^4_{\rm crt}+1)}\,.\nonumber
\end{eqnarray}

The estimated value of $b_0$ is $-0.0302993$. It seems that in the case of de
Sitter horizon for $b<b_0=-0.0302993$ there are not any horizon except the
mentioned one.

In the case of $b<0$ we have an interesting phenomenon due to the fact that
the function~$f$ has two local extrema (one minimum and one maximum). For a
special value of $b_0$ they collapse to one inflection point. This results in
one horizon as a \so\ of a system of \e s
\beq{3.17}
\bal
f(x_{\rm crt},a_{\rm crt},b_0)&=0\\
\pp fx(x_{\rm crt},a_{\rm crt},b_0)&=0\\
\pp{^2f}{x^2}f(x_{\rm crt},a_{\rm crt},b_0)&=0
\eal
\end{equation}
One finds $x_{\rm crt}\approx2.30331$, $a_{\rm crt}\approx2.8446$, $b_0
=-0.0302993$.

The points $x_{\rm crt}\approx2.30331$, $a_{\rm crt}\approx2.8446$ correspond
to the minimum of the \f\ $b(x)$ defined by the first \e\ \eqref{3.16}.

In Fig.\ 6 we give plots for $x_{\rm crt}(b)$ and $a_{\rm crt}(b)$ for $b>0$
and for $b<0$. In the case of $b<0$ we have two branches $x_{\rm crt}$ and
$a_{\rm crt}$.

\medskip
\def\opis{Plots of $x_{\rm crt}(b)$ and $a_{\rm crt}(b)$ for $b>0$
(Fig.~\the\nd A and Fig.~\the\nd B)\break and for $b<0$ (Fig.~\the\nd C and
Fig.~\the\nd D)}

\vbox{\vglue35pt
\obraz41
\grub0.2pt
\MT   0.000  100.000
\LT1200.000  100.000
\MT 1140.000 115.000 \LT 1200.000 100.000 \LT 1140.000 85.000
\cput(1200.000,25.000,b)
\MT 300.000   80.000
\LT 300.000  120.000
\cput(300.000,20.000,2)
\MT 500.000   80.000
\LT 500.000  120.000
\cput(500.000,20.000,4)
\MT 700.000   80.000
\LT 700.000  120.000
\cput(700.000,20.000,6)
\MT 900.000   80.000
\LT 900.000  120.000
\cput(900.000,20.000,8)
\MT1100.000   80.000
\LT1100.000  120.000
\cput(1100.000,20.000,10)
\MT 100.000    0.000
\LT 100.000  900.000
\MT 85.000 840.000 \LT 100.000 900.000 \LT 115.000 840.000
\rput(130.000,870.000,x_{\rm crt})
\MT  80.000  300.000
\LT 120.000  300.000
\lput(60.000,280.000,0.5)
\MT  80.000  500.000
\LT 120.000  500.000
\lput(60.000,480.000,1.0)
\MT  80.000  700.000
\LT 120.000  700.000
\lput(60.000,680.000,1.5)
\grub0.6pt
\MT  100.064  769.959
\LT  100.064  769.959
\LT  100.127  768.446
\LT  100.191  766.955
\LT  100.256  765.485
\LT  100.385  762.606
\LT  100.515  759.805
\LT  100.646  757.077
\LT  100.778  754.418
\LT  100.911  751.824
\LT  101.044  749.292
\LT  101.179  746.818
\LT  101.314  744.401
\LT  101.451  742.037
\LT  101.588  739.724
\LT  101.726  737.460
\LT  101.865  735.242
\LT  102.005  733.069
\LT  102.146  730.938
\LT  102.359  727.817
\LT  102.574  724.783
\LT  102.791  721.830
\LT  103.010  718.954
\LT  103.231  716.151
\LT  103.454  713.416
\LT  103.679  710.746
\LT  103.906  708.138
\LT  104.135  705.589
\LT  104.366  703.096
\LT  104.599  700.657
\LT  104.835  698.269
\LT  105.072  695.929
\LT  105.312  693.636
\LT  105.634  690.648
\LT  105.960  687.735
\LT  106.290  684.893
\LT  106.624  682.119
\LT  106.962  679.408
\LT  107.303  676.758
\LT  107.649  674.166
\LT  107.998  671.629
\LT  108.440  668.532
\LT  108.888  665.513
\LT  109.342  662.567
\LT  109.803  659.690
\LT  110.269  656.880
\LT  110.743  654.133
\LT  111.222  651.445
\LT  111.806  648.295
\LT  112.400  645.222
\LT  113.003  642.221
\LT  113.616  639.291
\LT  114.238  636.426
\LT  114.871  633.624
\LT  115.621  630.429
\LT  116.386  627.312
\LT  117.164  624.266
\LT  117.957  621.289
\LT  118.765  618.377
\LT  119.706  615.124
\LT  120.667  611.947
\LT  121.647  608.843
\LT  122.648  605.807
\LT  123.670  602.835
\LT  124.714  599.924
\LT  125.913  596.720
\LT  127.141  593.586
\LT  128.397  590.517
\LT  129.682  587.512
\LT  131.146  584.243
\LT  132.648  581.042
\LT  134.188  577.908
\LT  135.769  574.836
\LT  137.391  571.823
\LT  139.055  568.866
\LT  140.763  565.962
\LT  142.515  563.109
\LT  144.314  560.304
\LT  146.346  557.272
\LT  148.438  554.293
\LT  150.591  551.363
\LT  152.806  548.482
\LT  155.087  545.645
\LT  157.434  542.852
\LT  159.628  540.349
\LT  161.881  537.878
\LT  164.195  535.438
\LT  166.572  533.027
\LT  169.013  530.646
\LT  171.522  528.292
\LT  174.099  525.963
\LT  176.747  523.660
\LT  179.469  521.381
\LT  181.983  519.350
\LT  184.561  517.336
\LT  187.204  515.341
\LT  189.914  513.361
\LT  192.693  511.398
\LT  195.544  509.450
\LT  198.469  507.517
\LT  201.469  505.598
\LT  204.547  503.694
\LT  207.352  502.012
\LT  210.222  500.340
\LT  213.159  498.677
\LT  216.167  497.025
\LT  219.245  495.381
\LT  222.398  493.746
\LT  225.626  492.119
\LT  228.933  490.501
\LT  231.892  489.092
\LT  234.914  487.688
\LT  238.002  486.290
\LT  241.156  484.897
\LT  244.378  483.509
\LT  247.671  482.127
\LT  251.037  480.750
\LT  254.477  479.377
\LT  257.993  478.008
\LT  261.588  476.644
\LT  264.734  475.478
\LT  267.940  474.315
\LT  271.209  473.155
\LT  274.541  471.998
\LT  277.940  470.843
\LT  281.405  469.691
\LT  284.939  468.541
\LT  288.544  467.393
\LT  292.221  466.248
\LT  295.972  465.105
\LT  299.799  463.964
\LT  303.705  462.825
\LT  307.021  461.877
\LT  310.393  460.930
\LT  313.824  459.985
\LT  317.315  459.040
\LT  320.866  458.097
\LT  324.480  457.154
\LT  328.157  456.213
\LT  331.899  455.273
\LT  335.708  454.333
\LT  339.584  453.394
\LT  343.531  452.456
\LT  347.549  451.518
\LT  351.640  450.581
\LT  355.805  449.645
\LT  360.047  448.708
\LT  364.367  447.773
\LT  368.768  446.838
\LT  373.251  445.903
\LT  376.898  445.156
\LT  380.599  444.408
\LT  384.357  443.661
\LT  388.172  442.914
\LT  392.045  442.166
\LT  395.977  441.419
\LT  399.971  440.672
\LT  404.026  439.925
\LT  408.144  439.177
\LT  412.327  438.430
\LT  416.575  437.683
\LT  420.890  436.935
\LT  425.275  436.187
\LT  429.729  435.439
\LT  434.254  434.691
\LT  438.852  433.943
\LT  443.525  433.194
\LT  448.274  432.445
\LT  453.101  431.695
\LT  456.773  431.133
\LT  460.490  430.571
\LT  464.254  430.008
\LT  468.064  429.445
\LT  471.923  428.881
\LT  475.829  428.318
\LT  479.785  427.754
\LT  483.791  427.190
\LT  487.848  426.625
\LT  491.957  426.060
\LT  496.118  425.495
\LT  500.333  424.930
\LT  504.603  424.364
\LT  508.928  423.797
\LT  513.309  423.231
\LT  517.748  422.663
\LT  522.245  422.096
\LT  526.802  421.528
\LT  531.419  420.959
\LT  536.097  420.390
\LT  540.839  419.821
\LT  545.643  419.251
\LT  550.513  418.680
\LT  555.449  418.109
\LT  560.452  417.538
\LT  565.523  416.965
\LT  570.665  416.392
\LT  575.877  415.819
\LT  581.161  415.245
\LT  586.520  414.671
\LT  591.953  414.096
\LT  597.463  413.519
\LT  603.050  412.943
\LT  608.717  412.366
\LT  614.465  411.788
\LT  620.295  411.209
\LT  626.209  410.629
\LT  632.209  410.050
\LT  638.297  409.468
\LT  642.404  409.081
\LT  646.552  408.693
\LT  650.740  408.304
\LT  654.969  407.915
\LT  659.240  407.526
\LT  663.553  407.136
\LT  667.909  406.746
\LT  672.309  406.356
\LT  676.752  405.965
\LT  681.240  405.574
\LT  685.773  405.182
\LT  690.352  404.790
\LT  694.977  404.398
\LT  699.649  404.005
\LT  704.369  403.611
\LT  709.137  403.218
\LT  713.954  402.824
\LT  718.821  402.429
\LT  723.738  402.034
\LT  728.706  401.638
\LT  733.726  401.242
\LT  738.798  400.846
\LT  743.924  400.449
\LT  749.103  400.051
\LT  754.338  399.654
\LT  759.627  399.255
\LT  764.973  398.856
\LT  770.377  398.456
\LT  775.838  398.057
\LT  781.358  397.656
\LT  786.938  397.255
\LT  792.578  396.853
\LT  798.279  396.451
\LT  804.043  396.049
\LT  809.870  395.645
\LT  815.762  395.242
\LT  821.718  394.837
\LT  827.741  394.432
\LT  833.830  394.027
\LT  839.988  393.620
\LT  846.215  393.214
\LT  852.512  392.806
\LT  858.880  392.398
\LT  865.321  391.990
\LT  871.835  391.581
\LT  878.424  391.171
\LT  885.089  390.760
\LT  891.831  390.349
\LT  898.650  389.937
\LT  902.090  389.731
\LT  905.549  389.525
\LT  909.029  389.318
\LT  912.529  389.112
\LT  916.050  388.905
\LT  919.591  388.698
\LT  923.153  388.491
\LT  926.736  388.283
\LT  930.340  388.076
\LT  933.966  387.868
\LT  937.613  387.660
\LT  941.281  387.452
\LT  944.972  387.244
\LT  948.684  387.035
\LT  952.419  386.827
\LT  956.176  386.618
\LT  959.955  386.409
\LT  963.758  386.200
\LT  967.583  385.990
\LT  971.431  385.781
\LT  975.303  385.571
\LT  979.198  385.361
\LT  983.116  385.151
\LT  987.059  384.941
\LT  991.026  384.730
\LT  995.017  384.520
\LT  999.032  384.309
\LT 1003.072  384.098
\LT 1007.137  383.886
\LT 1011.227  383.675
\LT 1015.343  383.463
\LT 1019.484  383.251
\LT 1023.650  383.039
\LT 1027.843  382.827
\LT 1032.062  382.615
\LT 1036.307  382.402
\LT 1040.579  382.189
\LT 1044.877  381.976
\LT 1049.203  381.762
\LT 1053.556  381.549
\LT 1057.937  381.335
\LT 1062.345  381.121
\LT 1066.781  380.907
\LT 1071.246  380.692
\LT 1075.739  380.478
\LT 1080.261  380.263
\LT 1084.811  380.048
\LT 1089.392  379.833
\LT 1094.001  379.617
\LT 1098.640  379.402
\LT 1103.310  379.186
\kon x_{\rm crt}(b),\ b>0;
\obraz42
\grub0.2pt
\MT   0.000  100.000
\LT1200.000  100.000
\MT 1140.000 115.000 \LT 1200.000 100.000 \LT 1140.000 85.000
\cput(1200.000,25.000,b)
\MT 300.000   80.000
\LT 300.000  120.000
\cput(300.000,20.000,2)
\MT 500.000   80.000
\LT 500.000  120.000
\cput(500.000,20.000,4)
\MT 700.000   80.000
\LT 700.000  120.000
\cput(700.000,20.000,6)
\MT 900.000   80.000
\LT 900.000  120.000
\cput(900.000,20.000,8)
\MT1100.000   80.000
\LT1100.000  120.000
\cput(1100.000,20.000,10)
\MT 100.000    0.000
\LT 100.000  900.000
\MT 85.000 840.000 \LT 100.000 900.000 \LT 115.000 840.000
\rput(130.000,870.000,a_{\rm crt})
\MT  80.000  300.000
\LT 120.000  300.000
\lput(60.000,280.000,10)
\MT  80.000  500.000
\LT 120.000  500.000
\lput(60.000,480.000,20)
\MT  80.000  700.000
\LT 120.000  700.000
\lput(60.000,680.000,30)
\grub0.6pt
\MT  100.064  163.573
\LT  100.064  163.573
\LT  101.246  165.601
\LT  102.502  167.615
\LT  103.830  169.633
\LT  105.232  171.665
\LT  106.624  173.604
\LT  108.086  175.568
\LT  109.618  177.562
\LT  111.126  179.469
\LT  112.700  181.411
\LT  114.343  183.389
\LT  115.947  185.279
\LT  117.616  187.207
\LT  119.351  189.174
\LT  121.032  191.048
\LT  122.775  192.959
\LT  124.582  194.911
\LT  126.455  196.906
\LT  128.256  198.797
\LT  130.118  200.728
\LT  132.042  202.700
\LT  134.032  204.715
\LT  135.929  206.615
\LT  137.886  208.555
\LT  139.904  210.536
\LT  141.985  212.560
\LT  143.950  214.454
\LT  145.972  216.386
\LT  148.053  218.358
\LT  150.195  220.371
\LT  152.399  222.427
\LT  154.667  224.526
\LT  156.787  226.474
\LT  158.964  228.461
\LT  161.199  230.488
\LT  163.495  232.557
\LT  165.852  234.667
\LT  168.274  236.822
\LT  170.510  238.800
\LT  172.802  240.815
\LT  175.150  242.870
\LT  177.556  244.964
\LT  180.022  247.100
\LT  182.551  249.278
\LT  185.143  251.499
\LT  187.800  253.766
\LT  190.219  255.819
\LT  192.693  257.910
\LT  195.224  260.040
\LT  197.812  262.209
\LT  200.460  264.419
\LT  203.169  266.671
\LT  205.942  268.966
\LT  208.778  271.305
\LT  211.315  273.388
\LT  213.905  275.508
\LT  216.547  277.664
\LT  219.245  279.857
\LT  222.000  282.089
\LT  224.812  284.360
\LT  227.683  286.672
\LT  230.616  289.024
\LT  233.611  291.420
\LT  236.670  293.859
\LT  239.796  296.342
\LT  242.528  298.507
\LT  245.312  300.707
\LT  248.148  302.942
\LT  251.037  305.213
\LT  253.981  307.521
\LT  256.981  309.867
\LT  260.038  312.251
\LT  263.153  314.675
\LT  266.329  317.140
\LT  269.567  319.646
\LT  272.867  322.194
\LT  276.232  324.786
\LT  279.087  326.980
\LT  281.989  329.205
\LT  284.939  331.463
\LT  287.938  333.753
\LT  290.987  336.078
\LT  294.087  338.436
\LT  297.239  340.829
\LT  300.445  343.258
\LT  303.705  345.724
\LT  307.021  348.226
\LT  310.393  350.767
\LT  313.824  353.346
\LT  317.315  355.966
\LT  320.866  358.625
\LT  324.480  361.327
\LT  328.157  364.070
\LT  331.899  366.857
\LT  335.708  369.688
\LT  339.584  372.565
\LT  342.736  374.899
\LT  345.933  377.264
\LT  349.176  379.659
\LT  352.467  382.086
\LT  355.805  384.544
\LT  359.193  387.035
\LT  362.630  389.559
\LT  366.118  392.116
\LT  369.658  394.708
\LT  373.251  397.335
\LT  376.898  399.997
\LT  380.599  402.696
\LT  384.357  405.432
\LT  388.172  408.205
\LT  392.045  411.017
\LT  395.977  413.868
\LT  399.971  416.759
\LT  404.026  419.691
\LT  408.144  422.664
\LT  412.327  425.679
\LT  416.575  428.738
\LT  419.805  431.061
\LT  423.074  433.410
\LT  426.381  435.783
\LT  429.729  438.183
\LT  433.116  440.609
\LT  436.544  443.062
\LT  440.013  445.542
\LT  443.525  448.050
\LT  447.080  450.585
\LT  450.678  453.150
\LT  454.320  455.743
\LT  458.007  458.365
\LT  461.740  461.018
\LT  465.519  463.700
\LT  469.345  466.414
\LT  473.219  469.159
\LT  477.142  471.935
\LT  481.115  474.744
\LT  485.138  477.586
\LT  489.212  480.461
\LT  493.338  483.371
\LT  497.517  486.314
\LT  501.751  489.293
\LT  506.038  492.308
\LT  510.382  495.359
\LT  514.783  498.446
\LT  519.241  501.572
\LT  523.758  504.735
\LT  528.334  507.937
\LT  532.971  511.179
\LT  537.671  514.461
\LT  542.433  517.784
\LT  547.259  521.148
\LT  552.151  524.555
\LT  557.109  528.005
\LT  562.135  531.498
\LT  567.229  535.037
\LT  572.394  538.620
\LT  577.630  542.250
\LT  582.939  545.927
\LT  588.322  549.652
\LT  591.953  552.163
\LT  595.617  554.695
\LT  599.316  557.249
\LT  603.050  559.826
\LT  606.819  562.426
\LT  610.624  565.049
\LT  614.465  567.695
\LT  618.343  570.365
\LT  622.257  573.059
\LT  626.209  575.777
\LT  630.200  578.520
\LT  634.229  581.287
\LT  638.297  584.080
\LT  642.404  586.898
\LT  646.552  589.742
\LT  650.740  592.612
\LT  654.969  595.508
\LT  659.240  598.432
\LT  663.553  601.382
\LT  667.909  604.360
\LT  672.309  607.366
\LT  676.752  610.400
\LT  681.240  613.463
\LT  685.773  616.555
\LT  690.352  619.676
\LT  694.977  622.827
\LT  699.649  626.008
\LT  704.369  629.220
\LT  709.137  632.462
\LT  713.954  635.737
\LT  718.821  639.043
\LT  723.738  642.381
\LT  728.706  645.752
\LT  733.726  649.156
\LT  738.798  652.594
\LT  743.924  656.065
\LT  749.103  659.572
\LT  754.338  663.113
\LT  759.627  666.690
\LT  764.973  670.303
\LT  770.377  673.953
\LT  775.838  677.639
\LT  781.358  681.363
\LT  786.938  685.126
\LT  792.578  688.927
\LT  798.279  692.767
\LT  804.043  696.647
\LT  809.870  700.567
\LT  815.762  704.528
\LT  821.718  708.531
\LT  827.741  712.576
\LT  830.777  714.615
\LT  833.830  716.664
\LT  836.901  718.724
\LT  839.988  720.795
\LT  843.093  722.877
\LT  846.215  724.971
\LT  849.355  727.075
\LT  852.512  729.191
\LT  855.687  731.318
\LT  858.880  733.456
\LT  862.092  735.606
\LT  865.321  737.768
\LT  868.569  739.941
\LT  871.835  742.126
\LT  875.120  744.323
\LT  878.424  746.532
\LT  881.747  748.753
\LT  885.089  750.986
\LT  888.450  753.231
\LT  891.831  755.489
\LT  895.231  757.759
\LT  898.650  760.042
\LT  902.090  762.337
\LT  905.549  764.645
\LT  909.029  766.966
\LT  912.529  769.299
\LT  916.050  771.646
\LT  919.591  774.006
\LT  923.153  776.379
\LT  926.736  778.765
\LT  930.340  781.165
\LT  933.966  783.578
\LT  937.613  786.005
\LT  941.281  788.446
\LT  944.972  790.901
\LT  948.684  793.369
\LT  952.419  795.852
\LT  956.176  798.349
\LT  959.955  800.860
\LT  963.758  803.386
\LT  967.583  805.926
\LT  971.431  808.481
\LT  975.303  811.050
\LT  979.198  813.634
\LT  983.116  816.234
\LT  987.059  818.849
\LT  991.026  821.478
\LT  995.017  824.124
\LT  999.032  826.784
\LT 1003.072  829.461
\LT 1007.137  832.153
\LT 1011.227  834.861
\LT 1015.343  837.585
\LT 1019.484  840.325
\LT 1023.650  843.082
\LT 1027.843  845.855
\LT 1032.062  848.644
\LT 1036.307  851.450
\LT 1040.579  854.273
\LT 1044.877  857.113
\LT 1049.203  859.970
\LT 1053.556  862.845
\LT 1057.937  865.737
\LT 1062.345  868.646
\LT 1066.781  871.573
\LT 1071.246  874.518
\LT 1075.739  877.481
\LT 1080.261  880.462
\LT 1084.811  883.462
\LT 1089.392  886.480
\LT 1094.001  889.516
\LT 1098.640  892.572
\LT 1103.310  895.646
\kon a_{\rm crt}(b),\ b>0;
\vskip35pt
\obraz43
\grub0.2pt
\MT   0.000  100.000
\LT1200.000  100.000
\MT 1140.000 115.000 \LT 1200.000 100.000 \LT 1140.000 85.000
\cput(1200.000,25.000,b)
\MT 940.000   80.000
\LT 940.000  120.000
\MT 780.000   80.000
\LT 780.000  120.000
\cput(780.000,20.000,-0.010)
\MT 620.000   80.000
\LT 620.000  120.000
\MT 460.000   80.000
\LT 460.000  120.000
\cput(460.000,20.000,-0.020)
\MT 300.000   80.000
\LT 300.000  120.000
\MT 140.000   80.000
\LT 140.000  120.000
\cput(140.000,20.000,-0.030)
\MT1100.000    0.000
\LT1100.000  900.000
\MT 1085.000 840.000 \LT 1100.000 900.000 \LT 1115.000 840.000
\rput(1130.000,870.000,x_{\rm crt})
\MT1080.000  250.000
\LT1120.000  250.000
\rput(1140.000,230.000,2)
\MT1080.000  400.000
\LT1120.000  400.000
\rput(1140.000,380.000,4)
\MT1080.000  550.000
\LT1120.000  550.000
\rput(1140.000,530.000,6)
\MT1080.000  700.000
\LT1120.000  700.000
\rput(1140.000,680.000,8)
\grub0.6pt
\MT 1097.970  225.935
\LT 1097.970  225.935
\LT 1077.706  226.227
\LT 1057.515  226.524
\LT 1037.397  226.826
\LT 1017.352  227.133
\LT  997.381  227.445
\LT  977.482  227.761
\LT  957.657  228.083
\LT  937.906  228.412
\LT  918.228  228.746
\LT  898.625  229.085
\LT  879.096  229.432
\LT  859.641  229.784
\LT  840.260  230.143
\LT  820.955  230.509
\LT  801.725  230.883
\LT  782.570  231.264
\LT  763.491  231.655
\LT  744.488  232.053
\LT  725.561  232.459
\LT  706.710  232.876
\LT  687.937  233.301
\LT  669.241  233.738
\LT  650.623  234.185
\LT  632.083  234.644
\LT  613.622  235.115
\LT  597.075  235.551
\LT  580.592  235.996
\LT  564.174  236.453
\LT  547.821  236.924
\LT  531.534  237.408
\LT  515.314  237.905
\LT  499.160  238.416
\LT  483.074  238.945
\LT  467.055  239.490
\LT  451.106  240.054
\LT  435.225  240.637
\LT  419.415  241.242
\LT  403.675  241.870
\LT  388.008  242.524
\LT  374.142  243.128
\LT  360.334  243.756
\LT  346.586  244.411
\LT  332.897  245.094
\LT  319.269  245.809
\LT  305.704  246.561
\LT  292.201  247.351
\LT  278.763  248.188
\LT  267.059  248.963
\LT  255.407  249.782
\LT  243.808  250.653
\LT  233.909  251.446
\LT  224.052  252.291
\LT  214.237  253.195
\LT  206.091  254.002
\LT  197.977  254.864
\LT  189.895  255.796
\LT  183.454  256.599
\LT  177.036  257.465
\LT  172.238  258.164
\LT  167.454  258.913
\LT  162.684  259.722
\LT  157.929  260.608
\LT  154.768  261.249
\LT  151.615  261.943
\LT  148.469  262.698
\LT  146.899  263.105
\LT  145.332  263.534
\LT  143.766  263.991
\LT  142.203  264.478
\LT  140.643  265.004
\LT  139.085  265.577
\LT  137.530  266.211
\LT  135.978  266.928
\LT  134.429  267.764
\LT  132.885  268.801
\LT  131.345  270.298
\LT  130.501  272.022
\LT  130.425  272.642
\LT  130.425  272.853
\LT  130.501  273.483
\LT  131.338  275.310
\LT  132.855  276.990
\LT  134.368  278.212
\LT  135.878  279.233
\LT  137.386  280.134
\LT  138.892  280.954
\LT  140.396  281.714
\LT  141.898  282.427
\LT  143.399  283.102
\LT  144.899  283.746
\LT  146.397  284.363
\LT  149.391  285.535
\LT  152.381  286.637
\LT  155.367  287.685
\LT  159.840  289.175
\LT  164.308  290.589
\LT  168.770  291.942
\LT  174.713  293.672
\LT  180.649  295.334
\LT  188.062  297.335
\LT  196.949  299.652
\LT  207.308  302.266
\LT  219.139  305.169
\LT  232.446  308.358
\LT  250.191  312.531
\LT  278.319  319.053
\LT  302.060  324.545
\LT  321.401  329.051
\LT  337.811  332.913
\LT  352.771  336.478
\LT  367.774  340.102
\LT  381.319  343.424
\LT  394.906  346.810
\LT  407.023  349.880
\LT  419.179  353.011
\LT  429.850  355.806
\LT  440.554  358.655
\LT  451.295  361.565
\LT  462.073  364.537
\LT  471.342  367.138
\LT  480.642  369.792
\LT  489.974  372.502
\LT  499.339  375.271
\LT  508.739  378.102
\LT  516.600  380.512
\LT  524.486  382.970
\LT  532.400  385.478
\LT  540.341  388.040
\LT  548.312  390.657
\LT  556.312  393.333
\LT  564.343  396.070
\LT  572.407  398.872
\LT  580.504  401.741
\LT  588.636  404.683
\LT  595.167  407.090
\LT  601.722  409.549
\LT  608.301  412.059
\LT  614.906  414.625
\LT  621.536  417.248
\LT  628.193  419.932
\LT  634.878  422.680
\LT  641.592  425.493
\LT  646.646  427.648
\LT  653.412  430.585
\LT  658.507  432.837
\LT  663.620  435.134
\LT  668.751  437.477
\LT  673.901  439.867
\LT  679.071  442.306
\LT  684.260  444.797
\LT  689.470  447.343
\LT  694.701  449.944
\LT  699.954  452.603
\LT  705.229  455.324
\LT  710.527  458.108
\LT  715.848  460.959
\LT  721.194  463.879
\LT  726.564  466.873
\LT  731.960  469.943
\LT  737.383  473.093
\LT  741.013  475.239
\LT  744.656  477.424
\LT  748.311  479.649
\LT  751.979  481.915
\LT  755.660  484.223
\LT  759.355  486.576
\LT  763.064  488.975
\LT  766.786  491.421
\LT  770.523  493.917
\LT  774.275  496.463
\LT  778.042  499.062
\LT  781.824  501.717
\LT  785.623  504.429
\LT  789.437  507.200
\LT  793.268  510.034
\LT  797.116  512.931
\LT  800.982  515.897
\LT  804.865  518.932
\LT  806.814  520.476
\LT  810.725  523.623
\LT  812.688  525.226
\LT  814.656  526.848
\LT  816.629  528.491
\LT  818.606  530.154
\LT  820.589  531.839
\LT  822.577  533.546
\LT  824.570  535.276
\LT  826.568  537.028
\LT  828.572  538.804
\LT  830.581  540.604
\LT  832.596  542.429
\LT  834.616  544.279
\LT  836.642  546.154
\LT  838.674  548.057
\LT  840.712  549.986
\LT  842.755  551.945
\LT  844.805  553.931
\LT  846.861  555.948
\LT  848.923  557.994
\LT  850.992  560.072
\LT  853.067  562.181
\LT  855.148  564.324
\LT  857.236  566.500
\LT  859.331  568.711
\LT  861.433  570.959
\LT  863.542  573.242
\LT  865.658  575.565
\LT  867.782  577.925
\LT  869.913  580.327
\LT  872.051  582.769
\LT  874.197  585.255
\LT  876.351  587.785
\LT  878.513  590.361
\LT  880.683  592.983
\LT  882.862  595.655
\LT  885.049  598.376
\LT  887.244  601.150
\LT  889.449  603.978
\LT  891.662  606.861
\LT  893.885  609.802
\LT  896.117  612.803
\LT  898.359  615.866
\LT  900.610  618.993
\LT  902.871  622.187
\LT  905.143  625.451
\LT  907.425  628.786
\LT  909.718  632.197
\LT  912.022  635.686
\LT  914.338  639.255
\LT  916.664  642.910
\LT  919.003  646.653
\LT  921.354  650.488
\LT  923.717  654.419
\LT  926.093  658.451
\LT  928.482  662.588
\LT  930.885  666.835
\LT  933.301  671.198
\LT  935.732  675.683
\LT  938.177  680.295
\LT  940.638  685.041
\LT  943.114  689.927
\LT  945.606  694.962
\LT  948.115  700.154
\LT  950.641  705.511
\LT  953.184  711.043
\LT  955.746  716.761
\LT  958.327  722.677
\LT  960.927  728.800
\LT  963.548  735.147
\LT  966.190  741.730
\LT  968.854  748.567
\LT  971.541  755.674
\LT  974.251  763.073
\LT  976.986  770.783
\LT  979.747  778.829
\LT  982.535  787.238
\LT  985.352  796.040
\LT  988.198  805.267
\LT  991.075  814.959
\LT  993.986  825.157
\LT  996.931  835.910
\LT  999.912  847.274
\kon x_{\rm crt}(b),\ b<0;
\obraz48
\grub0.2pt
\MT   0.000  100.000
\LT1200.000  100.000
\MT 1140.000 115.000 \LT 1200.000 100.000 \LT 1140.000 85.000
\cput(1200.000,25.000,b)
\MT 940.000   80.000
\LT 940.000  120.000
\MT 780.000   80.000
\LT 780.000  120.000
\cput(780.000,20.000,-0.010)
\MT 620.000   80.000
\LT 620.000  120.000
\MT 460.000   80.000
\LT 460.000  120.000
\cput(460.000,20.000,-0.020)
\MT 300.000   80.000
\LT 300.000  120.000
\MT 140.000   80.000
\LT 140.000  120.000
\cput(140.000,20.000,-0.030)
\MT1100.000    0.000
\LT1100.000  150.000
\MT1100.000  220.000
\LT1100.000  900.000
\MT 1085.000 840.000 \LT 1100.000 900.000 \LT 1115.000 840.000
\rput(1130.000,870.000,a_{\rm crt})
\MT1080.000  250.000
\LT1120.000  250.000
\rput(1140.000,230.000,3)
\MT1080.000  400.000
\LT1120.000  400.000
\rput(1140.000,380.000,4)
\MT1080.000  550.000
\LT1120.000  550.000
\rput(1140.000,530.000,5)
\MT1080.000  700.000
\LT1120.000  700.000
\rput(1140.000,680.000,6)
\grub0.6pt
\MT 1097.970  275.864
\LT 1097.970  275.864
\LT 1077.706  275.013
\LT 1057.515  274.160
\LT 1037.397  273.306
\LT 1017.352  272.452
\LT  997.381  271.596
\LT  977.482  270.739
\LT  957.657  269.881
\LT  937.906  269.022
\LT  918.228  268.162
\LT  898.625  267.300
\LT  879.096  266.437
\LT  859.641  265.572
\LT  840.260  264.706
\LT  820.955  263.838
\LT  801.725  262.968
\LT  782.570  262.097
\LT  763.491  261.224
\LT  744.488  260.349
\LT  725.561  259.472
\LT  706.710  258.593
\LT  687.937  257.712
\LT  669.241  256.828
\LT  650.623  255.942
\LT  632.083  255.054
\LT  613.622  254.163
\LT  595.240  253.269
\LT  578.764  252.461
\LT  562.354  251.652
\LT  546.008  250.840
\LT  529.729  250.025
\LT  513.516  249.207
\LT  497.369  248.386
\LT  481.290  247.562
\LT  465.280  246.734
\LT  449.338  245.904
\LT  433.465  245.069
\LT  417.662  244.230
\LT  401.931  243.387
\LT  386.271  242.540
\LT  370.684  241.688
\LT  355.172  240.831
\LT  339.734  239.969
\LT  324.373  239.101
\LT  309.089  238.226
\LT  293.886  237.345
\LT  280.440  236.556
\LT  267.059  235.760
\LT  253.747  234.957
\LT  240.504  234.147
\LT  227.333  233.329
\LT  214.237  232.502
\LT  202.842  231.769
\LT  191.509  231.028
\LT  180.243  230.276
\LT  169.047  229.512
\LT  159.512  228.846
\LT  150.041  228.167
\LT  142.203  227.588
\LT  134.429  226.992
\LT  131.345  226.745
\LT  130.501  226.676
\LT  130.425  226.670
\LT  130.425  226.670
\LT  130.501  226.676
\LT  131.338  226.748
\LT  137.386  227.297
\LT  146.397  228.162
\LT  156.859  229.221
\LT  168.770  230.487
\LT  182.132  231.978
\LT  195.468  233.537
\LT  208.787  235.163
\LT  222.097  236.855
\LT  235.403  238.614
\LT  248.712  240.443
\LT  262.028  242.342
\LT  275.355  244.313
\LT  288.698  246.359
\LT  300.574  248.243
\LT  312.468  250.190
\LT  324.381  252.203
\LT  336.317  254.285
\LT  348.278  256.437
\LT  358.767  258.382
\LT  369.277  260.385
\LT  379.812  262.450
\LT  390.372  264.578
\LT  400.960  266.774
\LT  411.577  269.038
\LT  422.224  271.376
\LT  431.377  273.440
\LT  440.554  275.562
\LT  449.758  277.744
\LT  458.989  279.990
\LT  468.249  282.302
\LT  477.539  284.682
\LT  485.304  286.721
\LT  493.092  288.811
\LT  500.904  290.955
\LT  508.739  293.156
\LT  516.600  295.415
\LT  524.486  297.734
\LT  532.400  300.118
\LT  540.341  302.567
\LT  548.312  305.086
\LT  556.312  307.677
\LT  562.734  309.804
\LT  569.178  311.982
\LT  575.642  314.211
\LT  582.127  316.495
\LT  588.636  318.836
\LT  595.167  321.235
\LT  601.722  323.695
\LT  608.301  326.219
\LT  614.906  328.809
\LT  621.536  331.469
\LT  628.193  334.201
\LT  633.204  336.300
\LT  638.232  338.443
\LT  643.275  340.631
\LT  648.335  342.867
\LT  653.412  345.151
\LT  658.507  347.486
\LT  663.620  349.874
\LT  668.751  352.317
\LT  673.901  354.816
\LT  679.071  357.375
\LT  684.260  359.994
\LT  689.470  362.678
\LT  694.701  365.428
\LT  699.954  368.247
\LT  705.229  371.138
\LT  710.527  374.105
\LT  715.848  377.151
\LT  721.194  380.279
\LT  724.771  382.412
\LT  728.360  384.585
\LT  731.960  386.798
\LT  735.572  389.053
\LT  739.196  391.352
\LT  742.833  393.695
\LT  746.482  396.085
\LT  750.143  398.523
\LT  753.818  401.011
\LT  757.506  403.549
\LT  761.207  406.141
\LT  764.923  408.788
\LT  768.653  411.492
\LT  772.397  414.255
\LT  776.157  417.080
\LT  779.931  419.968
\LT  783.721  422.923
\LT  787.528  425.946
\LT  791.350  429.041
\LT  795.190  432.211
\LT  797.116  433.825
\LT  799.047  435.459
\LT  800.982  437.113
\LT  802.921  438.787
\LT  804.865  440.483
\LT  806.814  442.201
\LT  808.767  443.940
\LT  810.725  445.702
\LT  812.688  447.487
\LT  814.656  449.296
\LT  816.629  451.129
\LT  818.606  452.987
\LT  820.589  454.869
\LT  822.577  456.778
\LT  824.570  458.713
\LT  826.568  460.676
\LT  828.572  462.666
\LT  830.581  464.684
\LT  832.596  466.732
\LT  834.616  468.810
\LT  836.642  470.918
\LT  838.674  473.058
\LT  840.712  475.230
\LT  842.755  477.435
\LT  844.805  479.674
\LT  846.861  481.948
\LT  848.923  484.257
\LT  850.992  486.604
\LT  853.067  488.988
\LT  855.148  491.411
\LT  857.236  493.875
\LT  859.331  496.379
\LT  861.433  498.926
\LT  863.542  501.516
\LT  865.658  504.151
\LT  867.782  506.833
\LT  869.913  509.562
\LT  872.051  512.340
\LT  874.197  515.169
\LT  876.351  518.050
\LT  878.513  520.985
\LT  880.683  523.976
\LT  882.862  527.025
\LT  885.049  530.133
\LT  887.244  533.303
\LT  889.449  536.536
\LT  891.662  539.836
\LT  893.885  543.203
\LT  896.117  546.642
\LT  898.359  550.154
\LT  900.610  553.742
\LT  902.871  557.410
\LT  905.143  561.159
\LT  907.425  564.994
\LT  909.718  568.918
\LT  912.022  572.934
\LT  914.338  577.047
\LT  916.664  581.260
\LT  919.003  585.577
\LT  921.354  590.004
\LT  923.717  594.545
\LT  926.093  599.205
\LT  928.482  603.990
\LT  930.885  608.906
\LT  933.301  613.959
\LT  935.732  619.156
\LT  938.177  624.504
\LT  940.638  630.011
\LT  943.114  635.686
\LT  945.606  641.536
\LT  948.115  647.573
\LT  950.641  653.807
\LT  953.184  660.248
\LT  955.746  666.910
\LT  958.327  673.806
\LT  960.927  680.950
\LT  963.548  688.358
\LT  966.190  696.049
\LT  968.854  704.040
\LT  971.541  712.354
\LT  974.251  721.013
\LT  976.986  730.044
\LT  979.747  739.474
\LT  982.535  749.336
\LT  985.352  759.666
\LT  988.198  770.502
\LT  991.075  781.892
\LT  993.986  793.884
\LT  996.931  806.537
\LT  999.912  819.918
\kon a_{\rm crt}(b),\ b<0;}

\section{Dirac \e\ in the \NK{}}
In this section we deal with a generalization of a Dirac \e\ on~$P$. Thus we
consider spinor fields $\Ps,\ov\Ps$ on~$P$ transforming according to
$\Spin(1,4)$ (a~double covering group of $\SO(1,4)$---de Sitter group). We
want to couple these fields to gravity and electromagnetism. For
$\Ps$ and $\ov\Ps$ we have $\Ps,\ov\Ps:P\to\C^4$ and
\beq{4.1}
\bal
\Ps(\vf(g)p)&=\si(g^{-1})\Ps(p)\\
\ov\Ps(\vf(g)p)&=\ov\Ps(p)\si(g),
\eal
\end{equation}
where $\si\in\cL(\C^4)$, $p=(x,g_1)\in P$, $g,g_1\in\U(1)$.

On $E$ we define spinor ordinary fields $\psi,\ov\psi:E\to\C^4$. We suppose
that $\psi$ and~$\ov\psi$ are defined up to a phase factor and that
\beq{4.2}
\bal
\psi^f(x)&=\Ps(f(x))\\
\ov\psi{}^f(x)&=\ov\Ps(f(x))
\eal
\end{equation}
where $f:E\to P$ is a section of a bundle $\ul P$. In some sense spinor fields
on~$P$ are lifts of spinors on~$E$ (see Appendix C),
\beq{4.3}
\bal
\Ps(f(x))=\pi^\ast(\psi^f(x)),\quad&\psi^f=f^\ast \Ps\\
\ov\Ps(f(x))=\pi^\ast(\ov\psi{}^f(x)),\quad&\ov\psi{}^f=f^\ast \ov\Ps.
\eal
\end{equation}

Let us consider a different section of a bundle $\ul P$, $e:E\to P$. In this
case we have
$$
\gathered
\psi^e=e^\ast \Ps, \q \ov \psi=e^\ast\ov\Ps, \q
\psi{}^e(x)=\Ps(e(x)), \q \ov\psi{}^e(x)=\ov\Ps(e(x)),\\
\psi^e(x)=\psi^f(x)\exp\Bigl(\frac{ikq}{\hbar c}\,\chi(x)\Bigr), \q
\ov\psi{}^e(x)=\ov\psi{}^f(x)\exp\Bigl(-\frac{ikq}{\hbar c}\,\chi(x)\Bigr),
\endgathered
$$
where $kq$ is a charge of a fermion, $k=0,\pm1,\pm2, \dots$, for an electron $k=1$, $\chi$ is a gauge
changing \f.

Let us define an exterior gauge \dv\ $\gdv$ of the field~$\Ps$. One gets
\beq{4.4}
d\Ps=\z_\mu \Ps\t^\mu+\z_5\Ps\t^5
\end{equation}
and
\beq{4.5}
\bal
\gdv\Ps&=\hor d\Ps=\z_\mu\Ps\t^\mu\\
\gdv\ov\Ps&=\hor d\ov\Ps=\z_\mu\ov\Ps\t^\mu.
\eal
\end{equation}
Let $\g_\mu\in\cL(\C^4)$ be Dirac's matrices obeying the conventional
relations
\beq{4.6}
\{\g_\mu,\g_\nu\}=2\eta_\m
\end{equation}
(where $\eta_\m$ is a Minkowski tensor of signature $(---+)$) and let $B=B^+$
be a matrix \st
\beq{4.7}
\g^{\mu+}=B\g^\mu B^{-1}, \quad \ov\psi=\psi^+ B
\end{equation}
(the indices are raised by $\eta^\m$, an inverse tensor of $\eta_\m$), where
``$+$'' is a Hermitian conjugation, and
\beq{4.8}
\si_\m=\tfrac18[\g_\mu,\g_\nu].
\end{equation}
We define
$$
\g^5=\g^1\g^2\g^3\g^4 \in \cL(\C^4).
$$
One can easily check that
\beq{4.9}
\{\g_A,\g_B\}=2\ov g_{AB}
\end{equation}
where
\beq{4.10}
\begin{array}{rcl}
{}&\ov g_{AB}={\rm diag}(-1,-1,-1,+1,-1)&\\[2pt]
\hbox{and }&\g^A=(\g^\a,\g^5)&
\end{array}
\end{equation}
(the indices are raised by $\ov g{}^{AB}$, an inverse tensor of $\ov g_{AB}$).
We have
\beq{4.11}
\g^{5+}=B\g^5B^{-1} \qh{and} \ov\Ps=\PS^+B.
\end{equation}
So
\beq{4.12}
\g^{A+}=B\g^A B^{-1}.
\end{equation}

On the manifold $P$ we have an orthonormal \cd\ system $\t^A$ and we can
perform an infinitesimal change of the frame
\beq{4.13}
\bga
\t^{A'}=\t^A+\d \t^A=\t^A-\gd \ve,A,B,\t^B\\
\ve_{AB}+\ve_{BA}=0.
\ega
\end{equation}
If the spinor field $\Ps$ corresponds to $\t^A$ and $\Ps'$ to $\t^{A'}$ then
we get
\beq{4.14}
\bal
\Ps'&=\Ps+\d\Ps=\Ps-\ve^{AB}\wh\si_{AB}\Ps\\
\ov\Ps'&=\ov\Ps+\d\ov\Ps=\ov\Ps+\ov\Ps\wh\si_{AB}\ve^{AB}
\eal
\end{equation}
($\Ps$ and $\ov\Ps$ are Schouten $\si$-quantities (see Refs
\cite{433},~\cite{434}) where
\beq{4.15}
\wh\si_{AB}=\tfrac18[\g_A,\g_B].
\end{equation}
Notice that the \di\ of the spinor space for a $2n$-\di al space is $2^n$ and
it is the same for a $(2n+1)$-\di al one (in our case $n=2$).

We take a spinor field for a 5-dimensional space $P$ and assume that the
dependence on the 5th \di\ is trivial, i.e.\ Eq.~\eqref{4.1} holds. Taking a
section we obtain spinor fields on~$E$.

Let us introduce some new notions. We introduce a Levi-Civita symbol and a
dual Cartan's base
\bea{4.16}
{}&\ov\eta_{\a\b\g\d}, \qquad \ov\eta_{1234}=\sqrt{-\det(g_\(\a\b))}\\
&\ov\eta_\a=\frac1{2\cdot3}\,\ov\t^\d\wedge\ov\t^\g\wedge\ov\t^\b
\ov\eta_{\a\b\g\d} \label{4.17}\\
&\ov\eta=\frac14\ov\t^\a\wedge\ov\eta_\a.
\end{eqnarray}
We define
\beq{4.19}
\bal
\eta_\a&=\pi^\ast (\ov\eta_\a)\\
\eta&=\pi^\ast(\ov\eta)
\eal
\end{equation}
We rewrite here a Riemannian part of the \cn\ \eqref{1.15} introducing the \ct\
$\l=\frac{2\sqrt{G_N}}{c^2}$,
\beq{4.20}
\gd\wt w{},A,B,=\left(
\begin{array}{c|c}
\pi^\ast\bigl(\gd{\wt{\ov w}}{},\a,\b,)+\frac\l2\pi^\ast(\gd F,\a,\b,)\t^5\, &\,
\frac\l2 \pi^\ast(\gd F,\a,\g,\ov \t{}^\g)\\[2pt]\hline
\noalign{\vskip2pt}
-\frac\l2\pi^\ast (F_{\b\g}\ov \t{}^\g) & 0
\end{array}
\right)
\end{equation}
(see Refs \cite{98}, \cite{420}).

Let us consider exterior covariant \dv s of spinors $\Ps$ and $\ov\Ps$,
\beq{4.21}
\bal
\wt D\Ps&=d\Ps+\gd\wt w{},A,B,\dg\wh\si{},A,B,\Ps\\
\wt D\ov\Ps&=d\ov\Ps-\gd\wt w{},A,B,\Ps \dg\wh\si{},A,B,
\eal
\end{equation}
\wrt the Riemannian \cn\ $\gd \wt w{},A,B,$.

Now we introduce a \dv\ $\cD$, i.e.\ an exterior ``gauge'' \dv\ of a new
kind. This \dv\ may be treated as a generalization of minimal coupling scheme
between spinor and \elm c field on~$P$,
\beq{4.22}
\bal
\cD\Ps&=\hor D\Ps\\
\cD\ov\Ps&=\hor D\ov\Ps.
\eal
\end{equation}
We get
\beq{4.23}
\bal
\cD\Ps&=\wt{\ov\cD}\Ps-\frac\l8\,\gd F,\a,\mu,[\g_\a,\g_5]\Ps\t^\mu\\
\cD\ov\Ps&=\wt{\ov\cD}\,\ov\Ps+\frac\l8\,\gd F,\a,\mu,\ov\Ps[\g_\a,\g_5]\t^\mu
\eal
\end{equation}
where
\beq{4.24}
\bal
\wt{\ov\cD}\Ps&=\gdv \Ps+\pi^\ast (\gd{\wt{\ov w}}{},\a,\b,)\dg\si,\a,\b,\Ps\\
\wt{\ov\cD}\ov\Ps&=\gdv \ov\Ps-\pi^\ast (\gd{\wt{\ov w}}{},\a,\b,)
\ov\Ps\dg\si,\a,\b,.
\eal
\end{equation}

The \dv\ $\wt{\ov\cD}$ is a covariant \dv\ \wrt both $\pi^\ast(\gd{\wt{\ov
w}}{},\a,\b,)$ and ``gauge'' at once. It introduces an interaction between
\elm c and \gr al fields with Dirac's spinor in a classical well-known way
($\wt{\ov\cD}\Ps =\hor\wt{\ov D}\Ps$).

In Dirac theory we have the following Lagrangian for a spinor $\frac12$-spin
field on~$E$:
\beq{4.25}
\cL(\psi,\ov\psi,d)=i\,\frac{\hbar c}2\bigl(\ov\psi\, \ov l\wedge d\psi
+d\ov\psi\wedge \ov l\psi\bigr)+mc^2\ov\psi\psi\ov\eta
\end{equation}
where $\ov l=\g_\mu\ov\eta{}^\mu$.

Let us lift Lagrangian on a manifold $P$. We pass from spinors $\psi$ and
$\ov \psi$ to $\Ps$ and $\ov\PS$ and from the \dv\ $d$ to $\gdv$ or
to~$\wt{\ov \cD}$. This is a classical way. Moreover, we have to do with a
theory which unifies gravity and \elm sm and in order to get new physical
effects we should pass to our new \dv~$\cD$. Simultaneously we pass from
$\ov\eta$ to~$\eta$ and from $\ov l$ to $\pi^\ast(\ov l)=l$.

In this way one gets
\beq{4.26}
\cL_D(\Ps,\ov\Ps,\cD)=\frac{i\hbar c}2\bigl(\ov\PS l\wedge\cD\Ps
+\cD\ov\Ps\wedge l\Ps\bigr)+mc^2\ov\Ps \Ps\eta.
\end{equation}
Using formulae \eqref{4.23} one obtains
\beq{4.27}
\cL_D(\Ps,\ov\Ps,\cD)=\cL_D(\Ps,\ov\Ps,\wt{\ov \cD})
-i\,\frac{2\sqrt {G_N}}c\,\hbar F_\m \ov\Ps\g_5\si^\m \Ps\eta
\end{equation}
where
\beq{4.28}
\cL_D(\Ps,\ov\Ps,\wt{\ov \cD})=\frac{i\hbar c}2\bigl(\ov\Ps l\wedge
\wt{\ov\cD} \Ps+\wt{\ov\cD}\ov\Ps\wedge l\Ps\bigr)+mc^2\ov\Ps \Ps\eta.
\end{equation}

Now we should go back to a \spt\ $E$ (see Appendix~C) and we get the
following Lagrangian
\bg{4.29a}
\cL_D(\psi,\ov\psi,\cD)=\cL_D(\psi,\ov\psi,\wt{\ov\cD})
-i\,\frac{2\sqrt{G_N}}c\,\hbar F^\m\ov\psi\g_5\si_\m \psi\\
\cL_D(\psi,\ov\psi,\wt{\ov\cD})=\frac{i\hbar c}2 \bigl(\ov\psi\,\ov l
\wedge \wt{\ov\cD}\psi+\wt{\ov\cD}\psi\wedge \ov l\psi\bigr)
+mc^2\ov\psi \psi\ov \eta.\label{4.30a}
\end{gather}

We get a new term
\beq{4.29}
-i\,\frac{2\sqrt{G_N}}c\,\hbar F^\m\ov\psi \g_5\si_\m\psi.
\end{equation}
It is an interaction of the \elm c field with an anomalous dipole electric
moment. For such an anomalous interaction it reads
\beq{4.30}
i\,\frac{d_{kk}}2 \,F^\m \ov\psi\g_5\si_\m\psi.
\end{equation}
Our anomalous moment reads
\beq{4.31}
d_{kk}=-\frac{4\sqrt{G_N}}c\,\hbar=-\frac{4l\pl}{\sqrt\a}\,q
\simeq -7.56784835\times 10^{-32}\,{\rm[cm]}q
\end{equation}
where $l\pl$ is a \Pl\ length
$$
l\pl=\sqrt{\frac{\hbar G_N}{c^3}}\simeq1.61199\times10^{-35}{\rm m},
$$
$q$ is an elementary charge and
$$
\a=\frac{e^2}{\hbar c}\simeq\frac1{137}
$$
is a fine structure \ct.

This term can be also rewritten in a different way,
\beq{4.32}
-\frac2{\La_p}\,(\hbar^3c^5)^{1/2}F^\m \ov\psi\g_5\si_\m\psi
\end{equation}
where
\beq{4.33}
\bga
\La_p=m_pc^2\simeq 1.2209\times10^{19}{\rm GeV}\\
m_p=2.1765\times10^{-8}{\rm kg}
\ega
\end{equation}
are \Pl\ energy scale and \Pl\ mass. Thus we get a term which probably gives
us a trace of New Physics on a \Pl\ energy scale. This term is
nonrenormalizable in Quantum Field Theory and it is of 5 order in mass units
(i.e.\ $c=\hbar=1$) divided by an energy (mass) scale.

The term \eqref{4.30} can be written in a very convenient way
\beq{4.34}
d_{kk}\ov\psi\bigl(\b(\os\Si\cdot \os E+i\os\a\os B)\psi\bigr)
\end{equation}
where
\bg{4.35}
\b=\left(\begin{matrix} I & 0 \\ 0 & -I\end{matrix}\right), \quad
\os \a=\left(\begin{matrix} 0 & \os\si \\ \os\si & 0 \end{matrix}\right), \q
\os\g=\b\os\a\\
\os\Si=-\g^5\os \a=\g^4\g^5\os\g=\b\g^5\os \g \label{4.36}\\
\os\si=(\si_x,\si_y,\si_z), \label{4.37}
\end{gather}
$I$ is the identity matrix $2\times2$ and $\os\si$ are Pauli matrices. $\os E$ is an
electric field and $\os B$ is a magnetic field. In this way our term
introduces an anomalous dipole electric interaction and also an anomalous
magnetic dipole interaction. Of course the magnetic interaction is negligible
in comparison to ordinary magnetic moment interaction of an electron. One can
easily calculate this anomalous magnetic moment of an electron in terms of
Bohr magneton getting
$$
\frac4{\sqrt \a}\Bigl(\frac{m_{\rm e}}{m_{\rm p}}\Bigr)\mu_B=19.188
\times10^{-21}\mu_B,
$$
where $m_{\rm e}$ is a mass of an electron and $\mu_B=\frac{q\hbar}{2m_e}$ is
a Bohr magneton. From
the physical point of view the most important is the electric dipole moment
(EDM). So we see that using spinors $\Ps$ and $\ov\Ps$ and a \dv\ $\wt{\ov
\cD}$ in the Kaluza--Klein Theory we have achieved an additional \gr al-\elm
c effect. It is just an existence of a dipole moment of a fermion, which
value is determined by fundamental \ct s (only!).
This is another ``interference effect'' between \elm c and \gr al fields in
our unified field theory.
Thirring also has achieved in his paper \cite{421} a dipole electric moment
of fermion of the same order. In his theory a minimal rest mass of fermion is
of order of a \Pl\ mass. Thus his theory cannot describe a fermion from the
Standard Model. The anomalous moment in Thirring's theory depends on a mass
of a fermion. In order to get $d_{kk}$ of order $10^{-32}\,{\rm[cm]}q$ this must
be of a \Pl\ mass order. Otherwise the value of $d_{kk}$ can be smaller. (In
reality W.~Thirring obtains two types of anomalous Pauli terms---electric and
magnetic of the same order.)

In our case mass $m$ may be arbitrary, e.g.\ $m=0$. Thus we can consider also
massless fermions. We can also consider chargeless fermions, i.e.\ for $k=0$.
It is also worth noticing that Thirring's quantities $\Ps$
and~$\ov\Ps$ have nothing to do with our spinor fields $\Ps$ and~$\ov\Ps$ for
a mysterious Thirring's quantity $\vf$ which is absent in our theory (it
appears also in Thirring's definition of a parity operator). We develop the
theory considered here also in ordinary Kaluza--Klein Theory and in the
Kaluza--Klein theory with a torsion (see Refs \cite{98}, \cite{422},
\cite{423}). Someone develops a theory using our spinors $\Ps$ and~$\ov\Ps$
getting also anomalous electric dipole moments (see Refs \cite{424},
\cite{425}). We develop a similar approach for a Rarita--Schwinger field (see
Ref.~\cite{426}). In the case of the \NK{} we consider also a different
approach (see \cite{427}, \cite{428}). However now we consider the present as
appropriate.

Let us consider operations of reflection defined on a manifold~$P$. To define
them we choose first a local \cd\ system on~$P$ in such a way that we pass
from $\t^A$ to $dx^A$ (see Section~1), i.e.\ $(\pi^\ast(dx^\a),dx^5)$. In
this way
\beq{4.38}
x^A=(x^\a,x^5),\quad x^\a=(\os x,t).
\end{equation}
Then
\beq{4.39}
\Ps(p)=\Ps(x^A)=\Ps\bigl((\os x,t),x^5\bigr)
\end{equation}
and we define transformations: space reflection $P$ (do not confuse with a
manifold~$P$), time reversal $T$, charge reflection~$C$ and combined
transformations $PC$, $\t=PCT$,
\beq{4.40}
\Ps^C(x^\a,x^5)=C\Ps^\ast (x^\a,-x^5),
\end{equation}
where $C^{-1}\g_\mu C=-\g_\mu^\ast$.

Taking a section $f$ we get
\beq{4.41}
(\psi^f)^C (x^\a)=C\psi^{f\ast}(x^\a)
\end{equation}
and a charge changes the sign. The \rf\ $x^5\to -x^5$ as a charge \rf\ has
been already suggested by J.~Rayski (see Ref.~\cite{429}). For the space \cd\
\rf\ we have
\beq{4.42}
\Ps^P(x^\a,x^5)=\g^4 \Ps(-\os x,t,x^5).
\end{equation}
Taking a section $f$ we obtain
\beq{4.43}
(\psi^f)^P(\os x,t)=\g^4\psi^f(-\os x,t),
\end{equation}
i.e.\ a normal parity operator on~$E$.

This contrasts with Thirring's definition of the parity operator (Thirring
was forced to change the definition of the parity operator on 5-\di al space
and he could not obtain a normal parity operator on~$E$). The \tf\ of
time-reversal $T$ is defined by
\beq{4.44}
\Ps^T(\os x,t,x^5)=C^{-1}\g^1\g^2\g^3\Ps^\ast(\os x,-t,-x^5).
\end{equation}
Taking a section $f$ we get
\beq{4.45}
(\psi^f)^T(\os x,t)=C^{-1}\g^1\g^2\g^3(\psi^f)^\ast(\os x,-t)
\end{equation}
and a charge does change sign, i.e.\ a normal time-reversal operator on a
spsce-time.

To define a \tf\ $\t=PCT$ we write
\beq{4.46}
\Ps^\t (\os x,t,x^5)=-i\g^5\Ps(-\os x,-t,-x^5).
\end{equation}
Taking a section $f$ we get
\beq{4.47}
(\psi^f)^\t (\os x,t)=-i\g^5\psi^f(-\os x,-t)
\end{equation}
and a charge changes the sign.
The \tf\ $PC$ is as follows
\beq{4.48}
\Ps^{PC} (\os x,t,x^5)=\g^4C\Ps^\ast(-\os x,t,x^5).
\end{equation}
Taking a section $f$ we have
\beq{4.49}
(\psi^f)^{PC} (\os x,t)=\g^4C(\psi^f)^\ast(-\os x,t)
\end{equation}
and a charge changes a sign.

It is clear now that the \tf s obtained by us do not differ from those known
from the literature.

The additional term in Lagrangian \eqref{4.27} breaks $PC$ or $T$ symmetries
as in Thirring's theory (see Ref.~\cite{421}), but Thirring defines the
operator $PC$ in a different way. This can be easily seen by acting on both
sides of Eq.~\eqref{4.29} with the operator defined by Eq.~\eqref{4.48}. Of
course this breaking is very weak and it cannot be linked to $CP$-breaking
term in Cabbibo--Kobayashi--Maskava matrix. From this breaking due to
$\d_{PC}$-phase, which is responsible for $PC$ nonconservation in $K^0,\ov
K{}^0$ mesons decays and also for $D^0,\ov D{}^0$, $B_s,\ov B_s$, $B^0, \ov
B{}^0$ and so on, see Ref.~\cite{430n}, we can get a dipole electric moment of
an electron of order $8\times10^{-41}\,{\rm [cm]}q$ (if there is not New
Physics beyond SM, see Ref.~\cite{430}). This is because all Feynman diagrams
which induce EDM of electron vanish to three loops order.

According to Ref.~\cite{430} electron EDM
\beq{4.46d}
d_e=\biggl(\frac{g^2_w}{32\pi^2}\biggr)
\biggl(\frac{m_e}{M_w}\biggr)
\biggl[\ln\frac{\La^2}{M_W^2}+O(1)\biggr]\,d_W
\end{equation}
where
\beq{4.47d}
d_W=J\biggl(\frac{g_W^2}{32\pi^2}\biggr)\biggl(\frac q{2M_W}\biggr)
\frac{m_b^4m_s^2m_c^2}{M_W^2}
\end{equation}
is EDM for a $W$ boson, $\La$ is an energy scale for a New Physics
(beyond~SM),
$$
J=s_1^2s_2s_3c_1c_2c_3\sin\d_{CP}=2.96\times10^{-5}
$$
(see Ref.~\cite{432n}) is a Jarlskog invariant, $m_b,m_s,m_c$ are masses of quarks (we suppose the
existence of three families of fermions in~SM) and $s_i=\sin\t_i$,
$c_i=\cos\t_i$, $i=1,2,3$.

EDMs of an electron $d_e$ and quarks can induce EDMs of paramagnetic and
diamagnetic atoms
\bea{4.48d}
d_{\rm para}&\sim& 10\a^2Z^3d_e\\
d_{\rm dia}&\sim& 10Z^2\Bigl(\frac{R_N}{R_A}\Bigr)^2 \wt d_q. \label{4.49d}
\end{eqnarray}

For Thalium (Tl) and for Mercury (Hg) one gets
\bea{4.50}
d_{\rm Tl}&=&-585d_e\\
d_{\rm Hg}&=&7\times10^{-3}e(\wt d_u-\wt d_d)+10^{-2}d_e. \label{4.51}
\end{eqnarray}
For a neutron
$$
d_{\rm n}=(1.4\mp0.6)(d_d-0.25d_u)+(1.1\pm0.5)q(\wt d_d+0.5\wt d_u)
$$
where $d_d,d_u$ are EDM of quarks and $\wt d_d,\wt d_u,\wt d_q$ are color EDM
operators (see Ref.~\cite{431} and references cited therein). Recently we
have an upper bound on EDMs (see Ref.~\cite{432} and references cited
therein)
$$
|d_{\rm n}|<2.9\times 10^{-26}\,{\rm[cm]}q, \quad
|d_e|<1.6 \times 10^{-27}\,{\rm [cm]}q, \q
d({}^{199}{\rm Hg})<3.1\times 10^{-29}\,{\rm[cm]}q.
$$
In the case of $\t$-term in QCD we have also $d_n=3\times 10^{-16}
\t\,{\rm [cm]}q$ (see Ref.~\cite{431}).

Recently there has been a significant progress in obtaining an upper limit on
the EDM of an electron by using a polar molecule thorium monoxide (ThO). The
authors of Ref.~\cite{448} obtained an upper limit on~$d_e$,
\beq{4.58}
|d_e|<8.7 \times 10^{-29}{\rm[cm]}q.
\end{equation}
This is only of three orders of magnitude bigger than our result (see
Eq.~\eqref{4.31}). From the other side there is also a progress in
calculation of SM prediction of EDM for an electron coming from a phase
$\d_{CP}$ of CKM matrix. This calculation gives us the so called
\ti{equivalent} EDM (see Ref.~\cite{449}),
\beq{4.59}
d_e^{\rm equiv} \sim 10^{-38}{\rm[cm]}q,
\end{equation}
which is bigger of three orders of magnitude than the result from
Ref.~\cite{430n}.  Moreover, still smaller of six orders than our result. The
parameter~$\t$ from QCD is unknown and has no influence on EDM of an
electron. The existence of EDM of an electron coming from Kaluza--Klein
theory can help us in understanding of an asymmetry of matter-antimatter in the
Universe. This EDM moment which breaks PC and T~symmetry in an explicit way
can have an influence on the surviving of an annihilation matter with
antimatter following Big Bang.

It is interesting to notice that EDM from Kaluza--Klein Theory is the same
for a muon (a $\mu$ meson) and a tauon (a $\tau$ meson)
as for an electron. We get the same value for flavour
states of neutrinos. Due to this, EDM of this value can influence
oscillations of neutrinos species (see Ref.~\cite{450}).

To be honest, we write down a different, however trivial, coupling of spinor
fields $\Ps$ and~$\ov\Ps$ in Kaluza--Klein. This is a coupling to a \cn\ of
the form
\beq{4.60}
\gd\wh w{},A,B,=\left(
\begin{array}{c|c}
\pi^\ast\bigl(\gd{\wt{\ov w}}{},\a,\b,)\, &\,0\\[2pt]\hline
0 &\, 0
\end{array}
\right).
\end{equation}
In this way $\Ps$ and $\ov\Ps$ are transforming according to ${\rm SL}(2,\C)$
and new phenomena are absent, i.e.\ we have to do with Lagrangian
\eqref{4.28}.

Let us come back to neutrino oscillations in the presence of EDM. Let us
write a Lagrangian for three neutrino species neglecting \gr al field:
\bml{4.61}
\cL_D(\Ps_\l,\ov\Ps_\l,d)=\sum_{\l=\a,\b,\g}\biggl(\frac{i\hbar c}2
\bigl(\ov\Ps_\l l\wedge d\Ps_\l+ d\ov\Ps_\l\wedge l\Ps_\l\bigr)
+i\,\frac{d_{kk}}2 \,F^\m \ov\Ps_\l \g_5\si_\m \Ps_\l\biggr)\\
{}+\sum_{\l,\l'=\a,\b,\g}c^2\ov\Ps_\l m_{\l\l'}\Ps_{\l'}\eta.
\end{multline}
Despite the smallness of $d_{kk}$ its interaction with a strong electric and
magnetic fields can result in sizeable effects (see Eq.~\eqref{4.34}).
$m_{\l\l'}$~is a mass matrix for neutrinos which is not diagonal. In
particular $\a=e$, $\b=\mu$, $\g=\tau$.

Let us consider mass eigenstates of our neutrinos $\Ps_a$, $a=1,2,3$ (see
\cite{450})
\beq{4.62}
\PS_\l=\sum_{a=1,2,3}U_{\l a}\Ps_a.
\end{equation}
The unitary matrix $U=(U_{\l a})$ diagonalizes the mass matrix $\ov
m=(m_{\l\l'})$. The eigenvalues of the mass matrix are called $m_a$, $a=1,2,3$.
\begin{gather}\label{4.63}
\left(\begin{matrix}
m_1 & 0 & 0\\ 0 &m_2& 0\\ 0 &0 &m_3
\end{matrix}\right)=U^+ \,\ov m\,U\\
U=\left(\begin{matrix}
c_{12}c_{13} & s_{12}c_{13} & s_{13}e^{-i\d}\\
-s_{12}c_{23}-c_{12}s_{23}s_{13}e^{i\d} &
c_{12}c_{23}-s_{12}s_{23}s_{13}e^{i\d} & s_{23}c_{13}\\
s_{12}s_{23}-c_{12}c_{23}s_{13}e^{i\d} &
-c_{12}s_{23}-s_{12}c_{23}s_{13}e^{i\d} & c_{23}c_{13}
\end{matrix}\right)
\label{4.64}
\end{gather}
where $c_{ij}=\cos\t_{ij}$, $s_{ij}=\sin\t_{ij}$, the angles $\t_{ij}
\in[0,\frac\pi2]$, $\d\in[0,2\pi]$ is a Dirac CP violation phase (see
Refs.~\cite{430n}, \cite{450}, $i$ means $\l$---flavour, $j$ means $a$---mass
eigenstate).

In the new spinor variables the Lagrangian \eqref{4.61} reads
\beq{4.65}
\cL_D(\Ps_a,\ov\Ps_a,d)=\sum_{a=1,2,3}\biggl(\frac{i\hbar c}2 \bigl(\ov\Ps_a l\wedge d\Ps_a
+d\ov\Ps_a\wedge l\Ps_a\bigr) + \ov\Ps_a M_a \Ps \eta\biggr),
\end{equation}
where
\beq{4.66}
M_a=m_ac^2 + i\,\frac{d_{kk}}2 F^\m \g_5\si_\m = m_ac^2+d_{kk}\b
\bigl(\os\Si\cdot \os E+ i\os\a \cdot\os B\bigr)
\end{equation}
(see Eq.~\eqref{4.34}).

Using initial conditions for mass eigenstates
\bg{4.67}
\Ps_a(\os r,t=0)=\Ps_a^{(0)}(\os r) \\
\Ps_\la^{(0)}(\os r) = U_{\la a}\Ps_a^{(0)}(\os r)\label{4.68} \\
\Ps_a^{(0)}(\os r) =(U^{-1})_{a\la}\Ps_\la^{(0)}(\os r) \label{4.69}
\end{gather}
we can solve an initial value problem for linear \e s corresponding to the
Lagrangian \eqref{4.65}, finding an evolution in time of fields $\Ps_a$ (they
do not couple). Afterwards using \eqref{4.62} and \eqref{4.69} we find
oscillations of three neutrino flavours under an influence of magnetic and
electric fields due to additional term coming from Kaluza--Klein Theory.
Field \e s for $\Ps_a$ (Euler--Lagrange \e s for Lagrangian \eqref{4.65}) are
given in the following Hamilton form
\beq{4.70}
i\hbar c\,\frac{\pa \PS_a}{\pa t}=H_a\Ps_a,\quad a=1,2,3,
\end{equation}
where
\bg{4.71}
H_a=c\os\a\cdot\os p+\b m_ac^2-d_{kk}\bigl(\os\Si\cdot\os E+i\os\a\cdot
\os B\bigr)\\
\os p=-i\hbar\os\nabla.\label{4.72}
\end{gather}
Thus eventually one gets
\beq{4.73}
i\hbar c\,\frac{\pa \Ps_a}{\pa t}=-i\hbar c(\os\a\cdot\os\nabla)\Ps_a
+m_ac^2 \b\Ps_a- d_{kk}(\os\Si\cdot\os E+i\os \a\cdot\os B)\Ps_a, \q
a=1,2,3.
\end{equation}

Equations \eqref{4.73} are typical Dirac--Pauli \e s. Moreover, they have a
term which explicitly breaks PC \tf. We suppose $\os E={\rm const}$, $\os
B={\rm const}$. For Eqs~\eqref{4.73} are linear the general \so s are
expressed by the Fourier integral
\bml{4.74}
\Ps_a(\os r,t)=\int \frac{d^3\os p}{(2\pi)^{3/2}} \,e^{i\os p\cdot\os r}\\
{}\times \sum_{\z=\pm1}\Bigl[a_a^{(\z)} u_a^{(\z)}(\os p)
\exp\bigl(-iE(+)_a^{(\z)}t\bigr) + b_a^{(\z)}v_a^{(\z)}(\os p)
\exp\bigl(-i E(-)_a^{(\z)}t\bigr)\Bigr]
\end{multline}
where $a_a^{(\z)},b_a^{(\z)}$ are arbitrary \cf s, $u_a^{(\z)},v_a^{(\z)}$
are base spinors \st
\bea{4.75}
H_a u_a^{(\z)}&=&E(+)_a^{(\z)}u_a^{(\z)}\\
H_a v_a^{(\z)}&=&E(-)_a^{(\z)}v_a^{(\z)}.\label{4.76}
\end{eqnarray}
In the classical situation
\beq{4.77}
E(+)_a^{(\z)}=-E(-)_a^{(\z)}
\end{equation}
and $\z=\pm1$ describes different polarization states of the fermions $\Ps_a$
(see Refs \cite{1*}, \cite{**}). In our case $E(+)_a^{(+1)}$, $E(+)_a^{(-1)}$,
$E(-)_a^{(-1)}$, $E(-)_a^{(+1)}$ are roots of the polynomial of the fourth
order
\beq{4.78}
\det(H_a(\os p)-IE_a)=0, \q a=1,2,3,
\end{equation}
where $I$ is the identity matrix $4\times4$ and
\beq{4.79}
H_a(\os p)=c\os \a\cdot\os p+\b m_ac^2-d_{kk}\bigl(\os\Si\cdot\os E
+i\os\a\cdot\os\b\bigr), \q a=1,2,3.
\end{equation}

Spinors $u_a^{(\z)},v_a^{(\z)}$ are eigenvectors corresponding to those
eigenvalues. They are orthogonal.
Using formulae \eqref{4.35}--\eqref{4.37} one transforms Eqs
\eqref{4.78}--\eqref{4.79} into
\beq{4.80a}
H_a=\left(\begin{matrix}
m_ac^2I - d_{kk}(\os E\cdot\os \si) &&(c\os p-id_{kk}\os B)\cdot\os\si\\
(c\os p-id_{kk}\os B)\os\si &&d_{kk}(\os E\cdot\os \si)-m_ac^2I
\end{matrix}\right), \q a=1,2,3,
\end{equation}
and
\beq{4.81a}
\det\left(\begin{matrix}
(m_ac^2-E_a)I - d_{kk}(\os E\cdot\os \si) &&(c\os p-id_{kk}\os B)\cdot\os\si\\
(c\os p-id_{kk}\os B)\os\si &&d_{kk}(\os E\cdot\os \si)-(m_ac^2+E_a)I
\end{matrix}\right)=0, \q a=1,2,3,
\end{equation}
where $I$ is the $2\times2$ identity matrix.

Using explicit forms of Pauli matrices
\beq{4.82a}
\si_x=\left(\begin{matrix} 0 & 1 \\ 1 & 0 \end{matrix}\right)\q
\si_y=\left(\begin{matrix} 0 & -i \\ i & 0 \end{matrix}\right)\q
\si_z=\left(\begin{matrix} 1 & 0 \\ 0 & -1 \end{matrix}\right)
\end{equation}
one eventually gets
\beq{4.83a}
\hskip-6pt\det\left(\begin{matrix}
\gathered m_ac^2-E_a\\{}-d_{kk}E_z\endgathered &&
-d_{kk}(E_x-iE_y)&&
cp_z-id_{kk}B_z&&
\gathered c(p_x-ip_y)\\{}-id_{kk}(B_x-iB_y)\endgathered \\
\noalign{\vskip8pt}
-d_{kk}(E_x+iE_y)&&
\gathered m_ac^2-E_a\\{}+d_{kk}E_z\endgathered &&
\gathered c(p_x+ip_y)\\{}-id_{kk}(B_x+iB_y)\endgathered&&
-cp_z+id_{kk}B_z\\
\noalign{\vskip8pt}
cp_z-id_{kk}B_z&&
\gathered c(p_x-ip_y)\\{}-id_{kk}(B_x-iB_y)\endgathered&&
\gathered -m_ac^2-E_a\\{}+d_{kk}E_z\endgathered &&
d_{kk}(E_x-iE_y)\\
\noalign{\vskip8pt}
\gathered c(p_x+ip_y)\\{}-id_{kk}(B_x+iB_y)\endgathered&&
-cp_z+id_{kk}B_z&&
d_{kk}(E_x+iE_y)&&
\gathered -m_ac^2-E_a\\{}+d_{kk}E_z\endgathered
\end{matrix}\right)=0.
\end{equation}

Using initial conditions we can determine
\cf s $a_a^{(\z)}$ and~$b_a^{(\z)}$, i.e.\ we expand $\Ps_a^{(0)}(\os r)$
into Fourier integral
\beq{4.80}
\Ps_a^{(0)}(\os r)=\int \frac{d^3\os p}{(2\pi)^{3/2}}\,e^{i\os p\cdot\os r}
\sum_{\z=\pm1}\bigl[a_a^{(\z)}u_a^{(\z)}(\os p)+b_a^{(\z)}v_a^{(\z)}(\os p)
\bigr], \q a=1,2,3.
\end{equation}
We can consider several possibilities of neutrino flavour oscillations
supposing e.g.
\beq{4.81}
\Ps_\a^{(0)}(\os r)=\xi(\os r) \qh{and}
\Ps_\b^{(0)}(\os r)=\Ps_\g^{(0)}(\os r)=0.
\end{equation}
In this way
\beq{4.82}
\Ps_a^{(0)}(\os r)=U_{a\a}\xi(\os r)
\end{equation}
which can be considered as initial conditions for oscillations.

Moreover, this problem is beyond the scope of this paper and will be
considered elsewhere.

Let us notice that our generalization of a minimal coupling scheme
Eq.~\eqref{4.22} induces a new \cn\ on~$P$.
\bg{4.83}
\gd\check w{},A,B,=\hor(\gd\wt w,A,B,)\\
\qh{or} \gd\check w{},A,B,=\left(
\begin{array}{c|c}
\pi^\ast\bigl(\gd{\wt{\ov w}}{},\a,\b,)\, &\,
\frac\l2 \pi^\ast(\gd F,\a,\g,\ov \t{}^\g)\\[2pt]\hline
\noalign{\vskip2pt}
-\frac\l2\pi^\ast (F_{\b\g}\ov \t{}^\g) & 0
\end{array}
\right).\label{4.84}
\end{gather}
This \cn\ is metric but with non-vanishing torsion. Properties of this \cn\
have been extensively examined (also in the case of nonabelian Kaluza--Klein
Theories) in Ref.~\cite{0}.

Let us consider the following problem. What would it mean for Physics if
someone measured an  EDM for an electron of the value $d_{kk}=-\frac{4l_{\rm
pl}}{\sqrt\a}q$ as predicted in this section?
It would mean the fifth dimension is a reality in the sense of a 5-dimensional
Minkowski space.

An experiment which measures such a quantity strongly supports an idea of
rotations around the fifth axis in this space (the fifth dimension is a
space-like). This EDM exists only due to these rotations. Otherwise spinor
fields couple to a \cn\ \eqref{4.60} and there is not a new effect.

Even $P$ is a 5-\di al manifold, the additional fifth \di\ is not necessarily
of the same nature as the remaining four \di s, in particular three space \di
s. This \di\ is a gauge \di\ connected to the \elm c field. Moreover, we can
develop this theory using Yang--Mills' fields and also Higgs' fields using
\di al reduction procedure, expecting some additional effects. It means we
can expect something as ``travelling'' along additional \di s. This
perspective would have a tremendous importance for Physics and Technology.

Simultaneously an existence of an EDM of an electron has also very great
impact on our understanding of PC and~T symmetries breaking. This is also
very important.

Thus a mentioned measurement with an answer: \ti{Yes}, would have very
important physical, technological and even philosophical implications.

\section*{Conclusions}
In the paper we consider four problems:
\begin{enumerate}
\item Charge \cfn\ in the \NK{}.
\item Gravito-\elm c waves \so s in this theory.
\item An influence of a cosmological constant on a spherically-\s\ static \so.
\item Dirac \e s in \NK{}.
\end{enumerate}
There are some further prospects:
\begin{enumerate}
\item To find similar conditions for \cfn\ (of colour) in a nonabelian version in
the theory.
\item To find similar gravito--Yang--Mills waves.
\item To find spherical and cylindrical waves in the theory.
\end{enumerate}

Finally, we give some remarks. There are some misunderstandings connecting
Kaluza--Klein Theory, Einstein's Unified Field Theory, \eu\nos\ \eu\gr\
Theory (NGT), \eu\nos\ Kaluza--Klein Theory (NKKT), \eu\nos\ Jordan--Thiry
Theory (NJTT).

{\bf1.} First of all we comment a \ct\ $\la=\frac{2\sqrt{G_N}}{c^2}$. The \ct\ $\la$
appeared as a free parameter in this theory. Moreover in order to get
Einstein \e s with \elm c sources known from GR it is fixed and it is not free
any more. Why is there not a Planck's length?
I explain it shortly. The Kaluza theory is classical for a paper published by him
is classical as a classical paper in the scientific literature. It is also classical
for this theory is not quantum. For this we cannot get here a Planck's \ct.
This is simply for we need a Planck's \ct\ in order to construct the Planck's
length. \Pl's \ct\ is absent in Kaluza theory for this theory is classical
(non-quantum). The \Pl's length appeared in the further development done by
O.~Klein. O.~Klein considered a Klein--Gordon \e\ in 5-dimensional extension.
The \Pl's \ct\ is present in Klein--Gordon \e.
This \e\ can be considered as an \e\ for a classical scalar field. In
Kaluza--Klein theory \Pl's length appears as a scale of length.

\def\tr{transformation}
{\bf2.}
The classical Kaluza theory as a realistic unified field theory
has been abandoned by 1950's. Moreover,
due to some mathematical investigations a deep structure has been discovered
behind the theory. Let us describe it shortly. First of all it happens that
behind Maxwell theory of \elm sm there is a principal fibre bundle over a
\spt\ with a structural group $\U(1)$ and a \cn\ defined on this bundle is an
\elm c field. Gauge \tr, four-\pt, the first pair of Maxwell \e\ obtained a
clear geometrical meaning in terms of a fibre bundle approach (see Section~1
and Appendix~A).

It happens also that a classical Kaluza theory is a theory of metrized (in a
natural way) \elm c fibre bundle (see Ref.~\cite{97}).

This is a true unification of the two fundamental principles of invariance in
physics: a gauge invariance principle and a \cd\  invariance principle.

In Section 2 of Ref.~\cite{98} a classical KKT in this setting has been described
(see also the last two lines of page 576 with a fixing of the \ct~$\la$).

Moreover this paper is devoted to the KKT with torsion in such a way that we
put in the place of GR the Einstein--Cartan theory obtaining new features the
so-called ``interference effects'' between gravity and \elm sm going to some
effects which are small, moreover in principle measurable in experiment.
The \NK{} has been constructed using ideas and mathematical formalism similar
to those from Ref.~\cite{98}, i.e.\ to Kaluza--Klein Theory with torsion.

{\bf3.}
Let us consider Einstein Unified Field Theory. A.~Einstein started this
theory in 1920's. In 1950 he came back to this theory describing it in
Appendix~II of the fifth edition of his famous book \ti{The Meaning of
Relativity} (see Ref.~\cite{99}).

It is worth to mention that there are many versions of this theory. The
oldest Einstein--Thomas theory and after that Einstein--Strauss theory,
Einstein--Kaufmann theory. There are also two approaches, weak and strong
field \e s. The Einstein Unified Field Theory can be also considered as a
real theory and Hermitian theory. A slight deviation is the so-called
Bonnor's Unified Field Theory. In all of these approaches there are two
fundamental notions: \nos\ affine \cn\
$\gd\G,\la,\mu\nu,\ne\gd\G,\la,\nu\mu,$ and the \nos\ metric $g_{\mu\nu}
\ne g_{\nu\mu}$. \eu\cn\ and metric can be real or Hermitian. In this theory
there is also a second \cn\ $\gd W,\la,\mu\nu,\ne\gd W,\la,\nu\mu,$. \eu\cn\
$\gd\G,\la,\mu\nu,$ is a so-called constrained \cn, $\gd W,\la,\mu\nu,$ is called
unconstrained. All of these approaches have no free parameters. Some
parameters which appear in \so s of field \e s are \ti{integration \ct s}.

What was an aim to construct such theories? The aim was to find a unified
theory of gravity and \elm sm in such a way that GR and Maxwell theory appear
as some limit of the theory. This approach ended with fiasco. It was
impossible to obtain a Lorentz force.
It was impossible to obtain a Coulomb law too.

One can find all references to all versions of Einstein Unified Field Theory
in Refs \cite{1}, \cite{3}, \cite{4}, \cite{5} and we will not quote them here. Moreover, it
is worth to mention that A.~Einstein considered this theory as a theory of an
extended \gr. Moreover, there is a reference of A.~Einstein's idea to treat
this theory as a theory of an extended gravity only. A.~Einstein published a
paper on it in \ti{Scientific American} (the only one Einstein's paper in
this journal, see Ref.~\cite{Ein}).

Geometrical--mathematical properties of Einstein Unified Field Theory have
been described in a book by Vaclav Hlavat\'y (see Ref.~\cite{6d}).

In those times A.~Einstein started a program of geometrization of physics. Some
notions of this program have been described in Ref.~\cite{100}.

There is also an approach to this theory going in a different direction. It
has been summarized in the book by A.~H.~Klotz (see Ref.~\cite{101}).

{\bf 4.}
Let us comment NGT (\eu\nos\ \eu\gr al Theory) by J.~W.~Moffat.\break
J.~W.~Moffat
reinterpreted Einstein Unified Field Theory as a theory of a pure \gr al
field (see Ref.~\cite{5a}). He introduced material sources to the formalism.
Moreover, he introduced in his theory an additional universal \ct.
He and his co-workers developed this idea getting many interesting
results which are in principle testable by astronomical observations in the
Solar System and beyond. He was using both real and Hermitian theory. Simultaneously
he developed a formalism with two \cn s $\gd\G,\la,\mu\nu,$ and $\gd
W,\la,\mu\nu,$.

{\bf5.}
Let us comment the \NK{and the \eu\nos\ Jordan--Thiry }. I~posed and
developed these theories using the \nos\ metrization of an \elm c fibre
bundle using differential forms formalism as in Ref.~\cite{98}.

Early results concerning the \NK{} have been published in Refs \cite{102},
\cite{103}.

The final result of the theory with some developments has been published in
Ref.~\cite4. The paper contains also an extension to the \eu\nos\ Jordan--Thiry
Theory with a scalar field $\Ps$ (or~$\rho$). In order to get a pure \NK{} it
is enough to put ${\Ps=0}$ (or~${\rho=1}$). All new features as some
``interference effects'' between \elm c fields and \gr\ have been quoted in
Introduction of Ref.~\cite4. The theory has no free parameters except integration \ct s in
\so s.

It is possible to get an extension of the theory to the non-Abelian case. In
this case we have one free parameter. Moreover, this parameter can be fixed
by a \co ical \ct. The final version of this theory can be found in
Ref.~\cite3.  In Ref.~\cite1 one can find also an extension to the case with
Higgs' field and spontaneous symmetry breaking. In the last case there are
three free parameters which can be fixed by a \co ical \ct\ and scales of
masses.

{\advance\baselineskip by-0.35pt
I do not refer in my paper to the paper Ref.~\cite{104}, for the authors are using completely different approach (it is
better to say \ti{three approaches}). This approach is far away from
investigations in my work. Moreover, in future both approaches can meet and we
will shake hands. The only one point which is now common is a starting point,
a classical Kaluza Theory. We do not refer to Ref.~\cite{105}.
This paper deals with some problems
in NGT. However, NGT considered by them has only a little touch with NGT
considered here. They introduced a mass for skew-\s\ field
$B_{\mu\nu}=-B_{\nu\mu}$ (in our notation it is $g_\[\mu\nu]$). Moreover, the
$g_\[\mu\nu]$ can obtain a mass in a linear \ap ion of \eu\nos\ Non-Abelian
Kaluza--Klein Theory due to a \co ical \ct\ and it is not necessary to
introduce a mass term. It seems that this is a completely different
approach (see Ref.~\cite{105}). For a cure of NGT by a \co ical \ct\ see also
Ref.~\cite{106}.

Let us notice the following fact. Einstein's Unified Field Theory has been
abandoned as a realistic unified theory
for it has been proved using EIH (Einstein--Infeld--Hoffman) method
that there is not a Lorentz force term and Coulomb like law.

These are
disadvantages of Einstein Unified Field Theory but not NGT. This works now
for our advantage, for we do not see any term like Lorentz force and
Coulomb-like law in \gr al physics (I~do not mean a Newton \gr al law which
can be obtained in Einstein Unified Field Theory).
Someone said: ``{\it it is clever to use advantages, moreover, more clever is
to use disadvantages''} and this is a case. Moreover, in the \NK{} we get
Lorentz force term from $(N+4)$-dimensional (5-dimensional in an \elm c case)
geodetic \e s (see Refs~\cite1, \cite2, \cite3, \cite{4}).

All additional notions in the \NK{\JT\ } have been described in \cite1. We
get from $(N+4)$-\di al theory ($N=n+n_1$) four-\di al \e s due to an
invariance of a \nos\ metric and a \cn\ \wrt the right action of the group
(in the \elm c case this is a biinvariance of the action of a group $\U(1)$).

Let us notice also the following fact. \eu\e s obtained in the \NK{\JT\ } are
different from these in pure NGT. Due to this we can obtain nonsingular \so s
of field \e s in the \elm c case. These \so s possess a nonsingular metric
$g_\(\a\b)$ and nonsingular electric field. The asymptotic behaviour is as in
the case of Reissner--Nordstr\"om \so\ (see Refs \cite1, \cite4 and Section~3).
This is impossible to get in pure NGT.

{\bf 6.} In Section 2 we consider gravito-\elm c waves. They have nothing to
do with gravito-\elm sm in General Relativity. The notion of a wave is not so
easy as in Halliday's textbook. It is more general, see Ref.~\cite{Tr}.
A~wave carries an information. Roughly speaking, it can be modulated. It
means, it should possess an arbitrary \f\ of e.g.\ $(z-t)$. In the case of
nonlinear waves we use Riemann invariants (see e.g. Ref.~\cite{Kal}) and the
wave possesses an arbitrary \f\ of a Riemann invariant. Moreover, a
gravitational wave is more complicated, e.g.\ there is not a plane \gr al
wave. The \gr al field of such a wave is zero (the curvature tensor induced
by a metric describing a plane wave is zero). Moreover, we can consider
generalized plane waves (see Ref.~\cite{Zak}). \eu\gr al waves considered
in~\cite{Zak} are not only \gr al waves in a linear \ap ion. There are here
exact \so s of Einstein \e s which can describe a very strong \gr al field.
So we consider in Section~2 wave \so s of \NK{} in the sense of the mentioned
definition of a wave. Field \e s in the \NK{} describe \gr al and \elm c
fields. The \so s depend on arbitrary \f s of $(z-t)$. In the limit of zero
skewon and zero \elm c fields we get generalized plane waves known from the
book by Zakharov. Thus those \so s are gravito-\elm c waves. We use some
achievements in Einstein Unified Field Theory as results in a pure gravity
(see Refs \cite{7}--\cite{11}). The \elm c wave has remarkable properties to have
both invariants of an \elm c field $S=F_\m F^\m$, $P=F_\m F^{\ast\m}$ equal
to zero. The \elm c field for gravito-\elm c field wave has this property.

}\section*{Appendix A}
\def\theequation{A.\arabic{equation}}
\setcounter{equation}0
In the appendix we describe the notation and definitions of geometric
quantities used in the paper. We use a smooth principal bundle which is an
ordered sequence
\beq{A.0}
\ul P=(P,F,G,E,\pi),
\end{equation}
where $P$ is a total bundle manifold, $F$ is typical fibre, $G$, a Lie group,
is a structural group, $E$~is a base manifold and $\pi$ is a projection. In our
case $G=\U(1)$, $E$~is a \spt, $\pi:P\to E$.
We have a map $\vf:P\times G\to P$ defining an
action of~$G$ on~$P$. Let $a,b\in G$ and $\ve$~be a unit element of the
group~$G$, then $\vf(a)\circ \vf(b)=\vf(ba)$, $\vf(\ve)=\id$, where $\vf(a)p
=\vf(p,a)$. Moreover, $\pi\circ\vf(a)=\pi$. For any open set $U\subset E$ we
have a local trivialization $U\times G\simeq \pi^{-1}(U)$. For any $x\in E$,
$\pi^{-1}(\{x\})=F_x \simeq G$, $F_x$ is a fibre over~$x$ and is equal to~$F$.
In our case we suppose $G=F$, i.e.\ a Lie group $G$ is a typical fibre.
$\o$~is a 1-form
of \cn\ on~$P$ with values in the algebra of~$G$, $\mathfrak G$. In the case
of $G=\U(1)$ we use a notation $\a$ (an \elm c \cn). Lie algebra of $U(1)$
is~$R$. Let $\vf'(a)$ be a
tangent map to $\vf(a)$ whereas $\vf^\ast(a)$ is the contragradient
to~$\vf'(a)$ at a point~$a$. The form $\o$ is a form of ad-type, i.e.
\beq{A.1}
\vf^\ast(a)\o=\ad{a^{-1}}'\o,
\end{equation}
where $\ad{a^{-1}}'$ is a tangent map to the internal automorphism of the
group~$G$
\beq{A.2}
\ad a(b)=aba^{-1}.
\end{equation}
In the case of $\U(1)$ (abelian) the condition \eqref{A.1} means
\beq{A.3}
\mathop{\cL}_{\z_5}\a=0,
\end{equation}
where $\z_5$ is a Killing vector corresponding to one generator of the group
$\U(1)$. Thus this is a vector tangent to the operation of the group $\U(1)$
on~$P$, i.e.\ to $\vf_{\exp(i\chi)}$, $\chi=\chi(x)$, $x\in E$,
$\mathop{\cL}\limits_{\z_5}$ is a Lie \dv\ along $\z_5$.
We may introduce the distribution (field) of linear elements $H_r$, $r\in P$,
where $H_r\subset T_r(P)$ is a subspace of the space tangent to~$P$ at a
point~$r$ and
\beq{A.4}
v\in H_r \iff \o_r(v)=0.
\end{equation}
So
\beq{A.5}
T_r(P)=V_r\oplus H_r,
\end{equation}
where $H_r$ is called a subspace of \ti{horizontal\/} vectors and $V_r$ of
\ti{vertical\/} vectors.
For vertical vectors $v\in V_r$ we have $\pi'(v)=0$. This means that $v$ is
tangent to the fibres.

Let
\beq{A.6}
v=\hor(v)+\ver(v),\quad \hor(v)\in H,\ \ver(v)\in V_r.
\end{equation}
It is proved that the distribution $H_r$ is equal to choosing a \cn~$\o$. We
use the operation $\hor$ for forms, i.e.
\beq{A.7}
(\hor\b)(X,Y)=\b(\hor X,\hor Y),
\end{equation}
where $X,Y\in T(P)$.

The 2-form of a curvature is defined as follows
\beq{A.8}
\O=\hor d\o=D\o,
\end{equation}
where $D$ means an exterior covariant \dv\ \wrt $\o$. This form is also of
ad-type.

For $\O$ the structural Cartant \e\ is valid
\beq{A.9}
\O=d\o+\tfrac12[\o,\o],
\end{equation}
where
\beq{A.10}
[\o,\o](X,Y)=[\o(X),\o(Y)].
\end{equation}
Bianchi's identity for $\o$ is as follows
\beq{A.11}
D\O=\hor d\O=0.
\end{equation}
The map $f:E\supset U\to P$ such that $f\circ \pi=\id$ is called a
\ti{section} ($U$ is an open set).

From physical point of view it means choosing a gauge. A~covariant \dv\
on~$P$ is defined as follows
\beq{A.12}
D\Ps=\hor d\Ps.
\end{equation}
This \dv\ is called a \ti{gauge \dv}. $\Ps$ can be a spinor field on~$P$.

In this paper we use also a linear \cn\ on manifolds $E$ and $P$, using the
formalism of differential forms. So the basic quantity is a one-form of the
\cn\ $\gd\o,A,B,$. The 2-form of curvature is as follows
\beq{A.13}
\gd\O,A,B,=d\gd\o,A,B,+\gd\o,A,C, \wedge \gd\o,C,B,
\end{equation}
and the two-form of torsion is
\beq{A.14}
\T^A=D\t^A,
\end{equation}
where $\t^A$ are basic forms and $D$ means exterior covariant \dv\ \wrt \cn\
$\gd\o,A,B,$. The following relations are established \cn s with generally met
symbols
\beq{A.15}
\bal
\gd\o,A,B,&=\gd\G,A,BC,\t^C\\
\T^A&=\tfrac12\gd Q,A,BC,\t^B\wedge \t^C\\
\gd Q,A,BC,&=\gd\G,A,BC,-\gd\G,A,CB,\\
\gd\O,A,B,&=\tfrac12 \gd R,A,BCD,\t^C \wedge \t^D,
\eal
\end{equation}
where $\gd\G,A,BC,$ are \cf s of \cn\ (they do not have to be \s\ in indices
$B$ and~$C$), $\gd R,A,BCD,$ is a tensor of a curvature, $\gd Q,A,BC,$ is a
tensor of a torsion in a holonomic frame. Covariant exterior derivation \wrt $\gd\o,A,B,$ is given
by the formula
\beq{A.16}
\bal
D\Xi^A&=d\Xi^A+\gd\o,A,C,\wedge \Xi^C\\
D\gd\Si,A,B,&=
d\gd\Si,A,B,+\gd\o,A,C,\wedge \gd\Si,C,B,-\gd\o,C,B,\wedge \gd\Si,A,C,.
\eal
\end{equation}
The forms of a curvature $\gd\O,A,B,$ and torsion $\T^A$ obey Bianchi's
identities
\beq{A.17}
\bal
{}&D\gd\O,A,B,=0\\
&D\T^A=\gd\O,A,B,\wedge \t^B.
\eal
\end{equation}
All quantities introduced here can be found in Ref.~\cite{KN}.

In this paper we use a formalism of a fibre bundle over a \spt~$E$ with an
\elm c \cn~$\a$ and traditional formalism of differential geometry for linear
\cn s on~$E$ and~$P$. In order to simplify the notation we do not use fibre
bundle formalism of frames over $E$ and~$P$. A~vocabulary connected geometrical
quantities and gauge fields (Yang--Mills fields) can be found in
Ref.~\cite{97}.

In Ref.~\cite{Wu} we have also a similar vocabulary (see Table~I, Translation
of terminology). Moreover, we consider a little different terminology. First
of all we distinguished between a gauge \pt\ and a \cn\ on a fibre bundle. In
our terminology a gauge \pt\ $A_\mu \ov\t{}^\mu$ is in a particular gauge $e$
(a~section of a bundle), i.e.
\beq{A.18}
A_\mu \ov\t{}^\mu=e^\ast\o
\end{equation}
where $A_\mu \ov\t{}^\mu$ is a 1-form defined on $E$ with values in a Lie
algebra $\mathfrak G$ of~$G$. In the case of a strength of a gauge field we have
similarly
\beq{A.19}
\tfrac12 F_\m \ov\t{}^\mu \wedge \ov\t{}^\nu=e^\ast\O
\end{equation}
where $F_\m \ov\t{}^\mu \wedge \ov\t{}^\nu$ is a 2-form defined on~$E$ with
values in a Lie algebra $\mathfrak G$ of~$G$.

Using generators of a Lie algebra $\mathfrak G$ of $G$ we get
\beq{A.20}
A=\gd A,a,\mu, \ov\t{}^\mu X_a=e^\ast \o \quad\hbox{and}\quad
F=\tfrac12\gd F,a,\m,\ov\t{}^\mu \wedge \ov\t{}^\nu X_a=e^\ast \O
\end{equation}
where
\beq{A.21}
[X_a,X_b]=\gd C,c,ab,X_c, \quad a,b,c=1,2,\dots,n, \ n=\dim G(=\dim \mathfrak G),
\end{equation}
are generators of $\mathfrak G$, $\gd C,c,ab,$ are structure \ct s of a Lie
algebra of~$G$, $\mathfrak G$, $[\cdot,\cdot]$ is a commutator of Lie algebra
elements.

In this paper we are using Latin lower case letters for 3-\di al space indices. Here
we are using Latin lower case letters as Lie algebra indices. It does not
result in any misunderstanding.
\beq{A.22}
\gd F,a,\m,=\pa_\mu\gd A,a,\nu,-\pa_\nu\gd A,a,\mu,+\gd C,a,bc,\gd A,b,\mu,
\gd A,c,\nu,.
\end{equation}
In the case of an \elm c \cn\ $\a$ the field strength~$F$ does not depend on
gauge (i.e.\ on a section of a~bundle).

Finally it is convenient to connect our approach using gauge \pt s $\gd
A,a,\mu,$ with usually met (see Ref.~\cite{Pok}) matrix valued gauge
quantities $A_\mu$ and $F_\m$. It is easy to see how to do it if we consider
Lie algebra generators $X_a$ as matrices. Usually one supposes that $X_a$ are
matrices of an adjoint representation of a Lie algebra~$\mathfrak G$, $T^a$
with a normalization condition
\beq{A.23}
{\rm Tr}(\{T^a,T^b\})=2\d^{ab},
\end{equation}
where $\{\cdot,\cdot\}$ means anticommutator in an adjoint representation.

In this way
\bea{A.24}
A_\mu&=&\gd A,a,\mu, T^a,\\
F_\m&=&\gd F,a,\m, T^a. \label{A.25}
\end{eqnarray}
One can easily see that if we take
\beq{A.26}
F_\m=\pa_\mu A_\nu - \pa_\nu A_\mu + [A_\mu,A_\nu]
\end{equation}
from Ref.~\cite{Pok} we get
\beq{A.26a}
F_\m=(\gd F,a,\m,)T^a,
\end{equation}
where $\gd F,a,\m,$ is given by \eqref{A.22}. From the other side if we take
a section $f$, $f:U\to P$, $U\subset E$, and corresponding to it
\bea{A.27}
\ov A=\gd\ov A,a,\mu, \ov\t{}^\mu X_a&=&f^\ast \o\\
\ov F=\tfrac12\gd\ov F,a,\m, \ov\t{}^\mu \wedge \ov\t{}^\nu X_a&=&f^\ast \O
\label{A.28}
\end{eqnarray}
and consider both sections $e$ and $f$ we get transformation from $\gd
A,a,\mu,$ to $\gd\ov A{},a,\mu,$ and from $\gd F,a,\m,$ to $\gd\ov F{},a,\m,$ in
the following way. For every $x\in U\subset E$ there is an element $g(x)\in
G$ such that
\beq{A.29}
f(x)=e(x)g(x)=\vf(e(x),g(x)).
\end{equation}
Due to \eqref{A.1} one gets
\bea{A.30}
\ov A(x)&=&\ad{g^{-1}(x)}'A(x)+{g^{-1}(x)}\,dg(x)\\
\ov F(x)&=&\ad{g^{-1}(x)}'F(x) \label{A.31}
\end{eqnarray}
where $\ov A(x),\ov F(x)$ are defined by \eqref{A.27}--\eqref{A.28} and
$A(x),F(x)$ by \eqref{A.20}. The formulae \eqref{A.30}--\eqref{A.31} give a
geometrical meaning of a gauge transformation (see Ref.~\cite{97}). In an
\elm c case $G=\U(1)$ we have similarly, if we change a local
section from $e$ to~$f$ we get
$$
f(x)=\vf(e(x), \exp(i\chi(x)))  \quad (f:U\supset E\to P)
$$
and $\ov A=A+d\chi$.

Moreover,
in the traditional approach (see Ref.~\cite{Pok}) one gets
\bea{A.32}
\ov A_\mu(x)&=&U(x)^{-1}A_\mu(x)U(x)+U^{-1}(x)\pa_\mu U(x)\\
\ov F_\m(x)&=&U^{-1}(x)F_\m U(x), \label{A.33}
\end{eqnarray}
where $U(x)$ is the matrix of an adjoint representation of a Lie group $G$.

For an action of a group $G$ on $P$ is via \eqref{A.1}, $g(x)$ is exactly a
matrix of an adjoint representation of~$G$. In this way
\eqref{A.30}--\eqref{A.31} and \eqref{A.32}--\eqref{A.33} are equivalent.

Let us notice that usually a Lagrangian of a gauge field (Yang--Mills field)
is written as
\beq{A.34}
\cL_{\rm YM} \sim {\rm Tr}(F_\m F^\m)
\end{equation}
where $F_\m$ is given by \eqref{A.25}--\eqref{A.26}. It is easy to see that
one gets
\beq{A.35}
\cL_{\rm YM} \sim h_{ab}\gd F,a,\m, F^{b\m}
\end{equation}
where
\beq{A.36}
h_{ab}=\gd C,d,ac, \gd C,c,bd,
\end{equation}
is a Cartan--Killing tensor for a Lie algebra $\mathfrak G$, if we remember
that $X_a$ in adjoint representation are given by structure \ct s $\gd
C,c,ab,$.

Moreover, in Refs \cite{1,3} we use the notation
\beq{A.39}
\O=\tfrac12 \gd H,a,\m,\t^\mu \wedge \t^\nu X_a.
\end{equation}
In this language
\beq{A.40}
\cL_{\rm YM}=\tfrac1{8\pi} h_{ab}\gd H,a,\m, H^{b\m}.
\end{equation}
It is easy to see that
\beq{A.41}
e^\ast (\gd H,a,\m,\t^\mu \wedge \t^\nu X_a)=\gd F,a,\m,{\ov\t}^\mu
\wedge {\ov \t}^\nu X_a.
\end{equation}
Thus \eqref{A.40} is equivalent to \eqref{A.35} and to \eqref{A.34}.
\eqref{A.34} is invariant to a change of a gauge. \eqref{A.40} is invariant
\wrt the action of a group~$G$ on~$P$.

Let us notice that $h_{ab}\gd F,a,\m, F^{b\m}=h_{ab}\gd H,a,\m, \gd H,b,\m,$,
even $\gd H,a,\m,$ is defined on~$P$ and $\gd F,a,\m,$ on~$E$. In the
non-abelian case it is more natural to use $\gd H,a,\m,$ in place of $\gd
F,a,\m,$.

\section*{Appendix B}
\def\theequation{B.\arabic{equation}}
\setcounter{equation}0

In this appendix we find a formula for $H_{\nu\mu}$ from Eq.~\eqref{1.41}.
In order to do this let us solve this \e\ perturbatively. According to
Ref.~\cite4 and Ref.~\cite{KM} one gets
\beq{B.1}
H_{\a\b}=\nad{(0)}H_{\a\b}+\d\nad{(1)}H_{\a\b}+\d\nad{(2)}H_{\a\b}+\ldots
\end{equation}
where $\nad{(0)}H_{\a\b}$ is $H_{\a\b}$ in zero order of expansion \wrt
$h_{\a\b}$ where
\beq{B.2}
g_{\a\b}=\eta_{\a\b}+h_{\a\b}=\eta_{\a\b}+h_{(\a\b)}+h_{[\a\b]}
=\eta_{\a\b}+h_{(\a\b)}+g_{[\a\b]}
\end{equation}
and $\d\nadd{(k)}H_{\a\b}$ is a $k$-th correction to $\nad{(0)}H_{\a\b}$. One
gets
\beq{B.3}
g^{\mu\si}g_{\nu\si}=\bigl(\eta^{\mu\si}+\d\nad{(1)}h{}^{\mu\si}
+\d\nad{(2)}h{}^{\mu\si}+\ldots \bigr)(\eta_{\nu\si}+h_{\nu\si})=\d^\mu_\nu.
\end{equation}
From \eqref{B.3} one gets
\bea{B.4}
\d \nad{(1)}h{}^\m&=&-\eta^{\mu\si}\eta^{\nu\b}h_{\b\si}\\
\d \nad{(2)}h{}^\m&=&-\eta^{\nu\b}\d\nad{(1)}h{}^{\mu\si}h_{\b\si}
=\eta^{\nu\b}\eta^{\mu\g}\eta^{\a\si}h_{\si\g}h_{\b\a} \label{B.5}
\end{eqnarray}
where $\d\nadd{(k)}h{}^\m$ are $k$-th corrections to $\eta^\m$ (a zero order
of an inverse tensor of $g_\m$, $\eta^\m$ is an inverse Minkowski tensor). We
get
\beq{B.6}
g^\m=\eta^\m-\eta^{\mu\si}\eta^{\nu\b}h_{\b\si}+\eta^{\mu\g}\eta^{\nu\b}
\eta^{\a\si}h_{\b\a}h_{\si\g}.
\end{equation}
Eq.~\eqref{1.16} can be rewritten in a more convenient form
\beq{B.7}
H_{\b\si}-g^{\g\d}\bigl[g_{[\b\d]}H_{\g\si}+g_\[\g\si]H_{\b\d}\bigr]
=F_{\b\si}-2g_\[\d\si]g^{\d\g}F_{\b\g}.
\end{equation}
Using Eq.~\eqref{B.7} and writing
\bea{B.8}
H_{\b\a}&=&A_{\a\b}+B_{\a\b}\\
A_{\a\b}&=&A_{\b\a},\quad B_{\a\b}=-B_{\b\a} \label{B.9}
\end{eqnarray}
we can easily prove that $A_{\a\b}=0$. This means that $H_{\a\b}=-H_{\b\a}$
if $F_{\a\b}=-F_{\b\a}$. Using Eqs \eqref{B.6} and \eqref{B.7} one gets
\beq{B.10}
\bal
\nad{(0)}H_{\a\b}&=F_{\a\b}\\
\d\nad{(1)}H_{\b\si}&=\eta^{\g\d}(h_\[\b\d]F_{\g\si}-h_\[\a\d]F_{\g\b})\\
\d\nad{(2)}H_{\b\si}&=\eta^{\g\d}\eta^{\rho\a}(h_\(\rho\g)(h_\[\si\d]
F_{\a\b}-h_\[\b\d]F_{\si\a}))
\eal
\end{equation}
and eventually
\beq{B.11}
\nad{(2)}H_{\b\a}=F_{\b\a}+(\eta^{\g\d}-h^\(\g\d))(h_\[\b\d]F_{\g\a}
-h_\[\a\d]F_{\g\b}),
\end{equation}
where
\beq{B.12}
h^\(\g\d)=\eta^{\a\d}\eta^{\b\g}h_\(\a\b).
\end{equation}
Eq.~\eqref{B.11} can be rewritten in the form
\beq{B.13}
\nad{(2)}H_{\nu\mu}=F_{\nu\mu}-\nad{(1)}{\wt g}{}^\(\tau\a)(g_\[\mu\tau]
F_{\a\nu}-g_\[\nu\tau]F_{\a\mu})
\end{equation}
where $\nad{(1)}g{}^\(\tau\a)$ is an inverse tensor for $g_\(\a\b)$ up to the
first order of expansion \wrt $h_\(\a\b)$. One can easily generalize this \e\
to any order~$k$ getting
\beq{B.14}
\nadd{(k)}H_{\nu\mu}=F_{\nu\mu}-\nadd{(k-1)}{\wt g}{}^\(\tau \a)
(g_\[\mu\tau]F_{\a\nu}-g_\[\nu\tau]F_{\a\mu}).
\end{equation}
Taking $k\to\infty$ we get formula \eqref{1.41}, where $\nadd{(\infty)}H_\m
=H_\m$, $\nadd{(\infty)}{\wt g}_\(\a\b)=\wt g_\(\a\b)$.

Let us get Eq.~\eqref{1.41} from a general formula in $n$-\di al
generalization of Einstein Unified Field Theory obtained by Hlavat\'y and
Wrede (see Refs \cite{6d} and \cite{Wr}). One gets
\bml{B.15}
\gd\G,N,WM,=\gd\wt\G{},N,WM,+\tfrac12\bigl(\dg K,WM,N, -2\dg k,[M\cdot,A,
K_{W]AB}k^{NB}\bigr)\\ +h^{NE}\bigl\{\dg K,E(W\cdot,A, k_{M)A}+
\dg k,C\cdot,B, \bigl[\dg k,(M\cdot,C, K_{W)AB}\dg k,E\cdot,A,
-K_{EAB}\dg k,(W\cdot,A, \dg k,M)\cdot,C,\bigr]\bigr\}
\end{multline}
where
\bea{B.16}
\g_{AB}&=&h_{AB}+k_{AB}\\
h_{AB}&=&h_{BA}, \quad k_{AB}=-k_{BA} \label{B.17}\\
K_{ABC}&=&-\wt\nabla_A k_{BC}-\wt\nabla_Bk_{CA}+\wt\nabla_C k_{AB},\label{B.18}
\end{eqnarray}
$\gd \wt\G{},N,WM,$ is a Levi--Civita \cn\ generated by $h_{AB}=\g_\(AB)$
($\g_\[AB]=k_{AB}$). $\wt\nabla_A$ is a covariant \dv\ \wrt the \cn\ $\gd
\wt\G{}, N,WM,$.

The \cn\ $\gd\G,N,WM,$ is a \so\ of the \e
\beq{B.19}
D\g_{A+B-}=D\g_{AB}-\g_{AD}\gd Q,D,BC,(\G)\t^C=0, \quad
A,B,C,D,N,M=1,2,\dots,n,
\end{equation}
where $D$ is an exterior covariant \dv\ \wrt a \cn~$\G$.
\beq{B.20}
h^{AB}h_{BC}=\gd \d,A,C,
\end{equation}
and all indices are raised by $h^{AB}$ (E.~Schr\"odinger was surprised that
it was possible to find a \so\ to \eqref{B.19} in a covariant form).
The formula \eqref{B.15} is more general than that from Refs \cite{6d,Wr}
for in Eq.~\eqref{B.15} $\gd\wt\G{},N,WM,$ are \cf s of a Levi--Civita \cn.
This \cn\ can be considered in nonholonomic frame. Thus $\gd\wt\G{},N,WM,$
can be \nos\ in indices $W$ and~$M$. In Refs \cite{6d,Wr} $\gd\wt\G{},N,WM,$
mean Christoffel symbols. Moreover, the proof is exactly the same as in Refs
\cite{6d,Wr}. The authors of \cite{6d,Wr} are using a natural nonholonomic
frame connected to the \nos\ tensor $\g_{AB}$ in order to find formula
\eqref{B.15}. Moreover, this nonholonomic frame has nothing to do with the
frame we consider. They are supposing $\det(\g_{AB})\ne0$ and
$\det(\g_{(AB)})\ne0$, which is equivalent to our assumptions \eqref{1.5} and
\eqref{1.6} (in the case $n=5$ for $\g_{AB}$ given by Eq.~\eqref{1.13}).
Let us notice there is not any constraint imposed on a torsion
of the \cn.

V.~Hlavat\'y and C.~R. Wrede were first to consider $n$-\di al generalization
of a geometry from Einstein Unified Field Theory with \nos\ real tensor
$\g_{AB}$.

Here we are using capital Latin indices as indices of many-\di al manifolds.
This does not result in any misunderstanding. In general, in non-Abelian theory,
even in the case with spontaneous symmetry breaking, $n=4+N+N_1$, where $N$
is the \di\ of a Lie group and $N_1$ is the \di\ of a homogeneous space (see
Refs \cite{1,3,5}).

In our case we have $n=5$ and $\g_{AB}$ is given by Eq.~\eqref{1.13}.
It is easy to see that
\beq{B.21}
\gd\G,5,\m,=H_\m
\end{equation}
(in a lift horizontal basis, unholonomic frame). Thus  it is enough to
calculate $\gd\G,5,\m,$. One gets
\beq{B.22}
\gd\G,5,\o\mu,=\gd{\wt\G}{},5,\o\mu, - \tfrac12 K_{\o\mu5}-\tfrac14
\bigl\{\dg K,5\o\cdot,\a, k_{\mu\a}+\dg K,5\mu\cdot,\a,k_{\o\a}-\dg k,\g\cdot,
\b,k_{\a\b}\dg k,\o\cdot,\a,\dg k,\mu\cdot,\g, -
\dg k,\g\cdot,\b, K_{5\a\b}\dg k,\mu\cdot,\a,\dg k,\o\cdot,\g,\bigr\}
\end{equation}
where all indices are raised by $h^{\a\b}$
\bg{B.23}
h^{\a\b}h_{\a\g}=\gd \d,\b,\g,\\
g_{\a\b}=h_{\a\b}+k_{\a\b}\label{B.24}\\
h_{\a\b}=g_\(\a\b), \quad k_{\a\b}=g_\[\a\b].\nonumber
\end{gather}
We are keeping notation from Refs \cite{6d,Wr}.

Moreover in Ref.~\cite{98} (see Eq.~\eqref{2.15}) Levi--Civita \cn\ \cf s for
$\gd\wt\G{},N,WN,$ generated by $\g_\(AB)=h_{AB}$ are calculated. One has
$\gd\wt\G{},\a,\b5,=\gd F,\a,\b,$, $\gd\wt\G{},5,\b\g,=F_{\b\g}$, $\gd\wt\G{},\b,
5\g,=\gd F,\b,\g,$, $\gd\wt\G{},\a,\b\g,=\gd\wt{\ov\G}{},\a,\b\g,$ for $\la=2$
($n=5$ in Kaluza--Klein Theory). It
is easy to see that they are not \s\ in indices (they are not Christoffel
symbols for a frame is not holonomic, it is a lift horizontal basis).
The remaining \cf s are zero.

Using these results one gets
\bea{B.25a}
\wt\nabla_\mu k_{\mu5}&=&-\gd F,\tau,\o, k_{\mu\tau}=-\wt\nabla_\o
k_{5\mu}\\
\wt\nabla_5 k_{\o\mu}&=&-\gd F,\tau,\mu, k_{\o\tau}-\gd F,\tau,\o, k_{\tau\mu}
\label{B.25b}\\
\wt\nabla_\mu k_{\tau\o}&=&\wt{\ov\nabla}_\mu k_{\tau\o} \label{B.25c}
\end{eqnarray}
where we use the fact that
\bg{B.25d}
k_{\mu5}=k_{55}=k_{5\mu}=0 \\
\pa_5 k_{\o\mu}=0. \label{B.25e}
\end{gather}
Eventually we get
\bea{B.25}
K_{\o\mu 5}&=&2(\gd F,\tau,\o,k_{\mu\tau}-\gd F,\tau,\mu, k_{\o\tau})\\
K_{5\a\b}&=&2(\gd F,\tau,\o,k_{\mu\tau}-\gd F,\tau,\mu, k_{\tau\o}) \label{B.26}\\
K_{\o5\mu}&=&2\gd F,\tau,\mu, k_{\tau\o}.\label{B.27}
\end{eqnarray}
The remaining $K_{ABC}$ are zero. One gets
\beq{B.28}
H_{\o\mu}=F_{\o\mu}-\gd F,\tau,\o, k_{\mu\tau}+\gd F,\tau,\mu, k_{\o\tau}.
\end{equation}
Coming back to our notation from the paper (i.e.\ $k_\m=g_\[\m]$, $h^{\tau\a}
=\wt g{}^\(\tau\a)$) we get
\beq{B.29}
H_{\o\mu}=F_{\o\mu}-\wt g{}^\(\tau\a)F_{\a\o}g_\[\mu\tau]+\wt g{}^\(\tau\a)
F_{\a\mu}g_\[\o\tau],
\end{equation}
i.e.\ Eq.~\eqref{1.41}. In this way we have a consistency in our theory
getting the same results from both methods.

\section*{Appendix C}
\def\theequation{C.\arabic{equation}}
\setcounter{equation}0
\def\rp{representation}
\def\SL{{\rm SL}}

In this paper we consider two kinds of spinor fields $\Ps,\ov\Ps$ and
$\psi,\ov\psi$ defined respectively on~$P$ and~$E$. Spinor fields $\Ps$
and~$\ov\Ps$ transform according to $\Spin(1,4)$ and $\psi,\ov\psi$ according
to $\Spin(1,3)\simeq{\rm SL}(2,\C)$. We have
\beq{C.1}
U(g)\Psi(X)=D^F(g)\Psi(g^{-1}X), \q X\in M^{(1,4)}, \ g\in\SO(1,4).
\end{equation}
$\SO(1,4)$ acts linearly in $M^{(1,4)}$ (5-\di al Minkowski space). The
Lorentz group $\SO(1,3)\subset\SO(1,4)$. $D^F$ is a representation of
$\SO(1,4)$ (de~Sitter group) \st after a restriction to its subgroup
$\SO(1,3)$ we get
\beq{C.2}
D^F{}_{|\SO(1,3)}(\La)=L(\La),
\end{equation}
where
\beq{C.3}
L(\La)=D^{(1/2,0)}(\La)\oplus D^{(0,1/2)}(\La)
\end{equation}
is a Dirac representation of $\SO(1,3)$. More precisely, we deal with \rp s
of $\Spin(1,4)$ and $\Spin(1,3)\simeq \SL(2,\C)$ (see Ref.~\cite{52}). In
other words, we want spinor fields $\Ps$ and~$\ov\Ps$ to transform according
to such a \rp\ of $\Spin(1,4)$ which is induced by a Dirac \rp\
of~$\SL(2,\C)$. The complex \di s of both \rp s are the same:~4. The same are
also Clifford algebras
\beq{C.4}
C(1,4)\simeq C(1,3)
\end{equation}
(see Refs \cite{53}, \cite{54}).

One gets (up to a phase)
\beq{C.5}
\Ps_{|\SL(2,\C)}=\psi.
\end{equation}
Spinor fields $\psi$ and $\ov\psi$ transform according to Dirac \rp,
$\ov\psi=\psi^+B$. Our matrices $\g_\mu$ and $\g_A$ are \rp s of $C(1,3)$
($C(1,4)$).  One can consider projective \rp s for $\Ps$ and~$\psi$, i.e.\
\rp s of $\Spin(1,3)\ot \U(1)$ and $\SL(2,\C)\ot \U(1)$. Moreover, we do not
develop this idea here.

In this paper we develop the following approach to spinor fields on~$E$ and
on~$P$. We introduce orthonormal frames on~$E$ ($dx^1$, $dx^2$, $dx^3$,
$dx^4$) and on~$P$ ($dX^1=\pi^\ast(dx^1)$, $dX^2=\pi^\ast(dx^2)$,
$dX^3=\pi^\ast(dx^3)$, $dX^4=\pi^\ast(dx^4)$, $dX^5$). Our spinors $\Ps$
on $(P,\g_\(AB))$ and $\psi$ on $(E,g_\(\a\b))$ are defined as complex bundles
$\C^4$ over~$P$ or~$E$ with homomorphisms $\rho:C(1,4)\to \cL(\C^4)$ (resp.\
$\rho:C(1,3)\to \cL(\C^4)$) of bundles of algebras over~$P$ (resp.~$E$) \st
for every $p\in P$ (resp.~$x\in E$), the restriction of~$\rho$ to the fiber
over~$p$ (resp.~$x$) is equivalent to spinor \rp\ of a Clifford algebra
$C(1,4)$ (resp.~$C(1,3)$), i.e.\ $D^F$ (resp.\ Dirac \rp, see Refs
\cite{55},~\cite{56}). (There is also a paper on a similar subject (see
Ref.~\cite{57}).) Spinor fields $\Ps$ and~$\psi$ are sections of these
bundles. There is also an approach to consider spinor bundles for~$\Ps$
and~$\psi$ as bundles associated to principal bundles of orthonormal frames
for $(P,\g_\(AB))$ or $(E,g_\(\a\b))$ (spin frames). Spinor fields $\Ps$ and
$\psi$ are sections of these bundles. In our case we consider spinor fields
$\Ps$ and~$\ov\PS$ transforming according to \eqref{4.13} and \eqref{4.14}.
In the case of $\psi$ and $\ov\psi$ we have
\beq{C.6}
\bga
\ov\t{}^{\a\prime}=\ov\t{}^\a+\d\ov\t{}^\a=\ov\t{}^\a-\gd\ve,\a,\b,
\ov\t{}^\b\\
\ov\ve_{\a\b}+\ov\ve_{\b\a}=0.
\ega
\end{equation}

If the spinor field $\psi$ corresponds to $\ov\t{}^\a$ and $\psi'$ to
$\ov\t{}^{\a\prime}$ we get
\beq{C.7}
\bal
\psi'&=\psi+\d\psi=\psi-\ov\ve{}^{\a\b}\si_{\a\b}\psi\\
\ov\psi{}'&=\ov\psi+\d\ov\psi=\ov\psi+\ov\psi\ov\ve{}^{\a\b}\si_{\a\b}.
\eal
\end{equation}
Spinor fields $\Ps$ and $\ov\Ps$ are $\psi$ and $\ov\psi$ in any section of a
bundle~$P$. Simultaneously we suppose conditions \eqref{4.2}.

Similarly as for $\Ps,\ov\PS$ one gets
\bg{C.8}
\bal
\wt{\ov D}\psi&=d\psi + \gd\wt{\ov w}{},\a,\b, \dg\si,\a,\b, \psi\\
\wt{\ov D}\,\ov\psi&=d\ov\psi - \gd\wt{\ov w}{},\a,\b, \ov\psi \dg\si,\a,\b,
\eal\\
\bal
\wt{\ov\cD}\psi=\hor\wt{\ov D}\psi=\gdv\psi
+ \gd\wt{\ov w}{},\a,\b,\dg\si,\a,\b,\psi\\
\wt{\ov\cD}\,\ov\psi=\hor\wt{\ov D}\,\ov\psi=\gdv\ov\psi
- \gd\wt{\ov w}{},\a,\b,\ov\psi\dg\si,\a,\b,.
\eal \label{C.9}
\end{gather}

\section*{Appendix D}
\def\theequation{D.\arabic{equation}}
\setcounter{equation}0
\def\tv#1{\wt{\ov#1}{}}
\newdimen\krop
\setbox0=\hbox{$\scriptstyle\a$}
\krop=\wd0
\def\cdt{\hbox to\krop{$\scriptstyle\hfil\cdot\hfil$}}
\def\lw#1 {\lower#1pt\hbox\bgroup$\scriptstyle}
\let\hp\hphantom
\def\eg{$\egroup}
\let\na\nabla
\def\LC{Levi-Civita}

In this paper we proceed a unification and geometrization of \gr al and \elm
c interactions. Moreover, there is an approach (see Refs \cite{67},
\cite{68}, \cite{69}, \cite{70}, \cite{71}, \cite{72}) which is going in a
different direction. In that direction, the authors of those papers are
transforming all possible alternative theories of \gr\ described by some
geometric notions, i.e.\ \cn s, torsions, metric tensors and also with
nonstandard Lagrangians, i.e.\ nonlinear Lagrangians
($f(R)$, $f(R^\m R_\m)$, $f\X1((\gd R,\a,\b\m,
\ve^{\m\l\rho}\gd R,\b,\a\l\rho,)^2\Y1)$ etc.) to GR with additional
``matter fields''. In our notation $f$ means an arbitrary \f, $R$ is a scalar
curvature, $R_\m$ is a Ricci tensor, $\gd R,\a,\b\m,$ means a curvature
tensor (no necessary Riemann--Christoffel tensor). One can consider also some
different Lagrangians, i.e.\ $f(g_\m,R_\m)$. In this way some unified theories,
even Einstein Unified Field Theory, can be transformed into GR (General
Relativity) plus some additional ``matter fields'', i.e.\ scalar fields,
vector fields and so on. This is possible of course by using Legendre \tf\
techniques to define a new metric (\s) tensor and a Levi-Civita \cn\
compatible with this tensor.

The interpretation of this new tensor and a new \cn\ can be complex. They
could not have clear physical interpretation. In the case of \NK{} considered
in this paper we proceed in a little different way.

Let us consider our \nos\ \cn\ $\gd\ov W{},\l,\mu,$ and $\gd\ov w{},\l,\mu,$ on
a \spt~$E$. Using results from Ref.~\cite{6d} we can write (see Section~1):
\beq{D.1}
\gd\ov W{},\l,\m,=\gd\ov\G{},\l,\m,+\tfrac13 \gd\d,\l,\mu,\ov W_\nu
\end{equation}
and
\beq{D.2}
\gd \ov\G{},\l,\m,=\gd\tv\G,\l,\m,+\gd\ov Q{},\l,\m,+\gd\D,\l,\m,
\end{equation}
where $\gd\tv\G,\l,\m,$ is a Levi-Civita \cn\ induced by $g_\(\a\b)$ on~$E$
and
\bg{D.3}
\gd\ov Q{},\nu,\g\mu,=\frac12 \X2(\dg K,\g\mu,\nu,
-2g_{\tl[\mu\cdt]}^{\hp{\tl[\mu}\a}K_{\lw2.5 \g\tp \a\b\eg}g^\[\nu\b]\Y2)\\
\gd\D,\nu,\g\mu,=\wt g{}^\(\nu\d)\X2\{K^{\hp{\d(\g}\a}_{\d(\g\cdt}
g_{\lw2.5 \[\mu)\a]\eg} + g^{\hp{[\rho}\b}_{[\rho\cdt]}\X2[g^{\hp{(\mu}\rho}
_{([\mu\cdt]}K_{\lw2.5 \g)\a\b\eg} g^{\hp{[\d}\a}_{[\d\cdt]}
- K_{\d\a\b} g^{\hp{([\g}\a}_{([\g\cdt]} g^{\hp{[\mu)}\rho}_{[\mu)\cdt]}
\Y2]\Y2\} \label{D.4}\\
g_{\a\b}=g_\(\a\b)+g_\[\a\b] \label{D.5}\\
K_{\a\b\g}=-\tv\na_\a g_\[\b\g] - \tv\na_\b g_\[\g\a]
+ \tv\na_\g g_\[\a\b]. \label{D.6}
\end{gather}
$\tv\na$ is a covariant \dv\ \wrt the Levi-Civita \cn\ $\gd\tv\G{},\a,\b\g,
(\gd\tv w{},\a,\b,)$. We have of course
\beq{D.7}
\wt g{}^\(\a\b)g_\(\a\g)=\d^\b_\g
\end{equation}
and
\beq{D.8}
g^{\hp{[\g}\b}_\[\g\cdt]=\wt g{}^\(\b\a)g_\[\g\a].
\end{equation}

Starting from the formula for a 2-form of a curvature:
$$
\gd\ov\O{},\a,\b,(\ov W)=d\gd\ov W{},\a,\b,+\gd\ov W{},\a,\g,\wedge
\gd\ov W{},\g,\b,
$$
one gets
$$
\gd\ov\O{},\a,\b,(\ov W)=\gd\tv\O{},\a,\b, - \tv\na_{[\d}
\X2(\gd\ov Q{},\a,|\b|\g], + \gd\D,\a,|\b|\g],\ov\t{}^\d\wedge \ov\t{}^\g\Y2)
-\tfrac23 \gd\d,\a,\b, \ov W_\[\d,\g] \ov\t{}^\d\wedge \ov\t{}^\g
$$
where $\gd\tv\O{},\a,\b,$ is a 2-form of a curvature for a \LC\ \cn\ $\gd\tv
w{},\a,\b,$ on~$E$.

From the formula above we can easily read a tensor of a curvature $\gd
\ov R{},\a,\b\d\g,(\ov W)$ and afterwards a Moffat--Ricci tensor as given
below:
\beq{D.9}
\ov R_{\b\d}(\ov\G)=\tv R_{\b\d}- \frac12\X2(\tv\na_\d
\gd\ov Q{},\d,\b\g, + \tv\na_\d \gd\tv\D,\d,\b\g,\Y2)
+\frac14\X2(\tv\na_\g \gd\D,\a,\b\a, - \tv\na_\a \gd\D,\a,\b\g,\Y2)
\end{equation}
where $\tv R_{\b\g}$ is a Ricci tensor for a \LC\ \cn\ generated by
$g_\(\a\b)$. We get also Moffat--Ricci tensor for $\gd\ov
W{},\l,\mu,$. One gets
\beq{D.10}
\ov R_{\b\mu}(\ov W)=\ov R_{\b\mu}(\ov\G)+\tfrac23 \ov W_\[\b,\mu].
\end{equation}
The final result reads
\beq{D.11}
\ov R_{\b\mu}=\tv R_{\b\g}-\tfrac12\tv\na_\d \gd\ov Q{},\d,\b\g,
+\tfrac14\tv\na_\g \gd\D,\a,\b\a, - \tfrac34\tv\na_\d \gd\D,\d,\b\g,
+\tfrac23\ov W_\[\b,\g].
\end{equation}

One can also write a scalar curvature
\beq{D.12}
\ov R(\ov W)=g^\(\b\g)\tv R_{\b\g}
-\tfrac12 g^\[\b\g]\tv\na_\d \gd\ov Q{},\d,\b\g,
-\tfrac34 g^\(\b\g)\tv\na_\d \gd\D,\d,\b\g,
+\tfrac14 g^{\b\g}\tv\na_\g \gd\D,\a,\b\a, + \tfrac23 g^\[\b\g]\ov W_\[\b,\g].
\end{equation}
Thus now we can write Einstein \e s
\beq{D.13}
\ov R_{\a\b}(\ov W)=8\pi \nad{em}T_{\a\b}
\end{equation}
in the following way
\beq{D.14}
\tv R_{\b\g} -\tfrac12 \tv\na_\d \gd\ov Q{},\d,\b\g,
-\tfrac34 \tv\na_\d \gd\D,\d,\b\g, +\tfrac14 \tv\na_\g \gd\D,\a,\b\a,
+ \tfrac23 \ov W_\[\b,\g]=8\pi \nad{em}T_{\b\g}.
\end{equation}
Taking \s\ and anti\s\ part of \e\ \eqref{D.14} one gets
\bg{D.15}
\tv R_{\b\g}=8\pi \nad{em}T_\(\b\g)
+\tfrac34 \tv\na_\d \gd\D,\d,\b\g, -\tfrac14 \tv\na_{(\g} \gd\D,\a,\b)\a, \\
-\tfrac12 \tv\na_\d \gd\ov Q{},\d,\b\g,+\tfrac14 \tv\na_{[\g}\gd\D,\a,\b]\a,
+ \tfrac23 \ov W_\[\b,\g]=8\pi \nad{em}T_\[\b\g]. \label{D.16}
\end{gather}
One can eliminate $\ov W_\mu$ from the theory using \eqref{D.16} and getting
\beq{D.17}
\tfrac14 \tv\na_{\tl[\g}\gd\D,\a,{\b]|\a|,\mu\tp},
-\frac12\tv\D_\d \ov Q_\[\b\g,\mu] = 8\pi \nad{em}T_{\tl[\b\g],\mu\tp}.
\end{equation}
Let us consider our second Maxwell \e, i.e.\ Eq.~\eqref{1.39} writing it in
a new way. One gets
\bml{D.18}
\tv\na_\mu F^{\a\mu}=\gd\D,\mu,\d\mu, F^{\d\a} - \gd\ov Q{},\a,\d\mu,F^{\d\mu}\\
{}+\ov\na_\mu \X2(g^{\a\b}g^{\mu\g}\wt g^\(\tau\rho)
\X1(F_{\rho\g}g_\[\b\tau] - F_{\rho\b}g_\[\g\tau]\Y1)\Y2)
+2g^\[\a\b]\ov\D_\mu(g^\[\nu\b]F_{\nu\b}).
\end{multline}
In this way we get Einstein \e s Eq.~\eqref{D.15} and second pair of Maxwell
\e s in~GR \eqref{D.18}.

Moreover, we get a supplementary condition Eq.~\eqref{D.17}. In this way our
unified theory is equivalent to GR plus additional ``matter fields''.
Moreover, our ``matter field'' has pure geometrical origin. We get additional
terms on the right-hand side of Einstein \e s (some additional terms for an
effective energy-momentum tensor). We get also Eq.~\eqref{1.23}
(i.e.\ ${\falg^{[\m]}}_{,\nu}=0$) which is a
field \e\ for skewon $g_\[\m]$. We get also additional currents on the
right-hand side of the second pair of Maxwell \e s. This is similar to the
case of Einstein--Cartan theory (see Ref.~\cite{73} and references cited
therein). In this case the theory
is described by a metric (\s) tensor and a metric \cn\ on a \spt, which can
have non-zero torsion. The external sources are energy-momentum tensor (not
necessary \s) and spin density. One gets the following \e s:
\bg{D.19}
R_\m-\tfrac12g_\m R=8\pi t_\m\\
\gd Q,\rho,\m, +\gd\d,\rho,\mu, -\gd\d,\rho,\nu,\gd Q,\si,\mu\si,=8\pi \gd
s,\rho,\m,. \label{D.20}
\end{gather}
$\gd Q,\rho,\m,$ is a tensor of torsion for a metric \cn, $R_\m$ and~$R$ are
Ricci tensor and a scalar curvature for a \cn, $t_\m$ is an energy-momentum
tensor, $\gd s,\rho,\m,$ is a spin density tensor. \eu\e\ \eqref{D.20} can be
solved getting a torsion (and a contorsion)
\beq{D.21}
\gd Q,\rho,\m,=8\pi \X1(\gd s,\rho,\m,+\tfrac12\gd \d,\rho,\mu, \gd s,\si,
\nu\si,+\frac12 \gd\d,\rho,\nu,\gd s,\si,\si\mu,\Y1).
\end{equation}

Finally, we get a \cn\ on a \spt\ and we can rewrite Eq.~\eqref{D.19} in the
following way:
\bg{D.22}
\wt R_\m- \tfrac12g_\m \wt R=8\pi T^{\rm eff}_\m\\
T^{\rm eff}_\m =T_\m-2\pi\X1(s_{\m\g}s^\g + 2s_{\mu\g\d}\gd s,\d\g,\nu,
+s_{\mu\g\d}\gd s,\g\d,\nu,\Y1) + g_\m\X1(s_\g s^\g -s_{\d\g\a}s^{\a\g\d}
-\tfrac12 s_{\d\g\a}s^{\g\a\d}\Y1) \label{D.23}\\
T_\m =t_\m +\tfrac12\pi \wt\na_\rho \X1(\dg s,\nu\mu,\rho,+s^{\hp{\nu}\rho}
_{\nu\hp{\rho}\mu}+s^{\hp{\mu}\rho}_{\mu\hp{\rho}\nu}\Y1)
\label{D.24}\\
s_\a=\gd s,\g,\a\g, \nn
\end{gather}
where $\wt R_\m$, $\wt R$, $\wt\na$ are Ricci tensor, scalar curvature and
covariant \dv\ \wrt the \LC\ \cn\ generated by $g_{\a\b}$. Formula
\eqref{D.24} gives us Belifante--Rosenfeld symmetrization of the canonical
energy-momentum tensor. On the right-hand side of Eq.~\eqref{D.22} we have
additional ``matter fields'', and on the left-hand side a typical
term---Einstein tensor in~GR.

We can perform a very similar procedure in the case of Moffat (see
Ref.~\cite{5a}) and
Einstein--Cartan--Moffat theory (see Ref.~\cite{74}) using both procedures
described above. In the case of Kaluza--Klein Theory with torsion (see Refs
\cite{98}, \cite{420}) a similar to Einstein--Cartan Theory method can be
applied. In all of these cases GR with additional ``matter fields''
uses the same metric tensor as in original theory, which is not true in the
case of Refs \cite{67}, \cite{68}, \cite{69}, \cite{70}, \cite{71},
\cite{72}. Thus we have no problems with the physical interpretation of a \s\
metric and we can consider our theory as GR plus additional ``matter
fields'', which have a geometrical interpretation. In this way we can
complete the Einstein programme of a geometrization of physical interactions
getting ``interference effects'' between \gr al and \elm c fields
and prove that GR is a distinguished \gr al theory among alternative theories
of \gr\ and unified field theories.

\section*{Acknowledgement}
I would like to thank Professor B. Lesyng for the opportunity to carry out
computations using Mathematica\TM~6\footnote{Mathematica\TM\ is the
registered mark of Wolfram Co.} in the Centre of Excellence
BioExploratorium, Faculty of Physics, University of Warsaw.
I would like to thank an anonymous referee for critical comments to improve
my paper. The SCOAP3 sponsorship of my paper is highly appreciated.
\goodbreak

\end{document}